Níckolas de Aguiar Alves

# Nonperturbative Aspects of Quantum Field Theory in Curved Spacetime

Níckolas de Aguiar Alves

# Nonperturbative Aspects of Quantum Field Theory in Curved Spacetime

Thesis presented to the Graduate Program in Physics at the Federal University of ABC as partial requirement to obtain the degree of Master in Physics.

Advisor: Prof. Dr. André Gustavo Scagliusi Landulfo

Santo André – SP
2023

Este exemplar foi revisado e alterado em relação à versão original, de acordo com as observações levantadas pela banca examinadora no dia da defesa, sob responsabilidade única do autor e com a anuência do orientador.

This copy was revised and altered with respect to the original version, according with the observations raised by the examining committee on the defense day, under the author's sole responsibility and with the advisor's consent.

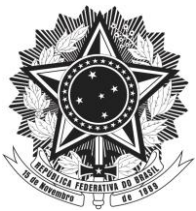

**MINISTÉRIO DA EDUCAÇÃO**
**Fundação Universidade Federal do ABC**
Avenida dos Estados, 5001 – Bairro Santa Terezinha – Santo André – SP
CEP 09210-580 · Fone: (11) 4996-0017

## FOLHA DE ASSINATURAS

Assinaturas dos membros da Banca Examinadora que avaliou e aprovou a Defesa de Dissertação de Mestrado do candidato, NÍCKOLAS DE AGUIAR ALVES realizada em 28 de Abril de 2023:

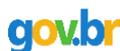
Documento assinado digitalmente
ANTONIO DUARTE PEREIRA JUNIOR
Data: 03/05/2023 05:33:55-0300
Verifique em https://validar.iti.gov.br

**Prof.(a) ANTONIO DUARTE PEREIRA JUNIOR**
UNIVERSIDADE FEDERAL FLUMINENSE

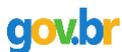
Documento assinado digitalmente
DANIEL AUGUSTO TUROLLA VANZELA
Data: 29/04/2023 01:00:01-0300
Verifique em https://validar.iti.gov.br

**Prof.(a) DANIEL AUGUSTO TUROLLA VANZELLA**
UNIVERSIDADE DE SÃO PAULO

**Prof.(a) GEORGE EMANUEL AVRAAM MATSAS**
UNIVERSIDADE ESTADUAL PAULISTA JÚLIO DE MESQUITA FILHO

**Prof.(a) MAURICIO RICHARTZ**
UNIVERSIDADE FEDERAL DO ABC

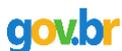
Documento assinado digitalmente
ANDRE GUSTAVO SCAGLIUSI LANDULFO
Data: 03/05/2023 15:07:48-0300
Verifique em https://validar.iti.gov.br

**Prof.(a) ANDRE GUSTAVO SCAGLIUSI LANDULFO**
UNIVERSIDADE FEDERAL DO ABC - Presidente

* Por ausência do membro titular, foi substituído pelo membro suplente descrito acima: nome completo, instituição e assinatura

Universidade Federal do ABC

This study was financed in part by the Coordenação de Aperfeiçoamento de Pessoal de Nível Superior—Brasil (CAPES)—Finance Code 001.

This study was financed in part by the São Paulo Research Foundation—Brasil (FAPESP)—Grant 2021/07372-7.

*I dedicate this thesis to two groups of people.*

*Firstly, to each one of the countless physicists, mathematicians, and philosophers whose efforts allowed me to reach this level of understanding within a single lifetime.*

*Secondly, most importantly, to those who come after me and gift me the honor of reading my words. I hope the blood, sweat, and tears I have invested in the following pages may spare you at least a few difficulties. May you harvest the dates I have planted.*

There are more things in heaven and earth, Horatio, than are dreamt of in your philosophy.

Hamlet, Prince of Denmark, in William Shakespeare's *The Tragedy of Hamlet, Prince of Denmark*, Act I, Scene V.

When it comes to atoms, language can be used only as in poetry. The poet, too, is not nearly so concerned with describing facts as with creating images and establishing mental connections.

Niels Bohr, as quoted by Heisenberg (1972, p. 41).

It takes more. Open your mind to the past—art, history, philosophy. And all this may mean something.

Captain Jean-Luc Picard (Sir Patrick Stewart), in *Star Trek: The Next Generation* Season 2, Episode 17: *Samaritan Snare*.

# Resumo

Teoria quântica de campos em espaços-tempos curvos é possivelmente o formalismo mais confiável com o qual se pode investigar efeitos quânticos na presença de campos gravitacionais intensos. No entanto, ela é comumente estudada por meio de tratamentos perturbativos. Nesta dissertação, desejamos usar o grupo de renormalização funcional—uma versão não-perturbativa do grupo de renormalização—como uma técnica para descrever fenômenos não-perturbativos em espaços-tempos curvos. O sistema escolhido é um detector de Unruh–DeWitt acoplado a um campo quântico escalar. Discutimos como formular um tal sistema em termos de uma ação e como calcular seu fluxo de renormalização para o caso de um detector inercial em um espaço-tempo plano, por simplicidade. Aprendemos, contudo, que os resultados são divergentes no limite em que o espaçamento energético do detector se anula. Possíveis abordagens alternativas são discutidas ao final.

Esta dissertação também apresenta uma revisão de teoria quântica de campos em espaços-tempos curvos através do formalismo algébrico, embora não assuma experiência prévia com análise funcional. Assim, ela fecha um buraco pedagógico na literatura. Ademais, também revisamos o grupo de renormalização funcional e derivamos a equação de Wetterich assumindo um conteúdo de campos genérico que pode incluir tanto campos bosônicos quanto fermiônicos. Uma tal dedução é dificilmente encontrada nas introduções pedagógicas disponíveis na literatura de física de altas energias.

Palavras-chave: teoria quântica de campos em espaços-tempos curvos, grupo de renormalização funcional, detectores de partículas.

# Abstract

Quantum field theory in curved spacetime is perhaps the most reliable framework in which one can investigate quantum effects in the presence of strong gravitational fields. Nevertheless, it is often studied by means of perturbative treatments. In this thesis, we aim at using the functional renormalization group—a nonperturbative realization of the renormalization group—as a technique to describe nonperturbative quantum phenomena in curved spacetimes. The chosen system is an Unruh–DeWitt particle detector coupled to a scalar quantum field. We discuss how to formulate such a system in terms of an action and how one can compute its renormalization group flow in the case of an inertial detector in flat spacetime, for simplicity. We learn, however, that the results are divergent in the limit in which the detector's energy gap vanishes. Possible workarounds are discussed at the end.

This thesis also presents a review of quantum field theory in curved spacetimes by means of the algebraic approach, although it assumes no previous experience with functional analysis. Hence, it fills a pedagogical gap in the literature. Furthermore, we also review the functional renormalization group and derive the Wetterich equation assuming a general field content that might include both bosonic and fermionic fields. Such a derivation is also hardly found in pedagogical introductions available in the high energy physics literature.

# Acknowledgments

Saruman believes that it is only great power that can hold evil in check, but that is not what I have found. I have found it is the small things, everyday deeds of ordinary folk, that keep the darkness at bay. Simple acts of kindness and love.

Gandalf (Sir Ian McKellen), in *The Hobbit: An Unexpected Journey*.

Who can say if I've been changed for the better? I do believe I have been changed for the better and because I knew you [...] I have been changed for good.

Glinda (Kristin Chenoweth) and Elphaba (Idina Menzel), in *Wicked: The Untold Story of the Witches of Oz*.

Writing acknowledgments is simultaneously incredibly similar and amazingly different from writing the remainder of this thesis. It is similar because each word appears on the screen due to the choices of my brain while also carrying a lot of feelings from my heart. It is different because it is way more difficult. This section was the first I started writing and as I type these words I'm still confident it will be the last one to be finished. You see, I think we humans may actually have a shot at understanding a thing or two about the Universe, but I'm not so confident understanding ourselves is that easy.

I speak no Greek, but I know the name Níckolas has its roots in the Greek words νίκη (nikē), meaning 'victory', and λαός (laos), meaning 'people'. When I was one-and-twenty I used to interpret it as "he who leads the people to victory". And I am four-and-twenty, and I read it as "he who is led by the people to victory". Quite naturally then, I nowadays see this thesis and everything herein as a result of years of help and support from numerous people, and I couldn't be prouder to be the one to bring those abstract notions of friendship and love into this physical form.

If I were given enough time, I'd be able to find a reason to thank each and every human being. Nevertheless, to avoid bureaucratic complications, I must complete this section within a deadline, which forces me to focus on a few people I particularly must mention. This does mean I chose to restrict this section to people who helped me in a more academic context. I apologize for any omissions—if you find it unfair your name is not here and feel you belong on these pages, you are correct.

First and foremost, I thank André G. S. Landulfo, my advisor. This project sounded terrific and terrifying in equal proportions from day one. In spite of that, even when I couldn't believe in myself, I felt that you did believe in me. I hope one day I can advise someone as well as you have advised me. Thank you for all of the discussions, encouragement, and inspiration.

A lot of what I know nowadays, and certainly a lot of the physicist I am today, is due to João C. A. Barata, Enrico Bertuzzo, Gustavo A. Burdman, and George E. A. Matsas. Through lectures, discussions, and day-to-day conversations you have not only taught me physics, but also introduced me to the physicist and professor I long to be. I also thank the late João E. Steiner, who helped me take my very first steps into scientific research. I wish I could knock on your door to thank you in person.

This thesis benefited a lot from the comments of Alex G. Dias, Breno Marques, Antônio D. Pereira Junior, and Vilson T. Zanchin, who composed my qualifying exam committee. Not only do I thank you for the helpful technical advice, but also for your very kind words, which were able to bring me peace and confidence when I really needed it. I also thank Antônio in particular for all of the fruitful and inspiring conversations we had—that gift at the Swieca Summer School was also very much appreciated.

A thesis defense is often a stressful process, but it was very pleasant to be able to discuss my work with Antônio D. Pereira Junior (once again!) and Daniel A. T. Vanzella, who composed my defense committee. Thank you for your kind words, for your support, and for the useful suggestions concerning this thesis and how to further develop it.

Many of the conversations I had with Lissa S. Campos, Ricardo Correa da Silva, and Thiago Raszeja happened during my undergraduate studies. Even so, you earned your place on these pages, for I had your guidance through each step I took along this path. Very often during my MSc, I had the joy of recalling our conversations and noticing how they suddenly were making much more sense. As I write these words, we are in four different countries, but I hope our paths cross again many times in the future.

I can hardly recall a time when I didn't know Pedro H. T. L. Tredezini and Beatriz Tucci, and to be fair I don't really want to. You have always been and will always be an inspiration to me. I'll forever hold our piccole particelle elementari in my heart and I hope we'll soon find the time to share courses and drinks again. Perhaps, this time, on the other side of the classroom.

The real treasure of a university is the friends we make along the way, and I couldn't be more thankful for the great conversations and discussions I had at our group's Journal Club. If I hadn't turned my camera off so often, you'd probably have noticed that I was smiling most of the time due to simply enjoying exploring the wonders of the universe with you.

I could never forget to thank our Iournal Cub, the Journal Club's undergraduate-level version. You guys gifted me the opportunity of revisiting and sharing with you each of the wonders of relativity, and there are very few presents better than spacetime itself. I can't wait to see the great histories that await you and to be able to cherish the fact I got to see the beginning of your worldlines.

Out of the great friends I have in our group, a few deserve special mention. I doubt this thesis would exist if it were not for me meeting João Lucas Miqueleto Reis. It is thanks to you that I found our group in the first place and if I know anything about the FRG it is because you helped me take my first steps.

A few friends work in areas a bit further away from mine, but were absolutely essential in helping me cope with the stresses of Academia during these last few years. I am deeply thankful for the support I have gotten from everyone at Karekatsuki, especially Melian Lopes Molinari, Wilson M. S. de Paula, and Pietro Raele. While I still haven't had the pleasure of meeting Gabrielle Weber in person, our online interactions and your always kind words have been a great source of peace and inspiration for me over the years.

As I mentioned at the beginning, this section is taking me a long time to write. I first added Betânia C. T. Backes to these paragraphs when we were still innocently flirting and crushing into each other. Back then you had already encouraged and inspired me in a myriad of different ways and kept me together through some quite difficult weeks. Nowadays we have been dating for a while and it feels as if I have known you my whole life. Furthermore, I think I no longer need to come up with excuses to give you a paragraph of your own. Thank you for proofreading my drafts and believing so much in me. Thank you for the poetry, for the sunshine, and for always reminding me of the things that matter. Most importantly, thank you for existing, and for all of the butterflies.

I still have a thousand fellow students to thank for trusting me enough to ask me a question or to talk to me about physics or life in academia. I hope I was able to help you at least a little bit. Those apparently silly conversations, or even questions at Stack Exchange, were often the greatest source of confidence and purpose I could wish for. They helped me stay calm as I tried to swim in the deep waters of research by making me remember how I got here in the first place.

To keep knowledge from humans is to chain their souls to the ground—a prison so tortuous that we are forced to deal with it by pretending it is not there at all. Therefore, I am deeply thankful for the financial support I received during the execution of this project. From March through November 2021 I was supported by the Coordenação de Aperfeiçoamento de Pessoal de Nível Superior (CAPES) through file 88887.600471/2021-00. From December 2021 to February 2023 I was supported by the São Paulo Research Foundation through grant 2021/07372-7. I thank you—both the institutions and the Brazilian people who sustain them—not only for the financial support, but for what it represents. I thank you for gifting me the freedom of knowledge.

I'm quite a LaTeX enthusiast and I want to also acknowledge some of the software behind this thesis. This document was written in LaTeX 2$_\varepsilon$ and typeset mostly in the version of EB Garamond provided by Duffner (2014). The monospaced font is Latin Modern Mono and the mathematics font is Garamond Math (Zhao and Zeng 2022). All illustrations were drawn with Ti*k*Z and Feynman diagrams were drawn with `tikz-feynman` (Ellis 2017).

Lastly, and perhaps most importantly, I thank you who gifted me your irretrievable time on Earth by reading my words. You see, different people have different relationships with death. As I write these words, one of the things I fear the most about it is the possibility I die with a thought, an idea, a point of view that I was not able to pass forward. I fear the effort that it took would have been in vain. Our time in this world is very limited to about a hundred years, and I was only able to write this thesis due to the efforts of thousands of ordinary people who wanted to push our knowledge a little bit further. Physics is difficult, and each new step comes at a price. As you read through these pages, do feel amazed at how gorgeous this arrogant Universe that loves to be noted is. However, also recall that each and every equation was not handed to humankind by Prometheus, but was rather conquered by people made of flesh and bone who could bleed, sweat, and cry. I am thankful to you, the reader, for valuing my own efforts with your time.

This thesis was written by standing on the shoulders of ordinary folk whose everyday deeds changed my world and kept darkness at bay. I, therefore, thank you, who are giants to me, for your acts of kindness and love. Were I offered the answers to all my questions about the Universe, I'd have to pass it, for it is far more enjoyable to search for them alongside you.

# Abbreviations

1PI  one-particle-irreducible

ASQG  asymptotically safe quantum gravity

CAR  canonical anticommutation relations

CCR  canonical commutation relations

EAA  effective average action

EFT  effective field theory

FRG  functional renormalization group

FRGE  functional renormalization group equation

GNS  Gelfand–Naimark–Segal

GR  general relativity

HEP  high energy physics

IR  infrared

KMS  Kubo–Martin–Schwinger

LHS  left-hand side

LPA  local potential approximation

μSC  microlocal spectrum condition

pAQFT  perturbative algebraic quantum field theory

QFT  quantum field theory

QFTCS  quantum field theory in curved spacetime

QG  quantum gravity

QM  quantum mechanics

RG  renormalization group

RHS  right-hand side

RQI  relativistic quantum information

SM  standard model

SR  special relativity

UV  ultraviolet

# Contents

# One

# Why Do Things Fall?

*We shall understand why it is interesting to consider the effects of gravity on quantum phenomena and why quantum field theory in curved spacetime is a fruitful framework. We shall also learn why the functional renormalization group is an interesting technique to be applied in these contexts and which sorts of problems it might help us uncover. Technical discussions are temporarily overlooked and delayed to the remaining chapters.*

Quantization of the gravitational field, which appears to be necessary for physical reasons, may be carried out without any new difficulties by means of a formalism fully analogous to that applied here.

Heisenberg and Pauli (1929), as translated by Stachel (1999, p. 527).

Although there has been a lot of work in the last fifteen years, I think it would be fair to say that we do not yet have a fully satisfactory and consistent quantum theory of gravity.

Hawking (1975).

I believe all physicists have experienced the widely spread belief that Physics is incomprehensibly difficult. While there is definitely truth in that belief—and the existence of unsolved problems in Physics is proof—it is also beautiful to notice how some truly deep questions are not only comprehensible, but simple enough to be asked by an arbitrarily chosen child. Why is the sky blue? How can airplanes fly? Why does water evaporate when heated up? How did the Universe begin? Can we go back in time? What is time? What are things made of? Why do they fall? These questions are easily asked, but hardly answered. Some of them are as old as humankind, but we still have not reached a satisfactory conclusion about them.

The Ancient Greeks, for example, already discussed what things are made of and why they fall. Aristotle proposed everything on Earth was made up of water, earth, fire, and air. Each of the four elements would have a tendency to move towards its natural place—upwards for fire and air, downwards for water and earth. They would also tend to move in straight lines, and would eventually get tired of moving. This provided an explanation of phenomena below the Moon, while the Heavens were ruled by the Physics of aether. This fifth element—or quintessence—was what the heavenly bodies were made of and it had the tendency to move in circles, explaining hence why planets and stars orbited the Earth (Russell 2004, Book I, Chap. 23).

Aristotle's theory provides a really good description of everyday phenomena (Rovelli 2015), but modern physicists usually have a different point of view when it comes to fundamental physics. Our cutting edge understanding of what things are made of corresponds to a list of elementary particles described by the so-called standard model (SM), which is reviewed, *e.g.*, on the book by Schwartz (2014). This model has a couple of advantages over the Aristotelian model. For example, it is capable of incredible quantitative predictions matching experiment to astonishingly high precision (Particle Data Group et al. 2022). It does lack a remarkable feature of the Aristotelian model, though: it cannot explain why things fall.

The best description of gravity currently known is general relativity (GR), discussed for example on the books by Hawking and Ellis (1973) and Wald (1984). This theory describes gravitational effects in terms of spacetime curvature, which is caused by the presence of matter—as Wheeler (2000, p. 235) has famously put it, "Spacetime tells matter how to move; matter tells spacetime how to curve". Just like the SM, GR has enormous success in describing experiments (Will 2014). Still, it lacks an impressive feature the Aristotelian model had: it cannot explain what things are made of.

At first glance, these facts seem merely to mean we now need different descriptions for different phenomena, which may seem like a perfectly reasonable conclusion. Nevertheless, this is not the case. The facts that the SM does not consider gravity and that GR does not consider the small-scale structure of matter mean that there are phenomena we are currently unable to understand. As good as our present-day theories may be, they are unreliable in situations in which we need to consider both the small-scale physics ignored by GR and the strong gravitational effects ignored by the SM. This happens, for example, very close to the Big Bang or within black holes. Given that—to our current knowledge—time literally starts at the Big Bang and literally ends inside black holes, it seems we fail to understand where we came from and where we are going to.

The issue deepens once we consider what each of these theories actually mean in modern physics. Whenever $x$ stands for position or $t$ for time in a physical equation, a spacetime structure is being assumed. In non-relativistic physics this assumption is often implicit, but it is there anyway. What GR truly represents in modern physics is our best knowledge of what spacetime *is*. It tells us what is the stage on which the Universe happens, and how this stage can become an actor in its own play. As for the SM, it is one of the greatest representatives of quantum mechanics (QM), which tells us the rules which all physical theories must abide by. Therefore, GR and QM are the two pillars supporting all known physics.

In spite of that, as Bronstein (1936a,b) first noticed, these two pillars are in some sense incompatible: once GR and QM are taken into account, it is not possible to determine a component of a field at some point $x$ to arbitrary precision. By "field" one means any function of spacetime, and these sorts of objects are among the main building blocks of both GR and the SM. To understand this impossibility, let us follow an adaptation of Bronstein's argument presented by Rovelli and Vidotto (2015, Sec. 1.2).

Suppose we want to determine the value of $\phi(x)$. In order to do so, we must first localize the point $x$ to some precision, say $L$. This can be done by placing a particle at $x$ and ensuring the particle's position has uncertainty $\Delta x < L$. Heisenberg's Uncertainty Principle then ensures the uncertainty in the particle's momentum must satisfy

$$\Delta p \geq \frac{\hbar}{2\Delta x} > \frac{\hbar}{2L}. \tag{1.0.1}$$

We know the mean value of $p^2$ is never smaller than $\Delta p^2$ (by the definition of $\Delta p$ in QM), and hence we may write

$$p^2 \geq \frac{\hbar^2}{4L^2}. \tag{1.0.2}$$

If we want to consider small uncertainties in the position, this will then naturally lead us to large momenta. In the ultra-relativistic limit, the particle's mass is negligible when compared to the particle's momentum, and hence we can write $E \approx pc$. This allows us to write

$$E^2 \geq \frac{\hbar^2 c^2}{4L^2},$$ (1.0.3)

which now incorporates QM and special relativity (SR).

We now bring in gravity. From GR, we know large concentrations of energy lead to the formation of an event horizon with Schwarzschild radius $R = \frac{2GM}{c^2}$, where $M$ is related to the energy $E$ through the famous formula $E = Mc^2$. Therefore, we can rewrite the Schwarzschild radius as[*]

$$R = \frac{2GE}{c^4}.$$ (1.0.4)

Hence, Eq. (1.0.3) states that the smaller the value of $L$, the larger the energy of the confined particle. Nevertheless, Eq. (1.0.4) tells us that the larger the energy of the confined particle, the larger the event horizon radius $R$. Eventually, for sufficiently small $L$, we shall have $L \leq R$. More specifically, if we impose $L \geq R$, the previous equations imply

$$\frac{L^2 c^8}{4G^2} \geq E^2,$$ (1.0.5a)

$$\geq \frac{\hbar^2 c^2}{4L^2},$$ (1.0.5b)

$$L^4 \geq \frac{\hbar^2 G^2}{c^6},$$ (1.0.5c)

$$L \geq \sqrt{\frac{\hbar G}{c^3}},$$ (1.0.5d)

which means $L$ must always be larger than the so-called Planck length $L_p \equiv \sqrt{\frac{\hbar G}{c^3}}$ if we want to be able to localize $x$. Otherwise, attempting to do so would form a black hole, preventing us from "peeking" at what is going on.

This is a pictorial argument, but it illustrates why understanding quantum gravity (QG) is more difficult than it is to understand other quantum theories. Notice this sketch doesn't lead to the same result for electromagnetism—if gravity did not exist, but electromagnetism still did, we would still be able to probe small scales by using neutral particles. However, everything is subject to gravity.

Nowadays, we do have a more solid understanding of gravity. We know it is not perturbatively renormalizable ('t Hooft and Veltman 1974; Goroff and Sagnotti 1986; reviewed by Percacci 2017, Secs. 3.5 and 3.6), we know how to use an effective field theory (EFT) perspective to obtain a quantum description of gravity (Donoghue 1994b,a; reviewed by Burgess 2004), and we know it is remarkably similar to many other familiar field theories (see, for example, Percacci 2023). We also have many candidate theories to a full description of QG (see, *e.g.*, Oriti 2009). Yet, no single theory can be claimed to be a fully satisfactory and consistent quantum theory of gravity.

Two styles exist within the attempt to fill this gap in human knowledge: a "top-down" approach and a "bottom-up" approach. The former consists in prescribing desirable features of a theory of QG—such as unitarity or the presence of a spin-2 field—to obtain a "goal theory" at high energies and use it to derive consequences at lower energies which should match what we already know. This is done, for example, within String Theory or Loop Quantum Gravity. The latter approach, bottom-up, consists in starting from what we already know and slowly attempting to climb up towards higher energies. This could be exemplified by studies relying on gravity as an EFT, by asymptotically safe quantum gravity (ASQG), or by the main framework of this thesis: quantum field theory in curved spacetime (QFTCS).

---

[*]Notice this is just an estimate. The properties of actual black holes do not depend only on mass, but also on angular momentum and charge.

## 1.1   Why Quantum Field Theory in Curved Spacetime?

QFTCS consists of the framework in which one considers the evolution of quantum fields upon a background composed of a fixed classical curved spacetime. In layman's terms, it can be though of as the theory of how elementary particles fall[*].

A first motivation for considering QFTCS is that it provides a reliable environment for studying quantum effects in the presence of gravity. While top-down approaches have the advantage of possibly leading to a fundamental and exact theory, they have the disadvantage of also depending on speculative assumptions. It is a high-risk, high-reward program. QFTCS, on the other hand, consists of gently modifying standard quantum field theory (QFT) in flat spacetime to adapt it to a curved background. The result is a language that relies only on our best-tested physical theories, without the need for any bold hypotheses, but which is still capable of providing astonishing results[†].

One may then argue, correctly, that QFTCS does not provide a fully satisfactory quantum description of gravity. While true, it is important to notice that this does not contradict the usefulness of the framework. For example, even though quantizing the electromagnetic field is a well-understood procedure nowadays, there are situations in which one can understand quantum effects in the presence of electromagnetic fields in a semiclassical approach. Indeed, an usual example of an application of time-dependent perturbation theory is to understand how an atom responds to the presence of an electromagnetic wave (Weinberg 2015, Chap. 6), and these investigations are far simpler to do than to consider quantum electrodynamics in full detail. Therefore, in some sense, the fact QFTCS does not quantize gravity should be understood as a feature of the framework, rather than simply a limitation. While it is true that we currently do not understand QG and doing it is a humongous open problem in theoretical physics, it should also be noted that having a fundamental theory and being able to use it to derive meaningful consequences are different things (*cf.* Anderson 1972). QFTCS allows us to comprehend quantum phenomena in the presence of gravitational fields without the need to deal with any further complications or degrees of freedom that are necessary in more complex theories. This is enough justification for this framework, at least within the range of phenomena accurately described by it.

Another usefulness of this approach is to obtain a deeper understanding of QFT itself. While the SM, for example, is typically formulated on an inertial reference frame in flat spacetime, we know spacetime is curved, and hence any theory in flat spacetime is an approximation. QFTCS allows us to dive into the foundations of QFT and understand what a QFT *is*, how it is formulated in other frames of reference, how to formulate it in the absence of any particular choice of coordinates or reference frame, which elements of the theory are indeed fundamental for its description, and so on.

---

[*]Although this analogy should be taken with a grain of salt—as we shall see on Section 2.4, the notion of elementary particle is not even well defined on a general curved spacetime.

[†]While QFTCS is our focus in this thesis, it is important to mention that seeing gravity as an EFT also has similar advantages in the sense of being a particularly "safe" approach.

QFTCS also provides a rich language to understand problems in which both the causal structure of spacetime and quantum effects should play a significant role. This happens, for example, in the field known as relativistic quantum information (RQI). Ever since the first few years of quantum theory, it has been noticed that QM and SR lead to curious predictions. The Einstein–Podolsky–Rosen "paradox" (Einstein, Podolsky, and Rosen 1935) is probably one of the best known examples. Modernly, these investigations have matured into a whole area of research trying to understand the role relativity and quantum effects can play when two observers try to communicate. In particular, examples from this field are some of the main motivations for this thesis, as we shall now see.

## 1.2  Why a Nonperturbative Approach?

As usual in many areas of physics, a great amount of the investigations done within QFTCS are carried out perturbatively. In this thesis, our goal is to introduce new techniques to the area so that nonperturbative results can be obtained. Our main focus will be on the example of a quantum field interacting with a particle detector.

### Relativistic Quantum Communication Channels

Within the field of RQI, a way of implementing a communication protocol between two observers is to give to each of them a qubit (*i.e.*, a two-level system) coupled to a relativistic quantum field. The observers can then convey information to one another by making measurements of their qubits, the effects of which are then carried by the field from sender to receiver.

A very general version of such a protocol was recently proposed by Landulfo (2016). The model will be described in further detail on Chapter 4, but some of its important features are:

i. it can be solved exactly due to the choice of coupling;

ii. the results obtained hold for a large class of spacetimes;

iii. it is not necessary to choose an arbitrary notion of "particles";

iv. the quantum state of the field can belong to a large class of "vacuum-like" states;

v. both sender and receiver can be assumed to have arbitrary trajectories through spacetime;

vi. both sender and receiver interact with the quantum field in only a bounded region of spacetime;

vii. the protocol allows for the transmission of classical information;

viii. observers that are not causally related are incapable of exchanging information.

In spite of these advantages, the model still has limitations. The very choice of coupling that allows the model to be solved exactly makes the transmission of quantum information impossible without the assistance of extra entanglement between the parts and does not allow one to harvest entanglement from the quantum field. Moreover, it prevents usage of the qubits as particle detectors.

Within QFTCS, and particularly within RQI, particle detectors are an omnipresent tool. Their basic concept is to couple a qubit to a quantum field in such a manner that the detector will get excited if it interacts with a "particle". This notion can be put in a detailed setting, as we shall do at the end of Section 2.4. The method can then be used to understand phenomena including, but not limited to, acceleration-induced thermality (Unruh and Wald 1984), neutrino oscillations (Torres et al. 2020), entanglement between quantum field modes (Pozas-Kerstjens and Martín-Martínez 2015), and has even been proposed as a substitute for clocks as the fundamental measurement apparatus in curved spacetimes (Perche and Martín-Martínez 2022). Hence, adapting the analysis carried out by Landulfo (2016) to a different coupling is of profound interest.

### Backreaction

A second motivation to study nonperturbative QFTCS is to obtain a deeper understanding of backreaction.

QFTCS *per se* means to consider quantum fields evolving on a fixed spacetime. Hence, the fields are treated as "test fields", meaning their stress-energy is assumed to have negligible effects upon the background geometry. This is, of course, an approximation, and it is interesting to wonder what happens once one considers how the fields affect the spacetime they are in. These effects are typically referred to as the backreaction of the fields on the background geometry.

Semiclassical gravity is an upgrade of QFTCS. In this new framework, one starts to consider self-consistent solutions of the Einstein equations and the quantum fields' equations of motion simultaneously. This can be done by considering, for example, the semiclassical Einstein equations,

$$G_{ab} = 8\pi \left[ T_{ab} + \omega(\hat{T}_{ab}) \right], \tag{1.2.1}$$

where $\omega(A)$ denotes the expectation value of the operator $A$ in the state $\omega$. $\hat{T}_{ab}$ is the stress-energy tensor of the quantum fields, $T_{ab}$ is the classical stress tensor, and $G_{ab}$ is the Einstein curvature tensor.

Notice that Eq. (1.2.1) has some limitations. In principle, one could try to analyze a given physical situation by using QFTCS, then compute $\omega(\hat{T}_{ab})$, renormalize it, and proceed iteratively. This procedure yields a perturbative approach. However, the approach will not make sense if "$T_{ab} \sim \omega(\hat{T}_{ab})$" in some sense, *i.e.*, if the quantum fields cannot be treated as mere perturbations upon a classical spacetime. Furthermore, it is difficult to understand how to deal with Eq. (1.2.1) in situations with "$\omega(\hat{T}_{ab}{}^2) \sim \omega(\hat{T}_{ab})^2$", *i.e.*, in which the quantum fluctuations are large. This happens, for example, for a compact quantum mass in a state which is a superposition of position eigenstates. The semiclassical Einstein equations claim that such a configuration would lead to the gravitational field of the mass positioned in between the two positions, which seems uncomfortable at best and even paradoxical at worst (see, *e.g.*, Wald 2020).

Given these limitations, it is interesting to wonder whether nonperturbative approaches could lead us to a better understanding of semiclassical gravity beyond the semiclassical Einstein equations. While this would not be a full theory of QG, it can represent a significant leap forward within a solid and reliable framework.

## 1.3   Why the Functional Renormalization Group?

There are many different nonperturbative approaches to studying QFT, such as exploiting the Dyson–Schwinger equations (Dyson 1949; Schwinger 1951a,b; reviewed by Itzykson and Zuber 1980, Sec. 10.1) or formulating the theory in a lattice. Our approach is to use the functional renormalization group (FRG), which means we shall obtain nonperturbative effects by considering the renormalization group (RG) flow computed to all orders in $\hbar$. This approach is reviewed in the books by Percacci (2017) and Reuter and Saueressig (2018), for example.

The RG is a manner of accounting for quantum corrections by promoting the coupling constants of a QFT to functions of scale. This encodes quantum effects in the "constants" and is extremely common within both high-energy physics (Schwartz 2014, Chap. 23) and statistical mechanics (Zinn-Justin 2007). Very often one computes this RG flow only up to a few orders in perturbation theory, but it is also possible to obtain nonperturbative results.

As it turns out, there are different ways of implementing this idea, and correspondingly there are many different functional renormalization group equations (FRGEs). We are interested in considering an FRGE often known as the "Wetterich equation" (Wetterich 1993; Bonini, D'Attanasio, and Marchesini 1993; Morris 1994). It is a flow equation for the effective average action, which itself is a notion of action suitable to a particular scale. In other words, the Wetterich equation describes how the action should change as one changes the scale of consideration for a theory. This has the advantage of involving an object which is familiar to high energy physics—*viz.*, the action—and also the Wetterich equation is a common tool within ASQG (Eichhorn 2019; Percacci 2017, 2023; Reuter and Saueressig 2018), meaning there is already literature on methods using this particular FRGE in curved spacetimes.

RG methods are particularly suitable for the problems we are interested in. When studying acceleration-induced thermality—one of the main sorts of effects considered in QFTCS—there is a natural energy scale dictated by the relevant temperature. Hence, it is simple to evolve the coupling constants to the scale of interest.

## 1.4 Thesis Structure

For the reasons previously explained, we are interested in applying FRG techniques to QFTCS. To do so in a concrete example, we shall work with the problem of understanding the nonperturbative RG flow of a Unruh–DeWitt detector. Here is the approach we shall take to achieve such goal in this thesis. The text assumes the reader to be somewhat experienced with GR and QFT. Most of the necessary material is covered in the books by Wald (1984) and Weinberg (1995, 1996), but other references are often suggested.

Chapter 2 reviews the main aspects of QFTCS. We shall do it through the so-called "algebraic approach", which provides a formulation of the theory independent of arbitrary choices of reference frame or coordinate system. It is rare to find discussions of the algebraic approach that do not rely on previous knowledge by the reader about functional analysis and other aspects of pure mathematics, and hence our discussion tries to avoid these prerequisites with the goal of filling a pedagogical gap in the literature. The chapter also discusses how the algebraic approach connects to other approaches—such as Fock space techniques or the Euclidean path integral approach used throughout most of this thesis—and derives and discusses the Unruh effect, one of the paradigmatic predictions of QFTCS.

Chapter 3 is concerned with the FRG techniques. We shall review some of the conceptual ideas behind the RG and derive the Wetterich equation for a theory with arbitrary field content. We also learn how to perform the calculations necessary to extract information from the FRGE and how to interpret these results.

Chapter 4 is the centerpiece of this project. It is concerned with the application of FRG techniques to Unruh–DeWitt particle detectors. We shall discuss the formulation of an Unruh–DeWitt detector in terms of a path integral and how to apply functional renormalization to such a system. Unfortunately, our calculations will lead to divergent results in the limit in which the detector's energy gap vanishes, at which point we shall discuss possible reasons and workarounds.

Chapter 5, at last, discusses the conclusions and future prospects.

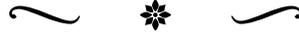

# Two

# Quantum Field Theory in Curved Spacetime

*We shall formulate quantum field theory in curved spacetime through the algebraic approach, a choice justified at the beginning of the chapter. The connection between this approach and others—such as the use of a Hilbert space—will also be clarified. The Unruh effect is derived in four different ways to illustrate the algebraic approach, the Euclidean techniques we shall need later on, the idea of a particle detector, and how one could obtain similar results with Fock space techniques.*

> Quantum field theory is—as its name suggests—the quantum theory of fields. "Particles" do not play any fundamental role in the formulation of quantum field theory.
>
> Wald ([2019](#)).

Perhaps the most important lesson of relativity is that, ironically, nothing relevant is relative. Many common expositions of the theory render this fact mysterious by focusing too heavily on changes of reference frames instead of their underlying geometrical structure. Nevertheless, it is a fact of physics, and perhaps of logic, that any well-defined experiment has the very same result for all observers and in any choice of coordinate system.

This choice of phrasing might seem to go against what is usual in many textbooks. Suppose, for example, that a certain particle is at rest with respect to Alice's reference frame, but in motion with respect to Bob's reference frame. One will find in textbooks the correct statement that, if Alice and Bob each measure the energy of the particle, they will find different results. Nevertheless, it is often overlooked that they are both performing different experiments in the sense their experimental apparatuses are set up differently. Indeed, one of the apparatuses is in motion with respect to the particle, while the other one is not. Hence, deep down, Alice and Bob are conducting different experiments. The only reason one would be prompted to think their measurements should match is the linguistic confusion of calling different properties of the particle by the same word: "energy".

This presents a conceptual difficulty when formulating general relativity (GR). One is faced with the fact that humans frequently use the same word for different concepts, and certain expositions might keep this ambiguity. For example, if one attempts at formulating relativity in terms of co-ordinate systems and how to change between them, then these sorts of linguistic confusions will be present in the formulation itself. Furthermore, one will be formulating the theory in terms of unphysical and arbitrary choices—after all, coordinate systems are not dependent on the physics, but on the physicist. Hence, such a formulation suffers from conceptual difficulties that may not affect the mathematics or the calculations if one is careful enough, but that make it dangerous for one to commit mistakes and misunderstand the meaning of the theory.

Nowadays, texts such as Hawking and Ellis (1973) and Wald (1984) avoid such issues by formulating GR in terms of its geometric meaning. For example, suppose again the situation in which Alice and Bob want to measure the energy of a particle. In geometric terms, this can be posed in the following manner. Let $p^a$ be the particle's four-momentum, $u^a$ be Alice's four-velocity, and $v^a$ be Bob's four-velocity. Then Alice desires to measure $-p_a u^a$, while Bob desires to measure $-p_a v^a$. Both of them desire to measure scalars, meaning their measurements cannot depend on choices of coordinates (as they should not). Furthermore, Alice is perfectly capable of performing the same experiment as Bob and measuring $-p_a v^a$, the only difference being that Alice will not call this quantity "energy". Quite naturally, when the same experiment (measuring $-p_a v^a$) is performed by different observers, they always obtain the same result.

Another way of noticing this fact is by setting up a simple mental experiment. Consider any experiment with results being either "yes" or "no". For example, this experiment could be "Is the particle's energy greater than or equal to 1 eV?". Suppose now that Bob attaches to his apparatus a bomb that will explode and destroy the laboratory if the experiment yields "yes". Day to day experience leads one to believe that both Bob and Alice will always agree on whether the bomb has or not gone off. In other words, changing observers cannot change the result of an experiment, even though it does change the words one uses to describe the results. They will not disagree on the results, but rather on the choice of calling the property being measured "energy". Alice might still refuse to call $-p_a v^a$ energy, but agrees on whether the bomb has or not gone off.

Similarly, consider any experiment which has a real number as a result. Suppose that the experiment is set up such that its result is shown on a computer screen. It would be extremely shocking if Alice and Bob happened to see different numbers on the same computer screen. The experiment never depended on the observer, but only on the experimental setup, which could have been made by either Alice or Bob. At least within classical physics, the role of the observer comes only in interpreting the results. One of them insists in calling the measured quantity by the name "energy", while the other refuses to do so.

As one could expect, quantum field theory in curved spacetime (QFTCS) inherits these linguistic difficulties from GR. Usual formulations of quantum field theory (QFT) in flat spacetime will very often depend heavily on coordinates and the Poincaré group (see Weinberg 1995, for example), and hence they can end up with the same conundrums of giving the same name to different objects. In flat spacetime, this is often excused. For example, the word "particle" is defined to have the same meaning by all inertial observers in Minkowski spacetime. Nevertheless, this will not be true in curved spacetimes or for non-inertial observers: the Unruh effect (Section 2.4) clearly shows that different observers may interpret the same physical state as possessing different quantities of "particles". Similarly, the word "vacuum" will turn out to be very subtle, since it typically is taken to mean "absence of particles".

Unfortunately, most expositions of QFT depend very heavily on the notions of particle and vacuum. Canonical quantization, for example, consists in formulating the theory by specifying its Hilbert space as the Fock space, which is inherently tied to a notion of particle and to a notion of vacuum. If these notions depend on the observer, then they are not due to physics, but due to the physicist. Hence, they are not fundamental and may end up obscuring important properties of the theory. While we can formulate QFTCS using a Fock space and exploit its properties to obtain interesting consequences, it is not always desirable to do so.

A second option would be to employ path integral techniques. Nevertheless, path integrals require a choice of a preferred vacuum state, which invariably introduces the physicist into the equations that ideally would involve only the physics. Furthermore, it is not obvious whether any state would be admissible on a path integral, especially in curved spacetimes. We shall discuss these issues in Section 2.5.

Path integrals can also involve one further difficulty. They are mathematically mysterious objects, due to the difficulty of understanding their convergence properties. Since this is a mathematical problem rather than a physical one, we shall overlook it in this text. However, it is not unusual to bypass these difficulties by formulating the theory in Euclidean signature—*i.e.*, one makes an analytical continuation to imaginary time to obtain expressions that are well-defined. In fact, an analytical continuation is necessary even in less rigorous approaches to make sense of the Gaussian integrals that come up (Peskin and Schroeder 1995, p. 286). This Euclidean approach introduces yet another limitation, since not all spacetimes admit such a continuation and, when they do, it might not be unique, as we are going to show in Section 2.5. Hence, this approach is also not always desirable.

These reasons lead authors such as Hollands and Wald (2015) to take another route, known as the "algebraic approach", which we now follow. We shall focus on what are the essential objects of QFTCS and make an effort to stick to them and avoid any other possible complications or prejudices. We are going to concern ourselves with the physics while avoiding the temptation of introducing our own choices of states, coordinates, and so on in our equations. While many discussions of this approach are fairly mathematically-inclined, we will ignore the mathematical details and proceed formally, instead of rigorously. In other words, while we shall borrow terms and notions commonly used in mathematical physics, we will keep the line of work common to theoretical physicists. This is not much different from how we would overlook the convergence of path integrals or the functional analytic details of Hilbert spaces had we chosen to follow other approaches.

It is important to point out that, even though some approaches have limitations when compared to others, there is not an ultimate approach we should always use. In fact, in this text we shall later employ Euclidean path integral methods, even though our initial formulation of QFTCS will be algebraic. The point is that different jobs require different tools. When formulating QFTCS, it is interesting to use a language that makes clear what are the theory's essential features, allowing us to have a clear picture of what we are doing. On the other hand, this language might turn out to be difficult to carry calculations with. Hence, once we are interested in doing actual computations, we will be opportunistic and use a more convenient approach.

## 2.1  What Constitutes a Physical Theory?

To introduce the algebraic approach, we shall motivate its use by following the discussions in the books by Alfsen and Shultz (2003, Chap. 6), Araki (1999, Chap. 1), and Strocchi (2008, Chap. 1). Our goal is to conclude that an algebraic structure is expected to arise in any physical theory, or at least in a wide class of physical theories. In more detail, we shall see how fairly general considerations about experiments with probabilistic outcomes lead to the notions of states and observables, as done by Araki (1999). Following Araki (1999) and Strocchi (2008), we will then argue that the observables have the algebraic structure of a Jordan algebra. At last, we shall consider the arguments given by Alfsen and Shultz (2003) to conclude it is natural to expect this Jordan algebra to be just the Hermitian operators on a larger $*$-algebra. Relevant definitions will be provided as they become necessary.

### States and Observables

Physics is an experimental science, and hence we will develop our description keeping this in mind as much as possible. While we may not provide details on how to set up specific experiments, it is important to remember that our goal is to describe results of experiments. The importance of this viewpoint can be exemplified by our previous example. While the meaning of the word "energy" depends on whether Alice or Bob is saying it, there is no ambiguity on the results of an experimental measurement.

Hence, let us begin by ensuring we are talking about experiments. In a laboratory conducting an experiment, one can identify four components that might be relevant:

  i. the physical system itself, which is the object of the measurement and was prepared in some way;

  ii. the experimental apparatus, used to perform the measurement on the system;

  iii. the observer conducting the experiment;

  iv. the environment, which is everything else.

We have previously argued that the observer should not be relevant on physical predictions, as long as "observer" is understood in the relativistic sense of the word. Quantum mechanics (QM) introduces some more nuance on the word "observer", and there is an ongoing discussion on their role in the theory, but we shall do as Araki ([1999](#)) and neglect any possible influences due to the observer. We shall also make the simplifying assumption that the environment is not playing a relevant role in the process. Therefore, we are assuming an experiment is completely defined by a physical system (including its detailed preparation) and by a measurement apparatus (also including its detailed preparation).

Let us then introduce a notation to "tag" the physical system and the apparatus. We will write $\omega$ to denote the physical system (and its detailed preparation) and $Q$ to denote the apparatus (and its detailed preparation). Hence, $\omega_1$ and $\omega_2$ may denote different systems, or the same system after undergoing different preparations, and so on. Similarly, $Q_1$ and $Q_2$ may mean completely different experimental apparatuses, or the same apparatus after undergoing two different preparations, and so on. Notice it is important to include the preparations of both the system and the apparatus in our discussion: there is no reason to expect *a priori* that different preparations of either will lead to the same experimental results.

To perform a measurement, we start with some isolated system $\omega$ and some isolated apparatus $Q$ which are known beforehand, including their detailed preparations. We then bring them together and let them interact until the apparatus sets in some state. For example, we wait until a needle is pointing somewhere, at which point we can read the result of the measurement. The possible results will be denoted by the symbols $p$, $q$, etc. These might be numbers or something different, such as the notion of "up" or "down".

This measurement can be performed multiple times by repeating or mimicking $\omega$ and $Q$. In some experiments, the same result will be obtained every single time. In other experiments, different results are obtained at each iteration. We shall make the hypothesis, however, that the results always approach some underlying probability distribution. This assumption is based on the practical justification that it has worked so far, and on the philosophical justification that it is difficult to conceive how to do physics in any other way. If experiments did not follow some sort of probability distribution, then perhaps we would not be able to do physics at all.

Suppose now we repeat the same experiment $N$ times and get the result $p$ a total of $n_p$ times. If there is indeed some underlying probability distribution, then the limit

$$w_\omega^Q(p) = \lim_{N \to \infty} \frac{n_p}{N} \tag{2.1.1}$$

exists. $w_\omega^Q(p)$ is then interpreted as the probability that measuring the system $\omega$ with the apparatus $Q$ yields the result $p$. As expected of a probability, notice that[*] $0 \le w_\omega^Q(p) \le 1$ and $\sum_p w_\omega^Q(p) = 1$.

The definition of the probabilities $w_\omega^Q(p)$ allows us to start discussing the notions of states and observables. Let us begin with states.

Suppose there are two system preparations $\omega_1$ and $\omega_2$ such that

$$w_{\omega_1}^Q(p) = w_{\omega_2}^Q(p) \tag{2.1.2}$$

for every apparatus $Q$ and for every result $p$. This is just a mathematical way of stating that these two different preparations $\omega_1$ and $\omega_2$ lead to the very same experimental results. Hence, it makes sense for us to consider them as being equivalent. We can then use Eq. (2.1.2) to define an equivalence relation on the space of all system preparations. The quotient of the space of all system preparations by this equivalence relation will be said to be the space of states. Therefore, a state is an equivalence class of system preparations. In other words, a state is a collection of system preparations that lead to the same experimental results. From now on, we will use $\omega$ to denote states rather than preparations.

In a similar fashion, let us suppose now two experimental apparatuses $Q_1$ and $Q_2$ such that

$$w_\omega^{Q_1}(p) = w_\omega^{Q_2}(p) \tag{2.1.3}$$

holds for every state $\omega$ and for every result $p$. Just as with the states, there is no point in distinguishing $Q_1$ and $Q_2$, since they both lead to the same experimental results. Hence, we shall also introduce an equivalence relation in the space of all apparatuses and proceed as with the states. We now get a space $\mathscr{A}$ of observables, which are the equivalence classes of apparatuses. In other words, an observable is a collection of experimental apparatuses that lead to the same experimental results. From now on, we will use $Q$ to denote observables rather than apparatuses.

### Functions of Observables

There is no *a priori* imposition on the possible results of an experiment. They could be concepts such as "up" and "down", "full" and "empty", and so on. They could be a color, as one would get if the scale on an ammeter was changed to a color gradient (Fig. 2.1 on the next page). While the collection of possible results can be fairly arbitrary, it is common and convenient to map them to real numbers. We shall also follow this convention from now on.

There is a lot of freedom in how we label our results. For example, suppose we want to measure the $z$-component of an electron's spin. In terms of real numbers, we can name them $+1$ and $-1$, $+\frac{\hbar}{2}$ and $-\frac{\hbar}{2}$, $1$ and $0$, and any other combination of two real numbers. The labels themselves are picked

---

[*]Some nuances arise when we consider infinite possible results. We will ignore them in our simplified treatment, but Araki (1999, Chap. 1) discusses this more carefully.

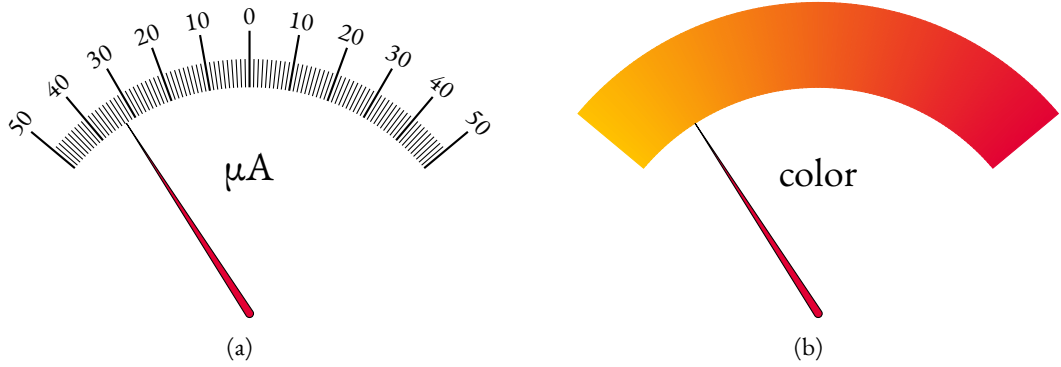

Figure 2.1: While it is typical to measure currents on an ammeter using real numbers as depicted on (a), this is
merely a convention. One could also make the unusual choice of labeling the ammeter's scale using
a color gradient and measure current using color, as depicted on (b).

only out of convention. We then notice that our present discussion has a curious feature: if we
measure the $z$-component of an electron's spin using an apparatus that yields the results $+1$ and $-1$
and later do the same experiment using an apparatus that yields $+\frac{\hbar}{2}$ and $-\frac{\hbar}{2}$ we cannot say that the two
apparatuses are measuring the same observable. After all, they gave different results and will never
yield the same probability distributions. Nevertheless, it is clear that the results of both experiments
are related by a simple change in scale. To account for this, we shall introduce the notion of function
of an observable.

Let $f : \mathbb{R} \to \mathbb{R}$ be some function. If $Q \in \mathcal{A}$, we define $f(Q) \in \mathcal{A}$ as the observable that yields $f(q)$
in any experiment in which $Q$ would yield $q$. In practice this can be achieved by simply rewriting the
numbers on the scale of an apparatus used to measure the observable $Q$. For example, in the case of
an analog apparatus used to measure an electron's spin in some direction, we can just use tape to stick
a piece of paper with the new values on top of the old ones.

$f$ does not need to be one-to-one, but if it fails to be, then $f(Q)$ is "coarser" than $Q$ in the sense
that it does not distinguish results as much as $Q$ does. As an example, consider once again the ammeter
on Fig. 2.1. Instead of mapping the results to a color gradient we could have chosen to simply map
the results to $-1$, $0$ or $+1$ depending only on the sign of the current. In this case, the change of scales loses
information about the experiment and $f(Q)$ is not able to distinguish results as much as $Q$ could.

A particularly interesting case is that in which $f(q) = 1$ for any $q$. In this case, we denote $f(Q) = \mathbb{1}$.
Since $f$ is constant, $\mathbb{1}$ does not depend on $Q$. If $f(q) = c$ for all $q$ where $c \in \mathbb{R}$, then we write $f(Q) = c\mathbb{1}$.
From an experimental point of view such a choice is hardly justified, but from a theoretical standpoint
it is only extremely inconvenient.

### Expectation Values

Before we discuss the algebraic structure of a theory, let us pick a state $\omega$ and an observable $Q \in \mathcal{A}$. We
define the expectation value of $Q$ in the state $\omega$ as

$$\omega(Q) = \sum_p p w_\omega^Q(p). \tag{2.1.4}$$

It is interesting to notice that if we know $\omega(Q)$ for every $Q$, we know $w_\omega^Q(p)$ for every $Q$ and $p$, since $w_\omega^Q(p) = \omega(\chi_p(Q))$, where $\chi_p$ is the function given by

$$\chi_p(q) = \begin{cases} 1, & \text{if } q = p, \\ 0, & \text{otherwise.} \end{cases} \tag{2.1.5}$$

Hence, knowing $w_\omega^Q(p)$ and knowing $\omega(Q)$ is interchangeable. Both of them can be seen as "coordinates" for the state $\omega$, since two states $\omega_1$ and $\omega_2$ are equal if, and only if, $\omega_1(Q) = \omega_2(Q)$ for all $Q$ and if, and only if, $w_{\omega_1}^Q(p) = w_{\omega_2}^Q(p)$ for all $Q$ and all $p$.

It is also interesting to point out that in QM we often need transition probabilities, but these can always be written in terms of expectation values. Indeed, this follows from the so-called polarization identity,

$$\langle \psi | \phi \rangle = \frac{1}{4}\left( \|\psi + \phi\|^2 - \|\psi - \phi\|^2 - i\|\psi + i\phi\|^2 + i\|\psi - i\phi\|^2 \right), \tag{2.1.6}$$

which can be proved by expanding the right-hand side (RHS). Notice that all terms on the RHS can be understood as expectation values. They might be expectation values in different states and concerning different observables, of course. Nevertheless, all probabilities in QM boil down to computing an expression of the form $\langle \psi | \phi \rangle$, and hence they all boil down to expectation values.

### Algebraic Structure of Observables

Suppose $Q_1$ and $Q_2$ are observables. We can define their sum as the observable $Q_1 + Q_2$ such that

$$\omega(Q_1) + \omega(Q_2) = \omega(Q_1 + Q_2) \tag{2.1.7}$$

for every state $\omega$. In experimental practice, $Q_1 + Q_2$ might not have a straightforward interpretation, meaning we might be expanding $\mathscr{A}$ beyond our original physical motivations.

Since we have already defined functions of observables, we can also define the product of $Q$ by a scalar: $c \cdot Q = f(Q)$, where $f(q) = cq$.

As one can check, these operations of addition and multiplication by a scalar turn $\mathscr{A}$ into a vector space.

We can still equip $\mathscr{A}$ with more structure in a natural way. For example, we are able to compute functions of elements of $\mathscr{A}$. In particular, given $Q \in \mathscr{A}$, $Q^2$ is well-defined by our previous prescriptions.

Using all of these structures we can see that $\mathscr{A}$ is naturally endowed with a product defined as

$$Q_1 \circ Q_2 = \frac{1}{2}\left[ (Q_1 + Q_2)^2 - Q_1^2 - Q_2^2 \right]. \tag{2.1.8}$$

This is known as the Jordan product. Notice that it is commutative by construction. Jordan, von Neumann, and Wigner (1934) have proven this product is weakly associative, *i.e.*,

$$(Q_1^2 \circ Q_2) \circ Q_1 = Q_1^2 \circ (Q_2 \circ Q_1). \tag{2.1.9}$$

If $Q_1$ and $Q_2$ are functions of a common underlying observable $Q$, then we are actually capable of expanding the definition of the Jordan product and write $Q_1 \circ Q_2 = \frac{1}{2}(Q_1 Q_2 + Q_2 Q_1)$. This expression makes sense because $Q_1 Q_2$ is actually of the form $f(Q)g(Q)$ for some $Q$ and some functions $f$ and $g$.

Hence, the expression actually means $Q_1 Q_2 = h(Q)$, where $h(q) = f(q)g(q)$. In this particular case, notice that $\circ$ is homogeneous, *i.e.*,

$$Q_1 \circ (\lambda Q_2) = (\lambda Q_1) \circ Q_2 = \lambda(Q_1 \circ Q_2) \tag{2.1.10}$$

for any real number $\lambda$. We shall follow Strocchi (2008, p. 20) and assume Eq. (2.1.10) for all $Q_1, Q_2 \in \mathscr{A}$.

Under these hypotheses, $\mathscr{A}$ then satisfies the following properties:

i. it is a real vector space;

ii. it has a bilinear product $\circ : \mathscr{A} \times \mathscr{A} \to \mathscr{A}$;

iii. $\circ$ is commutative;

iv. $\circ$ is weakly associative.

Such a structure is known as a Jordan algebra. It is also possible to show that it can be naturally turned into a normed space (Strocchi 2008, pp. 18–21), but we will overlook the topological properties to keep our discussion at an introductory level.

Jordan algebras often occur as the subspace of Hermitian operators acting on a Hilbert space. Indeed, if $\mathscr{H}$ is a Hilbert space and $Q_1$ and $Q_2$ are linear operators in $\mathscr{H}$, we can define their Jordan product as the symmetrized product $Q_1 \circ Q_2 = \frac{1}{2}(Q_1 Q_2 + Q_2 Q_1)$. Notice that if $Q_1$ and $Q_2$ are Hermitian, then so is $Q_1 \circ Q_2$. Weak associativity can be proved with a bit of algebra. Hence, this is a class of common examples of Jordan algebras. Notice this is relevant because it allows us to see how our construction of a natural algebraic structure for any physical theory is leading us to the Hilbert spaces of quantum mechanics[*].

Nevertheless, not all Jordan algebras are simply the Hermitian operators on some Hilbert space. The pathological example is the space of $3 \times 3$ Hermitian matrices with octonion entries (Alfsen and Shultz 2003, Theorem 4.6). We are then posed with the question of whether these pathological examples may or may not be of physical interest.

Most of the mathematical physics literature works not with Jordan algebras, but with $*$-algebras, which do resemble the operators in a Hilbert space in a well-defined sense. Since this leads to a convenient and powerful framework, we shall do the same. To justify this, we shall follow an argument given by Alfsen and Shultz (2003, pp. 196–197), but see also the argument by Streater (2007, Sec. 12.4).

When it comes to dealing with states and observables, Jordan algebras are sufficient. We are perfectly capable of using the structure of a Jordan algebra to obtain, for example, the probabilities we might be interested in. However, as Alfsen and Shultz (2003, pp. 196–197) argue, QM requires more than that from the observables, for they are also generators of symmetries. For example, momentum is not only the physical notion of momentum, but also the generator of translations. This means we also need the observables to have a Lie algebra structure, and this is not made possible by the Jordan product alone. However, suppose there is an associative product $\cdot$ such that $Q_1 \circ Q_2 = \frac{1}{2}(Q_1 \cdot Q_2 + Q_2 \cdot Q_1)$.

---

[*]It is also interesting to point out that classical mechanics also admits a formulation in terms of Hilbert spaces, often called Koopman–von Neumann mechanics (Mauro 2002). Our algebraic construction then seems to recover both quantum and classical theories. It is also possible to show that every commutative $C^*$-algebra can be understood as the collection of continuous functions on a phase space (Bratteli and Robinson 1987, Theorem 2.1.11; Landsman 2017, Theorem 3.1), which recovers our usual notion of observables in classical physics.

Then the commutator $[Q_1, Q_2] = Q_1 \cdot Q_2 - Q_2 \cdot Q_1$ can introduce the Lie algebra structure we need for the symmetry generators, solving our difficulty.

A convenient manner of implementing this is precisely by copying the algebraic structure of the linear operators acting on a Hilbert space. The technical term for such a structure is a ∗-algebra, which has the following properties

i. it is a complex vector space;

ii. it has a bilinear product $\cdot : \mathscr{A} \times \mathscr{A} \to \mathscr{A}$;

iii. $\cdot$ is associative;

iv. it has an antilinear operation $* : \mathscr{A} \to \mathscr{A}$ such that $(Q_1 \cdot Q_2)^* = Q_2^* \cdot Q_1^*$ and $Q^{**} = Q$.

Notice the ∗ operation mimics the Hermitian conjugate found in Hilbert spaces. It is often called an "involution".

Recall that we have been ignoring the topological properties for simplicity. Had we considered them, then we would be led to the structure of a $C^*$-algebra, which is a normed ∗-algebra with some convenient properties. More details can be found in the books by Alfsen and Shultz (2003), Araki (1999), and Strocchi (2008), among others.

$C^*$-algebras correspond to algebras of bounded operators, which intuitively means the observables have bounded values. This is different from what is traditionally done in the theoretical physics literature, since position, for example, is unbounded. Nevertheless, for practical experimentation aspects, we never measure unbounded values, and hence it is perfectly reasonable to work with bounded operators only. In any case, theoretical physicists not acquainted with functional analysis are more used to working with unbounded operators. Due to this and the fact that we want to avoid diving into discussions concerning norms and topological aspects, we can safely work with ∗-algebras only. Khavkine and Moretti (2015) take a similar stance within QFTCS.

## 2.2 Recovering Hilbert Spaces from ∗-Algebras

In the previous section, we explained why we expect all physical theories to be related to a ∗-algebra. Let us now see how we can recover the traditional notions of Hilbert spaces from these algebras. This is relevant within QFTCS because, as we will learn, there can be more than one Hilbert space associated to an algebra, and this lack of uniqueness is the origin of the difficulty to define what is meant by a "particle" in an observer-independent manner.

We now take the defining property of a physical theory to be a ∗-algebra $\mathscr{A}$. This gives us the observables, which are Hermitian operators in this algebra, *i.e.*, elements with $Q = Q^*$. Notice that not all Hermitian operators are physically significant—given any operator $Q$, $Q + Q^*$ is Hermitian. In ordinary QM, $Q$ could even be the unitary operator related to some symmetry transformation, and hence $Q + Q^*$ would not get the usual interpretation of observable we usually desire. Nevertheless, do notice this remark is not exclusive to the algebraic approach, since it also happens in the Hilbert space approach.

We have observables, but we still need states. Earlier, we noticed states can be seen as functions from the observables to the real line. Indeed, the expectation value of the observable $Q$ in the state $\omega$ was $\omega(Q)$. Furthermore, we defined the product of an observable by a scalar $\lambda$ and the addition of observables in such a way that

$$\omega(Q_1 + \lambda Q_2) = \omega(Q_1) + \lambda \omega(Q_2). \tag{2.2.1}$$

Hence, now that we are taking an algebra of observables as the defining feature of a theory, we will define the states as linear functionals on the algebra, *i.e.*, linear maps $\omega\colon \mathcal{A} \to \mathbb{C}$. They can yield complex values, since not every element in the algebra is a physical observable. Furthermore, we will require them to be normalized[*], $\omega(\mathbb{1}) = 1$, and to be positive, $\omega(Q^*Q) \geq 0$. Notice these impositions ensure we are mimicking the expectation values we already understand in QM.

Suppose now we are given an algebra $\mathcal{A}$ and a state $\omega\colon \mathcal{A} \to \mathbb{C}$. It is possible to perform a procedure known as the Gelfand–Naimark–Segal (GNS) construction (Gelfand and Naimark 1943; Segal 1947) that yields a Hilbert space $\mathcal{H}$ on which one can represent the algebra $\mathcal{A}$ by means of operators acting on $\mathcal{H}$. This Hilbert space does depend on $\omega$.

The word "represent" is used here in the sense of representation theory. Group representations are widely used in modern theoretical physics (see Zee 2016), and algebra representations are defined in a similar fashion. If $\mathcal{H}$ is a Hilbert space, $\mathcal{L}(\mathcal{H})$ the space of linear operators acting on $\mathcal{H}$, and $\mathcal{A}$ a ∗-algebra, a representation is a linear map $\pi\colon \mathcal{A} \to \mathcal{L}(\mathcal{H})$ that preserves the algebraic product and the involution, *i.e.*, $\pi(Q_1 Q_2) = \pi(Q_1)\pi(Q_2)$ and $\pi(Q^*) = \pi(Q)^*$. Hence, a representation of an algebra on a Hilbert space is a "copy" of the algebra in the operators acting on the Hilbert space. This copy may or may not be faithful: nothing prevents a representation from assigning the same operator on $\mathcal{L}(\mathcal{H})$ to different elements of $\mathcal{A}$. If the representation is one-to-one, it is said to be faithful.

The technical details of the GNS construction theorem vary depending on the assumptions taken. Khavkine and Moretti (2015, Theorem 5.1.13) discuss the theorem for unital ∗-algebras (*i.e.*, ∗-algebras which have a unit element), while most books on $C^*$-algebras discuss the $C^*$ case—which is the one usually meant when one speaks of the GNS construction. We will omit the details and focus on the physical meaning of the theorem.

Specifically, given a ∗-algebra satisfying some assumptions and some chosen state $\omega$ on said algebra, the GNS construction yields us

i. a Hilbert space $\mathcal{H}$;

ii. a representation $\pi\colon \mathcal{A} \to \mathcal{L}(\mathcal{H})$;

iii. a vector $|\omega\rangle \in \mathcal{H}$.

This triple has the property that $\omega(A) = \langle\omega|\pi(A)\omega\rangle$ for all $A \in \mathcal{A}$, and hence $|\omega\rangle$ represents $\omega$ as a vector in the Hilbert space. It also holds that[†]

$$\{\pi(A)\,|\omega\rangle\,;\, A \in \mathcal{A}\} = \mathcal{H}. \tag{2.2.2}$$

The GNS triple is also unique up to unitary isomorphism, *i.e.*, other triples with the same properties are related to the GNS triple by a unitary transformation. Do notice, however, that the GNS triple assumes a state underlying its construction, and different choices of states may lead to unequivalent representations.

---

[*]I implicitly assumed the algebra to have a unit $\mathbb{1}$, although this is not really necessary. See the book by Bratteli and Robinson (1987) for a more general treatment and more information.

[†]More precisely, the left-hand side (LHS) of Eq. (2.2.2) is dense on the Hilbert space.

Notice that the GNS construction provides the bridge between the abstract algebras we previously constructed and our usual notions of QM in a Hilbert space. However, it also allows us to start noticing some limitations of the Hilbert space approach.

Let us provide an example. It is a common misconception that a state is pure if it can be written as a vector on a Hilbert space, as opposed to a density matrix. Nevertheless, the GNS construction showed us that any state can be written as a vector on some Hilbert space. One can show that density matrices on a Hilbert space define algebraic states on the algebra $\mathscr{A} = \mathscr{B}(\mathscr{H})$ of bounded operators acting on the Hilbert space. Hence, the GNS construction allows us to pick any state, mixed or pure, and find some Hilbert space in which this state is a vector. A more careful way to define a pure state is by noticing that given a state $\omega$, there are always states $\omega_1$ and $\omega_2$ ($\omega_1 \neq \omega_2$) such that

$$\omega(A) = \lambda\omega_1(A) + (1-\lambda)\omega_2(A) \tag{2.2.3}$$

for all $A \in \mathscr{A}$, where $\lambda \in [0,1]$. At worst, take $\omega_1 = \omega$, any $\omega_2$, and $\lambda = 1$. A pure state is a state in which Eq. (2.2.3) holding for all $A \in \mathscr{A}$ implies $\lambda = 0$ or $\lambda = 1$. A mixed state is then simply a state that is not pure. For a pedagogical introduction to pure and mixed states from the point of view of the algebraic approach, see the review by Barata et al. (2021).

This raises an important question: if it is possible for the same state to be represented as a density matrix or as a vector, do we have any guarantee that all states are always represented? The answer is no. Some states might not be representable as density matrices in some Hilbert spaces. In fact, we define the folium of a state $\omega$ to be the collection of all algebraic states that can be written as a density matrix in the GNS representation induced by $\omega$. This is relevant because it means that working with Hilbert spaces keeps us from being able to see the whole picture. There are states that are not accessible from a given Hilbert space[*].

One might also wonder whether two representations of the same algebra $\mathscr{A}$ are always ensured to be equivalent. In other words, suppose $(\mathscr{H}_1, \pi_1)$ and $(\mathscr{H}_2, \pi_2)$, $\pi_i \colon \mathscr{A} \to \mathscr{L}(\mathscr{H}_i)$, are representations of $\mathscr{A}$. Is it always possible to find a unitary transformation $U \colon \mathscr{H}_1 \to \mathscr{H}_2$ such that $\pi(A)_2 = U\pi_1(A)U^*$? This is the most basic requirement for the two Hilbert spaces to yield the same theory. If we can't find such a transformation, then the inner products in a Hilbert space might have values different from those on the other Hilbert space. Hence, they can lead to different descriptions.

For quantum systems with a finite number of degrees of freedom, this question is answered by the Stone–von Neumann theorem (Hall 2013, Theorem 14.8), which ensures that all representations of the canonical commutation relations (CCR) for a system with finitely many degrees of freedom are equivalent[†]. However, the theorem fails in the case of infinitely many degrees of freedom, *i.e.*, for field theory. Therefore, we might get non-equivalent Hilbert spaces.

The issue with getting non-equivalent Hilbert spaces is we cannot tell which one is the correct description. In some situations, symmetry considerations might allow us to pick a preferred Hilbert space out of all possible ones. For example, in Minkowski spacetime there is a single Poincaré-invariant state, the Minkowski vacuum. Hence, it seems natural to pick the Hilbert space obtained by using the GNS construction with the Minkowski vacuum. This leads one to the usual treatment of QFT given, *e.g.*, by Weinberg ([1995](#)). Similar comments are applicable to QFT in stationary spacetimes such as Schwarzschild or De Sitter spacetimes (see Section [2.A](#)), but not in a general curved spacetime.

---

[*]There are situations in which one can always approximate a state arbitrarily well by density matrices in a given Hilbert space. This happens for a scalar quantum field in an arbitrary globally hyperbolic spacetime (Wald 1994, Theorem 4.5.2).

[†]Technically, the Stone–von Neumann theorem is a statement about the Weyl algebra, which corresponds to an exponentiated version of the CCR algebra (see Wald 1994, Chap. 2, for details). While the CCR algebra is merely a ∗-algebra, the Weyl algebra is a $C^*$-algebra, and hence it "behaves better" from a mathematical perspective.

Even in the occasion two different representations are equivalent, we should remark they can still have different interpretations. For example, in QFT, the Hilbert space is often taken to be a Fock space, which has a natural interpretation in terms of "particles". The notions of "particle" associated to the two equivalent Fock spaces might not be the same, and hence the unitary transformation will not preserve particle number. This is not an issue: we are interested in a quantum theory of fields, not of particles. However, it makes it clear that careless dependence on a Fock space might mix the actual physical content of the theory with misconceptions due to a belief in "particles" as fundamental entities.

## 2.3   Free Quantum Fields in Curved Spacetime

Given the previous motivations, we expect to be able to construct the algebra of observables for any theory we are interested in studying. However, this turns out to not be so easy. In spite of recent progress due to Brunetti et al. ([2022](#)) and Buchholz and Fredenhagen ([2020](#)), it is still not proven that any interacting quantum field theory admits an algebraic formulation[*]. In addition, while Hollands and Wald ([2010](#)) have put forth a proposal for how to define a quantum field theory nonperturbatively, the majority of the current knowledge is restricted to perturbative approaches.

Due to these reasons, we will focus on discussing the algebraic formulation of a free real scalar field on a globally hyperbolic spacetime. Further discussions can be found, for example, on the review by Hollands and Wald ([2015](#)) and references therein. This review is also our main reference for this section, alongside the book by Hack ([2016](#)). We should also add that this section assumes familiarity with the causal structures in GR (see, *e.g.*, Geroch [1970](#); Hawking and Ellis [1973](#); Pfäffle [2009](#); Wald [1984](#)).

In a curved spacetime, we take a real scalar field $\phi$ to satisfy the Klein–Gordon equation in the form

$$(\nabla_a \nabla^a - m^2)\phi = 0, \tag{2.3.1}$$

where $\nabla_a$ is the Levi-Civita connection associated with the Lorentzian metric $g_{ab}$ (signature $- + ++$). In order to construct the algebra, we are also interested in working with the Klein–Gordon equation with a source,

$$(\nabla_a \nabla^a - m^2)\phi = j, \tag{2.3.2}$$

where $j$ is some fixed arbitrary smooth function with compact support. We denote the space of smooth functions of compact support on spacetime $\mathcal{M}$ by $\mathscr{C}_0^\infty(\mathcal{M})$.

Let $\Sigma$ be a Cauchy surface for $\mathcal{M}$ with future-directed unit normal $n^a$. Let $f_0, f_1 \in \mathscr{C}_0^\infty(\Sigma)$. Then there is a unique function $\phi \in \mathscr{C}^\infty(\mathcal{M})$ such that the system

$$\begin{cases} (\nabla_a \nabla^a - m^2)\phi = j, \\ \phi|_\Sigma = f_0, \\ n^a \nabla_a \phi|_\Sigma = f_1 \end{cases} \tag{2.3.3}$$

---

[*]Given other formulations are typically not as mathematically sound, this does mean we still do not know whether interacting quantum field theories make mathematical sense at all.

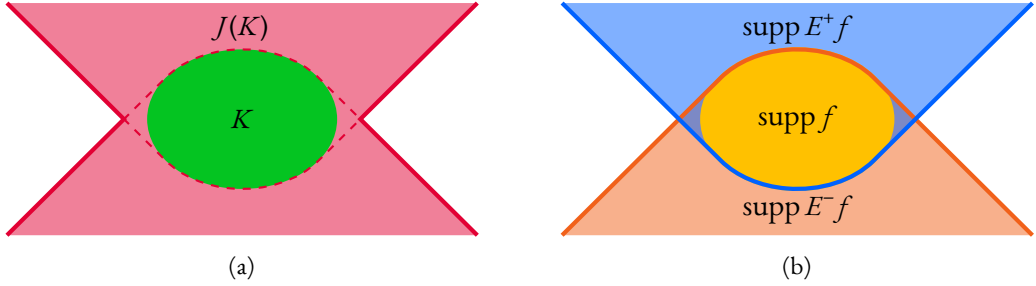

Figure 2.2: Causal diagrams illustrating the good causal behavior of solutions to the Klein–Gordon equation. The diagrams are drawn such that null geodesics are always at 45°. (a): if $K = \operatorname{supp} f_0 \cup \operatorname{supp} f_1 \cup \operatorname{supp} j$, then the solution $\phi$ for Eq. (2.3.3) on the facing page is such that $\operatorname{supp} \phi \subseteq J(K) = J^+(K) \cup J^-(K)$. In the picture, the dashed lines indicate the boundary of $J^\pm(K)$. (b): supports of the advanced and retarded propagators $E^-$ and $E^+$.

is satisfied (see, *e.g.*, Ginoux 2009, Theorem 3). This solution is causally well behaved in the sense that $\operatorname{supp} \phi \subseteq J^+(K) \cup J^-(K)$, where $K = \operatorname{supp} f_0 \cup \operatorname{supp} f_1 \cup \operatorname{supp} j$ (Ginoux 2009, Theorem 3). This is illustrated on Fig. 2.2a. It then follows that changing the value of, say, $f_0$ outside of $J^-(x)$—where $x$ lies to the future of $\Sigma$—will not affect the value of $\phi(x)$.

This good behavior allows us to define the advanced and retarded Green's operators (or propagators) for the Klein–Gordon equation. These are maps $E^\pm \colon \mathscr{C}_0^\infty(\mathscr{M}) \to \mathscr{C}^\infty(\mathscr{M})$ that map the source $j$ on Eq. (2.3.2) on the preceding page to a solution of the Klein–Gordon equation with boundary conditions defined in the following way. For the advanced propagator $E^-$, $E^- j$ is the unique solution with $\operatorname{supp} E^- j \subseteq J^-(\operatorname{supp} j)$. Similarly, the retarded propagator $E^+$ is defined such that $E^+ j$ is the unique solution with $\operatorname{supp} E^+ j \subseteq J^+(\operatorname{supp} j)$. In other words, the retarded propagator propagates a source to within the future light cone, while the advanced propagator propagates a source to within the past light cone (see Fig. 2.2b). The existence and uniqueness of these operators is assured for the Klein–Gordon equation in an arbitrary globally hyperbolic spacetime (Ginoux 2009, Corollary 6).

These operators might also be seen as distributional kernels. For example, we can write

$$(E^\pm j)(x) = \int E^\pm(x, y) j(y) \sqrt{-g} \, \mathrm{d}^4 y \,. \tag{2.3.4}$$

If we apply the Klein–Gordon operator to both sides of this equation, we see that

$$(\nabla_a \nabla^a - m^2) E^\pm(x, y) = \frac{1}{\sqrt{-g}} \delta^{(4)}(x, y), \tag{2.3.5}$$

where the occurrence of the metric determinant ensures coordinate-independence. Notice how these propagators are analogous to the retarded and advanced Green's function for Electrodynamics, for example. Notice also that in this way of writing we can express their support properties as

$$\operatorname{supp} E^\pm \subseteq \{(x, y) \in \mathscr{M} \times \mathscr{M}; x \in J^\pm(y)\}. \tag{2.3.6}$$

We are particularly interested in the Pauli–Jordan commutator function (or simply commutator function) $E$, defined through

$$E = E^+ - E^- \,. \tag{2.3.7}$$

Notice that the support properties of the advanced and retarded propagators imply that, as a distributional kernel,

$$\operatorname{supp} E \subseteq \{(x, y) \in \mathscr{M} \times \mathscr{M}; x \in J^+(y) \cup J^-(y)\}. \tag{2.3.8}$$

In Minkowski spacetime, translation symmetry implies $E(x, y) = E(x - y)$ and the commutator function can be written explicitly as (Bogoliubov and Shirkov [1982], p. 334)

$$E(x) = -\frac{1}{2\pi} \operatorname{sign}(x^0) \delta(-x^\mu x_\mu) + \frac{\operatorname{sign}(x^0) \Theta(-x^\mu x_\mu) J_1(m\sqrt{-x^\mu x_\mu})}{4\pi \sqrt{-x^\mu x_\mu}}, \tag{2.3.9}$$

where sign is the sign function, $\Theta$ is the Heaviside step function, $J_1$ is the Bessel function and $x^\mu x_\mu$ stands for the invariant interval. Notice that Bogoliubov and Shirkov ([1982], Sec. 18.1) define the Green's functions for the Klein–Gordon equation following a convention that differs from ours by a sign, leading to a sign difference between Eq. (2.3.9) and their expression.

One should note from Eq. (2.3.9) that the commutator function is a distribution, not a smooth function. For Minkowski spacetime, we can see it is singular on the light cone.

## Construction of the Algebra

With these preliminaries, we are ready to start defining the actual algebra we are interested in.

Firstly, we need to choose what are the observables we will consider. Within field theory, the field variables $\phi(x)$ are our most natural choice. However, some care must be taken: we know from QFT in flat spacetime that expressions such as $\phi(x)\phi(x)$ are ill-behaved and products of fields at the same spacetime point lead to infinities. This happens because mathematically quantum fields are not functions of spacetime, but operator-valued distributions. As such, we should actually smear them with test functions $f \in \mathscr{C}_0^\infty(\mathscr{M}, \mathbb{C})$. Hence, the fundamental variables we shall consider for now are

$$\phi(f) = \int \phi(x) f(x) \sqrt{-g} \, \mathrm{d}^4 x, \tag{2.3.10}$$

for $f \in \mathscr{C}_0^\infty(\mathscr{M}, \mathbb{C})$. We are not able to express quantities such as $\phi(x)\phi(x)$ yet, since working with distributions enforces us to work with linear expressions. However, we can later enlarge the algebra of observables to comprise such quantities and include, for example, the stress-energy-momentum tensor (see Section 2.B). We are working with complex test functions because we desire to get to an algebra over the complex numbers, as is required for quantum theory.

We start by defining our algebra $\mathscr{A} \equiv \mathscr{A}(\mathscr{M}, g_{ab})$ by having it be the $*$-algebra obtained by formally considering the identity together with all possible linear combinations, products, and involutions of objects of the form $\phi(f)$. Rigorously, this means we are considering the free $*$-algebra[*] generated by $\mathbb{1}$ and the elements $\phi(f)$ with $f \in \mathscr{C}_0^\infty(\mathscr{M}, \mathbb{C})$.

This procedure gives us an algebra, but it still does not reflect the physics we are interested in. For example, let $f, g \in \mathscr{C}_0^\infty(\mathscr{M}, \mathbb{C})$ and $c \in \mathbb{C}$. Then we know there are elements $\phi(f)$, $\phi(g)$, and $\phi(f + cg)$ in the algebra (we defined them to exist), but they are not related in any way. However, to match the properties we expect of smeared fields, we would like these quantities to respect

$$\phi(f + cg) = \phi(f) + c\phi(g). \tag{2.3.11}$$

---

[*] See, $e.g.$, the book by Geroch ([1985]) for more on free constructions.

We shall then enforce this by hand by taking appropriate quotients of the algebra. We are now left with an algebra that has this convenient property. This is the general procedure we shall follow to impose the remaining conditions we would like. We essentially force the structure to respect the desired properties by taking quotients at our convenience.

Next we would like to impose that the field satisfies the Klein–Gordon equation in the distributional sense. Using the known properties of distributional derivatives (see Strohmaier 2009), we see that this is expressed as

$$\phi\big((\nabla_a \nabla^a - m^2)f\big) = 0 \tag{2.3.12}$$

for all $f \in \mathscr{C}_0^\infty(\mathcal{M}, \mathbb{C})$.

Note that we could be quantizing a complex scalar field. Since we are interested in a real field, we impose that $\phi(x)$ is real. By looking at our expression for what is a smeared field, Eq. (2.3.10) on the preceding page, we see this translates to requiring

$$\phi(f)^* = \phi(\bar{f}), \tag{2.3.13}$$

where $\bar{f}$ is the complex conjugate of $f$.

At last, what we wrote so far could refer to a classical theory. All of the properties we gave would also be applicable to distributional solutions of the classical Klein–Gordon equation. Let us then enforce the theory is a quantum theory by imposing commutation relations. This happens by means of the commutator function,

$$[\phi(f_1), \phi(f_2)] = iE(f_1, f_2)\mathbb{1}, \tag{2.3.14}$$

where

$$E(f_1, f_2) = \int E(x, y)f_1(x)f_2(y)\sqrt{-g(x)}\, \mathrm{d}^4 x \sqrt{-g(y)}\, \mathrm{d}^4 y. \tag{2.3.15}$$

As commented by Hollands and Wald (2015) and discussed at length by Wald (1994), we could also choose to formulate the theory with a "symplectic smearing" based on test functions $f \in \mathscr{C}_0^\infty(\Sigma, \mathbb{C})$, with $\Sigma$ being a Cauchy surface. In this case, one sees Eq. (2.3.14) recovers the usual expression for the CCR. One should notice that $E(x, y)$ vanishes if $x$ and $y$ are spacelike related due to the support properties of the propagators, and hence Eq. (2.3.14) implements Einstein causality: operators on spacelike related regions commute.

It is also interesting to notice that if we introduce coordinates[*] $(t, \vec{x})$ on the spacetime manifold $\mathcal{M} = \mathbb{R} \times \Sigma$, then one can show that (Dimock 1980; Hack 2016, Theorem 2.2)

$$\partial_t E(t, \vec{x}; t', \vec{x}')\Big|_{t=t'} = -\delta^{(3)}(\vec{x}, \vec{x}') \quad \text{and} \quad E(t, \vec{x}; t', \vec{x}')\Big|_{t=t'} = 0, \tag{2.3.16}$$

the second equation being merely a statement that the commutator function vanishes on spacelike related events. Notice this means Eq. (2.3.14) implies

$$[\phi(t, \vec{x}), \phi(t, \vec{x}')] = 0, \tag{2.3.17}$$

$$[\partial_t \phi(t, \vec{x}), \partial_t \phi(t, \vec{x}')] = 0, \tag{2.3.18}$$

$$[\phi(t, \vec{x}), \partial_t \phi(t, \vec{x}')] = i\delta^{(3)}(\vec{x}, \vec{x}')\mathbb{1}, \tag{2.3.19}$$

---

[*]As mentioned in Appendix A, we write $\vec{x}$ for spatial points regardless of whether they actually have a vector space structure.

which are the usual equal-time commutation relations.

Notice Eq. (2.3.14) on the previous page also works as an *a posteriori* justification for our statement that the fields are actually operator-valued distributions. Since the commutator function is a distribution (as made evident by Eq. (2.3.9) on page 22), either the fields are distributional as well or we get to the contradictory statement that the commutator of two smooth quantities is somehow singular.

## Vacua

The algebra we have just constructed allows us to express all states in terms of correlation functions. Indeed, any element of the algebra is a linear combination of elements of the form $\phi(f_1)\cdots\phi(f_n)$ for some functions $f_i \in \mathscr{C}_0^\infty(\mathscr{M}, \mathbb{C})$. Hence, we can completely specify a state $\omega$ by providing the $n$-point functions

$$W_n(f_1, \dots, f_n) \equiv \omega(\phi(f_1)\cdots\phi(f_n)). \qquad (2.3.20)$$

In Lorentzian signature, these are sometimes referred to as "Wightman functions" (Streater and Wightman 2000).

As we have previously mentioned, the notions of "vacuum" and of "particle" become subtle within QFTCS. At this stage, it is interesting for us to give a well-defined meaning for these words so that we can relate the theory in curved spacetimes to the conventional formulations in flat spacetime. Which state deserves to be called "vacuum"? Is it even unique?

Within QFT in flat spacetime, the vacuum is often understood as the unique Poincaré invariant state. It can also be characterized by the fact it is the Hamiltonian's ground state. Nevertheless, both of these properties are problematic in curved spacetime. Poincaré symmetry is a property of flat spacetime, and hence there is no reason to expect it to be relevant, for example, in Schwarzschild spacetime. Similarly, "the Hamiltonian's ground state" is meaningless unless one has a preferred congruence of observers defining which Hamiltonian we mean. Hence, these properties are not convenient for us to use in a general spacetime. Instead, we need to focus on more general properties of the Minkowski vacuum which we can transport to other contexts.

Perhaps the most important feature of the Minkowski vacuum is that it allows us to employ Wick's theorem, which is at the basis of essentially any calculation performed in QFT in flat spacetime. Due to this, we define a state to be Gaussian (also known as "quasifree") if its correlation functions are such that, for any positive integer $n$ and any choice of functions $f_k \in \mathscr{C}_0^\infty(\mathscr{M}, \mathbb{C})$,

$$W_{2n-1}(f_1, \dots, f_{2n-1}) = 0, \qquad (2.3.21)$$

and

$$W_{2n}(f_1, \dots, f_{2n}) = \sum_{\text{pairings}} W_2(f_{i_1}, f_{i_2})\cdots W_2(f_{i_{2n-1}}, f_{i_{2n}}), \qquad (2.3.22)$$

which is just the usual notion we have of Wick's theorem. This is the definition given by Khavkine and Moretti (2015, Definition 5.2.22), but alternatively one can do as Hollands and Wald (2015) and provide a definition in terms of connected correlation functions, which we have not defined.

A convenient property of Gaussian states is that their GNS representations are always Fock spaces, with the Gaussian state being annihilated by all annihilation operators (Hollands and Wald 2015, Sec. 2; Khavkine and Moretti 2015, Theorem 5.2.24). However, these representations may or may not be irreducible. As one knows from group theory, it is always more convenient to work with irreducible representations for the sake of simplicity. It turns out that a state on a ∗-algebra leads to an irreducible GNS representation if, and only if, it is pure[*].

The Minkowski vacuum is pure and Gaussian. These properties seem to be (and are) fairly general and applicable to curved spacetimes, and hence they provide us with a convenient notion of what a vacuum is. A vacuum is a pure Gaussian state. Notice that, by construction of our definition, one can always find a Fock representation in which the vacuum is annihilated by all annihilation operators, and hence there is always some sense in which the vacuum is an absence of particles.

### Thermal States

While the vacuum is perhaps the most important state in QFT in flat spacetime, another important class of states are those in thermal equilibrium. In the algebraic approach, the notion of thermal equilibrium is given by means of the Kubo–Martin–Schwinger (KMS) condition (Kubo 1957; Martin and Schwinger 1959). Our discussion is inspired by the one given by Raszeja (2020, Sec. 2.3).

Suppose we have a system with finitely many degrees of freedom in contact with a thermal reservoir at inverse temperature $\beta = \frac{1}{T}$ (we take $k_B = 1$). Assuming the system to be in thermal equilibrium, its state is described by the density matrix (Kardar 2007, Eq. (6.86))

$$\rho = \frac{e^{-\beta H}}{Z}, \tag{2.3.23}$$

where $Z = \mathrm{tr}\left[e^{-\beta H}\right]$ is the partition function for the system and $H$ its Hamiltonian.

We take a finite system to provide us with some motivation and intuition while avoiding the difficulties that occur in larger systems (for a discussion, see Witten 2022). In this simplified case we can work with density matrices, but in field theory we will not always have a privileged Hilbert space. Therefore, we would like to obtain a purely algebraic condition that expresses thermal equilibrium.

Thermal equilibrium is a notion closely related to the time evolution of a system. For example, it can be characterized by stating that the entropy production rate of a system—which is a notion of time variation of entropy—vanishes (Tomé and Oliveira 2015). Hence, it is natural that to discuss equilibrium we need to consider some sort of time evolution.

In the context of finite quantum systems, this evolution is ruled by the Heisenberg equation of motion and, for a time-independent Hamiltonian, it is given by (Weinberg 2015, Eq. (3.6.6))

$$A(t) = e^{itH} A(0) e^{-itH}, \tag{2.3.24}$$

where $A(t)$ is some observable at time $t$ and we take $\hbar = 1$. Notice that we can also see this time evolution as a one-parameter group of automorphisms acting on the algebra $\mathcal{A}$. More specifically, we can write

$$\theta_t(A) = e^{itH} A e^{-itH} \tag{2.3.25}$$

to denote the time evolution of $A$ by an amount $t$. While unusual in ordinary QM, this notation will serve us well in the following.

Notice that the exponentials that occur on the definition of $\theta_t$ are similar to the exponential that occurs in the expression for a density matrix in thermal equilibrium, apart from the fact that $\rho$ involves a real exponential and $\theta_t$ involves imaginary exponentials. It is tempting, however, to consider an analytic continuation of $\theta_t$ to the complex $t$ plane. If this is possible, then notice that, given $A, B \in \mathcal{A}$, we have

$$\omega_\rho(BA) = \text{tr}[BA\rho], \tag{2.3.26a}$$

$$= \frac{1}{Z}\text{tr}[BAe^{-\beta H}], \tag{2.3.26b}$$

$$= \frac{1}{Z}\text{tr}[Ae^{-\beta H}B], \tag{2.3.26c}$$

$$= \frac{1}{Z}\text{tr}[Ae^{-\beta H}Be^{+\beta H}e^{-\beta H}], \tag{2.3.26d}$$

$$= \frac{1}{Z}\text{tr}[A\theta_{i\beta}(B)e^{-\beta H}], \tag{2.3.26e}$$

$$= \text{tr}[A\theta_{i\beta}(B)\rho], \tag{2.3.26f}$$

$$= \omega_\rho(A\theta_{i\beta}(B)). \tag{2.3.26g}$$

---

[*]In fact, this is the result for $C^*$-algebras (see Barata et al. 2021, Sec. 7). For ∗-algebras the result is a bit weaker (see Khavkine and Moretti 2015, Proposition 5.1.15).

In the previous expressions, $\omega_\rho$ is the state defined through $\omega_\rho(A) = \text{tr}[A\rho]$.

Eq. (2.3.26) leads one to the general expression

$$\omega(BA) = \omega(A\theta_{i\beta}(B)), \tag{2.3.27}$$

where $\theta_t$ can be any one-parameter group of automorphisms in the algebra, as long as it admits a suitable analytic extension. This property, known as the KMS condition, was used by Haag, Hugenholtz, and Winnink (1967) as a definition for equilibrium states in an algebraic setting.

It should be pointed out that while Eq. (2.3.27) is sufficient for $C^*$-algebras, general $*$-algebras are more subtle and require other additional conditions for the expectation values of the form $\omega(A_1 \cdots A_n)$ with $n > 2$ (see Hollands and Wald 2015). Nevertheless, our main focus is on Gaussian states, which are completely determined by the two-point function. Hence, for our purposes, Eq. (2.3.27) is enough. We say that a state satisfying Eq. (2.3.27) is a KMS state for the one-parameter group of automorphisms $\theta_t$ at inverse temperature $\beta$.

There is an important difference between how we defined vacua and how we defined KMS states. Notice that a vacuum for us is a state that is Gaussian and pure. These properties are related exclusively to the state and the algebra of observables, and are independent of any other choice or input from the physicist. Hence, given a state on an algebra of observables, one can immediately say whether it is a vacuum. KMS states, on the other hand, depend on the choice of a group of automorphisms. In other words, it depends on a choice of time evolution. Two different choices of time evolution might disagree on whether a given state is a KMS state. Hence, thermal equilibrium depends on something in addition to the state itself. In QFTCS this is relevant because different observers will have different definitions of time evolution. Hence, a state might be a KMS state for an observer, but not for another. Furthermore, two different observers might agree that a state is a KMS state, but disagree on what is its temperature. This is well illustrated by the Unruh effect, which shows that different observers can perceive the same vacuum state as having different temperatures depending on their acceleration.

## 2.4 Four Paths to the Unruh Effect

Once the basis of the theory has been outlined, let us consider the case of acceleration-induced thermality in Minkowski spacetime. We shall do this discussion in four different ways, since each of them can play different roles in explaining the physics going on. Furthermore, since experimental probing of the Unruh effect has only recently begun (the first claim of a direct observation has been made by Lynch et al. 2021), the existence of many paths to the same conclusion helps us to understand why it must be true.

Before we get to the actual calculations, let us explain what we are about to do. The Fulling–Davies–Unruh effect (Davies 1975; Fulling 1973; Unruh 1976), often called simply the Unruh effect, is the result that "for a [...] quantum field in its vacuum state in Minkowski spacetime, an observer with uniform acceleration $a$ will feel that he is bathed by a thermal distribution of quanta of the field at temperature $T$ given by $k_B T = \frac{\hbar a}{2\pi c}$" (Unruh and Wald 1984). It consists of a prediction made with QFTCS methods in flat spacetime, and challenges one's usual understanding of the meaning of "particles", since different observers are shown to have different particle interpretations of the same physical state. More details will be given in the following discussion, and even more can be found in the review by Crispino, Higuchi, and Matsas (2008).

In all of the following approaches, we consider Minkowski spacetime, $\mathcal{M} = (\mathbb{R}^4, \eta_{ab})$. The line element is given in Cartesian coordinates by

$$ds^2 = -dt^2 + dx^2 + dy^2 + dz^2. \tag{2.4.1}$$

When analyzing the Unruh effect, our interest will not be in the entire Minkowski spacetime, but rather on a region known as the right Rindler wedge (Rindler 1966). We will denote it by

$$R = \left\{ (t, x, y, z) \in \mathbb{R}^4; x > |t| \right\}. \tag{2.4.2}$$

This region can be understood as a globally hyperbolic spacetime in its own right, and it is particularly useful to mimic some properties found in black hole spacetimes. Many properties of the Rindler spacetime are reviewed, for example, in the books by Ellis and Williams (2000, Sec. 4.3) and Rindler (2006, Sec. 12. 4). Since the literature on this spacetime is vast, we shall state some of its properties without proof.

It is convenient for our purposes to cover the Rindler spacetime using the so-called Rindler coordinates (Rindler 1966, 2006, Sec. 12.4). We define them through

$$t = r \sinh a\eta \quad \text{and} \quad x = r \cosh a\eta, \tag{2.4.3}$$

for constant $a > 0$. They are illustrated on Fig. 2.3 on the next page. This definition leads to the line element

$$ds^2 = -a^2 r^2 \, d\eta^2 + dr^2 + dy^2 + dz^2. \tag{2.4.4}$$

Radar coordinates (Minguzzi 2005), which use $r = a^{-1} e^{a\xi}$, are also common in the literature.

Notice that surfaces of constant $\eta$ are Cauchy surfaces. Furthermore, the spacetime is static with Killing field $\left( \frac{\partial}{\partial \eta} \right)^a$. Since a four-dimensional spacetime has at most ten Killing fields and this is a section of Minkowski spacetime—which has its Killing fields as Poincaré transformations—we know that this Killing field is somehow related to Poincaré transformations. It turns out it is simply the generator of boosts along the $x$ direction. It also happens to be proportional to the four-velocities of observers with constant proper acceleration. In fact, the parameter $a$ introduced earlier is the proper acceleration of the observers moving along the locus with $\eta_{ab} \left( \frac{\partial}{\partial \eta} \right)^a \left( \frac{\partial}{\partial \eta} \right)^b = -1$.

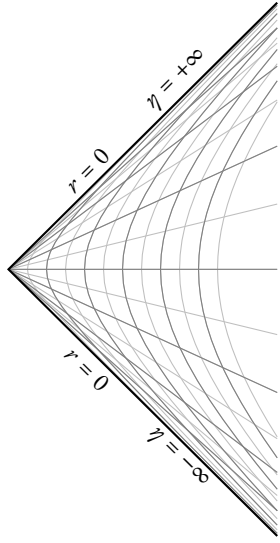

Figure 2.3: Depiction of how the Rindler coordinates given on Eq. (2.4.3) on the previous page cover the right Rindler wedge. The hyperbolae are curves of constant $r$, while the straight lines are curves of constant $\eta$.

The orbits induced by Lorentz boosts on Minkowski spacetime are illustrated on Fig. 2.4 on the facing page.

### Algebraic Approach

Our first derivation follows the algebraic spirit we have been establishing so far. We follow the discussion given by Hollands and Wald (2015, Sec. 2.2.a).

To obtain the desired QFT, we can simply consider the algebra of observables $\mathcal{A}(\mathcal{M})$, but now restrict it to only (linear combinations, products, and involutions of) observables of the form $\phi(f)$ with supp $f \subseteq R$ (a condition that implies supp $f \cap \partial R = \emptyset$). Through this restriction, we get to the subalgebra $\mathcal{A}(R) \subseteq \mathcal{A}(\mathcal{M})$.

If $\omega$ is a state on $\mathcal{A}(\mathcal{M})$, it is also a state on $\mathcal{A}(R)$—after all, we are simply considering less observables. Hence, the Minkowski vacuum defines a state on $\mathcal{A}(R)$. We desire to characterize it.

For simplicity, let us assume a massless field. In this case, we know that the two-point function is given by

$$W_2(x_1, x_2) = \underset{\varepsilon \to 0^+}{\text{w-lim}} \frac{1}{4\pi^2(x_1^\mu - x_2^\mu - i\varepsilon T^\mu)(x_\mu^1 - x_\mu^2 - i\varepsilon T_\mu)}, \tag{2.4.5}$$

which can be derived using the expressions given by Bogoliubov and Shirkov (1982, App. V.2). w-lim is the weak limit discussed on Appendix A and $T^a$ stands for any future-directed timelike vector.

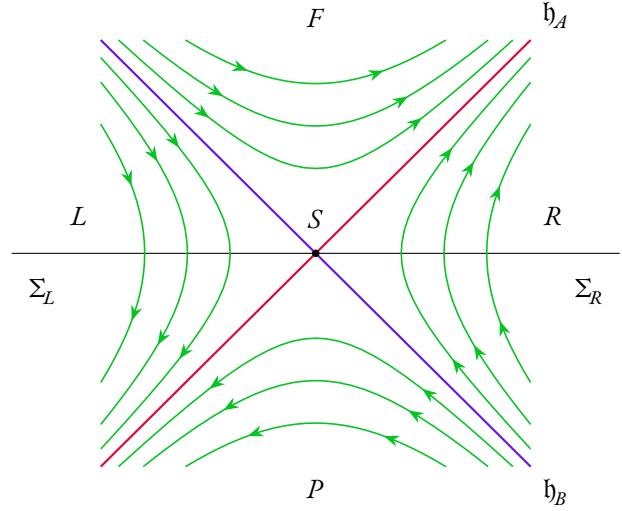

Figure 2.4: Orbits induced by Lorentz boosts on $1 + 1$-dimensional Minkowski spacetime. Notice that on the left and right Rindler wedges ($L$ and $R$, respectively) the orbits are timelike, while they are spacelike on the remaining wedges. On the null hypersurfaces $\mathfrak{h}_A = \{(t, x, y, z) \in \mathbb{R}^4; t = x\}$ and $\mathfrak{h}_B = \{(t, x, y, z) \in \mathbb{R}^4; t = -x\}$ that separate the wedges the orbits are also null. The spacelike submanifold $S = \mathfrak{h}_A \cap \mathfrak{h}_B$ is comprised of fixed points of the isometry orbits. $\Sigma_R$ (resp. $\Sigma_L$) is a Cauchy surface for the right (left) Rindler wedge.

Taking it to be $\left(\frac{\partial}{\partial t}\right)^a$, we find that, in Cartesian coordinates,

$$W_2(x_1, x_2) = \underset{\varepsilon \to 0^+}{\text{w-lim}} \frac{1}{4\pi^2 \left[(x_1 - x_2)^2 + (y_1 - y_2)^2 + (z_1 - z_2)^2 - (t_1 - t_2 - i\varepsilon)^2\right]}. \tag{2.4.6}$$

These expressions hold for the entire Minkowski spacetime, not only on the right Rindler wedge. In particular, notice they mean there are correlations between the left and right Rindler wedges, where the left wedge is defined as $L = \{(t, x, y, z) \in \mathbb{R}^4; x < -|t|\}$. When restricting the Minkowski vacuum to the right Rindler wedge, we drop these correlations, meaning the state can no longer be pure[*].

It turns out the state is not only mixed, but also a KMS state at inverse temperature $\beta = \frac{2\pi}{a}$ for the isometry $\eta \mapsto \eta + \tau$, where $\tau$ is some arbitrary parameter. In other words, it is thermal with respect to the time-evolution prescribed by accelerated observers, with temperature proportional to the acceleration.

To see this in the algebraic approach, we must show that the state satisfies the KMS condition. The Minkowski vacuum is Gaussian, and hence we essentially want to show that (for a more careful discussion, see Hollands and Wald 2015)

$$\omega(\phi(f)\theta_{\tau+i\beta}(\phi(g))) = \omega(\theta_\tau(\phi(f))\phi(g)), \tag{2.4.7}$$

where $\theta$ denotes the isometry $\eta \mapsto \eta + \tau$. To do so, one can use Eqs. (2.4.3) and (2.4.6) on page 27 and on this page to show that

$$W_2(x_1, \theta_{\tau+i\beta}(x_2)) = W_2(\theta_\tau(x_2), x_1). \tag{2.4.8}$$

[*]This follows from the Reeh–Schlieder theorem. See, for example, the discussion given by Clifton and Halverson (2001, Sec. 3).

This can be shown by direct calculation.

As noted by Sewell (1982), this is a special case of a more general theorem by Bisognano and Wichmann (1975, 1976). A discussion on this more technical result is given by Haag (1996, Sec. V.4.1). Since it holds in the more general context of axiomatic QFT, it also applies to interacting field theories.

### Fock Representation Approach

A second approach to deriving the Unruh effect employs the ideas we briefly discuss in Section 2.A on page 47. Namely, it exploits the fact that there is a natural choice of Fock representation of the CCR on a stationary spacetime. This also happens to be the most common derivation of the Unruh effect, and hence I shall only mention it briefly. More details can be found, for example, in the discussions by Crispino, Higuchi, and Matsas (2008), Mukhanov and Winitzki (2007, Chap. 8), and Wald (1994, Chap. 5), among others. The general technique is based on the papers by Hawking (1975) and Wald (1975), which were originally concerned not with the Unruh effect, but rather with the Hawking effect, which we shall briefly discuss at the end of this section.

The idea is the following. A stationary spacetime admits a natural notion of vacuum, as discussed in Section 2.A and references therein. The GNS representation of this vacuum gives rise to a Fock space. Hence, in spacetimes where two stationary Killing fields are available, we can try to relate these two Fock representations. This is the case for Minkowski spacetime, since we can choose to work in the Fock space $\mathscr{F}_M$ induced by the Minkowski vacuum or in the Fock space $\mathscr{F}_L \otimes \mathscr{F}_R$ induced by the Rindler vacuum (see Section 2.A) on each of the Rindler wedges.

In this situation it is possible to relate the creation and annihilation operators in the different Fock spaces. This is known as a Bogoliubov transformation (Bogoliubov 1947). Using this Bogoliubov transformation it is possible to formally obtain an $S$-matrix $S \colon \mathscr{F}_M \to \mathscr{F}_L \otimes \mathscr{F}_R$, which translates one representation into the other. After expressing the Minkowski vacuum in $\mathscr{F}_L \otimes \mathscr{F}_R$ one can trace out the modes in $\mathscr{F}_L$. This leaves a state in $\mathscr{F}_R$ formally described as a thermal density matrix at temperature $T = \frac{a}{2\pi}$.

This derivation is particularly interesting because it provides a first glance on how the notion of particle depends on the particular choice of representation. In the presence of more than one admissible representation, there is more than one admissible notion of particle and they will not coincide in general. The same state might have none or infinitely many particles depending on the chosen representation.

The approach sketched above was introduced by Wald (1975) when working on the Hawking effect, before the Unruh effect was actually discovered. The earlier work by Hawking (1975) was able to conclude thermality by simply computing the expectation value of the number of particles in the new representation without actually computing the full state.

### Euclidean Path Integral Approach

Given that much of the remaining chapters of this thesis will rely on Euclidean path integral techniques, it is instructive for us to also discuss the Unruh effect in this formulation. Our calculation mostly follows the original one due to Unruh and Weiss (1984), but Crispino, Higuchi, and Matsas (2008, Sec. II.I) also present a summarized version. We shall postpone a more technical discussion of Euclidean path integral techniques to Section 2.5.

Using Euclidean path integrals, we intend to show that the equality[*]

$$\langle 0_M | \mathcal{T} \phi(x_1) \cdots \phi(x_n) | 0_M \rangle = \frac{\operatorname{Tr}\left[e^{-\beta H_R} \mathcal{T} \phi(x_1) \cdots \phi(x_n)\right]}{\operatorname{Tr}\left[e^{-\beta H_R}\right]} \tag{2.4.9}$$

holds for all events $x_1, \ldots, x_n \in R$ when $\beta = \frac{2\pi}{a}$. In the previous expression, $H_R$ is the Rindler Hamiltonian, i.e., the generator of translations with respect to proper time for an accelerated observer with acceleration $a$. This is a different way of saying that $H_R$ generates translations with respect to $\eta$, or alternatively that it is $a$ times the boost generator.

Let us begin by writing $H_R$ down explicitly. The general action for a scalar field in Minkowski spacetime is

$$S[\phi] = -\int \left[\frac{1}{2} \nabla_a \phi \nabla^a \phi + V(\phi)\right] \sqrt{-g} \, d^4 x, \tag{2.4.10}$$

where $g$ is the determinant of the metric—we don't write it as $\eta$ to avoid confusion with the Rindler coordinate $\eta$. We chose to write the action in this way because it allows us to simply see how it will occur in non-inertial coordinates. Notice also that this time we added a general interaction potential $V(\phi)$, and hence this derivation is not restricted to free or massless fields.

Using Eq. (2.4.4) on page 27, we can see that the action can be written in Rindler coordinates as

$$S[\phi] = \int_{r>0} \left[\frac{1}{2(ar)^2}\left(\frac{\partial \phi}{\partial \eta}\right)^2 - \frac{1}{2}\left(\frac{\partial \phi}{\partial r}\right)^2 - \frac{(\vec{\nabla}_\perp \phi)^2}{2} - V(\phi)\right] ar \, d\eta \, d^2 x_\perp, \tag{2.4.11}$$

where we are writing $\vec{x}_\perp = (y, z)$.

There is a difference between Eqs. (2.4.10) and (2.4.11) that should be pointed out. While Eq. (2.4.10) is integrating over modes over the entire Minkowski spacetime, Eq. (2.4.11) is written in a coordinate system that is only defined on the right Rindler wedge. Hence, in writing Eq. (2.4.11), we are already assuming we are only paying attention to what happens on the right Rindler wedge.

In Rindler coordinates, the momentum canonically conjugate to $\phi$ is given by

$$\pi = \frac{\partial S}{\partial\left(\partial_\eta \phi\right)} = \frac{\partial_\eta \phi}{ar} \equiv \frac{1}{ar} \frac{\partial \phi}{\partial \eta}. \tag{2.4.12}$$

Therefore, we get to the Hamiltonian

$$H_R = \int_{r>0} \left[\frac{\pi^2}{2} + \frac{1}{2}\left(\frac{\partial \phi}{\partial r}\right)^2 + \frac{(\vec{\nabla}_\perp \phi)^2}{2} + V(\phi)\right] ar \, dr \, d^2 x_\perp. \tag{2.4.13}$$

We then consider the partition function

$$Z_R(\beta) \equiv \operatorname{Tr}\left[e^{-\beta H_R}\right]. \tag{2.4.14}$$

As a path integral, it can be written as (Năstase 2020, Chap. 71)

$$Z_R(\beta) = \int_{\phi(0)=\phi(\beta)} \exp\left(-\int_{r>0}\int_0^\beta ar\left[\frac{\pi^2}{2} + \frac{1}{2}\left(\frac{\partial \phi}{\partial r}\right)^2 + \frac{(\vec{\nabla}_\perp \phi)^2}{2} + V(\phi)\right] - i\pi \frac{\partial \phi}{\partial \tau} \, d\tau \, dr \, d^2 x_\perp\right) \mathcal{D}\phi \, \mathcal{D}\pi, \tag{2.4.15}$$

---

[*]Notice that the time ordering operator $\mathcal{T}$ can be regarded as a coordinate-independent object. If $x \in J^+(y)$, this happens in all coordinate systems, and hence the action of $\mathcal{T}$ on $\phi(x)\phi(y)$ also does. If $x$ and $y$ are spacelike related, then $\phi(x)$ and $\phi(y)$ commute and their ordering is irrelevant.

where $\phi(0) = \phi(\beta)$ means the integral runs over field configurations with periodic boundary conditions in $\tau$ with period $\beta$.

The integral over $\pi$ is Gaussian. It can be solved by noticing that

$$\frac{ar\pi^2}{2} - i\pi\frac{\partial\phi}{\partial\tau} = \frac{ar}{2}\left(\pi - \frac{i}{ar}\frac{\partial\phi}{\partial\tau}\right)^2 + \frac{1}{2ar}\left(\frac{\partial\phi}{\partial\tau}\right)^2. \tag{2.4.16}$$

Therefore, up to a superfluous normalization factor, one has

$$Z_R(\beta) = \int_{\phi(0)=\phi(\beta)} \exp\left(-\int_{r>0}\int_0^\beta \frac{1}{2ar}\left(\frac{\partial\phi}{\partial\tau}\right)^2 + ar\left[\frac{1}{2}\left(\frac{\partial\phi}{\partial r}\right)^2 + \frac{(\vec{\nabla}_\perp\phi)^2}{2} + V(\phi)\right]\mathrm{d}\tau\,\mathrm{d}r\,\mathrm{d}^2x_\perp\right)\mathcal{D}\phi. \tag{2.4.17}$$

Notice this expression can be understood in terms of the Euclidean, finite-temperature version of Eq. (2.4.11) on the preceding page. Namely,

$$Z_R(\beta) = \int_{\phi(0)=\phi(\beta)} e^{-S_{RE}^\beta[\phi]}\,\mathcal{D}\phi, \tag{2.4.18}$$

where the subscripts "$RE$" stand for "Rindler" and "Euclidean". Notice that if we had chosen other coordinate systems—such as an inertial coordinate system—the Euclidean action could be different (see the discussion in Section 2.5).

We are free to perform the integral in the exponent of Eq. (2.4.17) in whichever way we see fit. In particular, we can perform a change of variables according to

$$t_E = r\sin a\tau \quad\text{and}\quad x_E = r\cos a\tau. \tag{2.4.19}$$

While these are inspired by our definition of Rindler coordinates, notice we are not changing to a new coordinate chart on the manifold. We are only making a change of variables in the integral. One can then show that

$$\int_{r>0}\int_0^\beta \frac{1}{2ar}\left(\frac{\partial\phi}{\partial\tau}\right)^2 + ar\left[\frac{1}{2}\left(\frac{\partial\phi}{\partial r}\right)^2 + \frac{(\vec{\nabla}_\perp\phi)^2}{2} + V(\phi)\right]\mathrm{d}\tau\,\mathrm{d}r\,\mathrm{d}^2x_\perp$$
$$= \int_A \frac{1}{2}\left(\frac{\partial\phi}{\partial t_E}\right)^2 + \frac{1}{2}\left(\frac{\partial\phi}{\partial x_E}\right)^2 + \frac{1}{2}(\vec{\nabla}_\perp\phi)^2 + V(\phi)\,\mathrm{d}t_E\,\mathrm{d}x_E\,\mathrm{d}^2x_\perp, \tag{2.4.20}$$

where the integration region $A$ is illustrated on Fig. 2.5 on the facing page. Notice that Eq. (2.4.19) can only be single-valued if $\beta a \leq 2\pi$.

Consider now the case $\beta = \frac{2\pi}{a}$. We can then write

$$Z_R\left(\frac{2\pi}{a}\right) = \int \exp\left(-\int \frac{1}{2}\left(\vec{\nabla}_4\phi\right)^2 + V(\phi)\,\mathrm{d}^4x_E\right)\mathcal{D}\phi, \tag{2.4.21}$$

where we dropped the condition $\phi(0) = \phi(\beta)$, because it is now automatically implemented by the new variables. Notice, however, that the RHS of Eq. (2.4.21) is merely the generating functional at zero source for the theory in inertial coordinates. Hence,

$$Z_R(\beta) = \int e^{-S_{IE}[\phi]}\,\mathcal{D}\phi, \tag{2.4.22}$$

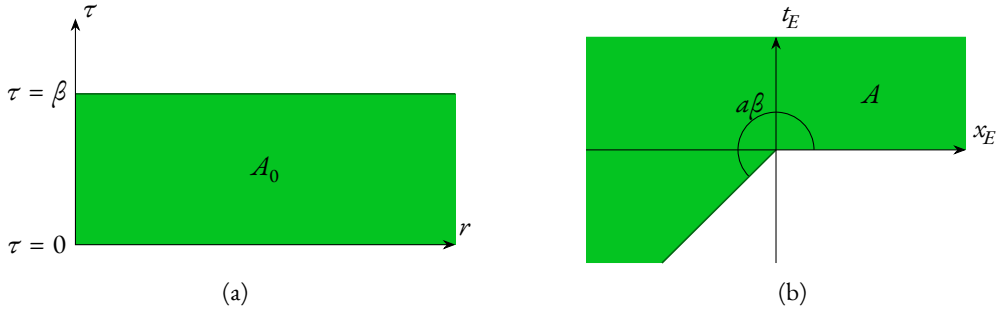

Figure 2.5: Integration region before ($A_0$) and after ($A$) the coordinate transformation done on Eq. (2.4.19) on the preceding page. Based on Figure 1 of the paper by Unruh and Weiss (1984).

where "$iE$" stands for "inertial" and "Euclidean".

These ideas can be generalized in a straightforward manner to a generating functional in the presence of a source, $Z[J]$. In this case, functional derivatives with respect to the source allow us to obtain the $n$-point correlation functions. We then find that

$$\langle 0_M | \phi(x_E^1) \cdots \phi(x_E^n) | 0_M \rangle = \frac{\text{Tr}\big[e^{-\beta H_R} \phi(x_E^1) \cdots \phi(x_E^n)\big]}{\text{Tr}\big[e^{-\beta H_R}\big]}. \tag{2.4.23}$$

Notice this is not Eq. (2.4.9) on page 31. Eq. (2.4.23) is an equality among correlation functions on a spacetime of Euclidean signature. However, the RHS of Eq. (2.4.9) on page 31 can be obtained from the RHS of Eq. (2.4.23) by means of the analytic continuation $\tau = i\eta$—this is how we went from Eq. (2.4.14) on page 31 to Eq. (2.4.15) on page 31. Similarly, the LHS of Eq. (2.4.9) on page 31 can be obtained from the LHS of Eq. (2.4.23) under $t_E = it$ (see Section 2.5). Nevertheless, as one might notice from Eqs. (2.4.3) and (2.4.19) on page 27 and on the facing page, it turns out that $\tau = i\eta$ and $t_E = it$ are actually the same analytic continuation. Hence, Eq. (2.4.23) implies Eq. (2.4.9) on page 31, concluding our proof.

### Particle Detector Approach

Another interesting approach for deriving the Unruh effect is to employ a particle detector. This allows us to obtain a different point of view on the effect, and will also provide us with useful experience for when we start working more closely with particle detectors in Chapter 4.

We shall consider an Unruh–DeWitt detector (DeWitt 1979; Unruh 1976). This is a two-level detector that can be excited or de-excited through interactions with the quantum field, similar to how an ammonia molecule can flip states upon interaction with an external electric field (see Feynman, Leighton, and Sands 2011, Chap. 9). Intuitively, the detector will flip from the ground state to the excited state when it absorbs a "particle", and will decay when it emits a "particle". Pictorially, we are considering a "particle in a box" that can interact with the field. For example, we are carrying around an electron in a box and use it to measure properties of the electromagnetic field. Further details are given by Unruh and Wald (1984) and Wald (1994, Sec. 3.3). Our discussion follows the review given by Burbano, Perche, and Torres (2021) and also draws from the seminal works by DeWitt (1979) and Unruh (1976).

We have already discussed at length how to describe a quantum field. For the detector, we shall consider a two-level quantum system—*i.e.*, a qubit— with free Hamiltonian

$$H_\Omega = \frac{\Omega}{2}\sigma^z, \tag{2.4.24}$$

where $\sigma^z$ is the Pauli matrix and $\Omega$ is a constant with dimension of energy. It represents the energy gap between the ground and excited states of the detector. Since we want the excited state to have an energy larger than that of the ground state, we assume $\Omega > 0$. Notice that $H_\Omega$ generates translations with respect to the detector's proper time.

We shall also introduce an interaction between the quantum field and the detector. We write, in the interaction picture,

$$H_{\mathrm{int}} = \varepsilon\sigma^x(\tau) \otimes \phi(z(\tau)), \tag{2.4.25}$$

where $\varepsilon$ is a coupling constant, $\sigma^x(\tau)$ is the Pauli matrix (which evolves in the interaction picture), $\phi$ is the quantum field, and $z(\tau)$ denotes the detector's worldline. Hence, we are prescribing a pointlike interaction between detector and field along the detector's worldline. This interaction could be more complex to allow us to turn the detector on and off, or to allow for the detector to have spatial degrees of freedom. We shall illustrate these more elaborate choices in Chapter 4, but this simple model is sufficient for our present purposes.

The quantum field also evolves with some Hamiltonian $H_\phi$. This Hamiltonian evolves the field along inertial time, and hence we need to introduce a correction factor to account for the evolution with respect to the detector's proper time. This is merely a factor of $\frac{\mathrm{d}t}{\mathrm{d}\tau}$, since

$$i\frac{\mathrm{d}}{\mathrm{d}\tau} = i\frac{\mathrm{d}t}{\mathrm{d}\tau}\frac{\mathrm{d}}{\mathrm{d}t} = \frac{\mathrm{d}t}{\mathrm{d}\tau}H_\phi. \tag{2.4.26}$$

At the end of the day, we have the Hamiltonian

$$H = \frac{\mathrm{d}t}{\mathrm{d}\tau}H_\phi + \frac{\Omega}{2}\sigma^z + \varepsilon\sigma^x(\tau) \otimes \phi(z(\tau)). \tag{2.4.27}$$

Let us then compute the excitation probability for the detector. Consider the system's initial state is $|g, 0\rangle = |g\rangle \otimes |0\rangle$, where $|g\rangle$ denotes the detector's ground state and $|0\rangle$ denotes the vacuum. We are mainly interested in the Minkowski vacuum, but most of our calculation also works for other states and spacetimes. We want to compute the probability that the system undergoes a transition to some state $|e, \phi\rangle = |e\rangle \otimes |\phi\rangle$, where $|e\rangle$ is the detector's excited state and $|\phi\rangle$ is an arbitrary field state. Hence, we are first trying to compute the amplitude

$$A_{g\to e}(\phi) = \langle e, \phi|U_{\mathrm{int}}|g, 0\rangle, \tag{2.4.28}$$

where $U_{\mathrm{int}}$ is the time-evolution operator in the interaction picture.

To compute this expression, we begin by writing the time-evolution operator as a Dyson series (Weinberg 2015, Eq. (8.7.13))

$$U_{\mathrm{int}}(\tau', \tau) = \mathcal{T}\exp\left(-i\int_\tau^{\tau'} H_{\mathrm{int}}(\tau_1)\,\mathrm{d}\tau_1\right), \tag{2.4.29a}$$

$$= \mathbb{1} + \sum_{n=1}^{+\infty}\frac{(-i)^n}{n!}\int_\tau^{\tau'}\cdots\int_\tau^{\tau'}\mathcal{T}\big(H_{\mathrm{int}}(\tau_1)\cdots H_{\mathrm{int}}(\tau_n)\big)\,\mathrm{d}\tau_1\cdots\mathrm{d}\tau_n, \tag{2.4.29b}$$

where $\mathcal{T}$ is the time-ordering operator.

Using Eq. (2.4.25) on the preceding page on the Dyson series, we find that

$$U_{\text{int}}(\tau', \tau) = \mathbb{1} + \sum_{n=1}^{+\infty} \frac{(-i\varepsilon)^n}{n!} \int_\tau^{\tau'} \cdots \int_\tau^{\tau'} \mathcal{T}\big[\big(\sigma^x(\tau_1) \cdots \sigma^x(\tau_n)\big) \otimes \big(\phi(\tau_1) \cdots \phi(\tau_n)\big)\big] \, d\tau_1 \cdots d\tau_n, \quad (2.4.30)$$

where we have now adopted the simplified notation $\phi(\tau) \equiv \phi(z(\tau))$.

Let us then consider the amplitude we are interested in. When computing $\langle e, \phi | U_{\text{int}} | g, 0 \rangle$, we can immediately see the identity drops out, since the two states are orthogonal. Hence, we are left with

$$A_{g \to e}(\phi; \tau, \tau') = \sum_{n=1}^{+\infty} \frac{(-i\varepsilon)^n}{n!} \int_\tau^{\tau'} \cdots \int_\tau^{\tau'} \langle e, \phi | \mathcal{T}\big[\big(\sigma^x(\tau_1) \cdots \sigma^x(\tau_n)\big) \otimes \big(\phi(\tau_1) \cdots \phi(\tau_n)\big)\big] | g, 0 \rangle \, d\tau_1 \cdots d\tau_n. \tag{2.4.31}$$

Notice that

$$\langle e, \phi | \mathcal{T}\big[\big(\sigma^x(\tau_1) \cdots \sigma^x(\tau_n)\big) \otimes \big(\phi(\tau_1) \cdots \phi(\tau_n)\big)\big] | g, 0 \rangle = \langle e | \mathcal{T}\big(\sigma^x(\tau_1) \cdots \sigma^x(\tau_n)\big) | g \rangle \, \langle \phi | \mathcal{T}\big(\phi(\tau_1) \cdots \phi(\tau_n)\big) | 0 \rangle. \tag{2.4.32}$$

At this stage, we cannot simplify the $n$-point function, but we can proceed with our calculation for the detector factor.

Let us begin by noticing that we can write $\sigma^x$ in terms of ladder operators as

$$\sigma^x = \sigma^+ + \sigma^-, \tag{2.4.33}$$

where

$$\sigma^\pm = \frac{\sigma^x \pm i\sigma^y}{2}. \tag{2.4.34}$$

The advantage of noticing this is that

$$\sigma^+ |g\rangle = |e\rangle, \quad \sigma^+ |e\rangle = 0, \quad \sigma^- |g\rangle = 0, \quad \text{and} \quad \sigma^- |e\rangle = |g\rangle, \tag{2.4.35}$$

which imply

$$\sigma^x |g\rangle = |e\rangle \quad \text{and} \quad \sigma^x |e\rangle = |g\rangle. \tag{2.4.36}$$

We then consider the slightly more complicated case where the Pauli matrix is being evolved in time, since we are working in the interaction picture. We then have

$$\sigma^x(\tau) |g\rangle = \exp(i\tau H_\Omega)\sigma^x \exp(-i\tau H_\Omega) |g\rangle, \tag{2.4.37a}$$

$$= \exp(i\tau H_\Omega)\sigma^x \exp\left(\frac{+i\tau\Omega}{2}\right) |g\rangle, \tag{2.4.37b}$$

$$= \exp\left(\frac{+i\tau\Omega}{2}\right)\exp(i\tau H_\Omega)\sigma^x |g\rangle, \tag{2.4.37c}$$

$$= \exp\left(\frac{+i\tau\Omega}{2}\right)\exp(i\tau H_\Omega) |e\rangle, \tag{2.4.37d}$$

$$= \exp\left(\frac{+i\tau\Omega}{2}\right)\exp\left(\frac{+i\tau\Omega}{2}\right) |e\rangle, \tag{2.4.37e}$$

$$= \exp(+i\tau\Omega) |e\rangle. \tag{2.4.37f}$$

An analogous calculation leads to

$$\sigma^x(\tau) |e\rangle = \exp(-i\tau\Omega) |g\rangle. \tag{2.4.38}$$

Therefore, we find that

$$\langle e | \sigma^x(\tau_1) \cdots \sigma^x(\tau_n) | g \rangle = \begin{cases} e^{i\Omega(\tau_1 - \tau_2 + \tau_3 - \cdots + \tau_n)}, & \text{if } n \text{ is odd,} \\ 0, & \text{if } n \text{ is even.} \end{cases} \quad (2.4.39)$$

In the time-ordered case, we get a similar result, but we must order the terms in the exponential correctly. Hence, we shall simply denote

$$\langle e | \mathcal{T}(\sigma^x(\tau_1) \cdots \sigma^x(\tau_{2n+1})) | g \rangle = \mathcal{T} e^{i\Omega(\tau_1 - \tau_2 + \tau_3 - \cdots + \tau_{2n+1})}, \quad (2.4.40)$$

and the expression vanishes if there is an even number of insertions.

Bringing all of this back to the Dyson series, we find that

$$A_{g \to e}(\phi; \tau, \tau') = \sum_{n \text{ odd}} \frac{(-i\epsilon)^n}{n!} \int_\tau^{\tau'} \cdots \int_\tau^{\tau'} \langle \phi | \mathcal{T}(\phi(\tau_1) \cdots \phi(\tau_n)) | 0 \rangle \, \mathcal{T} e^{i\Omega(\tau_1 - \cdots + \tau_n)} \, d\tau_1 \cdots d\tau_n. \quad (2.4.41)$$

The probability for the transition happening between the instants $\tau$ and $\tau'$ is then

$$p_{g \to e}(\tau, \tau') = \int \left| A_{g \to e}(\phi; \tau, \tau') \right|^2 \mathcal{D}\phi, \quad (2.4.42)$$

where we are integrating the field's state out, since we are looking only at the detector. Using the resolution of the identity written as $\int |\phi\rangle\langle\phi| \, \mathcal{D}\phi = \mathbb{1}$, we find that

$$p_{g \to e}(\tau, \tau') = \sum_{n, m \text{ odd}} \epsilon^{n+m} \frac{(-i)^{n-m}}{n! m!} \int_\tau^{\tau'} \cdots \int_\tau^{\tau'} \langle 0 | \mathcal{T}(\phi(\tau_1') \cdots \phi(\tau_m'))^\dagger \mathcal{T}(\phi(\tau_1) \cdots \phi(\tau_n)) | 0 \rangle \times$$
$$\times \, \mathcal{T} e^{i\Omega(\tau_1 - \cdots + \tau_n)} \mathcal{T} e^{-i\Omega(\tau_1' - \cdots + \tau_m')} \, d\tau_1 \cdots d\tau_n \, d\tau_1' \cdots d\tau_m'. \quad (2.4.43)$$

Up to leading order, we have

$$p_{g \to e}(\tau, \tau') = \epsilon^2 \int_\tau^{\tau'} \int_\tau^{\tau'} \langle 0 | \phi(\tau_1') \phi(\tau_1) | 0 \rangle \, e^{-i\Omega(\tau_1' - \tau_1)} \, d\tau_1 \, d\tau_1'. \quad (2.4.44)$$

Notice this expression means the probability of excitation is given by a Fourier transform of the two-point function.

So far, we did not need to specify the details of the state $|0\rangle$, the worldline of the detector, and not even the spacetime we are working in. This exhibits how useful particle detectors can be in a myriad of situations. Our case of interest concerns the Minkowski vacuum in Minkowski spacetime. For an inertial detector, the proper time $\tau$ would coincide with inertial time. In this case, we know the two-point function only has contributions due to positive frequencies (this follows from Eq. (2.4.6) on page 29), and hence the probability will vanish for $\Omega > 0$, which is our case of interest. Hence, an inertial detector will not detect any particles in the Minkowski vacuum, as expected. Nevertheless, notice that accelerated detectors have different frequency decompositions and, as a consequence, may lead to non-vanishing excitation probabilities.

Let us then specify the detector's worldline. This can be done naturally in Rindler coordinates, with which we specify the worldline as

$$z^\mu(\tau) = \left( \tau; \frac{1}{a}, 0, 0 \right). \quad (2.4.45)$$

We took $r = \frac{1}{a}$ because this corresponds to the worldline of the observer with proper acceleration $a$ defining Rindler coordinates (Rindler 1966, 2006, Sec. 12.4). We also took the coordinates $y(\tau) = z(\tau) = 0$ for simplicity, but they could have been given any other constant value without altering the following results.

Using Eqs. (2.4.3), (2.4.5) and (2.4.45) on page 27, on page 28 and on the preceding page, we find that

$$\langle 0_M | \phi(\tau_1')\phi(\tau_1) | 0_M \rangle = \operatorname*{w-lim}_{\varepsilon \to 0^+} \frac{a^2}{4\pi^2 \left[ (\cosh(a\tau_1') - \cosh(a(\tau_1 - i\varepsilon)))^2 - (\sinh(a\tau_1') - \sinh(a(\tau_1 - i\varepsilon)))^2 \right]},$$
(2.4.46)

where we chose to align the arbitrary future-directed timelike vector $T^a$ of Eq. (2.4.5) on page 28 along the $\left( \frac{\partial}{\partial \eta} \right)^a$ direction, for this simplifies the expression. The previous equation can then be further simplified using the properties of hyperbolic functions to get to

$$\langle 0_M | \phi(\tau_1')\phi(\tau_1) | 0_M \rangle = \operatorname*{w-lim}_{\varepsilon \to 0^+} \frac{-a^2}{16\pi^2 \sinh^2\left( \frac{1}{2} a \left( \tau_1' - \tau_1 - i\varepsilon \right) \right)}.$$
(2.4.47)

We can then notice that the probability of excitation is

$$p_{g \to e}(\tau, \tau') = -\frac{a^2 \varepsilon^2}{16\pi^2} \lim_{\varepsilon \to 0^+} \int_\tau^{\tau'} \int_\tau^{\tau'} \frac{\exp\left( -i\Omega(\tau_1' - \tau_1) \right)}{\sinh^2\left( \frac{a}{2}(\tau_1' - \tau_1 - i\varepsilon) \right)} \, d\tau_1' \, d\tau_1 \,,$$
(2.4.48a)

$$= -\frac{a\varepsilon^2}{8\pi^2} \lim_{\varepsilon \to 0^+} \int_\tau^{\tau'} \int_{a(\tau - \tau_1)/2}^{a(\tau' - \tau_1)/2} \frac{\exp\left( -\frac{2i\Omega\eta}{a} \right)}{\sinh^2(\eta - i\varepsilon)} \, d\eta \, d\tau_1 \,,$$
(2.4.48b)

where we defined $\eta = \frac{a(\tau_1' - \tau_1)}{2}$.

Let us then define the rate of excitation through

$$R_{g \to e} = \lim_{\substack{\tau' \to +\infty \\ \tau \to -\infty}} \frac{p_{g \to e}(\tau, \tau')}{\tau' - \tau},$$
(2.4.49a)

$$= -\frac{a\varepsilon^2}{8\pi^2} \lim_{\varepsilon \to 0^+} \int_{-\infty}^{+\infty} \frac{\exp\left( -\frac{2i\Omega\eta}{a} \right)}{\sinh^2(\eta - i\varepsilon)} \, d\eta \,.$$
(2.4.49b)

This Fourier transform can be computed using the residue theorem. One finds

$$R_{g \to e} = \frac{\varepsilon^2 \Omega}{2\pi \left( e^{\frac{2\pi\Omega}{a}} - 1 \right)}.$$
(2.4.50)

If we did the same calculations for the $|e, \phi\rangle \to |g, 0_M\rangle$ transition, we would get

$$R_{e \to g} = \frac{\varepsilon^2 \Omega}{2\pi \left( 1 - e^{-\frac{2\pi\Omega}{a}} \right)},$$
(2.4.51)

which is the same result with $\Omega \to -\Omega$. Notice then that this implies

$$\frac{R_{g \to e}}{R_{e \to g}} = e^{-\frac{2\pi\Omega}{a}},$$
(2.4.52)

meaning the detector satisfies the detailed balance (see Tomé and Oliveira 2015) at inverse temperature $\beta = \frac{2\pi}{a}$. This is a hallmark of a system in thermal equilibrium.

Given how abstract our previous approaches can be, it is interesting to notice how "experimental" this derivation is. While we used a simplified model for a particle detector, many physical systems can be understood as detectors. For example, a thermometer. One can even take it further and understand a steak as a particle detector, in which case the Unruh effect will present itself as a cooking method.

### Hawking Effect

Historically, the Unruh effect was preceded by a similar prediction in a completely different scenario: the Hawking effect (Hawking 1974, 1975).

In the Hawking effect, one considers a spacetime in which a star collapses to a black hole, as illustrated on Fig. 2.6 on the next page. For simplicity, we will consider only the case of spherically symmetric, uncharged collapse, although Hawking also treated the case in which the final black hole is a general Kerr–Newman black hole.

It was found that a static observer at infinity would see particles coming from the Schwarzschild black hole with a thermal spectrum at a temperature given by

$$T_H = \frac{1}{8\pi M}. \tag{2.4.53}$$

At the time, it was already expected that some sort of radiation would be emitted by rotating black holes, but it was surprising that Hawking's calculation also predicted the emission of radiation by Schwarzschild black holes (a historical account is given by Almeida 2021).

Soon after Hawking's calculation, Unruh (1976) investigated other aspects of the Hawking effect, with two of his conclusions being

  i.  the previously discussed Unruh effect;

 ii.  that the Hawking effect could be derived in empty Schwarzschild spacetime, rather than on a collapse spacetime, allowing for a simpler analysis. The presence of the star is exchanged by the imposition of boundary conditions for the field in the past horizon of Schwarzschild spacetime.

In quantum parlance, "imposing boundary conditions on the quantum field" means to select a quantum state. Hence, the state of the quantum field in Schwarzschild spacetime required for the observation of the Hawking effect is known as the Unruh vacuum.

It is particularly interesting that the Unruh vacuum presents a singular behavior near the past event horizon of Schwarzschild spacetime[*] (see Birrell and Davies 1982, Chap. 8, and references therein). This might seem troublesome at first, but notice there is no issue: the region in which the vacuum is singular does not exist physically. The presence of a collapsing star means there is no actual past event horizon, and hence the state is indeed nonsingular throughout the whole physical spacetime.

One might, however, wonder whether there is a state that is nonsingular throughout the whole maximally-extended Schwarzschild spacetime. There is indeed, and it is known as the Hartle–Hawking vacuum (Hartle and Hawking 1976). This state is stationary and leads to a prediction similar to the Hawking effect. Namely, a static observer will see particles coming from the black hole and from infinity with a thermal spectrum. This is an instance of the so-called Unruh effect in curved spacetimes (see Wald [1994](#), Sec. 5.3, for a discussion).

---

[*]In the language of Section 2.B, it fails to be Hadamard due to its behavior near the past event horizon.

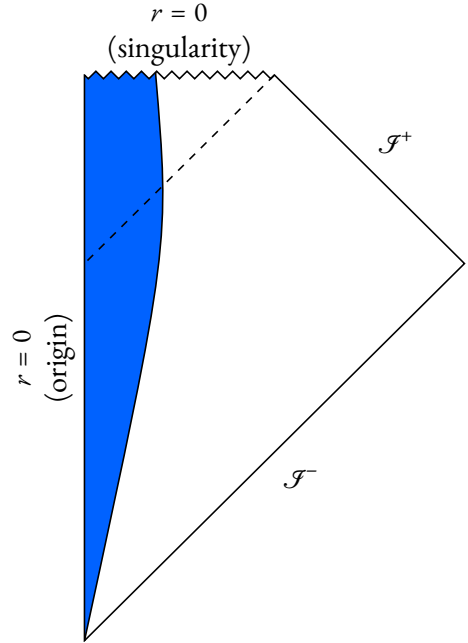

Figure 2.6: Penrose diagram for a spacetime with a collapsing star (shaded area), as considered by Hawking ([1975](#)). $\mathscr{I}^+$ (resp. $\mathscr{I}^-$) denotes the future (past) null infinity. It must be mentioned that this is a classical diagram—quantum effects would lead the black hole to evaporate, as discussed in more detail on the book by Wald ([1994](#), Chap. 7).

Some QFTCS discussions will derive these predictions in curved spacetime by considering the behavior of a static observer and argue with the equivalence principle that, since the observer has an uniform acceleration, the Unruh effect should apply, and hence the static observer should see the emission of particles with a thermal spectrum. While this argument does, *e.g.*, lead to the correct expression for the temperature, it leaves some important conceptual gaps unaddressed. For example, it does not account for the difference between the predictions given by the Unruh and Hartle–Hawking vacua, and it also makes it difficult to understand why one expects to see thermal radiation coming from a black hole, but not from regular spherical bodies, such as planets.

The argument is problematic because it fails to take into account important global properties. For example, the Hartle–Hawking vacuum is not physically acceptable in a collapsing star spacetime, for it predicts the emergence of modes from a white hole region. If this region does not exist in the physical spacetime, the Hartle–Hawking vacuum is not a reasonable model of physical reality. Similarly, the Unruh vacuum is unacceptable in empty Schwarzschild spacetime because it is singular at the past event horizon. Both of these vacua are inadmissible in a planetary spacetime because they predict field modes falling down a future event horizon which simply does not exist if a planet is present. Hence, in a planetary spacetime, a physically reasonable state would be, *e.g.*, the Boulware vacuum (Boulware [1975](#)). The Boulware vacuum is singular on both the future and past event horizons—which is irrelevant in a planetary spacetime—but predicts no particles will be seem by a static observer (see

Birrell and Davies 1982, Chap. 8, and references therein).

This example showcases two important aspects of the Hawking and Unruh effects. Firstly, that the presence of a horizon plays a relevant role. In Minkowski spacetime, the horizon is also present, for accelerated observers experience an event horizon similar to that of a black hole (see Ellis and Williams 2000, Sec. 4.3; Rindler 2006, Sec. 12.4). In full generality, the relevant structure is known as a bifurcate Killing horizon (see Wald 1994, Chap. 5, for details). Secondly, it shows our choice of vacuum in curved spacetime might lead to strikingly different predictions. Just as in ordinary QM and in the entirety of physics, in QFTCS it is meaningless to make any predictions without specifying the physical states to which they apply.

Some more discussion about the properties of the different vacua of Schwarzschild spacetime can be found on the books by Birrell and Davies (1982, Chap. 8) and Frolov and Novikov (1998, Sec. 11.2).

## 2.5   Euclidean Approach

In discussing the Unruh effect, we have also performed a calculation using an Euclidean path integral. In the following chapters, we shall also make extensive use of Euclidean path integrals. How does this approach relate to the algebraic approach we have been favoring so far?

As mentioned in the first few paragraphs of this chapter, the algebraic approach seems natural to formulate QFTCS in full generality. It does not assume preferred vacuum states, observers, CCR representations, special symmetries, and so on. Euclidean path integrals, on the other hand, consist of performing path integral computations on a Riemannian manifold that should be understood as an analytic continuation of a Lorentzian spacetime. One then analytically continues the results back to the physical spacetime. These procedures might not be possible for all spacetimes. Furthermore, path integrals assume a preferred state with respect to which the path integral is computed, and hence it is important for us to wonder about which states can be represented by a path integral.

The simplest case is, of course, given by the Minkowski vacuum in Minkowski spacetime. In this case, the Euclidean path integral approach can be formulated in a mathematically rigorous manner (at least in some cases) and then shown to recover other approaches in Minkowski spacetime, as discussed by Glimm and Jaffe (1987, Chaps. 6 and 19). At least on a formal level, the idea is the following. Using an Euclidean path integral, one can write correlation functions, such as

$$S_n(x_1, \ldots, x_n) = \int \exp\left(-S[\phi]\right)\phi(x_1)\cdots\phi(x_n)\,\mathscr{D}\phi\,. \tag{2.5.1}$$

The Euclidean $n$-point functions are sometimes referred to as "Schwinger functions" (Schwinger 1958). These functions can then be analytically continued back to Lorentzian signature to obtain the time-ordered $n$-point functions. The analytic continuations happen by means of a procedure often called "Wick rotation": one first defines Euclidean (or imaginary) time $t_E$ through $t = -it_E$, and later returns to Lorentzian signature by replacing $t_E = it$ back. To illustrate it, consider a scalar field in Minkowski spacetime. We have

$$S_n(x_1, \ldots, x_n) = \int \exp\left(-\int \frac{1}{2}\partial_a\phi\partial^a\phi + V(\phi)\,\mathrm{d}^4x_E\right)\phi(x_1)\cdots\phi(x_n)\,\mathscr{D}\phi\,, \tag{2.5.2a}$$

$$\rightarrow \int \exp\left(-i\int \frac{1}{2}\partial_a\phi\partial^a\phi + V(\phi)\,\mathrm{d}^4x\right)\phi(x_1)\cdots\phi(x_n)\,\mathscr{D}\phi\,, \tag{2.5.2b}$$

$$= \langle 0_M | \mathcal{T} \phi(x_1) \cdots \phi(x_n) | 0_M \rangle. \tag{2.5.2c}$$

The sign on the specification $t = -i t_E$ is particularly important. It is related to the fact that the Gaussian integrals used to evaluate the Lorentzian path integral on Eq. (2.5.2) on the facing page assume that time has a small imaginary part, *i.e.*, one makes the replacement $t \to t(1 - i\varepsilon)$ and takes the limit $\varepsilon \to 0^+$ at the end of the calculation to ensure the Gaussian formulae apply (Peskin and Schroeder 1995, p. 286). Incidentally, this very same $i\varepsilon$ prescription is what ensures we are computing the Feynman propagator rather than, *e.g.*, the two-point function. Hence, it is essential that we perform the Wick rotation according to $t = -i t_E$, or we could get other time-orderings for the $n$-point functions (for more details, see Hartman, Jain, and Kundu 2016). Notice the replacement $t_E = it$ is meant to be understood as $t_E = \lim_{\theta \to \frac{\pi}{2}^-} e^{i\theta} t$, or one will get the wrong $i\varepsilon$ prescription.

The first alarming issue of this prescription arises due to GR alone, without the need to consider quantum theory. We wish to make an analytic continuation by considering a notion of "imaginary time", but not even "real time" has a straightforward definition in a general curved spacetime. Hence, our first goal is to understand under which conditions we can give meaning to such a procedure.

## Imaginary Time in Curved Spacetime

Let us attempt at making an Euclidean continuation of De Sitter spacetime, for the sake of having an example (Percacci 2017, Sec. 5.2; Visser 2017). De Sitter spacetime can be covered with many different coordinates, some of which are reviewed by Spradlin, Strominger, and Volovich (2001). Let us mention a few possibilities with the cosmological constant set to $\Lambda = 3$ for simplicity. We can use global coordinates, which have

$$\mathrm{d}s^2 = -\mathrm{d}t^2 + \cosh^2 t \left( \frac{\mathrm{d}r^2}{1 - r^2} + r^2 \, \mathrm{d}\Omega^2 \right), \tag{2.5.3}$$

where $\mathrm{d}\Omega^2$ is the usual line element on the 2-sphere. These coordinates have spherical spatial sections. Alternatively, we could use planar spatial sections, which yield

$$\mathrm{d}s^2 = -\mathrm{d}t^2 + e^{-2t} \left( \mathrm{d}r^2 + r^2 \, \mathrm{d}\Omega^2 \right). \tag{2.5.4}$$

We can also choose to use hyperbolic spatial sections, which lead to

$$\mathrm{d}s^2 = -\mathrm{d}t^2 + \sinh^2 t \left( \frac{\mathrm{d}r^2}{1 + r^2} + r^2 \, \mathrm{d}\Omega^2 \right). \tag{2.5.5}$$

Lastly, a fourth interesting choice is to employ static coordinates, in which case the line element is

$$\mathrm{d}s^2 = -(1 - r^2) \, \mathrm{d}t^2 + \frac{\mathrm{d}r^2}{1 - r^2} + r^2 \, \mathrm{d}\Omega^2. \tag{2.5.6}$$

Naively, we would expect to be able to obtain an Euclidean manifold "analytically related" to De Sitter space by simply applying the prescription $t = -i t_E$. If we carry this out with Eqs. (2.5.3)

to (2.5.6) on the previous page, we find

$$ds^2 = dt_E{}^2 + \cos^2 t_E \left( \frac{dr^2}{1 - r^2} + r^2 \, d\Omega^2 \right), \tag{2.5.7}$$

$$ds^2 = dt_E{}^2 + e^{2it_E} \left( dr^2 + r^2 \, d\Omega^2 \right), \tag{2.5.8}$$

$$ds^2 = dt_E{}^2 - \sin^2 t_E \left( \frac{dr^2}{1 + r^2} + r^2 \, d\Omega^2 \right), \tag{2.5.9}$$

$$ds^2 = (1 - r^2) \, dt_E{}^2 + \frac{dr^2}{1 - r^2} + r^2 \, d\Omega^2 . \tag{2.5.10}$$

Eqs. (2.5.7) and (2.5.10) are indeed Riemannian metrics—in fact, they are just different coordinate choices on the 4-sphere. Nevertheless, Eq. (2.5.9) is Lorentzian with the opposite sign convention, and Eq. (2.5.8) is not even a real metric. If our naive prescription leads to these strange results, then how can we know when we will be able to make such a continuation procedure and whether it will work?

One way of dealing with this issue in a lot of generality is to give up on making a continuation on time and instead making an analytic continuation of the metric itself. In essence, the idea starts with the fact that, while any manifold admits a Riemannian metric, the existence of a Lorentzian metric is equivalent to the existence of a non-vanishing (timelike) vector field $t^a$ (Hawking and Ellis 1973, pp. 38–40). Using such a timelike vector field, we can then define an Euclidean metric $g_{ab}^{(E)}$ from the Lorentzian metric $g_{ab}^{(L)}$ through

$$g_{ab}^{(E)} = g_{ab}^{(L)} - 2 \frac{t_a \, t_b}{g_{cd}^{(L)} \, t^c t^d}. \tag{2.5.11}$$

This is guaranteed to provide a Riemannian metric for any choice of everywhere timelike vector field $t^a$. This approach is discussed in more detail by Candelas and Raine (1977), Gray (2016), Percacci (2017), and Visser (2017), and it is particularly important within quantum gravity. It is important to mention that this is not only an algebraic definition: the metric is being analytically continued to a complex metric. This is essential for us to be able to analytically continue the quantum theory back to Lorentzian signature later on. For further detail, see the previously mentioned references.

A problem with the prescription of Eq. (2.5.11) is that while it always yields a Riemannian metric and it is coordinate-independent, there is no guarantee that it will yield the same metric for different choices of vector field $t^a$. Nevertheless, our purposes are sufficiently restricted for this to not be a problem. Our interest is within QFTCS, which is usually concerned with globally hyperbolic spacetimes. This means we can exploit the fact that the spacetimes we are interested in always have the topology $\mathbb{R} \times \Sigma$ for some spatial manifold $\Sigma$ (Hawking and Ellis 1973, Proposition 6.6.8; Wald 1984, Theorem 8.3.14). Hence, the spacetime topology already picks up a preferred notion of time that we can use to define an Euclidean continuation. More specifically, the existence of a global time function $t$ is assured (Wald 1984, Theorem 8.3.14), and hence we have a preferred time direction given by $t_a = \nabla_a t$. Notice that in our example with De Sitter spacetime this notion of time corresponds to the one on Eq. (2.5.3) on the preceding page, which employs coordinates covering the whole manifold.

Static coordinate charts will also always yield a Riemannian metric. Indeed, if we have a static spacetime and are employing static coordinates, the line element will have the general form (Wald

$$\mathrm{d}s^2 = -V^2(x^1, x^2, x^3)\,\mathrm{d}t^2 + \sum_{\mu,\nu=1}^{3} h_{\mu\nu}(x^1, x^2, x^3)\,\mathrm{d}x^\mu\,\mathrm{d}x^\nu\,. \tag{2.5.12}$$

Notice the prescription $t = -it_E$ will always lead to a Riemannian metric in this case. More specifically, it leads to

$$\mathrm{d}s^2 = +V^2(x^1, x^2, x^3)\,\mathrm{d}t_E^2 + \sum_{\mu,\nu=1}^{3} h_{\mu\nu}(x^1, x^2, x^3)\,\mathrm{d}x^\mu\,\mathrm{d}x^\nu\,. \tag{2.5.13}$$

Notice that, in the language of Eq. (2.5.11) on the preceding page, this means we are picking $t^a$ to be the static Killing field.

Therefore, when working with globally hyperbolic spacetimes and with static spacetimes we are able to perform analytic continuations in a safe manner. It is also interesting to remark that the perspective of Eq. (2.5.11) on the facing page becomes particularly important in spacetimes that are stationary, but not static. Indeed, the naive prescription $t = -it_E$ would lead to imaginary off-diagonal terms—as can be seen on the Kerr metric, for example—but the metric continuation procedure still works.

As the upshot of this discussion, we see that there is some subtleties when attempting to analytically continue a spacetime. Nevertheless, we have a clear prescription for our goals. For example, we are capable of "Euclideanizing" Schwarzschild, De Sitter, and Rindler spacetimes.

### Analytic Continuation of Quantum States

Let us then assume we are working on a spacetime which does admit a reasonable analytic continuation. We then proceed to write Eq. (2.5.1) on page 40 in this spacetime. Once we analytically continue the Schwinger functions back to Lorentzian signature, we will end up with time-ordered $n$-points functions. A question arises: with respect to which quantum state are these $n$-point functions computed? This question can be overlooked when working with inertial observers in Minkowski spacetime, for it assumes the existence of a privileged Poincaré invariant state—the Minkowski vacuum—which is always the state of interest. Nevertheless, this need not be the case in curved spacetime, where a preferred state cannot be taken for granted. Recall, for example, our discussion about the Hawking effect in Section 2.4. We mentioned there are three particularly interesting vacuum choices in Schwarzschild spacetime: the Unruh vacuum, the Hartle–Hawking vacuum, and the Boulware vacuum. How can we distinguish among these states on a path integral?

Firstly, let us notice that to characterize a Gaussian state in a previously chosen spacetime it suffices to prescribe its Feynman propagator. Indeed, the spacetime structure fixes the expectation value of the commutator of fields (Eq. (2.3.14) on page 23), while the Feynman propagator can be used to obtain the expectation value of the anticommutator (as one can conclude from the formulae given by Fulling 1989, pp. 74–81; Wald 1994, Sec. 5.3). Using both expectation values, one obtains the two-point function, which is enough to fully characterize a Gaussian state.

Therefore, our goal is to distinguish among the possible Feynman propagators. In the Euclidean approach, the Feynman propagator is obtained from an analytic continuation of the two-point Schwinger function, meaning we want to understand how different two-point Schwinger functions might arise in the Euclidean spacetime.

The Feynman propagator $F$ relates to the time-ordered product of fields through

$$F(x, y) = i\omega(\mathcal{T}\phi(x)\phi(y)) \tag{2.5.14}$$

and it solves the equation

$$\Delta_x F(x, y) = \hat{\delta}(x, y),\tag{2.5.15}$$

where the subindex $x$ denotes the variable with respect to which the derivatives are to be carried out and $\Delta = -\nabla_a \nabla^a + m^2$. As a consequence, the two-point Schwinger function will satisfy[*]

$$\Delta_x^E S(x, y) = \hat{\delta}(x, y),\tag{2.5.16}$$

where $\Delta^E$ is the Euclidean analogue of $\Delta$. $S$ is still merely a Green's function. Hence, to understand the different possible vacua is to understand the possible Green's functions for the operator $\Delta^E$. However, just as we need the metric to be real analytic to perform an analytic continuation, we also need the states to be sufficiently well-behaved. On intuitive grounds, we can expect that stationary states in stationary spacetimes can be analytically continued to imaginary time, but there is no *a priori* reason to expect this to be possible for every state.

As discussed by Wald ([1979](#)), the uniqueness of two-point Schwinger functions can then be understood in terms of whether the operator $\Delta^E : \mathscr{C}_0^\infty(\mathcal{M}) \to L^2(\mathcal{M})$ is essentially self-adjoint[†], where $L^2(\mathcal{M})$ denotes the space of square-integrable functions[‡] on $\mathcal{M}$. This property holds for manifolds with "negligible boundary" (Gaffney [1951](#)). Among other examples, this includes complete Riemannian manifolds (Gaffney [1954](#); see also Grigor'yan [2009](#), Sec. 11.3, for a direct proof). Hence, the two-point Schwinger function is unique for complete Riemannian manifolds.

To understand this, let us consider the case of Minkowski spacetime. Its natural Euclideanization in inertial coordinates is given by $\mathbb{R}^4$. Hence, the operator $\Delta^E$ is given by

$$\Delta^E = -\sum_{\mu=1}^{4} \frac{\partial^2}{\partial (x^\mu)^2} + m^2.\tag{2.5.17}$$

For simplicity, let us consider the massless case. Then the equation for the Schwinger function is

$$\nabla_4^2 S(x, y) = -\hat{\delta}(x, y),\tag{2.5.18}$$

which means we simply desire to find the Green's function for Poisson's equation in $d = 4$. This is similar to an Electrostatics problem in $d = 4$ and hence we can use arguments common in Electrostatics (see, *e.g.*, Wald [2022](#), pp. 15–16) to argue that there is a unique function $S(x)$ such that $S(x, y) = S(x-y)$ (ensured by translation symmetry) that solves Eq. ([2.5.18](#)) and vanishes at infinity sufficiently fast. This unique solution is given by

$$S(x) = \frac{1}{4\pi^2 x^\mu x_\mu}.\tag{2.5.19}$$

---

[*]The coefficient on the RHS of Eq. ([2.5.16](#)) might seem wrong at first glance, but one should also remember to analytically continue the Dirac delta. This procedure is more easily done if one writes Eq. ([2.5.16](#)) as $\int \Delta_x S(x, y) \, \mathrm{d}^4 x_E = 1$ to get rid of the delta and notice the difference in coefficients comes from the Jacobian due to the Wick rotation.

[†]While physicists often just call an operator with the formal property $A = A^\dagger$ "Hermitian", mathematicians are more careful about the domain in which such an equality holds. "Essential self-adjointness" is one of the definitions that occur and it is discussed by essentially any book on functional analysis. See the book by Reed and Simon ([1980](#), Sec. VIII.2) for further discussion.

[‡]The action for a free field involves the integral of the square of the field, often with operator insertions. We expect at least the classical action to be finite, and hence it is reasonable to work with square-integrable functions. Otherwise, we could run into classical configurations with infinite energy.

The Wick rotation of this expression matches the expressions for $\langle 0_M | \mathcal{T}\phi(x)\phi(y) | 0_M \rangle$ given by Bogoliubov and Shirkov (1982, App. V.2).

$\mathbb{R}^4$ is complete as a Riemannian manifold, and hence the results due to Gaffney (1951, 1954) and Wald (1979) ensure the Schwinger function is unique. However, we can also interpret this by making an analogy with Electrostatics. $S(x, y)$ can be understood as if it was the scalar potential at the Euclidean event $x$ due to a point charge at $y$. Since we are considering $\Delta^E$ as an operator on $L^2(\mathcal{M})$, we want $\int S(x, y)f(y)\,\mathrm{d}^4 y$ to be square-integrable when $f$ is square-integrable, and hence $S(x)$ must vanish at infinity. This means we cannot add a solution of Laplace's equation to $S(x)$ to obtain a different possible solution, because solutions to the harmonic equation can only vanish at infinity if they vanish everywhere. Furthermore, since we are considering the problem in the whole Euclidean spacetime, we cannot use "image charges" outside of the region of interest to modify the solution.

The situation would change if we were interested in working with Rindler spacetime, for example. It can be shown that the Euclideanization of the metric given on Eq. (2.4.4) on page 27 still describes a piece of $\mathbb{R}^4$, but not the whole $\mathbb{R}^4$. Hence, we no longer have a complete manifold. From the Electrostatics perspective, this means there are other regions where we can put "image charges" and hence obtain other possible choices of Schwinger functions. This will then, of course, change the boundary conditions on the Schwinger functions. If we want to consider Rindler spacetime with the Minkowski vacuum, we are supposed to work with the Schwinger function given on Eq. (2.5.19) on the preceding page. If we were to add "image charges" outside of the analytically continued Rindler spacetime, we would be able to change the Schwinger function and obtain a different state upon analytic continuation back to Lorentzian signature.

This is the essential feature of how one can select different Gaussian states on Euclidean spacetimes: by selecting different boundary conditions on the two-point Schwinger function.

## States Expressible as Path Integrals

We then get to the most important question of this section: which states can be written in terms of an Euclidean path integral? In other words, for which states can the expression

$$\omega(\mathcal{T}\phi(x_1)\cdots\phi(x_n)) \stackrel{t\to -it_E}{=} \int \exp\left(-S[\phi]\right)\phi(x_1)\cdots\phi(x_n)\,\mathcal{D}\phi \tag{2.5.20}$$

be given meaning?

From our previous discussions, we expect to be capable of Euclideanizing a spacetime when it presents some sort of distinguished timelike vector. For example, a timelike Killing field, or $\nabla^a t$ for some global time function $t$. Since the Euclideanization procedure is, at its essence, an analytic continuation, we do not expect to gain much information about regions of spacetime that are not already real-analytic.

We also expect to be able to continue states that are somehow analytic in the time parameter being analytically continued. Hence, it is expected that we will manage to obtain sensible results for stationary states in static spacetimes, for example. In fact, Wald (1979) has shown the equivalence between the Euclidean definition of the Feynman propagator using the Schwinger function continuation and the natural definition under some conditions.

Hence, it seems reasonable that we can write expressions such as Eq. (2.5.20) at least for stationary states in static spacetimes. Based on our discussions about the analytic continuation of quantum states, we expect the selection of different possible stationary states to happen by means of the imposition of different boundary conditions on the fields being integrated over. Hence, at least in principle, one could distinguish the Minkowski and Rindler vacua, or the Unruh, Hartle–Hawking, and Boulware vacua, by their boundary conditions.

Whether the integral on Eq. (2.5.20) on the previous page is actually capable of reproducing the correct results in arbitrary spacetimes is far from obvious. While the weight $\exp(-S[\phi])$ is adequate for the Minkowski vacuum, it could happen that other states require more complicated expressions. Indeed, for interacting theories, Eq. (2.5.20) on the preceding page is a consequence of the Gell-Mann–Low formula (Gell-Mann and Low 1951; Peskin and Schroeder 1995, Eq. (4.31)), which allows one to relate Wightman functions in the free and interacting theories. As mentioned by D'Angelo et al. (2022, Remark 2.1) and Fredenhagen and Rejzner (2015), this formula is a property of the Minkowski vacuum, but there is no guarantee it will apply in more generality. As a consequence, the path integral expression of correlation functions might also fail.

Indeed, consider the derivation of the expression for correlation functions in terms of path integrals given by Peskin and Schroeder (1995, pp. 283–284). To obtain Eq. (2.5.1) on page 40 (or rather its Lorentzian version), the authors explicitly make use of the fact that the Minkowski vacuum is the state of minimum energy. This is not always available in general spacetimes, but it is in stationary spacetimes.

Our main interests in this thesis are to consider effects that can be attributed to free fields in stationary spacetimes. For example, we did not need to consider interacting fields to discuss the Unruh and Hawking effects, apart from coupling it to a Unruh–DeWitt detector. Hence, it seems reasonable that we can treat the quantum field as if it was completely free, and hence be able to write expressions such as Eq. (2.5.20) on the preceding page. Our calculations in Chapter 4 will ultimately be carried out only in Minkowski spacetime, where these limitations do not occur. Nevertheless, they can be relevant for future generalizations.

## Advantages of the Euclidean Approach

Given the limitations we just presented for the Euclidean approach in comparison with the algebraic approach, one might wonder why we still plan on using it on the following chapters.

As mentioned in this chapter's introduction, the main point is that different approaches have different strengths. For example, one will have a hard time searching for a GR book that writes the Riemann tensor using Dirac notation, just like one will have difficulty to find a QM textbook writing states using the abstract index notation that is common in relativity. None of the two choices of notation is inferior to the other. Rather, they possess different strengths and are particularly useful in different contexts even though, deep down, they are simply different notations for writing linear and multilinear algebra.

The difference between the algebraic and the Euclidean approaches can be thought in the same manner. The algebraic approach has the advantage of being abstract, which implies two things:

i. it allows for one to formulate QFTCS in a very general language, in which it is possible to think about many theories at the same time while focusing only on the essential features of the framework;

ii. it might make actual computations more difficult, since describing specific cases involves adding much more detail to the formalism.

Hence, it is an extremely useful approach to understand what QFTCS is, but it is not particularly convenient for carrying out computations. Other techniques, such as the use of Fock spaces or particle detectors, might provide easier methods.

The Euclidean approach, on the other hand, lacks on generality. We could only make sense of it for a handful of spacetimes and states. Nevertheless, it is extremely similar to well-known methods of QFT in flat spacetime and allows for more convenient ways of performing computations. While path integrals are often mathematically ill-defined, they are a wonderful device to perform calculations easily. Furthermore, the use of Euclidean signature makes the study of thermal phenomena particularly natural, since thermal equilibrium is mapped into periodicity in imaginary time.

In summary, we decided to employ the algebraic approach thus far because it allowed us to keep track of what are the essential features of the system being studied and other questions of relevance. We could clearly see that choices of coordinates, preferred observers, or the notion of "particle" are completely irrelevant from a fundamental point of view, while also noticing some importance that should be given to other properties that could go easily unnoticed. If our discussions used path integrals from the start, we might not have noticed the important matter of asking ourselves what is the actual physical state of the quantum field we are considering in our computations.

Now that our conceptual discussion of QFTCS is done and we plan on performing calculations, the algebraic approach is no longer so convenient. We can keep it in mind when interpreting results, but it will be easier to employ Euclidean methods when calculating renormalization group flows. In particular, when probing the Unruh effect, for example, we would be able to naturally consider the parameters of the detector at the scale determined by the Unruh temperature.

At the risk of contradicting the first statement in this chapter, it is relevant to notice that in doing physics the most appropriate approach is always relative to the physicist's goal.

## 2.A Fock Representations in Stationary Spacetimes

Many spacetimes of interest end up being stationary or asymptotically stationary. For example, the Kerr–Newman family, De Sitter spacetime, Minkowski spacetime and many more are stationary. When working in these spacetimes, one can exploit the available stationary symmetry when doing QFTCS. In this appendix, we will describe a qualitative and "handwaving" understanding of the role of symmetry. More detailed expositions can be found in the discussions by Khavkine and Moretti (2015, Sec. 5.2.7), Panangaden (2013), and Wald (1994, Sec. 4.3) and in the original papers by Ashtekar and Magnon (1975) and Kay (1978).

Let us begin by considering quantum fields in Minkowski spacetime. In Minkowski spacetime, it is common to discuss about creating and annihilating particles on a given state by using ladder operators. These operators are defined by means of the Fourier decomposition of the Klein–Gordon field. Namely, one can write

$$\phi(x) = \frac{1}{(2\pi)^{\frac{3}{2}}} \int \left( a_{\vec{p}} e^{ip\cdot x} + a_{\vec{p}}^{\dagger} e^{-ip\cdot x} \right) \frac{\mathrm{d}^3 p}{\sqrt{2\omega_{\vec{p}}}}, \qquad (2.A.1)$$

where

$$\omega_{\vec{p}} = +\sqrt{\left\| \vec{p} \right\|^2 + m^2} \qquad (2.A.2)$$

and we choose conventions such that

$$[a_{\vec{p}}, a_{\vec{q}}^\dagger] = \delta^{(3)}(\vec{p} - \vec{q}). \tag{2.A.3}$$

Notice that Eq. (2.A.1) on the previous page decomposes the field $\phi$ in terms of positive and negative frequencies. In other words, in terms of solutions associated with positive and negative energies. This difference between positive and negative energies is then used to define the creation and annihilation operators.

It should be remarked that ortochronous Poincaré transformations never flip the sign of the time-component of a four-vector. Hence, all inertial observers in Minkowski spacetime always agree on the sign of the energy of a given particle. As a consequence, all inertial observers agree on the decomposition given on Eq. (2.A.1) on the preceding page. It follows that they all agree that all annihilation operators annihilate the Minkowski vacuum and they all agree on how many particles there are in a given state of the quantum field.

From this discussion, we can already conclude that, in some sense, particles are an "energy-dependent concept". The separation between positive and negative energy is literally the way we usually define the ladder operators used in QFT in flat spacetime to create and annihilate particles. Loosely speaking, if two observers have different notions of what is energy, they might have two different notions of what is a particle.

The natural question to ask is then: what is energy? Intuitively, we can understand energy as being the Noether charge associated with time-translation symmetry—*i.e.*, energy is the conserved quantity induced by a timelike Killing field.

We now know what to expect. In a stationary spacetime, we have a timelike Killing field. Hence, in some sense we have an available notion of energy. This notion of energy can then be used to induce a preferred notion of particles, which leads us to a natural choice of Fock space. Formally, one defines the vacuum $|0\rangle$ by imposing it is annihilated by all annihilation operators and defines every other state in the Fock space by applying creation operators. See the previously mentioned references for a more rigorous approach.

There is, however, an interesting issue with this discussion. Consider Minkowski spacetime once again. The Minkowski vacuum is the unique Poincaré invariant state. Nevertheless, if we chose to restrict our attention to the right Rindler wedge as we often did in Section 2.4, then we would also be dealing with a different stationary spacetime and would be able to construct a vacuum that is invariant under the boost symmetry. This is known as the Rindler vacuum. Do these two states coincide?

They do not. By construction, the notion of time employed in the definition of the Rindler vacuum is the notion of proper time of an accelerated observer. On the other hand, Minkowski vacuum is built upon the notion of time as defined by inertial observers. Hence, the Rindler vacuum corresponds to the quantum state of the field in which an accelerated observer would see no particles. The Minkowski vacuum corresponds to the state of the field in which an inertial observer would see no particles. The Unruh effect proves that these two states do not coincide.

There are still more striking differences. The boost symmetries of Minkowski spacetime have a geometric structure known as a "bifurcate Killing horizon". Roughly speaking, this means the Killing field becomes null on a pair of crossing hypersurfaces, as depicted on Fig. 2.4 on page 29. More detailed definitions can be found, *e.g.*, in the discussions by Kay and Wald (1991, Sec. 2) and Wald (1994, Sec. 5.2). Other examples of spacetimes with Killing horizons are Schwarzschild, De Sitter, Schwarzschild–De Sitter, and Kerr spacetimes, among others. It was shown by Kay and Wald (1991) that spacetimes with such a structure admit at most one quasifree state that is both stationary and Hadamard. A Hadamard state is a state which allows the renormalization of nonlinear observables, such as the stress-energy-momentum tensor, and hence all physical states must be Hadamard states (see Section 2.B for more details).

Since both the Minkowski and Rindler vacua are stationary quasifree states, one of them must fail to be Hadamard. It does happen that the Rindler vacuum has an unphysical build up of energy near the null hypersurfaces $\mathfrak{h}_A$ and $\mathfrak{h}_B$ (see Birrell and Davies 1982, Eq. (6.157)).

Notice that this uniqueness result does not imply existence, and our earlier statement about existence of stationary states does not imply they are Hadamard. In fact, Kay and Wald (1991) have also shown there are no stationary Hadamard states on Kerr or Schwarzschild–De Sitter spacetimes.

## 2.B Introducing Nonlinear Observables

### Stress-Energy-Momentum Tensor

As we noted when constructing the algebra for free fields in Section 2.3, there is a mathematical difficulty in dealing with products of fields at the same point in spacetime, such as $\phi(x)\phi(x)$. Nevertheless, one often wants to consider these expressions. For example, the stress-energy-momentum tensor for a minimally coupled scalar field is given by (Wald 1984, Eq. (4.3.10))

$$T_{ab} = \nabla_a \phi \nabla_b \phi - \frac{1}{2} g_{ab} \left( \nabla_c \phi \nabla^c \phi + m^2 \phi^2 \right), \tag{2.B.1}$$

which involves the product of fields at the same point. Hence, if we desire to make sense of the very physical quantity that is the stress tensor, we must find a way of dealing with these issues.

This "bad behavior" in the ultraviolet (UV) is not exclusive to curved spacetimes. QFT in flat spacetimes also has these same difficulties. For example, let us consider the Hamiltonian as defined by an inertial observer in Minkowski spacetime. From Eq. (2.B.1) we see it is given by

$$H = \frac{\dot{\phi}^2}{2} + \frac{(\vec{\nabla}\phi)^2}{2} + \frac{m^2 \phi^2}{2}, \tag{2.B.2}$$

where the dot denotes a time derivative in the inertial observer's Cartesian coordinate system. Suppose we want to compute the expectation value of this Hamiltonian at some state $|\psi\rangle$. We desire to compute the quantity

$$\langle \psi | H | \psi \rangle = \frac{1}{2} \langle \psi | \dot{\phi}^2 + (\vec{\nabla}\phi)^2 + m^2 \phi^2 | \psi \rangle. \tag{2.B.3}$$

Nevertheless, this will lead to the presence of an infinite constant. Weinberg (2015, Sec. 11.6) writes this constant explicitly for the case of the electromagnetic field and attributes it to the vacuum fluctuations of the field. This is the main idea behind the concept of normal ordering, in which one decides to consider the quantity

$$\langle \psi | {:} H {:} | \psi \rangle \equiv \langle \psi | H | \psi \rangle - \langle 0 | H | 0 \rangle, \tag{2.B.4}$$

which intuitively means that one is interested not in the absolute energy of the field in the state $|\psi\rangle$, but only on the difference in energy between this state and the vacuum $|0\rangle$. The operator $:H:$ is said to be normal-ordered, and this term can also be defined in terms of manually moving the annihilation operators that occur in the expression for $H$ to the right (Weinberg 1995, p. 200). In many books, including the one by Weinberg (2015), this presentation is shortly followed by the comment that this treatment is reasonable as long as gravitation is not being taken into account. This, however, is not our case.

Wald (1977) dealt with this issue by introducing an axiomatic prescription to renormalize the stress energy tensor. One lists the properties desired for the expectation value of the stress tensor and then proceeds to discuss the existence and uniqueness of a prescription satisfying these properties. The original axioms given by Wald (1977) have later been adapted, for one of them cannot be satisfied (Wald 1978) and another one (the causality axiom) could eventually be generalized for a larger class of spacetimes by employing an algebraic philosophy. Our discussion will consider the axioms as presented by Wald (1994).

Firstly, notice that we can write the stress tensor in terms of a product of fields $\phi(x)\phi(x')$. Namely, we may write Eq. (2.B.1) on the preceding page in the form

$$T_{ab} = \lim_{x' \to x} \left\{ \left[ \nabla_a \nabla'_b - \frac{1}{2} g_{ab} \left( \nabla_c \nabla'^c + m^2 \right) \right] \phi(x)\phi(x') \right\}. \tag{2.B.5}$$

Therefore, we can relate the problem of computing $\omega(T_{ab})$ to the simpler task of computing $\omega(\phi^2)$. Our first requirement of a renormalized stress tensor is that this is consistent in a sense similar to how normal ordering works in flat spacetime—namely, our renormalization prescription allows one to compute the difference in energy (and so on) between any two "well-behaved" states.

Next, the stress tensor is local, and hence so should be its expectation value. Wald (1977) originally implemented this requirement by assuming the existence of natural Hilbert spaces in the far future and past—an assumption that is reasonable when the spacetime is asymptotically stationary, for example. Nevertheless, it is possible to provide a more general statement by employing the parlance of the algebraic approach.

In addition to these, we expect the expectation value of the stress tensor to be locally conserved, just like its classical counterpart. Hence, it should respect a continuity equation.

Lastly, the renormalized stress tensor in the Minkowski vacuum should vanish identically.

With these ideas in mind, we get to the four Wald axioms:

i. whenever $D(x, x') = \omega_1(\phi(x)\phi(x')) - \omega_2(\phi(x)\phi(x'))$ is a smooth function, the expression

$$\omega_1({:}T_{ab}{:}) - \omega_2({:}T_{ab}{:}) = \lim_{x' \to x} \left\{ \left[ \nabla_a \nabla'_b - \frac{1}{2} g_{ab} \left( \nabla_c \nabla'^c + m^2 \right) \right] D(x, x') \right\} \tag{2.B.6}$$

    holds;

ii. the expectation value $\omega({:}T_{ab}{:})$ depends locally on $\omega$ (see Wald 1994, p. 89, for details);

iii. $\nabla^a \omega(T_{ab}) = 0$ holds for all states;

iv. in Minkowski spacetime, $\langle 0_M | {:}T_{ab}{:} | 0_M \rangle = 0$.

These axioms are enough to provide a unique renormalization prescription up to local curvature terms. Indeed, suppose one has two renormalization prescriptions, which we shall denote through ${:}T_{ab}{:}_A$ and ${:}T_{ab}{:}_B$, both of them satisfying the four Wald axioms. The first axiom ensures that given two states $\omega_1$ and $\omega_2$ with $\omega_1(\phi(x)\phi(x')) - \omega_2(\phi(x)\phi(x'))$ being smooth, we have

$$\omega_1({:}T_{ab}{:}_A) - \omega_1({:}T_{ab}{:}_B) = \omega_2({:}T_{ab}{:}_A) - \omega_2({:}T_{ab}{:}_B), \tag{2.B.7}$$

which implies the difference

$$t_{ab} = \omega(:T_{ab}:_A) - \omega(:T_{ab}:_B) \tag{2.B.8}$$

is independent of the particular state of the field. The second axiom then ensures that $t_{ab}$ can depend only on the local geometry, and hence it must be constructed out of local curvature terms and of the metric tensor.

The third and fourth axioms then play the role of getting this prescription to a more convenient form. The third axiom implies $t_{ab}$ has to be conserved, hence restricting the allowed curvature terms. The fourth axiom takes care of an eventual cosmological constant term by setting it to $\Lambda = 0$, but could be modified to allow other possibilities.

The only natural length scale that we can consider at this stage is given by the field's mass, $m$. Assuming the ambiguities are continuous as $m \to 0$, we then have, on dimensional terms, that the possible contributions to the stress tensor are either $m^2 G_{ab}$ or the terms obtained by differentiating the Lagrangians[*] $R^2$ and $R_{ab}R^{ab}$. The first term can be absorbed in Newton's constant, while the other terms are ambiguities that cannot be resolved without further input from a theory of quantum gravity. Notice this is the same ambiguity that renders quantum GR one-loop non-renormalizable in the presence of matter (Percacci 2017, Sec. 3.5).

### Hadamard States

While the Wald axioms provide an uniqueness result on the renormalization of the stress tensor, we still could not establish existence. To do so, we can try to make sense of $\omega(\phi(x)\phi(x'))$ and then use Eq. (2.B.5) on the preceding page to obtain the stress tensor. Due to the fact that quantum fields are operator-valued distributions, this means we must still find a way of getting away with the issue of multiplying distributions.

When trying to multiply distributions, we know there are impossible cases. For example, one has

$$\delta(x)\delta(x) = \delta(0)\delta(x), \tag{2.B.9}$$

which leads to meaningless expressions whenever we try to integrate it against a smooth function. Nevertheless, there are situations in which we can multiply distributions. For example, any smooth function $f$ can be considered as a distribution, but we can still write

$$f(x)\delta(x) = f(0)\delta(x) \tag{2.B.10}$$

and get a sensible expression. We can even get two distributions that are not smooth functions and multiply them together, such as

$$\delta(x)\delta(x-1) = \delta(1)\delta(x-1) = 0. \tag{2.B.11}$$

One could argue that the only reason we have been able to give these examples is that they are cleverly designed to avoid getting two singularities at the same point. However, while that is a sufficient condition, it is not necessary. Consider the distributions

$$\varphi_{\pm}(x) = \underset{\varepsilon \to 0^+}{\text{w-lim}} \frac{1}{x \pm i\varepsilon}. \tag{2.B.12}$$

---

[*]One could also consider the Lagrangian $R^{abcd}R_{abcd}$, but it turns out that in four dimensions one can write this term in function of $R^2$, $R_{ab}R^{ab}$ and a topological invariant (see Percacci 2017, Sec. 2.4). However, topological invariants do not depend on the metric, only on the topology, and hence they will not contribute to the equations of motion.

Let us consider these distributions on $(-1, 1)$ and apply it to the function $b(x) = 1$. Then we have

$$\varphi_\pm(b) = \lim_{\varepsilon \to 0^+} \int_{-1}^{+1} \frac{1}{x \pm i\varepsilon} \, \mathrm{d}x = \lim_{\varepsilon \to 0^+} \left[ \mathrm{Log}(1 \pm i\varepsilon) - \mathrm{Log}(-1 \pm i\varepsilon) \right] = \mp i\pi, \qquad (2.B.13)$$

and hence we see $\varphi_\pm$ indeed diverge only in a distributional sense. Curiously, though, they admit being squared. Indeed, if we write

$$\varphi_\pm^2(x) = \operatorname*{w-lim}_{\varepsilon \to 0^+} \frac{1}{(x \pm i\varepsilon)^2}, \qquad (2.B.14)$$

which is the expected definition of "square" in this case, then we have

$$\varphi_\pm^2(b) = \lim_{\varepsilon \to 0^+} \int_{-1}^{+1} \frac{1}{(x \pm i\varepsilon)^2} \, \mathrm{d}x = \lim_{\varepsilon \to 0^+} \left[ -\frac{2}{1 + \varepsilon^2} \right] = -2, \qquad (2.B.15)$$

which is finite. Since the divergent behavior only occurs at $x = 0$ and our test function is finite there, we know we would get a finite result for any other test function that remains well-defined at $x = 0$.

One might then imagine $\varphi_\pm$ have some property that allows us to multiply them by other distributions. However, that is not quite the case. For example,

$$(\varphi_+ \varphi_-)(x) = \operatorname*{w-lim}_{\varepsilon \to 0^+} \frac{1}{x^2 + \varepsilon^2} \qquad (2.B.16)$$

fails to define a distribution. Indeed,

$$(\varphi_+ \varphi_-)(b) = \lim_{\varepsilon \to 0^+} \int_{-1}^{+1} \frac{1}{x^2 + \varepsilon^2} \, \mathrm{d}x = \lim_{\varepsilon \to 0^+} \frac{2}{\varepsilon} \arctan \varepsilon \to \infty. \qquad (2.B.17)$$

What is happening here? The trick is that $\varphi_\pm^2$ can be defined because we are also choosing distributions such that their singularities avoid each other, but now in a more subtle manner. While we are multiplying two distributions that are singular at the same point in space, their behaviors in Fourier space are compatible. An interesting property of Fourier transforms is that they are capable of translating singularities and smoothness into decay properties. Smooth functions have Fourier transforms that decay rapidly, faster than any polynomial. The Dirac delta—the stereotypical singular "function"—has a constant Fourier transform. This is exploited in the mathematical field of microlocal analysis to characterize the singularities of distributions in a more detailed manner, allowing one to identify in which directions of Fourier space a distribution is singular. Two distributions might happen to have "compatible" singular directions, in which case we are still able to give meaning to their product. More specifically, if the Fourier transform of one of the distributions does not decay rapidly in the direction $k$ in Fourier space, then the other distribution must decay sufficiently fast in the $-k$ direction to compensate it. This is known as Hörmander's criterion (Hörmander 2003, Theorem 8.2.10; Strohmaier 2009, Corollary 3).

To get a grasp of Hörmander's criterion, recall that the Fourier transform of a product is related to a convolution through a formula of the form

$$\mathscr{F}[f g](\xi) \propto \int \mathscr{F}[f](k) \mathscr{F}[g](\xi - k) \, \mathrm{d}^d k, \qquad (2.B.18)$$

where the proportionality constant depends on the choice of normalization of the Fourier transform $\mathscr{F}$. Hence, for $\mathscr{F}[fg](\xi)$ to remain finite, $\mathscr{F}[f](k)$ has to balance out the growth of $\mathscr{F}[g](\xi - k)$ and vice-versa[*]. In our previous example using the distributions $\varphi_\pm$ one has

$$\mathscr{F}[\varphi_\pm](k) \propto \mp i\Theta(\mp k), \tag{2.B.19}$$

and hence Hörmander's criterion was exactly what allowed or forbid us from multiplying them together. The constant of proportionality once again depends on the conventions taken for the Fourier transform.

These techniques also admit generalizations to manifolds (see Campos 2018, Sec. 2.3; Hörmander 2003, Chap. 8; Strohmaier 2009), even though Fourier analysis itself is not often available in curved spacetimes. The trick is that we can always multiply any distribution by a smooth function of compact support, which allows us to focus on a compact region. In this small region, we can use the fact that spacetime is locally flat to employ Fourier methods as if in flat space.

Using these techniques, one can summarize the divergence structure of distributions in an object known as the wavefront set. It is comprised of pairs $(x, k)$, where $x$ is a point in the manifold where the distribution is singular and $k$ is a singular direction in Fourier space. To determine the singular directions, one multiplies the distribution by smooth functions of compact support to employ Fourier analysis and check the decay properties. At the end of the day, we are left with an object $\mathrm{WF}(\phi) \subseteq T^*\mathscr{M}$ encoding the singularities of the distribution. More details are given in the review by Strohmaier (2009), for example.

As an example, let us consider the wavefront set of the two-point function in Minkowski spacetime. This is a bi-distribution (for it includes two fields), and hence it will be an object of the form $\mathrm{WF}(W_2) \subseteq T^*\mathscr{M} \times T^*\mathscr{M}$. It is given by (Strohmaier 2009, p. 118)

$$\mathrm{WF}(W_2) = \big\{ (x, k_a; y, -k_a); k_a \neq 0, k_a k^a = 0, (x - y)^\mu = \lambda k^\mu, \lambda \in \mathbb{R}, k_0 < 0 \big\}. \tag{2.B.20}$$

The first thing we can see from this wavefront set is that it violates Hörmander's criterion if we take $y \to x$. In that case, we will have singularities along opposite directions $\pm k_a$, and hence we will end up with an ill-defined object. Therefore, we cannot consider $W_2(x, x) = \omega(\phi(x)\phi(x))$ in this naive manner. However, as we know from QFT in flat spacetime, we are still able to perform normal-ordering. This comes by defining the object

$$\mathop{:}\phi(x)^2\mathop{:} \coloneqq \lim_{y \to x} \big[\phi(x)\phi(y) - W_2(x, y)\mathbb{1}\big]. \tag{2.B.21}$$

Notice we are carefully removing the singularities as we take the limit (although a rigorous treatment would require one to explain in which manner the limit should be taken).

The key behavior of the Minkowski vacuum and its wavefront set comes when we study what happens once we start playing around with $\mathop{:}\phi(x)^2\mathop{:}$. Notice, for example, that

$$\omega(\mathop{:}\phi(x)^2\mathop{::}\phi(y)^2\mathop{:}) = 2W_2(x, y)^2, \tag{2.B.22}$$

as one can find with an explicit calculation by employing the usual normal-ordering prescription—*i.e.*, by writing the fields using a Fourier decomposition and manually moving all annihilation operators to the right. We then see that to get an algebraic structure with the normal-ordered products we need it to be possible to square the two-point function. Eq. (2.B.20) on the preceding page is such that this is possible away from the coincidence limit[*]. Hence, the singularities of the Minkowski two-point function, while existent, are sufficiently well-behaved to allow us to deal with normal-ordered operators.

---

[*]If $f$ and $g$ are distributions of compact support, then their Fourier transforms are ensured to be smooth functions of polynomial growth (Strohmaier 2009, Theorem 5). As we will soon notice, distributions of compact support are sufficient for what we are interested in.

We then want to impose that physical states in general spacetimes should mimic these properties of the Minkowski vacuum. This is due to the following facts:

i. spacetime is locally flat, and hence it is reasonable to expect the UV behavior of physical states should somehow resemble the Minkowski vacuum;

ii. we desire to be able to extend the algebra of observables to nonlinear observables, and this can be done if the two-point functions are singular "in the right way".

This leads us to the microlocal spectrum condition (μSC). A state is said to satisfy the μSC—or, equivalently, to be a Hadamard state—if the wavefront set of its two-point function has the form (Khavkine and Moretti 2015, Eq. (5.76))

$$\mathrm{WF}(W_2) = \left\{ (x, p_a\, ; y, -q_a) \in T^*\mathcal{M} \times T^*\mathcal{M}; (x, p_a) \sim (y, q_a), p_a \rhd 0 \right\}. \qquad (2.\mathrm{B}.23)$$

In the previous expression, $(x, p_a) \sim (y, q_a)$ means there is a null geodesic $\gamma$ from $x$ to $y$ which is cotangent to $p_a$ at $x$ and $q_a$ is the parallel transport of $p_a$ from $x$ to $y$ along $\gamma$ with respect to the Levi-Civita connection. Furthermore, $p_a \rhd 0$ means $p_a$ is non-vanishing and future-directed, i.e., $p_a v^a \leq 0$ for all future-directed vectors $v^a$. Finally, we admit there are infinitely many null geodesics satisfying these requirements from $x$ to itself, corresponding to different covectors $p_a \in T^*_x\mathcal{M}$. Hence, the two-point function is singular along all future-directed null directions in the coincidence limit.

The μSC was introduced by Radzikowski (1992, 1996) and Radzikowski and Verch (1996) as an adaptation to the language of microlocal analysis of a previous local condition. The reviews by Campos (2018), Khavkine and Moretti (2015), and Strohmaier (2009) discuss in more details both approaches, how they relate to each other, and some of the historical advances made possible by employing methods from microlocal analysis.

The fact that we are restricting our interest to states with a well-defined singularity structure means we are capable of finding a bidistribution with a similar singularity structure, know as a Hadamard parametrix $H(x, x')$. This uses techniques introduced by Hadamard (1923), which explains the terms "Hadamard parametrix" and "Hadamard states". Using such a bidistribution, one can then get a notion of state-independent normal-ordering in curved spacetime: instead of subtracting the expectation value in a preferred vacuum state, one subtracts the Hadamard parametrix to obtain physically meaningful expressions. Further details can be found, e.g., in the review by Khavkine and Moretti (2015).

Only Hadamard states allow us to compute nonlinear observables, and hence they are the physically relevant states. Other states present singular behavior and will lead, for example, to the impossibility of computing physically sensible expressions for the stress-energy-momentum tensor.

Notice, however, that the region in which the state is singular might happen to be unphysical. This is the case for the Unruh and Boulware vacua on Schwarzschild spacetime mentioned at the end of Section 2.4. While they are not Hadamard states—the only Hadamard state on Schwarzschild spacetime is the Hartle–Hawking vacuum—their singular regions are unphysical for the situations in which they are interesting.  The Unruh vacuum models a quantum field in a collapsing star spacetime and is singular on the non-existing past event horizon.  Similarly, the Boulware vacuum is singular on the past and future event horizons, but this is not an issue since it is useful for modelling the quantum field in a stable star or planetary spacetime.  Therefore, while only Hadamard states are physical, there is no issue if the state turns out to be singular at an unphysical region of spacetime.  This is an issue with the simplifications we make when describing the spacetime, not with the state itself.

---

[*]At the coincidente limit, we need to renormalize the operator.

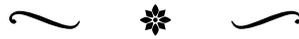

# Three

# Functional Renormalization Group

*We shall understand the general notions behind the renormalization group through the solutions of an ordinary differential equation. The case of field theory is introduced in its nonperturbative formulation, including a derivation of the Wetterich equation for arbitrary families of fields of arbitrary statistics. Methods for extracting information from the Wetterich equation are discussed in the end.*

> Therefore, conclusions based on the renormalization group arguments concerning the behavior of the theory summed to all orders are dangerous and must be viewed with due caution.
> So is it with all conclusions from local relativistic field theories.
>
> Bjorken and Drell (1965, p. 376).

> The set of transformations (50.22)–(50.24) is called the renormalization group. Rarely has there been a more pretentious name in the history of physics. It's like calling classical dynamics "the study of the Hamiltonian group of time translations". Nevertheless, that's what it's called.
>
> Coleman (2018, p. 1100).

The beauty and usefulness of theoretical physics lie in the fact that it allows for a few measurements to yield a myriad of predictions. By testing these new predictions against new measurements, one can then test the theory itself. The more the theory gets these predictions right, the more we trust it, and the more reasonable it is to assume it is an accurate model for how the Universe behaves. This is the most basic purpose of a physical theory: to gather, from a few inputs, sufficient information to make a plethora of predictions related to a certain class of phenomena. For example, the standard model (SM) takes 27 experimental inputs (Schwartz 2014, p. 641), but once these are fixed, one can make new predictions that can then be used to test the SM itself.

Interestingly, we are even able to use measurements at a certain scale and obtain results at other scales. In here, "scale" has a really wide meaning. In classical physics, we can use initial conditions to predict what happens at arbitrarily long times. Within high energy physics (HEP), one often uses measurements of the fine structure constant at low energies as inputs in precision tests of the SM (Particle Data Group et al. 2022, p. 178), a theory that works at far higher energies. There are, however, some caveats that one should consider.

One of the caveats has to do with the breakdown of perturbation theory. As pointed out in the book by Weinberg (1996, Chap. 18), $n$-loop amplitudes in quantum field theory (QFT) involving external momenta of order $q$ typically involve factors of the form $\alpha^n \left| \log\left(\frac{q^2}{m^2}\right) \right|^n$, where $m$ is the char-acteristic mass or energy scale of the theory and $\alpha$ is the perturbative parameter. This means that if $q^2 \gg m^2$, perturbation theory may become inconsistent even if $\alpha \ll 1$. Therefore, going to high energies can lead to issues.

The second caveat is more technical and can be understood as we investigate the first. As it turns out, regardless of perturbation theory, the behavior of a QFT under change of scale is far less trivial than what one has in classical theories. As pointed out in the pedagogical review by Delamotte (2004), the merge of quantum mechanics (QM) and special relativity leads to a new length scale in theories with mass scales: the Compton length $\frac{\hbar m}{c}$, which vanishes both for $\hbar \to 0$ and $c \to +\infty$. As a consequence, for length scales smaller than the Compton length, particle production and annihilation effects can kick in and significantly alter the behavior of the theory. These quantum fluctuations render a QFT harder to analyze than a classical theory would be.

## 3.1    Renormalization Group for a Differential Equation

Let us begin with the first caveat: the breakdown of perturbation theory at large scales. To do so, instead of using a field theoretical model, we shall consider a toy model presented by Delamotte (2004). Consider the initial value problem

$$\begin{cases} \dot{y}(t) = \epsilon y(t), \\ y(t_0) = r_0, \end{cases} \tag{3.1.1}$$

where $\epsilon \ll 1$ is a small parameter playing the role of a coupling constant. We know how to solve such a system exactly. The exact solution is given by

$$y(t) = e^{\epsilon(t-t_0)} r_0. \tag{3.1.2}$$

Nevertheless, suppose we did not know this. After all, we often cannot solve QFT calculations so easily. We could then try a "Dyson series approach". We notice that Eq. (3.1.1) implies

$$y(t) = r_0 + \int_{t_0}^{t} \dot{y}(t') \, dt', \tag{3.1.3a}$$

$$= r_0 + \epsilon \int_{t_0}^{t} y(t') \, dt', \tag{3.1.3b}$$

$$= r_0 + \epsilon \int_{t_0}^{t} \left[ r_0 + \epsilon \int_{t_0}^{t'} y(t'') \, dt'' \right] dt', \tag{3.1.3c}$$

$$= r_0 + r_0 \epsilon(t-t_0) + \epsilon^2 \int_{t_0}^{t} \int_{t_0}^{t'} y(t'') \, dt'' \, dt', \tag{3.1.3d}$$

$$= r_0 + r_0 \epsilon(t-t_0) + \frac{r_0 \epsilon^2 (t-t_0)^2}{2} + \epsilon^3 \int_{t_0}^{t} \int_{t_0}^{t'} \int_{t_0}^{t''} y(t''') \, dt''' \, dt'' \, dt', \tag{3.1.3e}$$

and so on. We could proceed writing the series explicitly and eventually notice it is just an exponential, but since we are interested in a perturbative approach, let us truncate the expression at order $\epsilon$. We learn that

$$y(t) = r_0 + \epsilon(t-t_0) + \mathcal{O}(\epsilon^2). \tag{3.1.4}$$

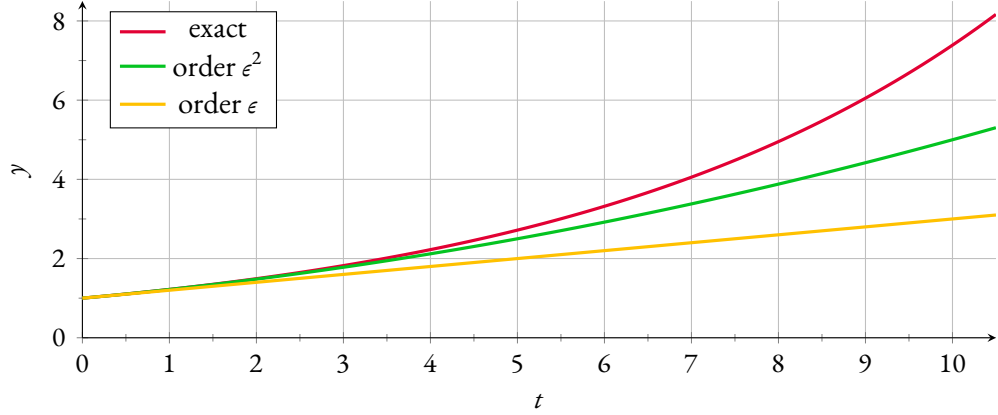

Figure 3.1: Comparison between the solutions to Eq. (3.1.1) on the facing page for $r_0 = 1$, $\varepsilon = 0.2$, and $t_0 = 0$. Notice that the exact solution (top) grows faster than both the order $\varepsilon^2$ approximation (middle) and the order $\varepsilon$ approximation (bottom). Even though $\varepsilon$ is small and the perturbative approach works at first, for long times there is a secular behavior that destroys the approximation.

This approximation is compared to the exact solution in Fig. 3.1, which also displays the solution up to order $\varepsilon^2$. In both cases, one notices that the approximation departs considerably from the exact solution at large values of $t$.

The issue at hand is that the perturbative expansion is not only on $\varepsilon$, but rather on $\varepsilon(t - t_0)$. At every order in perturbation theory $\varepsilon$ is accompanied by a term that diverges as $t \to +\infty$. Hence, at late times, the perturbative expansion will break down.

We then learn that a perturbative approach may become problematic if we attempt at using it to compute observables (such as $y(t)$) on scales too different from the experimental inputs we give (such as $y(t_0) = r_0$). While in principle we can pick an initial condition at whatever value of $t$ we desire, in practice one can improve approximation methods by keeping the scales close to each other.

Notice that this issue is similar to what can happen when solving a differential equation numerically using Euler's method (Burden, Faires, and Burden 2016, Sec. 5.2; Vesely 2001, Sec. 4.1.1), for example. While any initial condition is good in principle, in practice the approximations lead to an error build-up as one considers the behavior at late times.

### Beta Function

To improve our procedure, we would like to give $r_\tau = y(\tau)$ as input, where $\tau$ is some other finite time value intended to represent a different scale than $t_0$.

Up to order $\varepsilon$, we know

$$y(\tau) = r_\tau = r_0[1 + \varepsilon(\tau - t_0)] + \mathcal{O}(\varepsilon^2). \tag{3.1.5}$$

We can then solve for $r_0$ in terms of $r_\tau$ up to order $\varepsilon$. We find

$$r_0 = \frac{r_\tau}{1 + \varepsilon(\tau - t_0)} + \mathcal{O}(\varepsilon^2), \tag{3.1.6a}$$

$$= r_\tau[1 - \varepsilon(\tau - t_0)] + \mathcal{O}(\varepsilon^2). \tag{3.1.6b}$$

Hence, notice that

$$y(t) = r_0\big[1 + \varepsilon(t - t_0)\big] + \mathcal{O}(\varepsilon^2), \tag{3.1.7a}$$

$$= r_\tau\big[1 - \varepsilon(\tau - t_0)\big]\big[1 + \varepsilon(t - t_0)\big] + \mathcal{O}(\varepsilon^2), \tag{3.1.7b}$$

$$= r_\tau\big[1 + \varepsilon(t - \tau)\big] + \mathcal{O}(\varepsilon^2), \tag{3.1.7c}$$

and we can now express $y(t)$ directly in terms of $r_\tau$. This is useful if we are interested in values of $t$ closer to $\tau$ than to $t_0$. Nevertheless, this still leaves an issue: computing $r_\tau$ from $r_0$ still leads to an error that grows as the time difference $\tau - t_0$ increases.

A trick is to then use this procedure in infinitesimal steps. Instead of skipping from $t_0$ to $\tau$, we slide from $t_0$ to $t_0 + dt$. Or, more generally, we go from $t$ to $t + dt$, where $t$ can be any particular time instant. We then have

$$r_{t+dt} = r_t(1 + \varepsilon\, dt) + \mathcal{O}(\varepsilon^2). \tag{3.1.8}$$

Since we are doing this sliding of scale continuously, we can represent it as a differential equation. Notice that

$$dr_t = r_{t+dt} - r_t = \varepsilon\, dt + \mathcal{O}(\varepsilon^2), \tag{3.1.9}$$

and hence we can write the flow of $r_t$ as

$$\beta(r_t) \equiv \frac{dr_t}{dt} = \varepsilon r_t + \mathcal{O}(\varepsilon^2). \tag{3.1.10}$$

$\beta(r_t)$ is then said to be the beta function for the parameter $r_t$.

Incidentally, this is the differential equation we started with. For this simplified example, the sliding scale approach ends up being so good it leads us back to the exact solution. In a QFT calculation, the original problem would be more difficult and the beta function would provide a considerable simplification. Notice that Eq. (3.1.10) admits an expansion in $\varepsilon$ only, without influence of $t - t_0$. As a consequence, it can be used to correct our difficulties with perturbation theory in more complex problems.

## Renormalization Group Improvement

Another way of improving the behavior at a particular order is to notice that, at least in principle, the scale at which we give our experimental input should not really matter. For example, in the case of Eq. (3.1.1) on page 58, it is assured by the theory of ordinary differential equations that the initial value problem admits a unique solution. As a consequence, we can interpret the solution $y(t)$ in terms of a more general function $f \colon \mathbb{R} \times \mathbb{R} \to \mathbb{R}$. $f$ is the function that takes the initial condition $r_0$ and the time elapsed since the initial condition $t - t_0$, and gives the solution at $t$. Hence, $y(t) = f(r_0, t - t_0)$. In other words, $f(r_0, t - t_0)$ "propagates" $r_0$ to $r_t = y(t)$.

This strange reformulation is useful because we can now notice that

$$f(r_0, t - t_0) = f(f(r_0, \tau - t_0), t - \tau), \quad \text{for any } \tau, \tag{3.1.11}$$

which is just a fancy way of saying that the solution $y(t)$ could also be determined by imposing its value at $t = \tau$ rather than at $t = t_0$. Therefore, while Eq. (3.1.11) can be checked with the exact solution, notice that it holds as a matter of principle, not as a particular property of the exact solution. In fact, it must hold for any well-posed initial value problem, not only the one we are studying.

Eq. (3.1.11) on the facing page leads to a group structure. The elements are given by the functions $g_t(\cdot) = f(\cdot, t)$. They have the composition law $g_t \circ g_{t'} = f(f(\cdot, t'), t)$, which is associative as a consequence of associativity of the composition of functions. The neutral element is $g_0$, and the inverse of $g_t$ is $g_{-t}$.

From the point of view of the theory of differential equations, this is essentially a curiosity, and Eq. (3.1.11) on the preceding page may even be regarded as a trivial statement. However, we can use it to improve our perturbative solution.

At order $\epsilon$, we can approximate $f$ by the function

$$f_1(r_0, t - t_0) = r_0 + r_0 \epsilon (t - t_0), \tag{3.1.12}$$

and we then have $y(t) = f_1(r_0, t - t_0) + \mathcal{O}(\epsilon^2)$. Nevertheless, notice that, given $\tau \in \mathbb{R}$,

$$f_1(f_1(r_0, \tau - t_0), t - \tau) = f_1(r_0 + r_0 \epsilon (\tau - t_0), t - \tau) \tag{3.1.13a}$$

$$= r_0 + r_0 \epsilon (\tau - t_0) + (r_0 + r_0 \epsilon (\tau - t_0)) \epsilon (t - \tau) \tag{3.1.13b}$$

$$= r_0 + r_0 \epsilon (t - t_0) + r_0 \epsilon^2 (\tau - t_0)(t - \tau) \tag{3.1.13c}$$

$$= f_1(r_0, t - t_0) + r_0 \epsilon^2 (\tau - t_0)(t - \tau), \tag{3.1.13d}$$

and hence $f_1$ fails to satisfy Eq. (3.1.11) on the facing page at order $\epsilon^2$.

We know Eq. (3.1.11) on the preceding page must hold as a matter of principle. If it fails, this is due to perturbation theory. Hence, let us improve perturbation theory by imposing the validity of Eq. (3.1.11) on the facing page. We write

$$f_1^{\text{imp}}(r_0, t - t_0) \equiv r_0 + r_0 \epsilon (t - t_0) + r_0 \epsilon^2 G(t - t_0) \tag{3.1.14}$$

for some function $G$ that we wish to obtain. Notice that $G(0) = 0$ to ensure the initial condition $y(t_0) = r_0$. The superscript "imp" stands for "improved".

We now impose $f_1^{\text{imp}}$ satisfies Eq. (3.1.11) on the preceding page up to order $\epsilon^2$. Then it follows that

$$G(t - t_0) = G(\tau - t_0) + (\tau - t_0)(t - \tau) + G(t - \tau) + \mathcal{O}(\epsilon). \tag{3.1.15}$$

To solve this functional equation, we differentiate both sides with respect to $t_0$ and then set $t_0 = \tau$. We get

$$-G'(t - \tau) = -G'(0) - (t - \tau). \tag{3.1.16}$$

Denote $x \equiv t - \tau$. Then

$$G'(x) = x + G'(0). \tag{3.1.17}$$

This differential equation can be solved using the initial condition $G(0) = 0$—which $G$ must satisfy by hypothesis. We get

$$G(x) = \frac{x^2}{2} + ax, \tag{3.1.18}$$

for some constant $a$. One can then verify by direct substitution that this is a solution to Eq. (3.1.15).

Using this result, we find that

$$y(t) = r_0 + r_0 \epsilon (t - t_0) + \frac{r_0 \epsilon^2 (t - t_0)^2}{2} + a r_0 \epsilon^2 (t - t_0) + \mathcal{O}(\epsilon^2), \tag{3.1.19}$$

which gets the term of highest degree in $t - t_0$ correctly at the next order in perturbation theory. This is, of course, the most relevant term at order $\varepsilon^2$, and hence it represents an improvement over our previous solution.

We can once again make a comparison with the Euler method for solving differential equations numerically. Our present procedure in to impose that, as a matter of principle, Eq. (3.1.11) on page 60 must hold. Any violations of it are due to our approximation methods, and hence imposing its validity can improve our results. Similarly, the Euler method often fails to conserve energy, or to describe the system's evolution by means of a canonical transformation. Forcing the numerical method to respect the symplectic nature of time evolution can lead to improvements when solving Hamilton's equations numerically (see Vesely 2001, Sec. 4.2.5; Yoshida 1993).

## 3.2   Functional Renormalization Group

As with the differential equation we used as an example, in field theory calculations one often faces the issue of a perturbative expansion breaking down as the scale differences grow. Just as with the differential equation, a trick to obtain sensible results is to work with quantities at the correct scale. For example, while the fine structure constant is measured at low energies, when it enters SM calculations it is corrected to the right scale (Particle Data Group et al. 2022, p. 178). This corresponds to partially resumming the perturbative series (see Delamotte 2004), and hence provides an improvement to perturbation theory. Within QFT and statistical physics, the renormalization group (RG) is the name given to the idea of adapting a theory to different scales.

The fact the RG improves perturbation theory is one of our main motivations to study the functional renormalization group (FRG) and later apply it to Unruh–DeWitt detectors in Chapter 4. As we shall discuss, we want to perform a perturbative calculation involving the detector, but we are interested in nonperturbative regimes in the sense that we aim at considering detectors strongly coupled to a quantum field. Hence, instead of performing a perturbative expansion in the coupling between detector and field, one can perform an expansion in the detector's energy gap. A nonperturbative formulation of the RG will then be useful to allow us to improve the results by allowing them to flow to some scale of interest, such as the Unruh temperature for a detector probing the Unruh effect as described in Section 2.4. Since we want to consider strongly-coupled detectors, a perturbative formulation of the RG would not be adequate.

The perturbative RG is discussed in many references—such as the books by Hollowood (2013), Peskin and Schroeder (1995), Schwartz (2014), and Weinberg (1996)—and is not of much relevance to us. Therefore, we shall introduce the ideas directly by means of a nonperturbative approach. It is common to refer to a nonperturbative formulation of the RG as the FRG. It is also interesting to mention that, while our initial motivation for studying the RG was in improving perturbation theory, we shall arrive at a different formulation of QFT that allows us to study nonperturbative physics in an efficient manner.

Our discussion is mainly inspired by those due to Percacci (2017), Pereira (2016), and Reuter and Saueressig (2018). Nevertheless, our goal is to eventually describe a particle detector, which requires the use of anticommuting variables. Pedagogical references about the FRG usually derive the Wetterich equation assuming the fields to be bosonic, with an exception being the book by Kopietz, Bartosch, and Schütz (2010), which is a bit more oriented towards applications in condensed matter. Hence, we shall promptly introduce the concepts for a general field content that can include both bosonic and fermionic fields. While this may be a bit more difficult at first contact, it fills a pedagogical gap in the FRG literature from the HEP point of view.

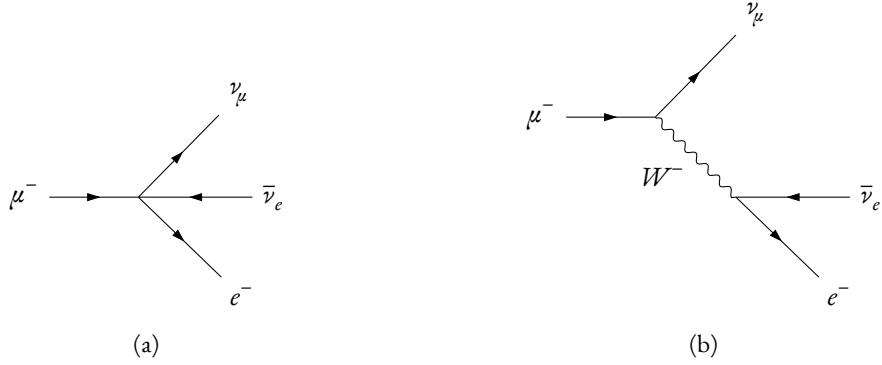

Figure 3.2: Diagrams for muon decay $\mu^- \rightarrow e^- + \bar{\nu}_e + \nu_\mu$. (a): in the Fermi theory of weak interactions; (b): in the SM. For the Fermi theory, the interaction between the four fermions happens at a single point, while for the SM it is mediated by the $W^-$ boson.

### Cutoffs and Effective Degrees of Freedom

The first description of weak interactions was given by Fermi (1933, 1934a,b). It consists of a four-fermion interaction and can be used to model decays such as $n^0 \rightarrow p^+ + e^- + \bar{\nu}_e$ and $\mu^- \rightarrow e^- + \bar{\nu}_e + \nu_\mu$. The muon decay, for example, is represented by the diagram on Fig. 3.2a. On the SM, however, the process is modelled differently. In this case, the theory describes the decay as being mediated by the $W^-$ boson, as depicted on the diagram given in Fig. 3.2b.

The $W$ boson has a mass $m_W = 80.377(12)$ GeV (Particle Data Group et al. 2022). At energies well below this threshold, it is not expected that the details of the interactions with the $W$ boson will be of any relevance, just like the detailed interactions in a water molecule are irrelevant when one is modeling the motion of the ocean. Hence, when performing calculations at low energies, we can simply use the Fermi theory, which provides a simpler description.

This is the point of view behind effective field theories (Burgess 2020). One can describe physics at a certain scale by considering only the degrees of freedom at that scale, without needing to bother about all of the details that happen at more fundamental scales. This decoupling of scales is the reason we can do classical physics without taking QM into account, and it lies at the heart of the RG.

To understand why, let us consider the general idea of ignoring the $W$ boson. How can we implement that in a field theory? By integrating out the undesired modes. For example, consider the generating functional for the sector of the SM we are interested in. It is given by

$$Z[J, K_+, K_-] = \int \exp\left(iS\left[\widehat{\Phi}, \widehat{W}^+, \widehat{W}^-\right] + \left\langle J, \widehat{\Phi} \right\rangle + \left\langle K_+, \widehat{W}^+ \right\rangle + \left\langle K_-, \widehat{W}^- \right\rangle\right) \mathcal{D}\widehat{\Phi} \, \mathcal{D}\widehat{W}^+ \, \mathcal{D}\widehat{W}^-, \quad (3.2.1)$$

where $\widehat{\Phi}$ denotes all other fields being considered (the muon, the electron, their neutrinos, etc), $J$ and $K_\pm$ denote classical sources, and $S$ denotes the theory's classical action. The relevant integrals and tensor contractions are understood in the inner products $\langle \cdot, \cdot \rangle$ between field and classical sources. We used a hat to clarify which fields are being integrated over. See Appendix A on page 137 for more information on this notation.

Suppose we are interested in probing only energy scales much smaller than $m_W$. In this case, the classical sources $K_\pm$ should be set to zero, for there is no sufficient energy around to excite on-shell $W$ bosons. Hence, we get

$$Z[J, 0, 0] = \int \exp\left(iS\left[\widehat{\Phi}, \widehat{W}^+, \widehat{W}^-\right] + \left\langle J, \widehat{\Phi}\right\rangle\right) \mathcal{D}\widehat{\Phi} \, \mathcal{D}\widehat{W}^+ \, \mathcal{D}\widehat{W}^- \,. \tag{3.2.2}$$

We may now perform the integral over the $W$ bosons and get to the generating functional

$$Z[J, 0, 0] = \int \exp\left(iS_{\text{eff.}}\left[\widehat{\Phi}\right] + \left\langle J, \widehat{\Phi}\right\rangle\right) \mathcal{D}\widehat{\Phi} \,, \tag{3.2.3}$$

where $S_{\text{eff.}}$ is an effective action. Rather than implementing the effects of the $W$ boson by means of the boson itself, we have "stored" the effects it could have on lower energy parameters. For example, integrating out $W$ leads to the occurrence of a term leading to the Fermi interaction depicted on Fig. 3.2a on the previous page (see Schwartz 2014, Sec. 29.4). In fact, the coupling constant of the Fermi interaction can be computed in terms of $m_W$ and other SM parameters (see Schwartz 2014, Eq. (29.73)).

It would be interesting to always work with a theory at the correct scale. Can we find a method to integrate out high energy modes without the need to integrate out a whole field at a time? In other words, is there a "continuous" way of doing this procedure?

Yes, there is. It lies at the basis of the formulation of the FRG that we shall implement. Nevertheless, to implement it, we will need to perform a Wick rotation and work in Euclidean spacetimes. The reason behind this necessity is technical convenience: we want to be able to have a clear way of defining whether a given four-vector $p^a$ lies above or below the scale we are interested in, and this is not possible in a straightforward manner in a Lorentzian manifold, since large energies might make up for large momenta and yield small values of $p^a p_a{}^*$. Nevertheless, as discussed in Section 2.5, asking for an Euclidean spacetime means we are restricting the available spacetimes, so this technical advantage does not come for free. Still, since the Unruh and Hawking temperatures often provide an adequate temperature scale at which we should consider a detector, it seems natural to use an Euclidean formulation for a particle detector.

We now wish to integrate out the modes above a certain scale $k$. How can that be achieved? With the $W$ boson, the basic idea was that its mass term would suppress the low-energy contributions. We can then generalize this idea by introducing an "artificial mass term". For example, for a scalar field in $d$ dimensions in Euclidean Minkowski spacetime, we can adapt the action $S[\phi]$ by adding to it a term

$$\Delta S_k[\phi] = \frac{1}{2} \int k^2 \phi^2(x) \, \mathrm{d}^d x \,. \tag{3.2.4}$$

Using this strategy, we can define a generating functional at the scale $k$ through

$$Z_k[J] = \int \exp\left(-S\left[\hat{\phi}\right] - \Delta S_k\left[\hat{\phi}\right] + \left\langle J, \hat{\phi}\right\rangle\right) \mathcal{D}\hat{\phi} \,, \tag{3.2.5}$$

where

$$\left\langle J, \hat{\phi}\right\rangle \equiv \int J(x)\hat{\phi}(x) \, \mathrm{d}^d x \,. \tag{3.2.6}$$

---

*In spite of this comment, D'Angelo et al. (2022) have recently proposed a variant of the Wetterich equation on Lorentzian manifolds by employing methods coming from perturbative algebraic quantum field theory (pAQFT). Nevertheless, the Euclidean formulation seems more adequate to study the systems in thermal equilibrium that one usually considers in QFTCS.

$Z_k$ is then defined through an integral that only takes into consideration the degrees of freedom at scales above $k$. Using this object, we shall soon be able to define an action suitable to be used at the scale $k$.

Using $k^2$ is actually not our only option. Instead of Eq. (3.2.4) on the facing page we can be somewhat more elaborate and work with

$$\Delta S_k[\phi] = \frac{1}{2} \int \phi(x) R_k(-\partial^2) \phi(x) \, \mathrm{d}^d x \propto \int \mathscr{F}[\phi](-p) R_k(p^2) \mathscr{F}[\phi](p) \, \mathrm{d}^d p \,, \tag{3.2.7}$$

where $\mathscr{F}[\phi]$ is the Fourier transform of $\phi$ and $R_k$ is a function of momentum that has some useful properties. It is typical to require that this function, often referred to as a cutoff or regulator, satisfies the following conditions (Percacci 2017, p. 130):

i.  for fixed $k$, $R_k(z)$ decays monotonically with $z$;

ii.  for fixed $z$, $R_k(z)$ grows monotonically with $k$;

iii.  $\lim_{k \to 0} R_k(z) = 0$, for any $z$;

iv.  for $z > k^2$, $R_k(z)$ vanishes sufficiently fast (usually exponentially fast);

v.  $R_k(0) = k^2$.

These conditions can be modified depending on the particular applications being considered.

Let us discuss why one typically wants these requirements. $R_k(z)$ should decay with $z$ because we want to suppress only the low-energy modes in the functional integral, but the high-energy modes should contribute as usual. This is also why we want $R_k(z)$ to vanish fast for $z > k^2$—the cutoff should not affect the high-energy modes. This will later protect the Wetterich equation from ultraviolet (UV) divergences.

$k$ determines the scale we are interested in. If we want to look at scales of larger $k$, the suppression must be stronger because we need to get rid of more modes. Hence, $R_k(z)$ is expected to grow with $k$. On the other hand, if $k = 0$, we want to integrate over all modes, and hence $\lim_{k \to 0} R_k(z) = 0$ is a reasonable requirement.

At last, $R_k(0) = k^2$ provides a normalization convention and protects the theory from infrared (IR) divergences.

Since both $R_k(z)$ and $z$ have dimension of mass squared, we can also define a dimensionless function $r$ through

$$R_k(z) = k^2 r\left(\frac{z}{k^2}\right). \tag{3.2.8}$$

In this manner, one can specify a cutoff by providing a "cutoff profile" $r(x)$. A few common choices are depicted in Fig. 3.3 on the next page.

We shall give particular emphasis to the cutoff profile

$$r(x) = (1 - x)\Theta(1 - x), \tag{3.2.9}$$

which is known as the Litim cutoff (Litim 2001). It is particularly convenient because it often allows to evaluate expressions explicitly.

When dealing with theories involving multiple fields we need to add cutoffs to all of them, but the general idea stays the same: we add quadratic terms to suppress low-energy modes. When working with multiple fields, or with fields that transform under some representation of a group, we will denote the cutoff matrix by $\mathscr{R}_k$. Hence, $R_k$ is always a scalar function, while $\mathscr{R}_k$ is a matrix in some suitable space. For example, the Litim cutoff for Dirac fermions is taken to be proportional to the $\gamma^\mu$ matrices (Litim [2001]). Using DeWitt's condensed notation (see Appendix A), we can write the cutoff term $\Delta S_k[\Phi]$ for a general collection of fields $\Phi$ as

$$\Delta S_k[\Phi] = \frac{1}{2} \Phi_\alpha \mathscr{R}_{k,\alpha\beta} \Phi_\beta, \tag{3.2.10}$$

where the indices $\alpha$ and $\beta$ run over field space, internal spaces, and spacetime. Alternatively, we shall often also write this expression as

$$\Delta S_k[\Phi] = \frac{1}{2} \langle \Phi, \mathscr{R}_k \Phi \rangle. \tag{3.2.11}$$

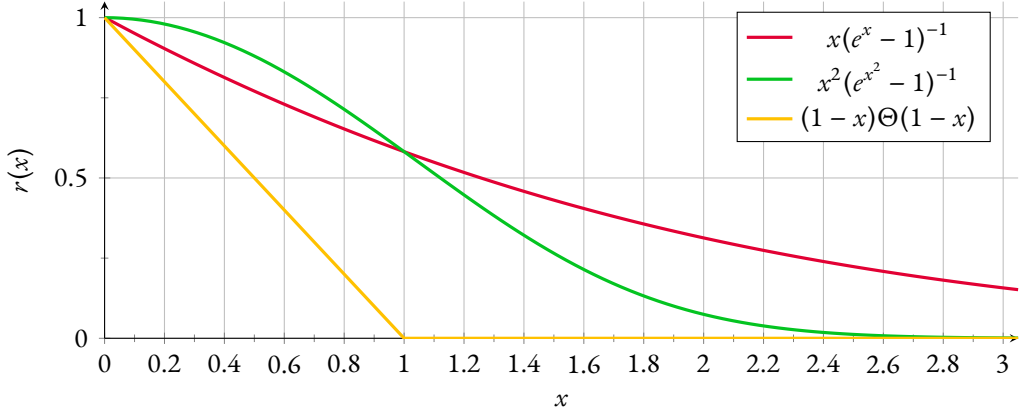

Figure 3.3: A few common cutoff profile choices.

This discussion can also be promptly generalized to curved spacetime. Details can be found, for example, in the book by Percacci ([2017], Sec. 6.6)

### Effective Average Action

With a clear concept of $Z_k[J]$, let us move on to obtain a notion of action appropriate to the scale $k$. To do this, we will exploit the other generating functionals available in QFT. Furthermore, from this point onward we will be working with a theory with a generic field content $\Phi$, which may include bosonic and fermionic fields $\Phi_i$. We shall extensively use the notation and results given in Section 3.A.

$Z_k[J]$ is often an unnecessarily complicated object. In full analogy with how we defined the connected generating functional in Section 3.A, we define $W_k[J]$ as

$$Z_k[J] = \exp W_k[J]. \tag{3.2.12}$$

$W_k[J]$ is an unusual object to work with in field theory, since it is a functional of the classical sources. As we did with the quantum action in Section 3.A, we now define a new object through

$$\tilde{\Gamma}_k[\Phi] \equiv \sup_J \left[ \langle J, \Phi \rangle - W_k[J] \right]. \tag{3.2.13}$$

This object resembles the quantum action as we defined in Section 3.A, but notice it still carries the regulator we added earlier when building $Z_k[J]$. Therefore, we now remove it by defining

$$\Gamma_k[\Phi] \equiv \tilde{\Gamma}_k[\Phi] - \Delta S_k[\Phi]. \qquad (3.2.14)$$

$\Gamma_k$ is referred to as the effective average action (EAA). It still depends on our choice of cutoff, but this dependence is analogous to the regularization scheme dependence that is usual in QFT. To discuss the EAA's meaning, let us first show a few of its properties that will be useful in order to understand it.

Firstly, consider the $k \to 0$ limit. In this case, by hypothesis, we will have $\Delta S_k \to 0$. Hence, we get $Z[J]$, and our definition of the EAA will reduce to the definition of the quantum action. Therefore,

$$\lim_{k \to 0} \Gamma_k = \Gamma. \qquad (3.2.15)$$

In other words, when we are not suppressing any modes, the EAA is simply the quantum action.

Next let us consider the other remarkable limit: $k \to \Lambda_{\mathrm{UV}}$, where $\Lambda_{\mathrm{UV}}$ is the UV cutoff for the theory. Let us denote by $\bar{J}$ the classical source $J$ that maximizes the expression in Eq. (3.2.13) on the facing page—notice it depends on $\Phi$. We can then write

$$Z_k[\bar{J}] = \exp W_k[\bar{J}], \qquad (3.2.16a)$$
$$= \exp\left(\langle \bar{J}, \Phi \rangle - \Delta S_k[\Phi] - \Gamma_k[\Phi]\right). \qquad (3.2.16b)$$

Therefore,

$$\exp\left(\langle \bar{J}, \Phi \rangle - \Delta S_k[\Phi] - \Gamma_k[\Phi]\right) = \int \exp\left(-S[\widehat{\Phi}] - \Delta S_k[\widehat{\Phi}] + \langle \bar{J}, \widehat{\Phi} \rangle\right) \mathscr{D}\widehat{\Phi}, \qquad (3.2.17a)$$
$$\exp\left(-\Gamma_k[\Phi]\right) = \int \exp\left(-S[\widehat{\Phi}] - \Delta S_k[\widehat{\Phi}] + \Delta S_k[\Phi] + \langle \bar{J}, \widehat{\Phi} - \Phi \rangle\right) \mathscr{D}\widehat{\Phi}. \qquad (3.2.17b)$$

Let us make a change of variables in the functional integral. We define $\widehat{\Xi} \equiv \widehat{\Phi} - \Phi$ and get

$$\exp\left(-\Gamma_k[\Phi]\right) = \int \exp\left(-S[\Phi + \widehat{\Xi}] - \Delta S_k[\Phi + \widehat{\Xi}] + \Delta S_k[\Phi] + \langle \bar{J}, \widehat{\Xi} \rangle\right) \mathscr{D}\widehat{\Xi}. \qquad (3.2.18)$$

Remember now that we defined $\Delta S_k$ as a term that is quadratic in the fields. Since it is only a quadratic term, we can express it as a Taylor expansion with the certainty that the series terminates, *i.e.*,

$$\Delta S_k[\Phi + \widehat{\Xi}] = \Delta S_k[\Phi] + \widehat{\Xi}_\alpha \frac{\partial \Delta S_k}{\partial \Phi_\alpha} + \frac{1}{2}\widehat{\Xi}_\alpha \frac{\overrightarrow{\partial} \Delta S_k \overleftarrow{\partial}}{\partial \Phi_\alpha \partial \Phi_\beta} \widehat{\Xi}_\beta, \qquad (3.2.19a)$$
$$= \Delta S_k[\Phi] + \widehat{\Xi}_\alpha \mathscr{R}_{k,\alpha\beta} \Phi_\beta + \Delta S_k[\widehat{\Xi}], \qquad (3.2.19b)$$
$$= \Delta S_k[\Phi] + \langle \widehat{\Xi}, \mathscr{R}_k \Phi \rangle + \Delta S_k[\widehat{\Xi}], \qquad (3.2.19c)$$

where the implicit sums and integrals over $\alpha$ and $\beta$ account for internal indices—such as spinorial or tensor indices, indices on a gauge group representation, etc—and for the spacetime dependency, just like in Section 3.A and Appendix A. The eventual fermionic character of the fields requires us to be careful with the sides from which we are differentiating each expression (for more on calculus of anticommuting variables, see DeWitt 2003, App. A).

We then get to

$$\exp\left(-\Gamma_k[\Phi]\right) = \int \exp\left(-S\big[\Phi + \hat{\Xi}\big] - \big\langle \hat{\Xi}, \mathscr{R}_k \Phi \big\rangle - \Delta S_k\big[\hat{\Xi}\big] + \big\langle \tilde{J}, \hat{\Xi} \big\rangle\right) \mathscr{D}\hat{\Xi}. \tag{3.2.20}$$

Had we done the same calculation for the quantum action, we would have found

$$\exp\left(-\Gamma[\Phi]\right) = \int \exp\left(-S\big[\Phi + \hat{\Xi}\big] + \big\langle \tilde{J}, \hat{\Xi} \big\rangle\right) \mathscr{D}\hat{\Xi}, \tag{3.2.21}$$

which is another route to seeing that $\lim_{k\to 0} \Gamma_k = \Gamma$.

We could follow the same arguments we did in Section 3.A when obtaining Eq. (3.A.16) on page 84 to conclude that

$$\tilde{J} = \mathbb{Z}\frac{\partial \tilde{\Gamma}_k}{\partial \Phi}, \tag{3.2.22}$$

where $\mathbb{Z}$ is the diagonal matrix with entries $+1$ for bosonic indices and $-1$ for fermionic indices defined on Eq. (3.A.12) on page 83. Notice Eq. (3.2.22) involves $\tilde{\Gamma}_k$, not $\Gamma_k$, for the EAA is not related to the connected generating functional simply by a Legendre transform: we also subtract the cutoff $\Delta S_k$ to get to it.

The general form of the cutoff $\Delta S_k$ is given on Eq. (3.2.11) on page 66. Notice then that

$$\tilde{J} = \mathbb{Z}\frac{\partial \Gamma_k}{\partial \Phi} + \mathbb{Z}\mathscr{R}_k\Phi. \tag{3.2.23}$$

If we use this expression on Eq. (3.2.20) alongside Eq. (3.A.14) on page 83, we find that

$$\exp\left(-\Gamma_k[\Phi]\right) = \int \exp\left(-S\big[\Phi + \hat{\Xi}\big] - \big\langle \hat{\Xi}, \mathscr{R}_k \Phi \big\rangle - \Delta S_k\big[\hat{\Xi}\big] + \Big\langle \mathbb{Z}\frac{\partial \Gamma_k}{\partial \Phi} + \mathbb{Z}\mathscr{R}_k\Phi, \hat{\Xi} \Big\rangle\right) \mathscr{D}\hat{\Xi}, \tag{3.2.24a}$$

$$= \int \exp\left(-S\big[\Phi + \hat{\Xi}\big] - \big\langle \hat{\Xi}, \mathscr{R}_k \Phi \big\rangle - \Delta S_k\big[\hat{\Xi}\big] + \Big\langle \mathbb{Z}\frac{\partial \Gamma_k}{\partial \Phi}, \hat{\Xi} \Big\rangle + \big\langle \hat{\Xi}, \mathscr{R}_k \Phi \big\rangle\right) \mathscr{D}\hat{\Xi}, \tag{3.2.24b}$$

$$= \int \exp\left(-S\big[\Phi + \hat{\Xi}\big] - \Delta S_k\big[\hat{\Xi}\big] + \Big\langle \mathbb{Z}\frac{\partial \Gamma_k}{\partial \Phi}, \hat{\Xi} \Big\rangle\right) \mathscr{D}\hat{\Xi}, \tag{3.2.24c}$$

$$= \int \exp\left(-S\big[\Phi + \hat{\Xi}\big] - \frac{1}{2}\big\langle \hat{\Xi}, \mathscr{R}_k \hat{\Xi} \big\rangle + \Big\langle \mathbb{Z}\frac{\partial \Gamma_k}{\partial \Phi}, \hat{\Xi} \Big\rangle\right) \mathscr{D}\hat{\Xi}. \tag{3.2.24d}$$

Let us now take the limit $k \to \Lambda_{\mathrm{UV}}$. In this case, we expect to have $\mathscr{R}_k \sim k^2$ up to some matrix structure. Indeed, in the scalar case we demanded that $R_k(0) = k^2$, and in the limit $k \to \Lambda_{\mathrm{UV}}$ we can safely assume $z \ll k^2$ for all modes, leading us to the behavior $R_k(z) \sim k^2$. The same idea applies to the more complicated case in which we are working with a matrix $\mathscr{R}_k$. This means we get to

$$\lim_{k\to\Lambda_{\mathrm{UV}}} \exp\left(-\Gamma_k[\Phi]\right) = \lim_{k\to\Lambda_{\mathrm{UV}}} \int \exp\left(-S\big[\Phi + \hat{\Xi}\big] - \frac{k^2}{2}\big\langle \hat{\Xi}, \hat{\Xi} \big\rangle + \Big\langle \mathbb{Z}\frac{\partial \Gamma_k}{\partial \Phi}, \hat{\Xi} \Big\rangle\right) \mathscr{D}\hat{\Xi}. \tag{3.2.25}$$

Notice $\exp\left(-\frac{k^2}{2}\big\langle \hat{\Xi}, \hat{\Xi} \big\rangle\right)$ is approaching a Dirac delta. Hence, in the limit we can write

$$\exp\left(-\Gamma_{\Lambda_{\mathrm{UV}}}[\Phi]\right) = \exp\left(-S[\Phi]\right), \tag{3.2.26}$$

$$\Gamma_{\Lambda_{\mathrm{UV}}}[\Phi] = S[\Phi]. \tag{3.2.27}$$

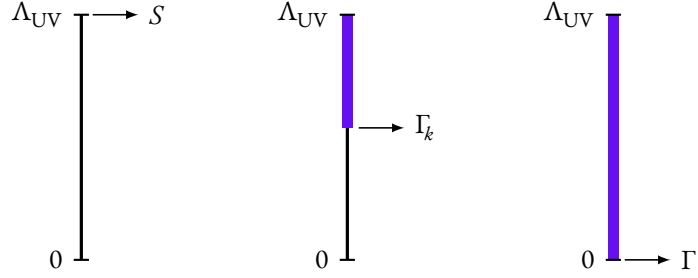

Figure 3.4: Integrated modes for the classical action $S$, for the EAA $\Gamma_k$, and for the quantum action $\Gamma$. The black thin lines illustrate all possible modes, while the thick purple lines illustrate which ones are integrated over in the path integrals defining each of the actions. The classical action can be understood as being at the UV cutoff $\Lambda_{UV}$ in the sense that it does not include corrections due to modes at any scales. On the other hand, the quantum action is associated with the IR cutoff at $k = 0$ because it includes the corrections from all quantum modes. The EAA interpolates between them by considering modes from the scale $k$ up to the UV cutoff $\Lambda_{UV}$.

Therefore, if we take $k \to 0$, the EAA becomes the quantum action. If we take $k \to \Lambda_{UV}$, the EAA becomes the classical action. We then see that the EAA is an interpolation between the classical and quantum actions.

This sheds some light on the meaning of the EAA: it is an action adequate for working at the scale $k$. This might seem counterintuitive at first, for it is unusual to think of the classical action as being a high-energy object, while the quantum action is a low-energy object. Why is this so?

This can be understood by thinking about our previous example concerning muon decay. If we want to deal with the theory at low energies, then we need to take the high-energy quantum effects into consideration by integrating them out. We are then left with an effective action adapted to work at low energies.

If we are working at $k \sim 0$, then all modes must be integrated out. In this case, we get to an action—the quantum action—that already accounts for all quantum corrections already at tree level. The effects of loops are now translated as new vertices, just like the effect of the $W$ boson had been accounted for by a new four-fermion vertex. On the other hand, if we are working at $k \sim \Lambda_{UV}$, we cannot integrate any modes out. Hence, we need to work with our bare action and manually consider all effects we are interested in. Similarly, $\Gamma_k$ is to be understood as an action suited for working at the scale $k$. This is illustrated in Fig. 3.4.

This should also explain the seemingly unusual choice of working with an arbitrary IR cutoff and integrating up rather than working with an UV cutoff and integrating down, as is usual in perturbative formulations of the Wilsonian RG. The reason is that to consider the full quantum theory we have to integrate over all modes up to the UV anyway. The FRG philosophy we are using is that if we are working at the scale $k$, we can first integrate over the modes with energies larger than $k$ and then use the resulting EAA on the path integral over lower energy modes.

### Wetterich Equation

To actually describe the RG flow, it is not convenient to work with this integral formulation. Instead, we can derive a differential equation describing how the EAA changes as one smoothly changes the scale of interest $k$. This will be a functional renormalization group equation (FRGE).

There are many possible FRGEs (for a review, see Kopietz, Bartosch, and Schütz 2010, Chap. 7). We are interested in working with the Wetterich equation. It was originally derived by Wetterich (1993), Bonini, D'Attanasio, and Marchesini (1993), and Morris (1994). However, Kopietz, Bartosch, and Schütz (2010, p. 193) mention other "essentially equivalent" expressions had been given earlier by Nicoll and Chang (1977), Nicoll, Chang, and Stanley (1974), and Weinberg (1978). The Wetterich equation is a functional differential equation for the EAA. Other FRGEs may involve other generating functionals or use different definitions for the effective action.

We begin with the definition of the EAA. Letting $\tilde{J}$ be the classical source that maximizes the supremum on Eq. (3.2.13) on page 66 we have

$$\Gamma_k[\Phi] = \langle \tilde{J}, \Phi \rangle - W_k[\tilde{J}] - \Delta S_k[\Phi], \tag{3.2.28a}$$

$$\partial_k \Gamma_k[\Phi] = -\partial_k W_k[\tilde{J}] - \partial_k \Delta S_k[\Phi]. \tag{3.2.28b}$$

Let us then compute each of these derivatives.

For the cutoff term we have

$$\partial_k \Delta S_k[\Phi] = \frac{1}{2} \langle \Phi, \partial_k \mathscr{R}_k \Phi \rangle, \tag{3.2.29a}$$

$$= \frac{1}{2} \mathrm{STr}[\partial_k \mathscr{R}_k (\Phi \otimes \Phi)], \tag{3.2.29b}$$

$$= \frac{1}{2} \mathrm{STr}[(\Phi \otimes \Phi) \partial_k \mathscr{R}_k], \tag{3.2.29c}$$

where STr denotes the supertrace (see DeWitt 2003, p. 989). This is similar to a regular trace, but it takes into account a negative sign in front of fermionic entries. One can write $\mathrm{Tr}[\mathbb{Z}A] = \mathrm{STr}[A]$, for example. It occurs on Eq. (3.2.29) because we had to exchange the positions of the two occurrences of $\Phi$ when writing the inner product as a trace. This is essentially a fancier version of the trick

$$\langle \psi | A \psi \rangle = \mathrm{tr}[A | \psi \rangle \langle \psi |] \tag{3.2.30}$$

commonly performed in QM.

Since the classical field respects $\Phi = \omega(\widehat{\Phi})$, where $\omega$ is the field's state, we can write Eq. (3.2.29) as

$$\partial_k \Delta S_k[\Phi] = \frac{1}{2} \mathrm{STr}[(\omega(\widehat{\Phi}) \otimes \omega(\widehat{\Phi})) \partial_k \mathscr{R}_k]. \tag{3.2.31}$$

For the connected generating functional, we have

$$\partial_k W_k[J] = \partial_k \log Z_k[J], \tag{3.2.32a}$$

$$= \frac{1}{Z_k[J]} \partial_k Z_k[J], \tag{3.2.32b}$$

$$= \frac{1}{Z_k[J]} \partial_k \int \exp\left(-S[\widehat{\Phi}] - \Delta S_k[\widehat{\Phi}] + \langle J, \widehat{\Phi} \rangle\right) \mathscr{D}\widehat{\Phi}, \tag{3.2.32c}$$

$$= -\frac{1}{Z_k[J]} \int \exp\left(-S[\widehat{\Phi}] - \Delta S_k[\widehat{\Phi}] + \langle J, \widehat{\Phi} \rangle\right) \partial_k \Delta S_k[\widehat{\Phi}] \mathscr{D}\widehat{\Phi}, \tag{3.2.32d}$$

$$= -\frac{1}{2} \frac{1}{Z_k[J]} \int \exp\left(-S[\widehat{\Phi}] - \Delta S_k[\widehat{\Phi}] + \langle J, \widehat{\Phi} \rangle\right) \mathrm{STr}[(\widehat{\Phi} \otimes \widehat{\Phi}) \partial_k \mathscr{R}_k] \mathscr{D}\widehat{\Phi}, \tag{3.2.32e}$$

$$= -\frac{1}{2} \omega\left(\mathrm{STr}[(\widehat{\Phi} \otimes \widehat{\Phi}) \partial_k \mathscr{R}_k]\right), \tag{3.2.32f}$$

$$= -\frac{1}{2} \operatorname{STr}\big[\omega\big(\widehat{\Phi} \otimes \widehat{\Phi}\big)\partial_k \mathcal{R}_k\big], \tag{3.2.32g}$$

where we used Eq. (3.2.29) on the preceding page.

Eqs. (3.2.28), (3.2.31) and (3.2.32) on the facing page now yield

$$\partial_k \Gamma_k[\Phi] = \frac{1}{2} \operatorname{STr}\big[\big(\omega\big(\widehat{\Phi} \otimes \widehat{\Phi}\big) - \omega\big(\widehat{\Phi}\big) \otimes \omega\big(\widehat{\Phi}\big)\big)\partial_k \mathcal{R}_k\big], \tag{3.2.33}$$

where we now identify the connected two-point function. We can then write this equation as

$$\partial_k \Gamma_k[\Phi] = \frac{1}{2} \operatorname{STr}\Big[\Big(\frac{\partial}{\partial J} \otimes \frac{\partial}{\partial J} W_k\Big)\partial_k \mathcal{R}_k\Big]. \tag{3.2.34}$$

Eq. (3.A.21) on page 85 implies

$$\Big(\frac{\partial}{\partial J} \otimes \frac{\partial}{\partial J} W_k\Big) = \Big(\frac{\vec{\partial}\, \tilde{\Gamma}_k\, \overleftarrow{\partial}}{\partial \Phi\, \partial \Phi}\Big)^{-1}, \tag{3.2.35a}$$

$$= \Big(\frac{\vec{\partial}\, \Gamma_k\, \overleftarrow{\partial}}{\partial \Phi\, \partial \Phi} + \frac{\vec{\partial}\, \Delta S_k\, \overleftarrow{\partial}}{\partial \Phi\, \partial \Phi}\Big)^{-1}, \tag{3.2.35b}$$

$$= \big(\Gamma_k^{(2)} + \mathcal{R}_k\big)^{-1}, \tag{3.2.35c}$$

where in the last line we defined

$$\Gamma_k^{(2)} \equiv \frac{\vec{\partial}\, \Gamma_k\, \overleftarrow{\partial}}{\partial \Phi\, \partial \Phi}. \tag{3.2.36}$$

Eqs. (3.2.34) and (3.2.35) now yield

$$\partial_k \Gamma_k = \frac{1}{2} \operatorname{STr}\big[\big(\Gamma_k^{(2)} + \mathcal{R}_k\big)^{-1}\partial_k \mathcal{R}_k\big]. \tag{3.2.37}$$

It is conventional and convenient to multiply this equation by $k$, and to define the adimensional RG time $t$ through $t = \log k$. This implies $\partial_t = k\partial_k$ and we find, at last,

$$\partial_t \Gamma_k = \frac{1}{2} \operatorname{STr}\big[\big(\Gamma_k^{(2)} + \mathcal{R}_k\big)^{-1}\partial_t \mathcal{R}_k\big]. \tag{3.2.38}$$

Eq. (3.2.38) is the Wetterich equation. From this point onward, whenever we talk about *the* FRGE, Eq. (3.2.38) is the expression being referred to.

Notice that this derivation was exact. At no point have we performed any sort of approximation. Therefore, Eq. (3.2.38) is an exact expression. Furthermore, our derivation did not assume any background spacetime structure or similar structure, the exception being that we did assume we could write the generating functional in accordance with the Gell-Mann–Low formula. This implies the possible limitations we mentioned at the end of Section 2.5.

Since we assumed the cutoff $\mathcal{R}_k(z)$ to vanish at large $z$, Eq. (3.2.38) is an expression involving UV finite terms. Note, however, that the FRGE provides the expression for $\partial_t \Gamma_k$, not the EAA directly. If we integrate the FRGE to obtain the full effective action, we will run into the usual UV divergences common in field theory.

It is also interesting to notice that the FRGE can be understood as a new formulation of QFT that is equivalent to the path integral formulation. It has, however, the advantage that a classical action would enter the formalism only by means of a boundary condition at $k = \Lambda_{\mathrm{UV}}$. Apart from this, the theory is "local in scale", with the "degree of locality" being determined by how well the cutoff $\mathcal{R}_k$ can screen the undesired modes.

It is interesting to compare the Wetterich equation to its one-loop analogue. In Euclidean signature, the one-loop approximation for the quantum action, $\Gamma^{\text{one-loop}}$, is given by (see Eq. (3.A.33) on page 86)

$$\Gamma^{\text{one-loop}} = S + \frac{1}{2}\,\mathrm{STr}\big[\log S^{(2)}\big]. \tag{3.2.39}$$

If we use this result with the EAA we get to (*cf.* Codello 2010, Eq. (2.20))

$$\Gamma_k^{\text{one-loop}} = S + \frac{1}{2}\,\mathrm{STr}\big[\log\big(S^{(2)} + \mathcal{R}_k\big)\big]. \tag{3.2.40}$$

Differentiating with respect to $t = \log k$ leads to

$$\partial_t \Gamma_k^{\text{one-loop}} = \frac{1}{2}\,\mathrm{STr}\big[\big(S^{(2)} + \mathcal{R}_k\big)^{-1}\partial_t \mathcal{R}_k\big]. \tag{3.2.41}$$

This expression has the same form as the Wetterich equation, up to the fact that the Wetterich equation writes $\Gamma_k^{(2)}$ instead of $S^{(2)}$ on the right-hand side (RHS). Hence, the FRGE improves the one-loop result to an exact result by imposing the action to flow with scale. This also means the exact flows computed with the Wetterich equation will resemble one-loop results, although they are now considering contributions from all orders.

## 3.3 Computing Nonperturbative Renormalization Flows

### Beta Functions

Let us prescribe a general expression for the EAA. Suppose it is given by

$$\Gamma_k[\Phi] = \sum_i g_i(k)\mathcal{O}_i[\Phi], \tag{3.3.1}$$

where the $\mathcal{O}_i$ are local operators, meaning they can involve only finitely many positive powers of derivatives. Notice that this ansatz then implies

$$\partial_t \Gamma_k[\Phi] = \sum_i \partial_t g_i(k)\mathcal{O}_i[\Phi], \tag{3.3.2}$$

where we assumed the operators not to depend on the scale $k$.

The quantities

$$\beta_i = \partial_t g_i(k) = k\partial_k g_i(k) \tag{3.3.3}$$

are the beta functions for the theory. Therefore, knowing $\partial_t \Gamma_k$ allows us to read off the beta functions, which earns $\partial_t \Gamma_k$ the name "beta functional".

Typically, each of the couplings $g_i$ will have a different mass dimension[*] $d_i$. It is then convenient to define the dimensionless couplings $\tilde{g}_i \equiv k^{-d_i} g_i$. Notice then that the dimensionless beta functions will be given by

$$\tilde{\beta}_i \equiv \partial_t \tilde{g}_i, \tag{3.3.4a}$$

_______________

[*]We take $G$ to be dimensional in this chapter.

$$= k\partial_k\big(k^{-d_i}g_i\big), \tag{3.3.4b}$$

$$= -d_i k^{-d_i}g_i + k\partial_k(g_i)k^{-d_i}, \tag{3.3.4c}$$

$$= -d_i\tilde{g}_i + k^{-d_i}\beta_i, \tag{3.3.4d}$$

where no summation is implied.

As one would expect, the dimensionless beta functions are dimensionless. Since they can only depend on the couplings, and we have also written the couplings in terms of a single scale $k$ and of the dimensionless couplings, we can conclude that the dimensionless beta functions do not depend explicitly on $k$. Indeed, they must remain dimensionless, but there are no other scales around to cancel the dimensions of $k$. Hence, the dimensionless beta functions describe an autonomous system of differential equations for the couplings. By solving this system we can then obtain the RG flow of the theory.

The focus on dimensionless couplings was explained by Weinberg (1979). Suppose we desire to compute some reaction rate $R$ with mass dimension $D$. $R$ could be a cross section, or a decay rate, for example. We expect, on dimensional grounds, that it will be given by an expression of the form

$$R = k^D f\Big(\frac{E}{k}, X, \tilde{g}(k)\Big), \tag{3.3.5}$$

where $k$ is the renormalization scale, $E$ is some physical energy scale characterizing the process, $X$ denotes the collection of all dimensionless quantities characterizing the process (energy ratios, angles, etc), and $\tilde{g}(k)$ denotes the collection of all dimensionless renormalized couplings at scale $k$.

The core idea of the RG is that the theory should be independent of the renormalization scale. Hence, we can pick whichever $k$ we prefer. Picking $k = E$ we get to

$$R = E^D f\big(1, X, \tilde{g}(E)\big). \tag{3.3.6}$$

We then see that if the dimensionless couplings diverge at a finite energy scale $E$—*i.e.*, if the theory has a Landau pole—this will likely lead to a divergence in measurable quantities. Had we used the dimensionful couplings, then their divergences would not need to mean anything, since it could be simply due to a high enough energy.

The fact that we assumed the operators $\mathcal{O}_i$ to be independent of scale can seem unusual if compared to expositions of the Wilsonian RG. In the Wilsonian approach, one often keeps the kinetic term always normalized by redefining the fields upon a change of scale (see, *e.g.*, Peskin and Schroeder 1995, Sec. 12.1). When working with the FRGE, though, it is more convenient to refrain from rescaling the fields and instead allow the kinetic term to have its own coupling constant. For example, we would write the EAA for a scalar field with a quartic interaction as

$$\Gamma_k = \int \frac{Z_k}{2}\partial_a\phi\partial^a\phi + \frac{m^2 Z_k}{2}\phi^2 + \frac{\lambda_k Z_k^2}{4!}\phi^4\, \mathrm{d}^d x. \tag{3.3.7}$$

This new coupling $Z_k$ can then be used to account for the wavefunction renormalization of the field $\phi$. Since it can be absorbed by a field redefinition, it is said to be "inessential", as opposed to "essential" couplings that cannot be redefined away (for further discussion, see Weinberg 1979, Sec. 16.3).

### Truncations

With the FRGE at hand, we would like to start computing nonperturbative RG flows. Nevertheless, recall that the FRGE can be understood as a formulation of QFT equivalent to the path integral formalism. Hence, exactly computing the flow should be about as easy as exactly computing the generating functional.

The reason we are working with the FRG is not because it will provide exact results—it will not. Rather, we choose to work with it because it provides a formulation of QFT with which we can work nonperturbatively. We still need to make approximations to actually be able to solve the equations, but these approximations do not need to assume the coupling constants to be small. Therefore, our results will be approximate, but nonperturbative.

We expect the EAA to have a form such as the one given in Eq. (3.3.1) on page 72. This typically involves infinitely many terms that are generated by the RG flow. As we shall shortly see in the example of a scalar theory—and later in Chapter 4 with a particle detector—even if we start with only a couple of couplings, the RG flow tends to generate infinitely many interactions (see Eq. (3.3.26) on page 78 and the discussion following it). This is similar to how integrating out the $W$ boson leads to the appearance of a quartic fermion interaction in the theory of weak interactions—as we integrate out the high-energy modes, new vertices appear to compensate for their effects.

Given that an infinite-dimensional theory space is fairly complicated, we shall approximate the EAA by truncating it. This truncation may or may not have only finitely many terms. Its main purpose is to offer a compromise between the infinitely difficult exact problem and the simplicity of possibly unrealistic theories. Ideally, the truncation will be able to keep the main features of the exact theory while still allowing us to perform computations. Approximating an infinitely difficult reality by a convenient simplification that keeps its main features is, perhaps, an excellent description of the entirety of theoretical physics. "Truncation" is merely the name we give to this procedure in the context we are working on.

Once we truncate the expression for the EAA, we can compute the beta functions by using it as an ansatz on the RHS of the FRGE. We can then compute the supertrace and obtain an expression for the beta functional $\partial_t \Gamma_k$. Nevertheless, this expression will typically not lie on the truncated theory space. Hence, we will need to also truncate the results and ignore the beta functions for couplings that lie outside of our original truncation.

As an example, consider an interacting scalar field in Minkowski spacetime. We shall calculate its RG flow soon. Suppose we start our computations assuming it has a $\phi^4$ interaction, but no other interaction terms. Then using the FRGE will lead to terms contributing to beta functions of other couplings, such as $\phi^6$ (see Eq. (3.3.26) on page 78). However, since our original truncation ignored the $\phi^6$ term, our calculations are incapable of predicting how this term affects its own RG flow or even the RG flows of the other couplings we were already considering. This means our results for this term are not reliable. Hence, we choose to drop this term entirely and consider only the flow of the terms that were in our original truncation. If we want to consider the flow of more terms, we should take those terms into consideration from the very beginning.

A good truncation will be able to mimic the main features of a theory, despite being considerably simpler. Ideally, the RG flow computed with the truncated ansatz will be similar to what one would obtain from the full theory space followed by a projection onto the truncated space, as illustrated in Fig. 3.5 on the next page.

It is not trivial to tell whether a given truncation is good or not. Since the actual theory does not depend on the choice of cutoff, a good sanity check to test the validity of a truncation is to use a few different cutoffs and see whether they lead to similar results. Another possibility is to use increasingly more complex truncations and test whether the results are stable under the addition of new terms. As an example, one might add different terms to the Einstein–Hilbert action to obtain different truncations for a quantum gravity (QG) theory and see whether the results are stable (see Eichhorn [2019](#), Table 2).

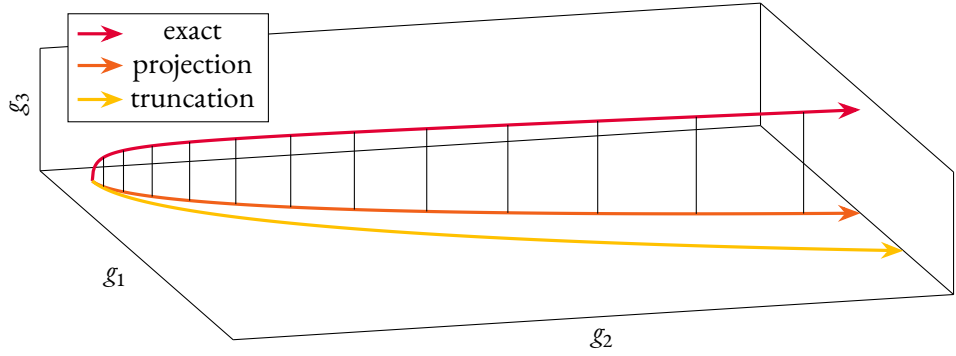

Figure 3.5: Illustration of a truncation of a three-dimensional theory space to a two-dimensional theory space. Ideally, the truncated trajectory, which is calculated with the approximate EAA, will be similar to the projected trajectory, which is computed with the full EAA and then projected onto the truncated theory space. This figure is based on Figure 2.3 of the book by Reuter and Saueressig ([2018](#)).

Further discussion about approximation schemes for the FRGE can be found in the review by Delamotte ([2012](#), Sec. 2.2.3) and in the book by Reuter and Saueressig ([2018](#), Sec. 2.2).

### Nonperturbative Flow of a Scalar Theory in Flat Spacetime

As an example, we can carry out the calculations for the nonperturbative RG flow for a theory with a single scalar field in flat spacetime. In this manner, we can get a feeling for the general procedure we shall follow for computing beta functions. We follow the discussion in the book by Percacci ([2017](#), Sec. 6.4), but see also the book by Reuter and Saueressig ([2018](#), Sec. 2.2.4).

We begin by prescribing an ansatz for the EAA. We shall work with the expression

$$\Gamma_k[\phi] = \int \frac{1}{2} \partial_a \phi \partial^a \phi + V_k(\phi) \, \mathrm{d}^d x \,, \tag{3.3.8}$$

which is known as the local potential approximation (LPA). In this approximation, we are ignoring the wavefunctiion renormalization—which could be considered by adding a variable coupling $Z_k$ to the kinetic term—and any derivative terms. $d$ denotes the spacetime dimension. Notice that we are already writing the Euclidean action.

For this theory, which includes a single bosonic field, the FRGE is simplified to the form

$$\partial_t \Gamma_k = \frac{1}{2} \operatorname{Tr}\left[ \left( \Gamma_k^{(2)} + R_k \right)^{-1} \partial_t R_k \right]. \tag{3.3.9}$$

We can keep the cutoff $R_k$ unspecified for the time being.

Let us begin by computing the Hessian $\Gamma_k^{(2)}$. Since there is only one bosonic field, it is simply given by

$$\frac{\delta^2 \Gamma_k}{\delta\phi(x)\delta\phi(y)} = -\partial^2 \hat{\delta}^{(d)}(x-y) + V_k''(\phi)\hat{\delta}^{(d)}(x-y), \qquad (3.3.10)$$

which is obtained by differentiating Eq. (3.3.8) on the preceding page. We shall denote the operator $-\partial^2$ by $\Delta$ in order to simplify notation and to match the formulae given in Section 3.B. The Dirac deltas in Eq. (3.3.10) are simply identities on the functional matrices labelled by the spacetime coordinates $x$ and $y$. As an operator, we can simply write

$$\Gamma_k^{(2)} = \Delta + V_k''(\phi). \qquad (3.3.11)$$

Therefore, we get to

$$\partial_t \Gamma_k = \frac{1}{2} \operatorname{Tr}\left[\frac{\partial_t R_k(\Delta)}{\Delta + R_k(\Delta) + V_k''(\phi)}\right]. \qquad (3.3.12)$$

It is useful to define $P_k(z) \equiv z + R_k(z)$ so that we may write

$$\partial_t \Gamma_k = \frac{1}{2} \operatorname{Tr}\left[\frac{\partial_t P_k(\Delta)}{P_k(\Delta) + V_k''(\phi)}\right]. \qquad (3.3.13)$$

To compute this functional trace, we shall resort to the heat kernel techniques described on Section 3.B. Using Eq. (3.B.20) on page 89 we can see that

$$\partial_t \Gamma_k = \frac{1}{2} \operatorname{Tr}\left[\frac{\partial_t P_k(\Delta)}{P_k(\Delta) + V_k''(\phi)}\right], \qquad (3.3.14a)$$

$$= \frac{1}{2(4\pi)^{\frac{d}{2}}} \sum_{n=0}^{+\infty} Q_{\frac{d}{2}-n}\left[\frac{\partial_t P_k}{P_k + V_k''(\phi)}\right] \int \operatorname{tr}[a_n(x)]\,\mathrm{d}^d x, \qquad (3.3.14b)$$

$$= \frac{1}{2(4\pi)^{\frac{d}{2}}} Q_{\frac{d}{2}}\left[\frac{\partial_t P_k}{P_k + V_k''(\phi)}\right] \int \operatorname{tr}[a_0(x)]\,\mathrm{d}^d x, \qquad (3.3.14c)$$

$$= \frac{V_d}{2(4\pi)^{\frac{d}{2}}} Q_{\frac{d}{2}}\left[\frac{\partial_t P_k}{P_k + V_k''(\phi)}\right], \qquad (3.3.14d)$$

where $V_d$ is the (infinite) volume of $d$-dimensional spacetime. It is cancelled by a similar infinite volume on the definition of $\partial_t \Gamma_k$, as we soon shall see[*]. Since we are working in flat spacetime with $\Delta = -\partial^2$ (i.e., without an endomorphism) the objects $a_n$ vanish for $n > 0$.

On Eq. (3.3.14) we have written the potential term outside of the spacetime integral. We could do this because in the LPA we are ignoring any derivative terms, and hence the spatial dependence of $\phi$ is irrelevant. This allowed us to treat it as a constant. This is similar to how one would compute the one-loop effective potential, for example (see Peskin and Schroeder 1995, Sec. 11.3; Weinberg 1996, Sec. 16.2).

---

[*]To avoid the weirdness of dealing with these infinite volumes one could mimic what is done in pAQFT and use a spacetime cutoff to the action that makes it an integral over a compact region of spacetime (see Rejzner 2016, Sec. 4.1). This would limit us to a finite spacetime volume and we could still argue that the effective potential flows as it does, since the final results would need to be independent of the choice of spacetime cutoff.

Notice our ansatz for the EAA implies

$$\partial_t \Gamma_k[\phi] = \partial_t \left( \int \frac{1}{2} \partial_a \phi \partial^a \phi + V_k(\phi) \, \mathrm{d}^d x \right), \tag{3.3.15a}$$

$$= \int \partial_t V_k(\phi) \, \mathrm{d}^d x, \tag{3.3.15b}$$

$$= V_d \partial_t V_k(\phi), \tag{3.3.15c}$$

where we once again used the hypothesis of a constant field $\phi$ to pull the potential out of the integral. This spacetime volume $V_d$ cancels the one on Eq. (3.3.14) on the facing page and leads us to

$$\partial_t V_k(\phi) = \frac{1}{2(4\pi)^{\frac{d}{2}}} Q_{\frac{d}{2}} \left[ \frac{\partial_t P_k}{P_k + V_k''(\phi)} \right]. \tag{3.3.16}$$

To proceed, let us consider a general even potential given by

$$V_k(\phi) = \sum_{n=0}^{+\infty} \frac{\lambda_{2n} \phi^{2n}}{(2n)!}. \tag{3.3.17}$$

It follows that

$$\beta_{2n} = \partial_t \lambda_{2n} = \left. \frac{\mathrm{d}^{2n}(\partial_t V_k(\phi))}{\mathrm{d}\phi^{2n}} \right|_{\phi=0}, \tag{3.3.18}$$

which means, under the light of Eq. (3.3.16), that

$$\beta_{2n} = \frac{1}{2(4\pi)^{\frac{d}{2}}} \left. \frac{\mathrm{d}^{2n}}{\mathrm{d}\phi^{2n}} Q_{\frac{d}{2}} \left[ \frac{\partial_t P_k}{P_k + V_k''(\phi)} \right] \right|_{\phi=0}, \tag{3.3.19a}$$

$$= \frac{1}{2(4\pi)^{\frac{d}{2}}} \left. \frac{\mathrm{d}^{2n}}{\mathrm{d}\phi^{2n}} Q_{\frac{d}{2}} \left[ \frac{\partial_t P_k}{P_k + \lambda_2 + \sum_{n=1}^{+\infty} \frac{\lambda_{2n+2} \phi^{2n}}{(2n)!}} \right] \right|_{\phi=0}. \tag{3.3.19b}$$

The first few beta functions will be given by (Percacci 2017, Eqs. (6.35) to (6.38))

$$\beta_0 = \frac{1}{2(4\pi)^{\frac{d}{2}}} Q_{\frac{d}{2}} \left[ \frac{\partial_t P_k}{P_k + \lambda_2} \right], \tag{3.3.20a}$$

$$\beta_2 = \frac{1}{2(4\pi)^{\frac{d}{2}}} \left[ -\lambda_4 Q_{\frac{d}{2}} \left[ \frac{\partial_t P_k}{(P_k + \lambda_2)^2} \right] \right], \tag{3.3.20b}$$

$$\beta_4 = \frac{1}{2(4\pi)^{\frac{d}{2}}} \left[ -\lambda_6 Q_{\frac{d}{2}} \left[ \frac{\partial_t P_k}{(P_k + \lambda_2)^2} \right] + 6\lambda_4^2 Q_{\frac{d}{2}} \left[ \frac{\partial_t P_k}{(P_k + \lambda_2)^3} \right] \right], \tag{3.3.20c}$$

$$\beta_6 = \frac{1}{2(4\pi)^{\frac{d}{2}}} \left[ -\lambda_8 Q_{\frac{d}{2}} \left[ \frac{\partial_t P_k}{(P_k + \lambda_2)^2} \right] + 30\lambda_4\lambda_6 Q_{\frac{d}{2}} \left[ \frac{\partial_t P_k}{(P_k + \lambda_2)^3} \right] - 90\lambda_4^3 Q_{\frac{d}{2}} \left[ \frac{\partial_t P_k}{(P_k + \lambda_2)^4} \right] \right]. \tag{3.3.20d}$$

If we wish to proceed, we must compute the $Q$-functionals, which depend on the choice of cutoff. However, this is similar to the dependence on renormalization scheme that occurs on perturbative treatments, so there is nothing to worry about.

The Litim cutoff (see Eqs. (3.2.8) and (3.2.9) on page 65) is particularly convenient because it allows us to obtain analytic results. Otherwise, we could have to resort to numerical methods. With the Litim cutoff we have

$$P_k(z) = z + R_k(z), \tag{3.3.21a}$$

$$= z + (k^2 - z)\Theta(k^2 - z) \tag{3.3.21b}$$

and

$$\partial_t P_k(z) = 2k^2 \Theta(k^2 - z). \tag{3.3.22}$$

Therefore,

$$Q_{\frac{d}{2}}\left[\frac{\partial_t P_k}{(P_k + \lambda_2)^l}\right] = \frac{1}{\Gamma\left(\frac{d}{2}\right)} \int_0^{k^2} \frac{2k^2 z^{\frac{d}{2}-1}}{(k^2 + \lambda_2)^l} \, dz, \tag{3.3.23a}$$

$$= \frac{2k^2}{\Gamma\left(\frac{d}{2}\right)(k^2 + \lambda_2)^l} \int_0^{k^2} z^{\frac{d}{2}-1} \, dz, \tag{3.3.23b}$$

$$= \frac{4k^{d+2}}{d\Gamma\left(\frac{d}{2}\right)(k^2 + \lambda_2)^l}, \tag{3.3.23c}$$

$$= \frac{2k^{d+2}}{\Gamma\left(\frac{d}{2} + 1\right)(k^2 + \lambda_2)^l}. \tag{3.3.23d}$$

For spacetime dimension $d$, the coupling $\lambda_{2n}$ has mass dimension $d_{2n} = 2n - (n-1)d$. Therefore, we may define the dimensionless couplings through

$$\tilde{\lambda}_{2n} = k^{-2n+(n-1)d} \lambda_{2n}. \tag{3.3.24}$$

Using this definition, we get

$$Q_{\frac{d}{2}}\left[\frac{\partial_t P_k}{(P_k + \lambda_2)^l}\right] = \frac{2k^{d+2-2l}}{\Gamma\left(\frac{d}{2} + 1\right)(1 + \tilde{\lambda}_2)^l}. \tag{3.3.25}$$

Once we employ Eq. (3.3.4) on page 72 and take all of these formulae into consideration, we get to the dimensionless beta functions for the Litim cutoff. The first few of them are (Percacci 2017, Eq. (6.46))

$$\tilde{\beta}_0 = -d\tilde{\lambda}_0 + c_d \frac{1}{1 + \tilde{\lambda}_2}, \tag{3.3.26a}$$

$$\tilde{\beta}_2 = -2\tilde{\lambda}_2 - c_d \frac{\tilde{\lambda}_4}{(1 + \tilde{\lambda}_2)^2}, \tag{3.3.26b}$$

$$\tilde{\beta}_4 = (d - 4)\tilde{\lambda}_4 + c_d \left[-\frac{\tilde{\lambda}_6}{(1 + \tilde{\lambda}_2)^2} + \frac{6\tilde{\lambda}_4^2}{(1 + \tilde{\lambda}_2)^3}\right], \tag{3.3.26c}$$

$$\tilde{\beta}_6 = (2d - 6)\tilde{\lambda}_6 + c_d \left[-\frac{\tilde{\lambda}_8}{(1 + \tilde{\lambda}_2)^2} + \frac{30\tilde{\lambda}_4\tilde{\lambda}_6}{(1 + \tilde{\lambda}_2)^3} - \frac{90\tilde{\lambda}_4^3}{(1 + \tilde{\lambda}_2)^4}\right], \tag{3.3.26d}$$

with

$$c_d = \frac{1}{(4\pi)^{\frac{d}{2}} \Gamma\left(\frac{d}{2} + 1\right)}. \tag{3.3.27}$$

Notice that we were able to easily compute the contributions of infinitely many couplings to the RG flow without the need to ever make a perturbative approximation. In particular, we notice the occurrence of denominators with factors of $1 + \tilde{\lambda}_2$. This is a nonperturbative contribution, for it corresponds to infinitely many powers of $\tilde{\lambda}_2$.

Eq. (3.3.26) on the facing page implies that if a single coupling $\lambda_{2n}$ with $n > 1$ is non-vanishing, then all other terms are going to be generated by the RG flow. Due to this, the quantum action for interacting theories will typically involve infinitely many terms—all the possible ones that respect the theory's symmetries.

Recalling the fact that $\tilde{\lambda}_2 = k^{-2}\lambda_2$, it is interesting to notice that the denominators with powers of $1 + \tilde{\lambda}_2$ tend to 1 for large energies ($k^2 \gg \lambda_2$), but suppress the contributions to the beta functions at low energies ($k^2 \ll \lambda_2$). This is a consequence of the fact that a field decouples at energies much smaller than its mass, just like happened with our $W$ boson example for the muon decay.

### Understanding the Renormalization Flow

Once we have computed the beta functions, we need to interpret them to obtain information about how the RG flow works. This is mainly a matter of studying the properties of a (usually nonlinear) dynamical system, a subject well-known in mathematics and introduced pedagogically in the book by Strogatz (2018), for example. We shall make a very short review of the main features that are relevant to field theory. More details can be found, *e.g.*, in the book by Weinberg (1996, Secs. 18.3 and 18.4), which inspires this section.

To study the behavior of a dynamical system, it is often interesting to consider its topological properties. For example, the existence of fixed points, which are points in theory space where the RG flow "stops". Mathematically, one has

$$\tilde{\beta}_i \Big|_{\tilde{g}^*} = 0, \tag{3.3.28}$$

where $\tilde{g}^*$ is the fixed point. These sorts of structures are interesting because they will often have specific behaviors such as attracting or repulsing RG trajectories, and hence they can be used to obtain a qualitative behavior of the RG flow.

The prototypical example is the Gaussian fixed point. In any theory, the point in theory space given by a free massless theory is a fixed point. Since this point in theory space is the absence of interactions, the RG flow will not be able to generate a flow that drags the theory away from it[*]. Some theories, such as quantum chromodynamics, flow to this fixed point in the UV, and by doing so earn themselves the title of "asymptotically free", since they become free theories in the limit of infinite energy.

Suppose now $\tilde{g}^*$ is a fixed point. Then it is convenient for us to define the variables

$$y_i = \tilde{g}_i - \tilde{g}_i^*. \tag{3.3.29}$$

---

[*]With the Wetterich equation this can be seen by noticing free theories are quadratic in the fields, and hence the RHS of the FRGE will not include any instances of the fields. Hence, the sole non-vanishing beta function will be that of the cosmological constant.

In this way, we can analyze how the RG flow behaves in the vicinity of the fixed point. In particular, we want to consider the linearization of the flow about the fixed point.

Close to the fixed point, the flow of $y_i$ is determined by the functions

$$\partial_t y_i = \partial_t \big( \tilde{g}_i(k) - \tilde{g}_i^\star \big), \tag{3.3.30a}$$

$$= \partial_t \tilde{g}_i(k) - 0, \tag{3.3.30b}$$

$$= \tilde{\beta}_i(\tilde{g}), \tag{3.3.30c}$$

$$= \tilde{\beta}_i(\tilde{g}^\star) + \left( \frac{\partial \tilde{\beta}_i}{\partial \tilde{g}_j} \right)_{\tilde{g}^\star} \big( \tilde{g}_j - \tilde{g}_j^\star \big) + \mathcal{O}\big( (\tilde{g} - \tilde{g}^\star)^2 \big), \tag{3.3.30d}$$

$$= \left( \frac{\partial \tilde{\beta}_i}{\partial \tilde{g}_j} \right)_{\tilde{g}^\star} y_j + \mathcal{O}\big( y^2 \big), \tag{3.3.30e}$$

where we used the fact that, by assumption, $\tilde{\beta}_i(\tilde{g}^\star) = 0$. Hence, for small $y$, we can linearize the RG flow and write it as the matrix equation

$$\partial_t y = My, \tag{3.3.31}$$

where $M$ has entries given by

$$M_{ij} = \left( \frac{\partial \tilde{\beta}_i}{\partial \tilde{g}_j} \right)_{\tilde{g}^\star}. \tag{3.3.32}$$

$M$ is constant, for the dimensionless beta functions cannot depend explicitly on $k$ and we are looking at a fixed value of the couplings.

The formal solution to Eq. (3.3.31) is given by

$$y(t) = e^{tM} y(0). \tag{3.3.33}$$

We can decompose this solution in terms of eigenvectors of $M$, each of them respecting[*]

$$\langle v_m | y(t) \rangle = e^{t\lambda_m} \langle v_m | y(0) \rangle, \tag{3.3.34}$$

where $v_m$ is an eigenvector associated to the eigenvalue $\lambda_m$ and summation over $m$ is not implied. We have three possible situations:

i.  $\mathrm{Re}[\lambda_m] < 0$, meaning the flow is dragged toward the fixed point along the direction of $v_m$ as $t$ grows. Hence, the direction $v_m$ is called UV-attractive or IR-repulsive.

ii.  $\mathrm{Re}[\lambda_m] = 0$, meaning the linear analysis is insufficient to determine whether the fixed point is attractive or repulsive, and one needs to consider the quadratic (or higher) terms. We say the direction $v_m$ is marginal.

iii.  $\mathrm{Re}[\lambda_m] > 0$, meaning the flow is dragged away from the fixed point along the direction of $v_m$ as $t$ grows. Hence, the direction $v_m$ is called UV-repulsive or IR-attractive.

---

[*]The case in which the eigenvectors of $M$ do not span the vector space is always associated with the presence of degenerate eigenvalues. We can then understand it as the limiting case of a nondegenerate problem (Strogatz 2018, Example 5.2.5; Weinberg 1996, footnote on p. 140).

UV-repulsive directions are also called irrelevant, since they can only reach the fixed point in the UV if $\langle v_m | y(0) \rangle = 0$. Hence, if one assume the theory does reach the fixed point in the UV—for example because this assumption seems reasonable to keep the theory physical at all scales and free of Landau poles—then the irrelevant directions provide predictions in the IR limit. One does not have to worry about the values of their couplings constants because they are already fixed by the flow, and hence they are irrelevant in this sense. Similarly, UV-attractive directions are also called relevant, because the values of their coupling constants still have to be fixed experimentally even if the theory flows toward the fixed point in the UV.

It is interesting to point out that these ideas are at the core of asymptotically safe quantum gravity (see Eichhorn 2019; Percacci 2017, 2023; Reuter and Saueressig 2018; Weinberg 1979), an approach to QG in which one assumes gravity can be described by a QFT that approaches a fixed point in the UV. Theories that approach a fixed point in the UV are said to be asymptotically safe, since they are safe from Landau poles. If the fixed point has only finitely many relevant directions, then there are only finitely many couplings to be fixed experimentally. In this case, even if the theory is perturbatively nonrenormalizable it still leads to finite results (due to being asymptotically safe) and retains its predictive power (since it has only finitely many free parameters).

In the vicinity of the Gaussian fixed point, we can see from Eq. (3.3.4) on page 72 that the dimensionless beta functions will behave approximately as[*]

$$\tilde{\beta}_i \approx -d_i \tilde{g}_i. \tag{3.3.35}$$

Hence, the matrix $M$ is already diagonal and has eigenvalues $-d_i$. As a consequence, the directions with $d_i < 0$ are irrelevant and those with $d_i > 0$ are relevant, while those with $d_i = 0$ are marginal. This is the same classification one usually finds in perturbative discussions about renormalizability (see Weinberg 1995, p. 503).

## 3.A  Path Integrals and Generating Functionals

In order to establish our notation and conventions for the definitions of the generating functionals commonly used in field theory, let us quickly review them in this appendix. A more detailed account can be found in the books by Peskin and Schroeder (1995, Sec. 11.5) and Weinberg (1996, Sec. 16.1). Some manipulations in this section will involve calculus of anticommuting numbers, which is well reviewed by DeWitt (2003, App. A).

Consider a theory with field content $\Phi$. This is to be understood as a "vector of fields", with each of the fields being given by $\Phi_i$, for some $i$. We shall denote by $\Phi_\alpha$ the value of some field component at some spacetime event, so that $\alpha$ carries the information about which field we are talking about, which entry in internal spaces (such as gauge spaces, for example), and the spacetime event. This is known as DeWitt's condensed notation (DeWitt 1964). Each of the fields may be either bosonic or fermionic. We define the generating functional of the theory $Z[J]$ and the connected generating functional $W[J]$ through

$$Z[J] = \exp\left(W[J]\right) = \int \exp\left(-S[\widehat{\Phi}] + \langle J, \widehat{\Phi} \rangle\right) \mathscr{D}\widehat{\Phi}, \tag{3.A.1}$$

---

[*]Technically, Eq. (3.3.35) ignores the linear terms that might occur on Eq. (3.3.4) on page 72 and which do occur, for example, on Eq. (3.3.26) on page 78. Nevertheless, the conclusion is correct. See the book by Percacci (2017, p. 173) for a more careful argument.

where we added a hat to $\Phi$ to make it clear that it is being integrated over. In the previous equation, we have also written the shorthand notation

$$\left\langle J, \widehat{\Phi} \right\rangle \equiv J_\alpha \widehat{\Phi}_\alpha, \tag{3.A.2}$$

where the Einstein summation convention also implies an integral over spacetime events. This notation is based on those used by DeWitt (1964) and Kopietz, Bartosch, and Schütz (2010, Chaps. 6 and 7).

Notice that

$$\frac{\partial Z}{\partial J_\alpha} = \int \exp\left(-S\left[\widehat{\Phi}\right] + \left\langle J, \widehat{\Phi} \right\rangle\right) \widehat{\Phi}_\alpha \, \mathscr{D}\widehat{\Phi}, \tag{3.A.3}$$

$$\frac{\partial^2 Z}{\partial J_\beta \partial J_\alpha} = \int \exp\left(-S\left[\widehat{\Phi}\right] + \left\langle J, \widehat{\Phi} \right\rangle\right) \widehat{\Phi}_\beta \widehat{\Phi}_\alpha \, \mathscr{D}\widehat{\Phi}, \tag{3.A.4}$$

and so on. Hence, the $n$-point correlation functions in the presence of the source $J$ are given by

$$\omega\left(\widehat{\Phi}_{\alpha_1} \cdots \widehat{\Phi}_{\alpha_n}\right) = \frac{1}{Z[J]} \frac{\partial^n Z}{\partial J_{\alpha_1} \cdots \partial J_{\alpha_n}} = \frac{\int \exp\left(-S\left[\widehat{\Phi}\right] + \left\langle J, \widehat{\Phi} \right\rangle\right) \widehat{\Phi}_{\alpha_1} \cdots \widehat{\Phi}_{\alpha_n} \, \mathscr{D}\widehat{\Phi}}{\int \exp\left(-S\left[\widehat{\Phi}\right] + \left\langle J, \widehat{\Phi} \right\rangle\right) \mathscr{D}\widehat{\Phi}}. \tag{3.A.5}$$

Let us then check the derivatives of $W[J]$. We will already introduce the notation

$$\omega_c\left(\widehat{\Phi}_{\alpha_1} \cdots \widehat{\Phi}_{\alpha_n}\right) \equiv \frac{\partial^n W}{\partial J_{\alpha_1} \cdots \partial J_{\alpha_n}}. \tag{3.A.6}$$

Our goal is now to understand its meaning. Notice that

$$\omega_c\left(\widehat{\Phi}_\alpha\right) = \frac{\partial W}{\partial J_\alpha}, \tag{3.A.7a}$$

$$= \frac{\partial}{\partial J_\alpha} \log Z[J], \tag{3.A.7b}$$

$$= \frac{1}{Z[J]} \frac{\partial Z}{\partial J_\alpha}, \tag{3.A.7c}$$

$$= \omega\left(\widehat{\Phi}_\alpha\right). \tag{3.A.7d}$$

Furthermore,

$$\omega_c\left(\widehat{\Phi}_\alpha \widehat{\Phi}_\beta\right) = \frac{\partial^2 W}{\partial J_\alpha \partial J_\beta}, \tag{3.A.8a}$$

$$= \frac{\partial}{\partial J_\alpha}\left(\frac{\partial}{\partial J_\beta} \log Z[J]\right), \tag{3.A.8b}$$

$$= \frac{\partial}{\partial J_\alpha}\left(\frac{1}{Z[J]} \frac{\partial Z}{\partial J_\beta}\right), \tag{3.A.8c}$$

$$= \frac{1}{Z[J]} \frac{\partial^2 Z}{\partial J_\alpha \partial J_\beta} - \frac{1}{Z[J]} \frac{\partial Z}{\partial J_\alpha} \frac{1}{Z[J]} \frac{\partial Z}{\partial J_\beta}, \tag{3.A.8d}$$

$$= \omega\left(\widehat{\Phi}_\alpha \widehat{\Phi}_\beta\right) - \omega\left(\widehat{\Phi}_\alpha\right)\omega\left(\widehat{\Phi}_\beta\right). \tag{3.A.8e}$$

If we were to write this expression diagrammatically, we would have

$$\omega_c\left(\widehat{\Phi}_\alpha\widehat{\Phi}_\beta\right) = \alpha \text{———} \beta \; - \; \alpha \text{———} \beta \;, \tag{3.A.9}$$

which means $\omega_c\left(\widehat{\Phi}_\alpha\widehat{\Phi}_\beta\right)$ is the two-point function minus its disconnected components. Hence, $\omega_c\left(\widehat{\Phi}_\alpha\widehat{\Phi}_\beta\right)$ is just the connected two-point function. The same ideas hold for more complicated diagrams (see Peskin and Schroeder 1995, pp. 379–380; Weinberg 1996, Sec. 16.1), and hence we interpret $W[J]$ as the generator of connected $n$-point correlation functions.

$Z[J]$ and $W[J]$ both have the same physical content. Nevertheless, $Z[J]$ generates all diagrams, while $W[J]$ generates only the connected ones. Hence, $W[J]$ is a simpler object, meaning it will often be more convenient to work with it. However, there is still another generating functional that is even simpler: the quantum action, or effective action.

The quantum action $\Gamma$ is a Legendre–Fenchel transform (see Wipf 2021, Sec. 5.3.1) of the connected generating functional,

$$\Gamma[\Phi] \equiv \sup_J \left[\langle J, \Phi\rangle - W[J]\right]. \tag{3.A.10}$$

Let us understand some of its properties.

Notice that if the maximum is achieved for some $\bar{J}$ (something usually assumed in the physics literature for simplicity), then it must satisfy

$$\frac{\partial}{\partial J_\alpha}\left(\langle J, \Phi\rangle - W[J]\right)\Big|_{J=\bar{J}} = 0, \tag{3.A.11a}$$

$$\Phi_\alpha = \frac{\partial W}{\partial J_\alpha}\Big|_{J=\bar{J}}, \tag{3.A.11b}$$

$$= \omega\left(\widehat{\Phi}_\alpha\right)_{\bar{J}}. \tag{3.A.11c}$$

Hence, $\Gamma[\Phi]$ is a functional of the classical fields $\Phi = \omega\left(\widehat{\Phi}\right)$.

To compute the behavior of the derivatives of the quantum action, let us first introduce the notation

$$\zeta_\alpha = \begin{cases} +1, & \text{if } \alpha \text{ is a bosonic index,} \\ -1, & \text{if } \alpha \text{ is a fermionic index.} \end{cases} \tag{3.A.12}$$

We will also introduce the $\mathbb{Z}$ matrix with elements

$$\mathbb{Z}_{\alpha\beta} = \delta_{\alpha\beta}\zeta_\alpha, \tag{3.A.13}$$

where summation is not implied. Since expressions with $\zeta_\alpha$ will often end up involving an odd number of indices, we will typically write the summations explicitly when necessary during the rest of this section.

As an example of how to use $\mathbb{Z}$, we can write

$$\langle J, \Phi\rangle = \int_\alpha J_\alpha\Phi_\alpha, \tag{3.A.14a}$$

$$= \int_\alpha \zeta_\alpha\Phi_\alpha J_\alpha, \tag{3.A.14b}$$

$$= \langle \Phi, \mathbb{Z}J\rangle, \tag{3.A.14c}$$

where we wrote the integral explicitly to avoid notational ambiguities due to the occurrence of $\zeta_\alpha$.

We now note that the definition of the quantum action implies

$$\frac{\partial \Gamma}{\partial \Phi_\alpha} = \frac{\partial}{\partial \Phi_\alpha} \big( \langle \bar{J}, \Phi \rangle - W[\bar{J}] \big), \tag{3.A.15a}$$

$$= \left\langle \frac{\partial \bar{J}}{\partial \Phi_\alpha}, \Phi \right\rangle + \left\langle \frac{\partial \Phi}{\partial \Phi_\alpha}, \mathbb{Z}\bar{J} \right\rangle - \frac{\partial}{\partial \Phi_\alpha} W[\bar{J}], \tag{3.A.15b}$$

$$= \left\langle \frac{\partial \bar{J}}{\partial \Phi_\alpha}, \Phi \right\rangle + \left\langle \frac{\partial \Phi}{\partial \Phi_\alpha}, \mathbb{Z}\bar{J} \right\rangle - \int_\beta \frac{\partial \bar{J}_\beta}{\partial \Phi_\alpha} \frac{\partial W}{\partial \bar{J}_\beta}, \tag{3.A.15c}$$

$$= \left\langle \frac{\partial \bar{J}}{\partial \Phi_\alpha}, \Phi \right\rangle + \left\langle \frac{\partial \Phi}{\partial \Phi_\alpha}, \mathbb{Z}\bar{J} \right\rangle - \int_\beta \frac{\partial \bar{J}_\beta}{\partial \Phi_\alpha} \Phi_\beta, \tag{3.A.15d}$$

$$= \left\langle \frac{\partial \bar{J}}{\partial \Phi_\alpha}, \Phi \right\rangle + \left\langle \frac{\partial \Phi}{\partial \Phi_\alpha}, \mathbb{Z}\bar{J} \right\rangle - \left\langle \frac{\partial \bar{J}}{\partial \Phi_\alpha}, \Phi \right\rangle, \tag{3.A.15e}$$

$$= \left\langle \frac{\partial \Phi}{\partial \Phi_\alpha}, \mathbb{Z}\bar{J} \right\rangle, \tag{3.A.15f}$$

$$= \zeta_\alpha \bar{J}_\alpha, \tag{3.A.15g}$$

where $\bar{J}$ maximizes the argument of the supremum on Eq. (3.A.10) on the preceding page and we wrote all integrals explicitly when necessary. In a more compact notation, we can write

$$\bar{J} = \mathbb{Z}\frac{\partial \Gamma}{\partial \Phi}. \tag{3.A.16}$$

The chain rule then allows us to write

$$\frac{\partial}{\partial \Phi_\alpha} = \int_\beta \frac{\partial \bar{J}_\beta}{\partial \Phi_\alpha} \frac{\partial}{\partial \bar{J}_\beta}, \tag{3.A.17a}$$

$$= \int_\beta \frac{\partial}{\partial \Phi_\alpha} \left( \zeta_\beta \frac{\partial \Gamma}{\partial \Phi_\beta} \right) \frac{\partial}{\partial \bar{J}_\beta}, \tag{3.A.17b}$$

$$= \int_\beta \frac{\partial^2 \Gamma}{\partial \Phi_\alpha \partial \Phi_\beta} \zeta_\beta \frac{\partial}{\partial \bar{J}_\beta}. \tag{3.A.17c}$$

Let us write this as a matrix expression. Define

$$\left[ \frac{\partial}{\partial \Phi} \otimes \frac{\partial}{\partial \Phi} \right]_{\alpha\beta} \equiv \frac{\partial^2}{\partial \Phi_\alpha \partial \Phi_\beta}. \tag{3.A.18}$$

Then we may write Eq. (3.A.17) as

$$\frac{\partial}{\partial \Phi} = \left( \frac{\partial}{\partial \Phi} \otimes \frac{\partial}{\partial \Phi} \Gamma \right) \mathbb{Z}\frac{\partial}{\partial \bar{J}}, \tag{3.A.19a}$$

$$= \frac{\overset{\rightarrow}{\partial}}{\partial \Phi} \Gamma \frac{\overset{\leftarrow}{\partial}}{\partial \Phi} \frac{\partial}{\partial \bar{J}}, \tag{3.A.19b}$$

where we used $\mathbb{Z}$ to transform one of the left-derivatives in a right-derivative. This was possible because $\Gamma$ is a commutative number, while the arguments of the derivatives are either commutative, in which case derivatives from both sides coincide, or anticommutative, in which case they differ by a sign (see DeWitt 2003, pp. 977–978).

Using this result, we find that

$$\Phi = \frac{\partial W}{\partial J}, \tag{3.A.20a}$$

$$\mathbb{1} = \frac{\partial}{\partial \Phi} \frac{\partial W}{\partial J}, \tag{3.A.20b}$$

$$= \frac{\vec{\partial} \Gamma \vec{\partial}}{\partial \Phi \partial \Phi} \left( \frac{\partial}{\partial J} \otimes \frac{\partial}{\partial J} W \right), \tag{3.A.20c}$$

and hence

$$\left( \frac{\partial}{\partial J} \otimes \frac{\partial}{\partial J} W \right) = \left( \frac{\vec{\partial} \Gamma \vec{\partial}}{\partial \Phi \partial \Phi} \right)^{-1}, \tag{3.A.21}$$

which means the second derivative of the quantum action is the inverse of the connected propagator, where care is due with the fermionic signs.

This is most of the mathematical machinery we need. Nevertheless, a few more comments about the quantum action are in place. We shall state these facts without proof, but more information can be found on the books by Peskin and Schroeder (1995, Sec. 11.5) and Weinberg (1996, Sec. 16.1).

Firstly, effects coming from all loop levels are taken into account by considering the tree-level diagrams coming from the quantum action. In this sense it replaces the classical action, since its "classical level" already includes all quantum corrections.

Secondly, the quantum action generates the one-particle-irreducible (1PI) diagrams, which are the connected diagrams that cannot be rendered disconnected by cutting any single internal line. 1PI diagrams can be used to write all connected diagrams, which can then be used to write all diagrams. Hence, $\Gamma[\Phi]$ is an object simpler than $W[J]$ or $Z[J]$. Despite being simpler, it still includes all of the physical content of the theory.

Finally, notice how these constructions resemble those found in statistical systems (Peskin and Schroeder 1995, Sec. 11.3). The generating functional $Z[J]$ is analogous to the partition function of a statistical system, $W[J]$ resembles the Helmholtz free energy, and at last $\Gamma[\Phi]$ resembles the Gibbs free energy.

## One-Loop Quantum Action

It is also interesting for us to derive an expression for the one-loop quantum action. We will adapt the discussion given by Codello (2010, App. B.2) to theories with fermions. The expression for the one-loop quantum action for a bosonic field is given in the books by Peskin and Schroeder (1995, Eq. (11.63)) and Weinberg (1996, Eq. (16.2.8)) for the case of Lorentzian signature.

We begin by noticing that Eqs. (3.A.14) and (3.A.16) on page 83 and on the preceding page, alongside the definitions of $W[J]$ and $\Gamma[\Phi]$, allow us to write

$$\exp(-\Gamma[\Phi]) = \int \exp\left( -S\big[\widehat{\Phi}\big] + \left\langle \widehat{\Phi} - \Phi, \frac{\partial \Gamma}{\partial \Phi} \right\rangle \right) \mathscr{D}\widehat{\Phi} . \tag{3.A.22}$$

It is convenient for us to define a fluctuation field $\widehat{\Xi} = \widehat{\Phi} - \Phi$. We can then write

$$\exp(-\Gamma[\Phi]) = \int \exp\left( -S\big[\widehat{\Xi} + \Phi\big] + \left\langle \widehat{\Xi}, \frac{\partial \Gamma}{\partial \Phi} \right\rangle \right) \mathscr{D}\widehat{\Xi} . \tag{3.A.23}$$

Our one-loop expansion consists in expanding the quantum action in powers of $\hbar$ and then dropping terms proportional to $\hbar^2$ or higher powers. Hence, let us restore $\hbar$ in our equations by writing

$$\exp\left(-\frac{1}{\hbar}\Gamma[\Phi]\right) = \int \exp\left(-\frac{1}{\hbar}S\big[\widehat{\Xi}+\Phi\big] + \frac{1}{\hbar}\left\langle\widehat{\Xi},\frac{\partial\Gamma}{\partial\Phi}\right\rangle\right)\mathscr{D}\widehat{\Xi}. \qquad (3.A.24)$$

If we redefine $\widehat{\Xi}\to\sqrt{\hbar}\widehat{\Xi}$, we get to

$$\exp\left(-\frac{1}{\hbar}\Gamma[\Phi]\right) = \int \exp\left(-\frac{1}{\hbar}S\big[\sqrt{\hbar}\widehat{\Xi}+\Phi\big] + \frac{1}{\sqrt{\hbar}}\left\langle\widehat{\Xi},\frac{\partial\Gamma}{\partial\Phi}\right\rangle\right)\mathscr{D}\widehat{\Xi}. \qquad (3.A.25)$$

Let us then perform two expansions. The first of them is to write the quantum action in the form

$$\Gamma[\Phi] = \sum_{n=0}^{+\infty} \hbar^n \Gamma_n[\Phi] \qquad (3.A.26)$$

(see Toms 2007, pp. 376–378, for a justification of the use of only integer powers of $\hbar$). The second is to Taylor expand the classical action as

$$S\big[\sqrt{\hbar}\widehat{\Xi}+\Phi\big] = S[\Phi] + \sqrt{\hbar}\left\langle\widehat{\Xi},\frac{\partial S}{\partial\Phi}\right\rangle + \frac{\hbar}{2}\left\langle\widehat{\Xi},\frac{\vec{\partial}S\vec{\partial}}{\partial\Phi\partial\Phi}\widehat{\Xi}\right\rangle + \mathcal{O}\big(\hbar^{\frac{3}{2}}\big), \qquad (3.A.27a)$$

$$= S[\Phi] + \sqrt{\hbar}\left\langle\widehat{\Xi},\frac{\partial S}{\partial\Phi}\right\rangle + \frac{\hbar}{2}\left\langle\widehat{\Xi},S^{(2)}\widehat{\Xi}\right\rangle + \mathcal{O}\big(\hbar^{\frac{3}{2}}\big). \qquad (3.A.27b)$$

Notice the positions of the entries are important, since some of the fields anticommute.

Notice then that we get to

$$\exp\left(-\frac{1}{\hbar}\Gamma_0[\Phi] - \Gamma_1[\Phi] + \mathcal{O}(\hbar)\right)$$
$$= \int \exp\left(-\frac{1}{\hbar}S[\Phi] - \frac{1}{\sqrt{\hbar}}\left\langle\widehat{\Xi},\frac{\partial S}{\partial\Phi}\right\rangle - \frac{1}{2}\left\langle\widehat{\Xi},S^{(2)}\widehat{\Xi}\right\rangle + \frac{1}{\sqrt{\hbar}}\left\langle\widehat{\Xi},\frac{\partial\Gamma_0}{\partial\Phi}\right\rangle + \mathcal{O}\big(\sqrt{\hbar}\big)\right)\mathscr{D}\widehat{\Xi}. \quad (3.A.28)$$

At leading order in $\hbar$, we find that

$$\Gamma_0[\Phi] = S[\Phi] + \text{constant}, \qquad (3.A.29)$$

which was expected. Hence, we can simplify the equation to

$$\exp(-\Gamma_1[\Phi] + \mathcal{O}(\hbar)) = \int \exp\left(-\frac{1}{2}\left\langle\widehat{\Xi},S^{(2)}\widehat{\Xi}\right\rangle + \mathcal{O}\big(\sqrt{\hbar}\big)\right)\mathscr{D}\widehat{\Xi}. \qquad (3.A.30)$$

We have reduced the problem to computing a Gaussian integral. Using the known expressions for Gaussian integrals, we find that to one-loop order (*cf.* Altland and Simons 2010, Eqs. (3.19) and (4.19); DeWitt 2003, Eq. (A.120))

$$\exp(-\Gamma_1[\Phi]) \propto \big[\text{SDet}\,S^{(2)}\big]^{-\frac{1}{2}}, \qquad (3.A.31)$$

where SDet denotes the superdeterminant (DeWitt 2003, p. 989). Therefore,

$$\Gamma_1[\Phi] = \frac{1}{2}\log\big(\text{SDet}\,S^{(2)}\big), \qquad (3.A.32a)$$

$$= \frac{1}{2}\,\text{STr}\big[\log S^{(2)}\big]. \qquad (3.A.32b)$$

Thus, to one-loop order, we may write

$$\Gamma^{\text{one-loop}} = S + \frac{\hbar}{2}\,\text{STr}\big[\log S^{(2)}\big]. \qquad (3.A.33)$$

## 3.B   Heat Kernel Techniques

When working with the FRGE, we often need to compute complicated functional traces. This appendix is intended to describe how to perform the calculations we shall encounter. It is based on the books by Percacci ([2017](#)) and Reuter and Saueressig ([2018](#), App. D), and on the paper by Wald ([1979](#)). One might also check the more extensive reviews by Avramidi ([2000](#)) and Vassilevich ([2003](#)).

Let us denote

$$\Delta = -g^{ab}\nabla_a\nabla_b + \mathbb{E}, \tag{3.B.1}$$

where $g_{ab}$ is the (Euclidean) spacetime metric, $\nabla_a$ is the Levi-Civita connection associated with $g_{ab}$, and $\mathbb{E}$ is some linear map that can act on both spacetime indices and eventual internal indices carried by the fields[*]. The endomorphism $\mathbb{E}$ could be, for example, a mass term or perhaps some curvature term. It is included to make the discussion a bit more general.

When working with the FRGE, we might find ourselves wishing to compute an expression of the form

$$\mathrm{Tr}[W(\Delta)], \tag{3.B.2}$$

for some arbitrary function $W(\Delta)$. How can we proceed to compute the functional trace?

Let $\widetilde{W}(s)$ denote the inverse Laplace transform of $W(z)$, i.e., suppose

$$W(z) = \int_0^{+\infty}\widetilde{W}(s)e^{-sz}\,\mathrm{d}s\,. \tag{3.B.3}$$

We may then write

$$\mathrm{Tr}[W(\Delta)] = \int_0^{+\infty}\widetilde{W}(s)\,\mathrm{Tr}\!\left[e^{-s\Delta}\right]\mathrm{d}s\,, \tag{3.B.4}$$

which reduces the problem to that of computing the simpler trace $\mathrm{Tr}\!\left[e^{-s\Delta}\right]$.

Assuming $\Delta$ is sufficiently well-behaved, we can write the action of $e^{-s\Delta}$ on some test function $f$ by means of an integral kernel. Namely, we can write

$$\left(e^{-s\Delta}f\right)(x) = \int K(s;x,y;\Delta)f(y)\sqrt{g}\,\mathrm{d}^d y\,, \tag{3.B.5}$$

where $\sqrt{g}\,\mathrm{d}^d y$ denotes the integration measure on the manifold. $K(s;x,y;\Delta)$ is known as the heat kernel, while $e^{-s\Delta}$ is also called the heat operator. Notice that if $f$ is not a scalar function—for example, it could be a tensor field or a matrix in some internal space—then $K(s;x,y;\Delta)$ might also have further structure.

Notice that the heat operator is a solution to the heat equation,

$$(\partial_s + \Delta)e^{-s\Delta} = 0. \tag{3.B.6}$$

It has the initial condition

$$e^{-s\Delta}\Big|_{s=0} = \mathbb{1}. \tag{3.B.7}$$

It then follows from Eq. ([3.B.5](#)) that the heat kernel satisfies

$$(\partial_s + \Delta)K(s;x,y;\Delta) = 0 \tag{3.B.8}$$

---

[*]It is possible to add a Yang–Mills connection to $\nabla_a$, but we refrain from doing so since we will not be considering gauge theories in this thesis.

and

$$\lim_{s \to 0} K(s; x, y; \Delta) = \tilde{\delta}^{(d)}(x, y). \tag{3.B.9}$$

In terms of the heat kernel, we can write the trace $\mathrm{Tr}\big[e^{-s\Delta}\big]$ as

$$\mathrm{Tr}\big[e^{-s\Delta}\big] = \int K(s; x, x; \Delta) \sqrt{g} \, \mathrm{d}^d x \,. \tag{3.B.10}$$

Hence, we have reduced the problem of computing $\mathrm{Tr}[W(\Delta)]$ to that of understanding the heat kernel.

### Early Time Expansion

Notice that the parameter $s$ must have dimension of length squared, since $\Delta$ has dimension of inverse length squared. Hence, if we manage to expand the heat operator for small values of $s$—"early times", since $s$ is similar to time in the heat equation—we will be highlighting the UV properties of the heat operator. In such a limit, we expect it to be similar to what one would get for flat space.

Therefore, in order to gain some intuition, let us consider the problem of obtaining the heat kernel in flat space for the operator

$$\Delta = -\tilde{\delta}^{ab} \nabla_a \nabla_b + m^2, \tag{3.B.11}$$

which typically occurs for a massive scalar field. Since we are working in flat space, we can write this operator in Cartesian coordinates as

$$\Delta = -\sum_{\mu=1}^{d} \frac{\partial^2}{\partial (x^\mu)^2} + m^2. \tag{3.B.12}$$

Therefore, the heat kernel can be seen as the solution to the initial value problem

$$\begin{cases} \left( \dfrac{\partial}{\partial s} - \displaystyle\sum_{\mu=1}^{d} \dfrac{\partial^2}{\partial (x^\mu)^2} + m^2 \right) K(s; x, y; \Delta) = 0, \\ K(0; x, y; \Delta) = \tilde{\delta}^{(d)}(x, y). \end{cases} \tag{3.B.13}$$

To solve the equation we can take a Fourier transform with respect to the spatial variables $x$. This also requires taking a Fourier transform for the initial condition. We are left with the initial value problem for a simple exponential function. If we solve it and then transform back, we end up with

$$K(s; x, y; \Delta) = \frac{1}{(2\pi)^d} \int \exp\left( i \partial_{\mu\nu} k^\mu (x^\nu - y^\nu) - k^2 s - m^2 s \right) \mathrm{d}^d k \,. \tag{3.B.14}$$

This is a Gaussian integral in $d$ dimensions. Using the known formulae for these sorts of integrals (see, *e.g.*, Altland and Simons 2010, Eq. (3.13)) we get to

$$K(s; x, y; \Delta) = \frac{1}{(4\pi s)^{\frac{d}{2}}} \exp\left( -\frac{(x^\mu - y^\mu)(x_\mu - y_\mu)}{4s} - m^2 s \right). \tag{3.B.15}$$

Hence, we expect to be able to write the heat kernel for a more general operator in the form

$$K(s; x, y; \Delta) = \frac{1}{(4\pi s)^{\frac{d}{2}}} \exp\left(-\frac{\sigma(x, y)}{4s}\right) \sum_{n=0}^{+\infty} A_n(x, y) s^n, \qquad (3.B.16)$$

where $\sigma(x, y)$ is the square of the geodesic distance between $x$ and $y$, and the coefficients $A_n(x, y)$ are generally matrices on internal spaces. Wald (1979) proved this expression[*] for scalar fields on non-compact manifolds with self-adjoint $\Delta$. It was already known for compact manifolds. One also has the boundary condition that $\lim_{y \to x} A_0(x, y) = \mathbb{1}$—the identity on the internal space—in order for Eq. (3.B.9) on the preceding page to be satisfied.

From Eqs. (3.B.10) and (3.B.16) on the facing page and on the current page we can see that

$$\mathrm{Tr}\left[e^{-s\Delta}\right] = \frac{1}{(4\pi s)^{\frac{d}{2}}} \sum_{n=0}^{+\infty} \int \mathrm{tr}[a_n(x)] \sqrt{g}\, \mathrm{d}^d x\, s^n, \qquad (3.B.17)$$

where $a_n(x) \equiv A_n(x, x)$ and tr denotes the algebraic trace over the finite-dimensional internal space. Eq. (3.B.17) is the early time expansion for the trace of the heat operator.

The matrices $a_n$ can be written in terms of curvature tensors and of the endomorphism $\mathbb{E}$. For dimensional reasons, notice that $a_n$ must be built out of terms with exactly $2n$ derivatives. These coefficients can be found in mathematical physics references such as those by Avramidi (2000) and Vassilevich (2003). We have, for example, the terms (Reuter and Saueressig 2018, App. D.1)

$$a_0 = \mathbb{1}, \qquad (3.B.18)$$

$$a_1 = \frac{R}{6} \mathbb{1} - \mathbb{E}, \qquad (3.B.19)$$

where $R$ is the Ricci scalar. Further terms involve more derivatives, meaning the expressions get progressively more complex.

## Traces of Functions of Operators

Let us then get back to the problem of finding $\mathrm{Tr}[W(\Delta)]$. From Eqs. (3.B.4) and (3.B.17) on page 87 and on this page we find

$$\mathrm{Tr}[W(\Delta)] = \int_0^{+\infty} \widetilde{W}(s)\, \mathrm{Tr}\left[e^{-s\Delta}\right] \mathrm{d}s, \qquad (3.B.20a)$$

$$= \sum_{n=0}^{+\infty} \int_0^{+\infty} \frac{\widetilde{W}(s) s^n\, \mathrm{d}s}{(4\pi s)^{\frac{d}{2}}} \int \mathrm{tr}[a_n(x)] \sqrt{g}\, \mathrm{d}^d x, \qquad (3.B.20b)$$

$$= \frac{1}{(4\pi)^{\frac{d}{2}}} \sum_{n=0}^{+\infty} \int_0^{+\infty} \widetilde{W}(s) s^{-\left(\frac{d}{2}-n\right)} \mathrm{d}s \int \mathrm{tr}[a_n(x)] \sqrt{g}\, \mathrm{d}^d x, \qquad (3.B.20c)$$

$$= \frac{1}{(4\pi)^{\frac{d}{2}}} \sum_{n=0}^{+\infty} Q_{\frac{d}{2}-n}[W] \int \mathrm{tr}[a_n(x)] \sqrt{g}\, \mathrm{d}^d x, \qquad (3.B.20d)$$

where we defined the $Q$-functionals through

$$Q_n[W] = \int_0^{+\infty} \widetilde{W}(s) s^{-n}\, \mathrm{d}s. \qquad (3.B.21)$$

---

[*]Up to the mathematical caveats that we are ignoring.

Eq. (3.B.20) on the previous page is the expression we wished to obtain.

Suppose $n > 0$. Then

$$Q_n[W] = \frac{1}{\Gamma(n)} \int_0^{+\infty} W(z) z^{n-1} \, \mathrm{d}z \,. \tag{3.B.22}$$

Indeed,

$$\frac{1}{\Gamma(n)} \int_0^{+\infty} W(z) z^{n-1} \, \mathrm{d}z = \frac{1}{\Gamma(n)} \iint_0^{+\infty} \widetilde{W}(s) z^{n-1} e^{-sz} \, \mathrm{d}z \, \mathrm{d}s \,, \tag{3.B.23a}$$

$$= \frac{1}{\Gamma(n)} \iint_0^{+\infty} \widetilde{W}(s) u^{n-1} s^{-n} e^{-u} \, \mathrm{d}u \, \mathrm{d}s \,, \tag{3.B.23b}$$

$$= \int_0^{+\infty} \widetilde{W}(s) s^{-n} \, \mathrm{d}s \,, \tag{3.B.23c}$$

$$= Q_n[W] \,. \tag{3.B.23d}$$

For $n \leq 0$, we cannot use the integral representation of the Gamma function we just applied, but we can obtain other formulas.

Notice that Eq. (3.B.3) on page 87 implies

$$W^{(k)}(z) = (-1)^k \int_0^{+\infty} s^k \widetilde{W}(s) e^{-sz} \, \mathrm{d}s \,, \tag{3.B.24}$$

and hence

$$Q_n[W] = (-1)^k Q_{n+k}[W^{(k)}] \,. \tag{3.B.25}$$

We are then able to choose a convenient value of $k$ and obtain $Q_n[W]$ for $n \leq 0$ by employing Eqs. (3.B.22) and (3.B.25). This leads to the general formula

$$Q_n[W] = \frac{(-1)^k}{\Gamma(n+k)} \int_0^{+\infty} W^{(k)}(z) z^{n+k-1} \, \mathrm{d}z \tag{3.B.26}$$

for any $k$ such that $n + k > 0$.

## Traces of Covariant Derivatives of Functions of Operators

Let us now consider a more complicated trace. Namely, suppose for some reason we got to an expression of the form

$$\mathrm{Tr}[\mathcal{O}W(\Delta)] \equiv \mathrm{Tr}\Big[\mathcal{O}^{a_1 \cdots a_n} \nabla_{a_1} \cdots \nabla_{a_n} W(\Delta)\Big] \,. \tag{3.B.27}$$

These sorts of expressions will occur in Section 4.3, when we are dealing with the Unruh–DeWitt detector. How can we deal with them?

Firstly, notice we can assume $\mathcal{O}^{a_1 \cdots a_n}$ to be completely symmetric. The antisymmetric parts can be rewritten in terms of curvature tensors, leading us back to the form of Eq. (3.B.27), although with $\mathcal{O}^{a_1 \cdots a_n}$ now being of a smaller rank.

The first step is to use the inverse Laplace transform, as we did before. Notice that

$$\mathrm{Tr}[\mathcal{O}W(\Delta)] = \int_0^{+\infty} \widetilde{W}(s) \, \mathrm{Tr}\Big[\mathcal{O}e^{-s\Delta}\Big] \, \mathrm{d}s \,, \tag{3.B.28}$$

and now we just have to study the trace of the heat operator. Notice that

$$\left(\mathcal{O}e^{-s\Delta}f\right)(x) = \int \mathcal{O}K(s;x,y;\Delta)f(y)\sqrt{g}\,\mathrm{d}^d y, \tag{3.B.29}$$

where the operator $\mathcal{O}$ acts on the $x$ coordinates. Hence, we get to

$$\mathrm{Tr}\left[\mathcal{O}e^{-s\Delta}\right] = \mathrm{tr} \int \mathcal{O}^{a_1\cdots a_n}\overline{\nabla_{(a_1}\cdots\nabla_{a_n)}K(s;x,y;\Delta)}\sqrt{g}\,\mathrm{d}^d x, \tag{3.B.30}$$

where the overline denotes the coincidence limit $y \to x$. The algebraic trace tr in front of the integral takes care of eventual contributions due to internal indices. To simplify notation, we shall also write

$$K_{a_1\cdots a_n} \equiv \overline{\nabla_{(a_1}\cdots\nabla_{a_n)}K(s;x,y;\Delta)}. \tag{3.B.31}$$

Eq. (3.B.30) provides an expression for $\mathrm{Tr}\left[\mathcal{O}e^{-s\Delta}\right]$ in terms of the heat kernel evaluated at $x \neq y$ until the coincidence limit is taken at the last minute. Hence, computation of traces by means such as the one provided by Eq. (3.B.30) are known as off-diagonal heat kernel techniques.

Using Eqs. (3.B.16) and (3.B.30) on page 89 and on this page we can obtain expressions for the traces of the form $\mathrm{Tr}\left[\mathcal{O}e^{-s\Delta}\right]$ (see Reuter and Saueressig 2018, App. D.2, for a bit more detail). The case of a flat manifold with vanishing $\mathbb{E}$ is of particular interest in Chapter 4. In this case, one has (Reuter and Saueressig 2018, Eq. (D.29))

$$K_{(a_1\cdots a_{2n})} = \frac{1}{(4\pi s)^{\frac{d}{2}}}\left(-\frac{1}{2s}\right)^n\left[\hat{\partial}_{a_1 a_2}\cdots\hat{\partial}_{a_{2n-1}a_{2n}} + \left(\frac{(2n)!}{2^n n!} - 1\right)\text{permutations}\right], \tag{3.B.32a}$$

$$= \frac{1}{(4\pi s)^{\frac{d}{2}}}\left(-\frac{1}{2s}\right)^n\frac{(2n)!}{2^n n!}\hat{\partial}_{(a_1 a_2}\cdots\hat{\partial}_{a_{2n-1}a_{2n})}. \tag{3.B.32b}$$

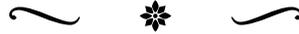

# Four

# Nonperturbative Unruh–DeWitt Detectors

*A motivation for the study of the nonperturbative renormalization group flow of a particle detector is given. We then review how to formulate an Unruh–DeWitt detector in terms of an action, and attempt to compute its functional renormalization group flow. Our approach will lead to issues on the gapless limit of the detector, and we shall discuss other possible ideas for further analyses.*

Have you guessed the riddle yet?

The Mad Hatter, in Lewis Carroll's *Alice's Adventures in Wonderland*, Chapter VII: A Mad Tea-Party.

Understanding the relationship between quantum mechanics (QM) and relativity is an issue nearly as old as QM itself. For example, the considerations raised by Einstein, Podolsky, and Rosen (1935) naturally lead to questions about whether quantum entanglement allows for superluminal exchange of information between two systems. As these questions evolved, they developed into an area of physics known as relativistic quantum information, which is concerned with how information can be exchanged between observers in spacetime.

When studying such problems, quantum field theory in curved spacetime becomes a particularly natural language since it is capable of encoding quantum effects while also taking into consideration the intricacies due to a curved spacetime. A quantum field can be used as a medium for communication with observers accessing it by using particle detectors such as the Unruh–DeWitt detector we described in Section 2.4. By performing a measurement on their detector, an observer can encode information in the field. Later, a second observer may retrieve this information by performing a new measurement on their detector. Within this paradigm, Landulfo (2016) proposed a communication protocol for two observers moving through a general globally hyperbolic spacetime and studied its features in a completely nonperturbative manner. We shall briefly review his model and findings as a motivation for the following sections.

One considers two observers, Alice and Bob, each possessing a two-level quantum system—a qubit. We assume them to be in some arbitrary globally hyperbolic spacetime on which there is a scalar quantum field, which is described by the methods outlined in Section 2.3. The field is assumed to interact with the qubits by means of the interaction-picture Hamiltonian (Landulfo 2016, Eq. (24))

$$H_{\text{int}}(t) = \sum_j \epsilon_j(t) \int_{\Sigma_t} \psi_j(t, \tilde{x}) \sigma_j^x(t) \otimes \phi(t, \tilde{x}) \sqrt{-g} \, d^3 x \qquad (4.0.1)$$

(*cf.* Eq. (2.4.25) on page 34). In this expression, $t$ is a global time function assured to exist on any globally hyperbolic spacetime (Wald 1984, Theorem 8.3.14), $\Sigma_t$ is the Cauchy surface at constant $t$, $\vec{x}$ denotes the coordinates on $\Sigma_t$, $j$ labels whether the qubit is Alice's or Bob's, $\epsilon_j(t)$ is the coupling between qubit $j$ and the field (allowed to have a time-dependence so the coupling can be turned on and off), $\psi_j(t, \vec{x})$ is a function modelling the spacetime extension of the qubit, $\sigma^x$ is the Pauli matrix, $\phi$ is the quantum field, and $g$ is the spacetime metric determinant[*]. While we use the notation $\vec{x}$ for a point in $\Sigma_t$, do notice that $\Sigma_t$ does not necessarily have a vector space structure. The total Hamiltonian of the composite system made of the field and qubits is then given in the interaction picture by

$$H(t) = H_\phi + H_{\mathrm{int}}(t). \qquad (4.0.2)$$

Notice this differs from the model we previously considered in Section 2.4 due to the absence of "gap terms" for the qubits (see Eq. (2.4.24) on page 34). We shall comment on this shortly.

As shown by Landulfo (2016), this model leads to a number of advantages. By using it, a communication protocol can be established with the following properties:

   i. it involves a model that can be exactly solved;

  ii. the results hold for a general globally hyperbolic spacetime;

 iii. one does not need to assume a representation of the canonical commutation relations;

  iv. the field can be assumed to be on an arbitrary Gaussian state;

   v. both sender and receiver are allowed to have arbitrary trajectories;

  vi. both sender and receiver only interact with the field on a bounded region of spacetime;

 vii. classical information can be transmitted between sender and receiver;

viii. causality is manifest, with spacelike-related observers being unable to communicate.

Nevertheless, the same simplification that allows for these advantages also brings a few limitations. The model is exactly solvable because it lacks a gap term such as the one shown on Eq. (2.4.24) on page 34. This limitation prevents the observers from transmitting quantum information without the assistance of extra entanglement between the parts, from harvesting entanglement from the vacuum (Simidzija, Jonsson, and Martín-Martínez 2018), and from using the qubits as particle detectors. Indeed, the absence of a gap term spoils the interpretation that the detector gets excited when it detects a particle, since there is no longer a difference between ground and excited states.

Due to these reasons, we are interested in modifying the model analyzed by Landulfo (2016) by now considering the Hamiltonian

$$H(t) = H_\phi + \sum_j \frac{\Omega_j}{2} \frac{\mathrm{d}\tau_j}{\mathrm{d}t} \sigma_j^z + H_{\mathrm{int}}(t). \qquad (4.0.3)$$

---

[*]Eq. (4.0.1) on the preceding page does include the determinant of the spacetime metric rather than the determinant of the induced metric on $\Sigma_t$. This can be understood by recalling that the Hamiltonian is not a scalar, but the action—which includes the time integral of the Hamiltonian—is.

This new Hamiltonian adds gap terms to the qubits considered on Eq. (4.0.2) on the preceding page. The factors $\frac{\mathrm{d}\tau_j}{\mathrm{d}t}$ are included because the gap terms generate translations with respect to the detector's proper times, but $H(t)$ should generate translations with respect to the global time function $t$. With this new Hamiltonian we shall be able to bypass some of the limitations imposed by the previous model. However, this will cost us the ability to solve the model exactly in an arbitrary spacetime.

We would still like to be able to consider qubits strongly coupled to the quantum field, since this is one of the main features of the analysis given by Landulfo (2016). Nevertheless, we will no longer be able to solve the model exactly. A possibility is to then perform a perturbative expansion not on the couplings to the field, but rather on the energy gaps of the detectors. As long as the gaps are non-vanishing, we should already be able to use the qubits as particle detectors and obtain other interesting results concerning communication, and hence there is no great loss in assuming the gaps to be small.

This perturbative treatment can be improved by considering the renormalization group (RG) flow of the field-detector system. As mentioned in Chapter 3, the RG flow can be used to partially resum the perturbative expansion and hence improve the results obtained by perturbative methods (see, *e.g.*, Delamotte 2004). Therefore, we are interested in employing functional renormalization group (FRG) techniques to obtain a nonperturbative understanding of the behavior of particle detectors coupled to quantum fields in curved spacetimes.

Notice that particle detectors are interesting tools to probe the Unruh and Hawking effects, which have well-defined temperature scales. Hence, they are often taken to be systems in thermal equilibrium. Since thermal equilibrium can be understood as periodicity in imaginary time (this is essentially the physical meaning of the Kubo–Martin–Schwinger condition mentioned in Section 2.3 on page 20), an Euclidean approach gains a natural interpretation and our Euclidean formulation of the FRG seems particularly interesting. The Unruh or Hawking temperature will then often provide the scale of interest at which we shall consider the theory's parameters.

Our goal in this chapter will then be to discuss the FRG flow of an Unruh–DeWitt detector. While ultimately we would like to obtain the flow for accelerated detectors in Minkowski spacetime or static detectors in Schwarzschild spacetime in order to discuss the Unruh and Hawking effects, we shall focus on the simpler case of an inertial detector in flat spacetime. As we will notice, this simple case is already complicated enough to require a careful analysis by itself. Generalizations for other trajectories, spacetimes, and the analysis of communication aspects are outside the scope of this thesis. Nevertheless, in the following sections we shall avoid making unnecessary simplifications, and hence many of our intermediate results hold in more generality.

## 4.1 Action for a Particle Detector

When working with the functional renormalization group equation (FRGE), we need to have an ansatz for the effective average action (EAA). Therefore, we would like to have an action formulation for the particle detector we will be working with. To obtain one we will follow the paper by Burbano, Perche, and Torres (2021), but first we will specify the model we are interested in studying in a Hamiltonian formulation.

### Hamiltonian Formulation

Since we are concerned only with the FRG analysis of the particle detector, we can simplify the Hamiltonian given by Eqs. (4.0.1) and (4.0.3) on page 93 and on page 94 by considering a single detector. This leaves us with

$$H(t) = H_\phi + \frac{\Omega}{2}\frac{\mathrm{d}\tau}{\mathrm{d}t}\sigma^z + H_{\mathrm{int}}(t), \tag{4.1.1}$$

where

$$H_{\mathrm{int}}(t) = \epsilon(t) \int_{\Sigma_t} \psi(t,\vec{x})\sigma^x(t) \otimes \phi(t,\vec{x})\sqrt{-g}\,\mathrm{d}^3x. \tag{4.1.2}$$

These Hamiltonians generate translations with respect to the global time function $t$. To generate translations with respect to proper time—which will be more interesting for our purposes—we can write

$$H(\tau) = \frac{\mathrm{d}t}{\mathrm{d}\tau}H_\phi + \frac{\Omega}{2}\sigma^z + \frac{\mathrm{d}t}{\mathrm{d}\tau}H_{\mathrm{int}}(\tau). \tag{4.1.3}$$

We shall also simplify the interaction term given by Eq. (4.1.2). Instead of working with the general case of a smeared particle detector, we will focus on pointlike detectors, which are commonly known as Unruh–DeWitt detectors (DeWitt 1979; Unruh 1976).

To consider a pointlike detector means to choose the smearing function $\psi$ on Eq. (4.1.2) to have the form

$$\psi(t,\vec{x}) = \frac{1}{\sqrt{-g}}\int \delta^{(d)}(x,z(\tau))\,\mathrm{d}\tau, \tag{4.1.4}$$

where the integral is taken over the detector's proper time and $z(\tau)$ denotes the detector's worldline (*cf.* Perche and Martín-Martínez 2022, Eq. (51); Poisson, Pound, and Vega 2011, Eq. (17.4)). We can then write the interaction Hamiltonian on Eq. (4.1.2) as

$$H_{\mathrm{int}}(t) = \epsilon(t) \iint_{\Sigma_t} \delta^{(d)}(x,z(\tau))\sigma^x(t) \otimes \phi(t,\vec{x})\,\mathrm{d}^{d-1}x\,\mathrm{d}\tau, \tag{4.1.5a}$$

$$= \epsilon(t) \int \delta\big(t - z^0(\tau)\big)\sigma^x(t) \otimes \phi(t,\vec{z}(\tau))\,\mathrm{d}\tau, \tag{4.1.5b}$$

$$= \epsilon(t) \int \frac{\delta\big(\tau - (z^0)^{-1}(t)\big)}{\left|\frac{\mathrm{d}z^0}{\mathrm{d}\tau}\right|}\sigma^x(t) \otimes \phi(t,\vec{z}(\tau))\,\mathrm{d}\tau, \tag{4.1.5c}$$

$$= \epsilon(t) \int \delta\big(\tau - (z^0)^{-1}(t)\big)\frac{\mathrm{d}\tau}{\mathrm{d}t}\sigma^x(t) \otimes \phi(t,\vec{z}(\tau))\,\mathrm{d}\tau, \tag{4.1.5d}$$

$$= \epsilon(t)\sigma^x(t) \otimes \phi(t,\vec{z}(\tau(t)))\frac{\mathrm{d}\tau}{\mathrm{d}t}. \tag{4.1.5e}$$

Eqs. (4.1.3) and (4.1.5) then yield

$$H(\tau) = \frac{\mathrm{d}t}{\mathrm{d}\tau}H_\phi + \frac{\Omega}{2}\sigma^z + \epsilon(\tau)\sigma^x(\tau) \otimes \phi(z(\tau)). \tag{4.1.6}$$

Notice that on Eq. (4.1.6) we wrote all factors in the interaction term as functions of $\tau$, despite the fact that on Eq. (4.1.5) they were written in terms of the global time function $t$. This is possible because the detector is being implicitly assumed to move on a causal trajectory, and hence its worldline intercepts each of the hypersurfaces of constant $t$ exactly once, as a consequence of the fact that these hypersurfaces are Cauchy surfaces (Wald 1984, Theorem 8.3.14). Therefore, $\tau$ and $t$ are in one-to-one correspondence.

### Action Formulation

Let us then figure out the action describing the detector. We do not need to consider the term on Eq. (4.1.6) on the preceding page describing the quantum field, since it will only amount to the usual action of a scalar field. We will then focus on the remaining terms, which involve the detector.

The problem of formulating the Unruh–DeWitt detector by means of an action was recently addressed by Burbano, Perche, and Torres (2021) with the goal of providing a path integral formulation of particle detectors. We shall follow their approach.

We begin by introducing the $\mathfrak{su}(2)$ ladder operators by[*]

$$\sigma^{\pm} = \frac{\sigma^x \pm i\sigma^y}{2}.\tag{4.1.7}$$

With this definition we can then write

$$\sigma^x = \sigma^+ + \sigma^-\tag{4.1.8}$$

and

$$\sigma^z = \sigma^+ \sigma^- - \sigma^- \sigma^+.\tag{4.1.9}$$

Therefore, the gap term becomes

$$H_\Omega = \frac{\Omega}{2}\left(\sigma^+ \sigma^- - \sigma^- \sigma^+\right).\tag{4.1.10}$$

Eq. (4.1.10) resembles the Hamiltonian for a fermionic harmonic oscillator (Nakahara 2003, Sec. 1.5.1). In particular, notice how

$$\{\sigma^+, \sigma^-\} = 1 \quad\text{and}\quad \{\sigma^\pm, \sigma^\pm\} = 0.\tag{4.1.11}$$

The path integral quantization of such a system is a well-understood problem in QM and is discussed, for example, by Nakahara (2003, Sec. 1.5). It requires us to employ Grassmann variables (which are reviewed by DeWitt 2003, App. A; Nakahara 2003, Sec. 1.5).

If our problem was simply to quantize a fermionic harmonic oscillator, we would introduce a pair of Grassmann variables $\bar{\theta}$ and $\theta$ which, upon quantization, would satisfy the canonical anticommutation relations (CAR)

$$\{\bar{\theta}, \theta\} = 1, \quad \{\theta, \theta\} = 0, \quad\text{and}\quad \{\bar{\theta}, \bar{\theta}\} = 0.\tag{4.1.12}$$

This is the same algebra given on Eq. (4.1.11). Hence, a path integral formulation in terms of Grassmann variables would yield the correct algebra we are expecting. Eq. (4.1.12) can also be seen to describe the complex Clifford algebra $\mathbb{C}\ell_2$ (see Vaz Jr. and Rocha Jr. 2016; Woit 2017, Chaps. 28 and 29) by defining

$$\chi_1 = \frac{\theta + \bar{\theta}}{\sqrt{2}} \quad\text{and}\quad \chi_2 = \frac{i(\theta - \bar{\theta})}{\sqrt{2}}\tag{4.1.13}$$

and noticing this implies

$$\{\chi_i, \chi_j\} = \delta_{ij}.\tag{4.1.14}$$

---

[*]Since these operators lie on the complexification of $\mathfrak{su}(2)$, they are actually elements of $\mathfrak{sl}(2, \mathbb{C})$ (Hall 2015, Sec. 3.6).

An Unruh–DeWitt detector, however, is more complicated than a fermionic harmonic oscillator. In terms of the Grassmann variables $\bar{\theta}$ and $\theta$, we would represent the gap term by "mimicking" Eq. (4.1.10) on the previous page with an action term similar to

$$S_\Omega \sim \frac{\Omega}{2}\big(\bar{\theta}\theta - \theta\bar{\theta}\big),\tag{4.1.15}$$

where the details still need to be worked out and we are ignoring the necessary integrals. For now, we are only interested in the algebraic structure. The extra difficulty in the Unruh–DeWitt detector comes from the fact we also need to somehow represent the interaction with the field. Our current proposal would look like

$$S_{\text{int}} \sim \epsilon\phi\big(\bar{\theta} + \theta\big).\tag{4.1.16}$$

This is not an acceptable action. Indeed, since $\phi$ is real and $\bar{\theta}$ and $\theta$ are Grassmann numbers, this is an a-number, but the action should be a c-number. Hence, this approach cannot be right.

To correctly represent the detector, we must find a manner to represent the CAR algebra in an "even manner", *i.e.*, in such a way that each term occurring in the action always involves the product of an even number of Grassmann fields. In this way, we can ensure the action will be a c-number.

We already know the terms in the action can be written in terms of the Clifford algebra $\mathbb{C}\ell_2$, but this requires the use of odd terms. Nevertheless, it is known within the theory of Clifford algebras that $\mathbb{C}\ell_n^+ \simeq \mathbb{C}\ell_{n-1}$, where $\mathbb{C}\ell_n^+$ represents the even part of $\mathbb{C}\ell_n$ and $\simeq$ denotes an isomorphism. This follows, for example, from the results given in the book by Vaz Jr. and Rocha Jr. (2016, Theorem 4.2 and 4.4). Therefore, $\mathbb{C}\ell_2 \simeq \mathbb{C}\ell_3^+$, and we can represent the necessary degrees of the freedom for the detector by employing a representation of $\mathbb{C}\ell_3$. Venturelli (2007) discusses this approach in the related problem of the spin-boson system in condensed matter physics.

When working with the FRGE, the representation provided by $\mathbb{C}\ell_3$ turns out to be inconvenient. The ansatz provided by the classical action is a bad truncation for the EAA because the interaction term is not capable of generating corrections to other terms, and hence we would have to supplement the ansatz with more coupling constants. To avoid this, we will reapply the result $\mathbb{C}\ell_n^+ \simeq \mathbb{C}\ell_{n-1}$ and work with a representation given by $\mathbb{C}\ell_4^+$. We then have the relations

$$\mathbb{C}\ell_4^+ \simeq \mathbb{C}\ell_3 \supseteq \mathbb{C}\ell_3^+ \simeq \mathbb{C}\ell_2.\tag{4.1.17}$$

This is the same approach chosen by Burbano, Perche, and Torres (2021).

$\mathbb{C}\ell_4$ is generated by four Grassmann variables (in addition to the identity, which is commutative). Hence, we shall work with two pairs of Grassmann variables: $\theta$, $\bar{\theta}$, $\eta$, and $\bar{\eta}$. Upon quantization they will respect the algebra given by

$$\{\bar{\theta},\theta\} = 1 \quad\text{and}\quad \{\bar{\eta},\eta\} = 1\tag{4.1.18}$$

while all other anticommutators vanish. In analogy with Eq. (4.1.13) on the preceding page, we may define

$$\chi_3 = \frac{\eta + \bar{\eta}}{\sqrt{2}} \quad\text{and}\quad \chi_4 = \frac{i(\eta - \bar{\eta})}{\sqrt{2}}\tag{4.1.19}$$

and obtain

$$\{\chi_i,\chi_j\} = \delta_{ij}\tag{4.1.20}$$

for $i, j = 1, 2, 3, 4$.

Since our original goal was to represent $\mathbb{C}\ell_2$, how can we now interpret these objects coming from $\mathbb{C}\ell_4$? Our trick is to see both the ground and excited states of the detector as excitations on some more fundamental "true vacuum". This idea is similar to how, deep down, the energy levels on a hydrogen atom are all excitations upon the Minkowski vacuum for the electron field—instead of interpreting the changes in energy levels as straightforward energy transitions, we can also interpret them as the annihilation of an electron at some energy level and the creation of an electron at the new energy level. A similar concept was employed by Torres et al. (2020) to understand neutrino oscillations. In their model, one describes a particle detector in terms of protons and neutrons, which are both seem as excitations upon an underlying "true vacuum" that represents the absence of nucleons.

From this point of view, we may then see the operators $\sigma^\pm$ as being shorthands for

$$\sigma^+ = a_e^\dagger a_g \quad \text{and} \quad \sigma^- = a_g^\dagger a_e, \tag{4.1.21}$$

where $a_g$, $a_g^\dagger$, $a_e$, and $a_e^\dagger$ are the ladder operators for the ground and excited states. In other words, $\sigma^+$, for example, acts on the detector by annihilating the ground state $|g\rangle$ and then creating the excited state $|e\rangle$. To see how this relates to our representation of $\mathbb{C}\ell_4$, notice that these new ladder operators satisfy

$$\{a_e^\dagger, a_e\} = 1 \quad \text{and} \quad \{a_g^\dagger, a_g\} = 1 \tag{4.1.22}$$

with all other anticommutators vanishing. This is the same algebra the Grassmann variables $\theta$, $\bar{\theta}$, $\eta$, and $\bar{\eta}$ satisfy, as shown on Eq. (4.1.18) on the preceding page. Hence, we can interpret these Grassmann variables as representing the ladder operators associated with the ground and excited states upon some more fundamental underlying vacuum. More specifically, $\bar{\theta}$, $\bar{\eta}$, $\theta$, and $\eta$ are associated with $a_e^\dagger$, $a_g^\dagger$, $a_e$, and $a_g$, respectively.

In terms of these new ladder operators, the gap term can be written as

$$H_\Omega(a_g^\dagger, a_g, a_e^\dagger, a_e) = \frac{\Omega}{2}\left(a_e^\dagger a_g a_g^\dagger a_e - a_g^\dagger a_e a_e^\dagger a_g\right), \tag{4.1.23a}$$

$$= \frac{\Omega}{2}\left(a_e^\dagger a_e - a_e^\dagger a_g^\dagger a_g a_e - a_g^\dagger a_g + a_g^\dagger a_e^\dagger a_e a_g\right), \tag{4.1.23b}$$

$$= \frac{\Omega}{2}\left(a_e^\dagger a_e - a_g^\dagger a_g\right). \tag{4.1.23c}$$

Similarly, the interaction term becomes

$$H_{\text{int}}(a_g^\dagger, a_g, a_e^\dagger, a_e) = \epsilon(\tau)\sigma^x(\tau) \otimes \phi(z(\tau)), \tag{4.1.24a}$$

$$= \epsilon(\tau)\left[\sigma^+(\tau) + \sigma^-(\tau)\right] \otimes \phi(z(\tau)), \tag{4.1.24b}$$

$$= \epsilon(\tau)\left[a_e^\dagger(\tau)a_g(\tau) + a_g^\dagger(\tau)a_e(\tau)\right] \otimes \phi(z(\tau)). \tag{4.1.24c}$$

We want then to obtain the Euclidean action for this theory. Given a Hamiltonian in terms of creation and annihilation operators, it is well-known how one can obtain the coherent state path integral (see Altland and Simons 2010, Sec. 4.2). Hence, we obtain (Burbano, Perche, and Torres 2021, pp. 6–7)

$$S = S_{\hat{\phi}} + \int \bar{\theta}\dot{\theta} + \bar{\eta}\dot{\eta} + H_\Omega(\bar{\eta}, \eta, \bar{\theta}, \theta) + H_{\text{int}}(\bar{\eta}, \eta, \bar{\theta}, \theta) \, \mathrm{d}\tau, \tag{4.1.25}$$

where the dots denote differentiation with respect to $\tau$, the terms in $H_\Omega$ and $H_{\text{int}}$ are understood as being evaluated on the detector's worldline, and $S_{\hat{\phi}}$ denotes the quantum field's action. More explicitly, we can write the classical action for the field-detector system as

$$S = S_\phi + \int \bar{\theta}\dot{\theta} + \bar{\eta}\dot{\eta} + \frac{\Omega}{2}(\bar{\theta}\theta - \bar{\eta}\eta) + \varepsilon(\tau)\phi(\tau)(\bar{\eta}\theta + \bar{\theta}\eta)\,\mathrm{d}\tau\,, \qquad (4.1.26)$$

where $\phi(\tau) \equiv \phi(z(\tau))$ and the Grassmann variables are all understood as being defined on the detector's worldline.

## 4.2    Setting Up the FRGE

We are now in position to discuss the ansätze that we will need when working with the FRGE for the field-detector system. Namely, we must choose a truncation for the EAA and a regulator. We shall also make a few preliminary calculations that will be necessary to compute the RG flow.

### Truncation Ansatz

Inspired by Eqs. (3.3.8) and (4.1.26) on page 75 and on the current page, we shall consider the ansatz

$$\Gamma_k = \int \frac{1}{2}\partial_a\phi\partial^a\phi + V_k(\phi)\,\mathrm{d}^d x + \int \bar{\theta}\dot{\theta} + \bar{\eta}\dot{\eta} + \frac{\Omega_k}{2}(\bar{\theta}\theta - \bar{\eta}\eta) + \varepsilon_k\chi(\tau)\phi(\tau)(\bar{\eta}\theta + \bar{\eta}\theta)\,\mathrm{d}\tau\,, \qquad (4.2.1)$$

where $\chi(\tau)$ is some previously chosen function modelling how the detector can be turned on and off. A few comments are in place.

Firstly, we are assuming the field to be subject to an arbitrary potential depending only on the field itself. This assumption is being made to keep the expression more general for the time being, and because it will allow us to take a look at whether field self-interactions can affect or be affected by the detector. Furthermore, assuming an interacting field will not affect most of our computations, and hence restricting our attention to free fields would not simplify them. In any case, we may later choose the potential to simply be a mass term, if we prefer[*].

Secondly, notice that on Eq. (4.1.26) the coupling between detector and field was responsible for turning the interaction on and off. The most natural generalization would then be to "promote" $\varepsilon(\tau)$ to a two-variable function $\varepsilon_k(\tau)$. Such a function lives in the tensor product space of the spaces of functions of $k$ and of $\tau$. Hence, it can be written as a sum of functions of the form $f(k)g(\tau)$, but will generally not have this form. We chose to make the simplifying assumption in which $\varepsilon_k(\tau) = \varepsilon_k\chi(\tau)$. This is not completely general from a mathematical point of view, but we expect it to be sufficient for our interests. A more general assumption also seems likely to lead to more difficulties than improvements. In any case, one can still wonder whether splitting $\chi(\tau) = \chi_1(\tau) + \chi_2(\tau)$ and working with two couplings $\varepsilon_k^1$ and $\varepsilon_k^2$ such that $\varepsilon_k(\tau) = \varepsilon_k^1\chi_1(\tau) + \varepsilon_k^2\chi_2(\tau)$ would lead to the same results. For now, our calculations will not depend on this simplifying assumption, so we will keep the shorthand $\varepsilon_k(\tau) = \varepsilon_k\chi(\tau)$ to save space, but it can be relevant later on. At the final stages of our calculations, though, we will make one more simplifying assumption and take $\varepsilon_k(\tau) = \varepsilon_k$, without any time dependence.

Thirdly, notice that we are assuming the spacetime manifold to be flat on Eq. (4.2.1). We will also implicitly assume our path integrals to be computed with respect to the Minkowski vacuum, so we can be comfortable with the self-interaction terms we are considering for the quantum field, in spite of our previous concerns mentioned in Section 2.5. Nevertheless, it seems reasonable to expect the following calculations to be generalizable for curved spacetimes, at least when considering stationart states on static spacetimes.

---

[*]I thank Prof. Alex Gomes Dias for the suggestion of considering the effects of field self-interactions.

Lastly, we are not considering the running of any derivative terms. Therefore, our approach will be similar to when we considered the behavior of a scalar field with the local potential approximation in Section 3.3. In particular, we can assume the fields to be constant throughout spacetime in the Wetterich equation.

### Hessian of the EAA

The right-hand side of the Wetterich equation involves the Hessian of the EAA. Hence, we will surely need to compute it. We shall do this now, because looking at its form will also help us choose appropriate regulators to work with.

To make the functional traces in the Wetterich equation clear, we would like to work only with fields defined over the whole spacetime. Otherwise, we would get uncomfortable expressions that would require us to integrate over proper time and spacetime variables separately. While it may be possible to deal with such expressions, they are surely not usual in the literature, and hence we shall avoid them in favor of more common techniques.

Therefore, we would like to understand the detector in terms of fields defined over spacetime, not only on the worldline. To do so, we will "promote" the Grassmann variables to spacetime fields. To keep the action unchanged we shall use Dirac deltas to keep the physical detector variables restricted to its worldline.

We will illustrate it first with the interaction term. Notice that we can write

$$\int \epsilon_k(\tau)\phi(\tau)(\bar{\theta}\eta + \bar{\eta}\theta)\,\mathrm{d}\tau = \int \epsilon_k(\tau)\phi(z(\tau))(\bar{\theta}\eta + \bar{\eta}\theta)\,\mathrm{d}\tau\,, \tag{4.2.2a}$$

$$= \int \epsilon_k(\tau)\phi(z(\tau))(\bar{\theta}\eta + \bar{\eta}\theta)\vartheta^{(d)}(x - z(\tau))\,\mathrm{d}^d x\,\mathrm{d}\tau\,, \tag{4.2.2b}$$

$$= \int \epsilon_k(\tau)\phi(x)(\bar{\theta}\eta + \bar{\eta}\theta)\vartheta^{(d)}(x - z(\tau))\,\mathrm{d}^d x\,\mathrm{d}\tau\,, \tag{4.2.2c}$$

where we used the fact that $\phi$ was originally defined on the whole spacetime, not only on the detector's worldline. If we now promote the other fields to fields on the whole spacetime and copy the steps we did on Eq. (4.1.5) on page 96, we can write

$$\int \epsilon_k(\tau)\phi(\tau)(\bar{\theta}\eta + \bar{\eta}\theta)\,\mathrm{d}\tau = \int \epsilon_k(t)\phi(x)(\bar{\theta}(x)\eta(x) + \bar{\eta}(x)\theta(x))\vartheta^{(d)}(x - z(\tau))\,\mathrm{d}^d x\,\mathrm{d}\tau\,, \tag{4.2.3a}$$

$$= \int \epsilon_k(t)\phi(x)(\bar{\theta}(x)\eta(x) + \bar{\eta}(x)\theta(x))\vartheta^{(d-1)}(\vec{x} - \vec{z}(t))\frac{\mathrm{d}\tau}{\mathrm{d}t}\,\mathrm{d}^d x\,. \tag{4.2.3b}$$

Since $t$ and $\tau$ are in one-to-one correspondence, there is nothing wrong with writing $\epsilon_k(t)$. We could do the same with the Grassmann fields, but it will be convenient to treat them as defined on spacetime without further simplifications.

Similarly, the gap term becomes

$$\int \frac{\Omega_k}{2}(\bar{\theta}\theta - \bar{\eta}\eta)\,\mathrm{d}\tau = \int \frac{\Omega_k}{2}(\bar{\theta}\theta - \bar{\eta}\eta)\vartheta^{(d-1)}(\vec{x} - \vec{z}(\tau))\frac{\mathrm{d}\tau}{\mathrm{d}t}\,\mathrm{d}^d x\,. \tag{4.2.4}$$

The kinetic terms, on the other hand, involve a time derivative and hence are a bit more subtle. Notice, however, that

$$\int \tilde{\theta} \partial_\tau \theta \, \mathrm{d}\tau \tag{4.2.5}$$

is parameterization invariant. Hence, since $\tau$ and $t$ are in a one-to-one correspondence, we have

$$\int \tilde{\theta}(\tau) \partial_\tau \theta(\tau) \, \mathrm{d}\tau = \int \tilde{\theta}(t) \partial_t \theta(t) \, \mathrm{d}t \,, \tag{4.2.6a}$$

$$= \int \tilde{\theta}(t, \vec{x}) \partial_t \theta(t, \vec{x}) \hat{\delta}^{(d-1)}(\vec{x} - \vec{z}(t)) \, \mathrm{d}t \, \mathrm{d}^{d-1}x \,, \tag{4.2.6b}$$

$$= \int \tilde{\theta}(x) \partial_t \theta(x) \hat{\delta}^{(d-1)}(\vec{x} - \vec{z}(t)) \, \mathrm{d}^d x \,, \tag{4.2.6c}$$

$$= \int \tilde{\theta}(x) \partial_\tau \theta(x) \hat{\delta}^{(d-1)}(\vec{x} - \vec{z}(t)) \frac{\mathrm{d}\tau}{\mathrm{d}t} \, \mathrm{d}^d x \,, \tag{4.2.6d}$$

$$= \int \tilde{\theta}(x) \frac{\mathrm{d}z^\mu}{\mathrm{d}\tau} \partial_\mu \theta(x) \hat{\delta}^{(d-1)}(\vec{x} - \vec{z}(t)) \frac{\mathrm{d}\tau}{\mathrm{d}t} \, \mathrm{d}^d x \,, \tag{4.2.6e}$$

$$= \int \tilde{\theta}(x) u^\mu \partial_\mu \theta(x) \hat{\delta}^{(d-1)}(\vec{x} - \vec{z}(t)) \frac{\mathrm{d}\tau}{\mathrm{d}t} \, \mathrm{d}^d x \,, \tag{4.2.6f}$$

$$= \int \tilde{\theta}(x) u^a \partial_a \theta(x) \hat{\delta}^{(d-1)}(\vec{x} - \vec{z}(t)) \frac{\mathrm{d}\tau}{\mathrm{d}t} \, \mathrm{d}^d x \,, \tag{4.2.6g}$$

where we defined $u^\mu = \frac{\mathrm{d}z^\mu}{\mathrm{d}\tau}$. Notice $u^a$ is the detector's four-velocity. It should be understood as a function of $\tau$ (or $t$), since it is defined only on the detector's worldline.

If we bring everything together we find that the EAA can be written as

$$\Gamma_k = \int \frac{1}{2} \partial_a \phi \partial^a \phi + V_k(\phi) \, \mathrm{d}^d x$$
$$+ \int \hat{\delta}^{(d-1)}(\vec{x} - \vec{z}(t)) \frac{\mathrm{d}\tau}{\mathrm{d}t} \left[ \tilde{\theta} u^a \partial_a \theta + \tilde{\eta} u^a \partial_a \eta + \frac{\Omega_k}{2} (\tilde{\theta}\theta - \tilde{\eta}\eta) + \epsilon_k(t)\phi(\tilde{\theta}\eta + \tilde{\eta}\theta) \right] \mathrm{d}^d x \,, \tag{4.2.7}$$

where all fields are now understood as being defined over spacetime.

To further simplify this expression, let us introduce the notation

$$\xi \equiv \begin{pmatrix} \theta \\ \eta \end{pmatrix} \tag{4.2.8}$$

and

$$\gamma_\tau = \hat{\delta}^{(d-1)}(\vec{x} - \vec{z}(t)) \frac{\mathrm{d}\tau}{\mathrm{d}t} \tag{4.2.9}$$

so that we may write

$$\Gamma_k = \int \frac{1}{2} \partial_a \phi \partial^a \phi + V_k(\phi) + \gamma_\tau \left[ \tilde{\xi}^\dagger u^a \partial_a \xi + \frac{\Omega_k}{2} \tilde{\xi}^\dagger \sigma_z \xi + \epsilon_k(t)\phi \tilde{\xi}^\dagger \sigma_x \xi \right] \mathrm{d}^d x \,, \tag{4.2.10}$$

where $\sigma_x$ and $\sigma_z$ are the Pauli matrices[*].

---

[*]I am now writing them as $\sigma_x$ and $\sigma_z$ rather than $\sigma^x$ and $\sigma^z$, because they are now not being considered as operators on the detector's Hilbert space, but merely as matrices.

Therefore, we have the derivative

$$\frac{\partial \Gamma_k}{\partial \Phi(x)} = \begin{pmatrix} -\partial^2 \phi(x) + V_k'(\phi) + \gamma_\tau \epsilon_k(t) \bar{\xi}^\dagger(x) \sigma_x \xi(x) \\ \gamma_\tau \left[ u^a \partial_a \bar{\xi}(x) - \frac{\Omega_k}{2} \sigma_z \bar{\xi}(x) - \epsilon_k(t) \phi(x) \sigma_x \bar{\xi}(x) \right] \\ \gamma_\tau \left[ u^a \partial_a \xi(x) + \frac{\Omega_k}{2} \sigma_z \xi(x) + \epsilon_k(t) \phi(x) \sigma_x \xi(x) \right] \end{pmatrix}, \tag{4.2.11}$$

where we took the convention $\Phi^\top = \left( \phi, \xi^\top, \bar{\xi}^\dagger \right)$. To make this computation, it is convenient to notice that

$$\int \bar{\xi}^\dagger \partial_\tau \xi \, \mathrm{d}\tau = \int \xi^\dagger \partial_\tau \bar{\xi} \, \mathrm{d}\tau, \tag{4.2.12}$$

as one can show using integration by parts and the fact $\xi$ is composed of Grassmann variables. It is also useful to notice that

$$\bar{\xi}^\dagger \sigma_x \xi = -\xi^\dagger \sigma_x \bar{\xi}, \tag{4.2.13}$$

and similarly for $\sigma_z$.

Eq. (4.2.11) implies the Hessian is

$$\frac{\bar{\partial} \Gamma_k \bar{\partial}}{\partial \Phi(x) \partial \Phi(x')} = \begin{pmatrix} -\partial^2 + V_k''(\phi) & \gamma_\tau \epsilon_k \bar{\xi}^\dagger \sigma_x & -\gamma_\tau \epsilon_k \xi^\dagger \sigma_x \\ -\gamma_\tau \epsilon_k \sigma_x \bar{\xi} & 0 & \gamma_\tau \left[ u^a \partial_a - \frac{\Omega_k}{2} \sigma_z - \epsilon_k \phi \sigma_x \right] \\ \gamma_\tau \epsilon_k \sigma_x \xi & \gamma_\tau \left[ u^a \partial_a + \frac{\Omega_k}{2} \sigma_z + \epsilon_k \phi \sigma_x \right] & 0 \end{pmatrix} \partial^{(d)} \left( x - x' \right), \tag{4.2.14}$$

where we were able to omit the spacetime dependence of the fields and of $\epsilon_k(t)$ due to the Dirac delta.

## Choice of Regulator

We choose to work with a regulator with the form

$$\mathcal{R}_k \left( x, x' \right) = \begin{pmatrix} R_k^\phi & 0 & 0 \\ 0 & 0 & -\gamma_\tau R_k^d \sigma_z \\ 0 & \gamma_\tau R_k^d \sigma_z & 0 \end{pmatrix} \partial^{(d)} \left( x - x' \right), \tag{4.2.15}$$

where this choice is motivated by the desire that the detector regulator matches a gap term, preventing divergences in the gapless limit. Another possible choice would be to match the kinetic terms—this was used by Jungnickel and Wetterich (1996) and Litim (2001) with chiral fermions to avoid spoiling chiral symmetry, for example.

For now, we do not need to specify our choices for $R_k^\phi$ and $R_k^d$ and we shall keep it like that for as long as possible. Nevertheless, when the time comes, we will use cutoffs defined through profiles according to

$$R_k^\phi(z) = k^2 r_\phi \left( \frac{z}{k^2} \right) \tag{4.2.16a}$$

and

$$R_k^d(z) = k r_d \left( \frac{z}{k^2} \right). \tag{4.2.16b}$$

A particularly interesting case is the Litim cutoff (Litim 2001), given by

$$R_k^\phi(z) = (k^2 - z) \Theta(k^2 - z) \tag{4.2.17}$$

and

$$R_k^d(z) = \left(k - \frac{z}{k}\right)\Theta(k^2 - z). \tag{4.2.18}$$

Eq. (4.2.17) on the preceding page is identical to the cutoff given by Eq. (3.2.9) on page 65 and Eq. (4.2.18) is the same cutoff, apart from a division by $k$ to keep dimensional consistency.

## 4.3    Renormalization Group Flow

Let us then proceed to use the Wetterich equation to find the beta functions for $\Omega_k$, $\epsilon_k$, and for the field self-couplings. While these calculations are not trivial, they are fairly algorithmic. We shall follow the procedure outlined by Benedetti et al. (2011).

### Regularized Propagator

As we can see on Eq. (3.2.38) on page 71, we need to compute the inverse of $\Gamma_k^{(2)} + \mathscr{R}_k$. Since $\Gamma_k^{(2)} + \mathscr{R}_k$ is the inverse of the connected propagator given by $W_k$, we can say that we need to compute the "regularized propagator" for the theory. While this is an arduous task, it is simplified by noticing our choice of cutoff on Eq. (4.2.15) on the preceding page is a sparse matrix. Hence, upon taking the supertrace of $(\Gamma_k^{(2)} + \mathscr{R}_k)^{-1}\partial_t\mathscr{R}_k$, only a handful of terms of $(\Gamma_k^{(2)} + \mathscr{R}_k)^{-1}$ will be relevant. Let us figure out which terms are these.

Suppose

$$(\Gamma_k^{(2)} + \mathscr{R}_k)^{-1} = \begin{pmatrix} \mathbf{A} & \mathbf{B} & \mathbf{C} \\ \mathbf{D} & \mathbf{E} & \mathbf{F} \\ \mathbf{G} & \mathbf{H} & \mathbf{I} \end{pmatrix}. \tag{4.3.1}$$

Then the Wetterich equation reads, for the regulator given on Eq. (4.2.15) on the previous page,

$$\partial_t\Gamma_k = \frac{1}{2}\,\mathrm{STr}\left[(\Gamma_k^{(2)} + \mathscr{R}_k)^{-1}\partial_t\mathscr{R}_k\right], \tag{4.3.2a}$$

$$= \frac{1}{2}\,\mathrm{STr}\left[\begin{pmatrix} \mathbf{A} & \mathbf{B} & \mathbf{C} \\ \mathbf{D} & \mathbf{E} & \mathbf{F} \\ \mathbf{G} & \mathbf{H} & \mathbf{I} \end{pmatrix}\begin{pmatrix} \partial_t R_k^\phi & 0 & 0 \\ 0 & 0 & -\gamma_\tau\partial_t R_k^d\sigma_z \\ 0 & \gamma_\tau\partial_t R_k^d\sigma_z & 0 \end{pmatrix}\right], \tag{4.3.2b}$$

$$= \frac{1}{2}\,\mathrm{STr}\left[\begin{pmatrix} \mathbf{A}\partial_t R_k^\phi & \mathbf{C}\gamma_\tau\partial_t R_k^d\sigma_z & -\mathbf{B}\gamma_\tau\partial_t R_k^d\sigma_z \\ \mathbf{D}\partial_t R_k^\phi & \mathbf{F}\gamma_\tau\partial_t R_k^d\sigma_z & -\mathbf{E}\gamma_\tau\partial_t R_k^d\sigma_z \\ \mathbf{G}\partial_t R_k^\phi & \mathbf{I}\gamma_\tau\partial_t R_k^d\sigma_z & -\mathbf{H}\gamma_\tau\partial_t R_k^d\sigma_z \end{pmatrix}\right], \tag{4.3.2c}$$

$$= \frac{1}{2}\,\mathrm{Tr}\left[\mathbf{A}\partial_t R_k^\phi\right] - \frac{1}{2}\,\mathrm{Tr}\left[\mathbf{F}\gamma_\tau\partial_t R_k^d\sigma_z\right] + \frac{1}{2}\,\mathrm{Tr}\left[\mathbf{H}\gamma_\tau\partial_t R_k^d\sigma_z\right], \tag{4.3.2d}$$

where $t$ represents the RG time $t = \log k$ (we will no longer make use of the spacetime's global time function in this chapter) and we omitted the omnipresent Dirac delta $\delta^{(d)}(x - x')$. We see then that we only need to compute three block entries of $(\Gamma_k^{(2)} + \mathscr{R}_k)^{-1}$. In fact, we do not even need to compute these entries exactly, for some of their components will lie outside the truncation we are working with. Hence, we will be able to perform a few simplifications during our computations.

To compute $(\Gamma_k^{(2)} + \mathcal{R}_k)^{-1}$ we will treat the functional matrices formally rather than writing their products as

$$(\mathrm{M}_1 \mathrm{M}_2)(x, z) = \int \mathrm{M}_1(x, y) \mathrm{M}_2(y, z)\, \mathrm{d}^d y, \tag{4.3.3}$$

which would be a more rigorous approach. Since some of the entries of $\Gamma_k^{(2)}$ are Grassmann-valued (see Eq. (4.2.14) on page 103), we must compute the inverse by using formulae for inverses of block matrices. These are given, for example, in the book by Bernstein (2018, Sec. 3.9). We can, however, exploit the fact that our truncation only needs terms that are at most quadratic in the Grassmann fields, for this means we do not need an exact inverse. It suffices to make a series expansion approach. This is, in fact, the way inverses in the Grassmann algebra are defined (DeWitt 2003, App. A).

To compute the block inverse, we will consider $\Gamma_k^{(2)} + \mathcal{R}_k$ as given by blocks according to

$$\begin{pmatrix} A & B \\ C & D \end{pmatrix} \equiv \begin{pmatrix} P_k + V_k''(\phi) & \gamma_\tau \epsilon_k \bar{\xi}^\dagger \sigma_x & -\gamma_\tau \epsilon_k \bar{\xi}^\dagger \sigma_x \\ -\gamma_\tau \epsilon_k \sigma_x \bar{\xi} & 0 & \gamma_\tau \left[ u^a \partial_a - \left( R_k^d + \frac{\Omega_k}{2} \right) \sigma_z - \epsilon_k \phi \sigma_x \right] \\ \gamma_\tau \epsilon_k \sigma_x \xi & \gamma_\tau \left[ u^a \partial_a + \left( R_k^d + \frac{\Omega_k}{2} \right) \sigma_z + \epsilon_k \phi \sigma_x \right] & 0 \end{pmatrix}, \tag{4.3.4}$$

where we introduced the shorthand

$$P_k(\Delta) \equiv \Delta + R_k^\phi(\Delta), \tag{4.3.5}$$

where $\Delta = -\partial^2$. We then notice that, if $\mathrm{M}_1$ and $\mathrm{M}_2$ are matrices,

$$(\mathrm{M}_1 + \mathrm{M}_2)^{-1} = \mathrm{M}_1^{-1} - \mathrm{M}_1^{-1}\mathrm{M}_2\mathrm{M}_1^{-1} + \mathrm{M}_1^{-1}\mathrm{M}_2\mathrm{M}_1^{-1}\mathrm{M}_2\mathrm{M}_1^{-1} - \mathrm{M}_1^{-1}\mathrm{M}_2\mathrm{M}_1^{-1}\mathrm{M}_2\mathrm{M}_1^{-1}\mathrm{M}_2\mathrm{M}_1^{-1} + \cdots. \tag{4.3.6}$$

As we mentioned, this formula is particularly useful when working with anticommuting variables (see DeWitt 2003, Eq. (A.5)).

We can then write

$$\begin{aligned} \left( \Gamma_k^{(2)} + \mathcal{R}_k \right)^{-1} &= \begin{pmatrix} A & 0 \\ 0 & D \end{pmatrix}^{-1} - \begin{pmatrix} A & 0 \\ 0 & D \end{pmatrix}^{-1} \begin{pmatrix} 0 & B \\ C & 0 \end{pmatrix} \begin{pmatrix} A & 0 \\ 0 & D \end{pmatrix}^{-1} \\ &\quad + \begin{pmatrix} A & 0 \\ 0 & D \end{pmatrix}^{-1} \begin{pmatrix} 0 & B \\ C & 0 \end{pmatrix} \begin{pmatrix} A & 0 \\ 0 & D \end{pmatrix}^{-1} \begin{pmatrix} 0 & B \\ C & 0 \end{pmatrix} \begin{pmatrix} A & 0 \\ 0 & D \end{pmatrix}^{-1} + \cdots, \end{aligned} \tag{4.3.7}$$

where the dots stand for higher terms in the series expansion that will involve at least three instances of Grassmann variables coming from $B$ or $C$ (and hence lie beyond our truncation). Notice then that we have

$$\left( \Gamma_k^{(2)} + \mathcal{R}_k \right)^{-1} = \begin{pmatrix} A^{-1} & 0 \\ 0 & D^{-1} \end{pmatrix} - \begin{pmatrix} 0 & A^{-1}BD^{-1} \\ D^{-1}CA^{-1} & 0 \end{pmatrix} + \begin{pmatrix} A^{-1}BD^{-1}CA^{-1} & 0 \\ 0 & D^{-1}CA^{-1}BD^{-1} \end{pmatrix} + \cdots. \tag{4.3.8}$$

By comparing this expression to Eqs. (4.3.1) and (4.3.2) on the preceding page, we see that the middle term does not contribute to the Wetterich equation. Hence, we only need to consider the remaining terms.

Eq. (4.3.4) allows us to see that

$$A^{-1} = \frac{1}{P_k + V_k''(\phi)} \tag{4.3.9}$$

and

$$D^{-1} = \frac{1}{\gamma_\tau \left[ (u^a \partial_a)^2 - \left( R_k^d + \frac{\Omega_k}{2} \right)^2 - \epsilon_k^2 \phi^2 \right]} \begin{pmatrix} 0 & u^a \partial_a + \left( R_k^d + \frac{\Omega_k}{2} \right) \sigma_z + \epsilon_k \phi \sigma_x \\ u^a \partial_a - \left( R_k^d + \frac{\Omega_k}{2} \right) \sigma_z - \epsilon_k \phi \sigma_x & 0 \end{pmatrix}.$$

$$(4.3.10)$$

With a little more algebraic work, one can also find that

$$A^{-1}BD^{-1}CA^{-1} = \frac{2\gamma_\tau \epsilon_k^2 \left[ \epsilon_k \phi \xi \bar{\xi}^\intercal \sigma_x \xi - \left( R_k^d + \frac{\Omega_k}{2} \right) \bar{\xi}^\intercal \sigma_z \xi \right]}{\left[ P_k + V_k''(\phi) \right]^2 \left[ (u^a \partial_a)^2 - \left( R_k^d + \frac{\Omega_k}{2} \right)^2 - \epsilon_k^2 \phi^2 \right]}.$$

$$(4.3.11)$$

It is a bit more complicated to compute $D^{-1}CA^{-1}BD^{-1}$. To do it, we begin by noticing that

$$CA^{-1}B = \frac{1}{P_k + V_k''(\phi)} \begin{pmatrix} -\gamma_\tau \epsilon_k \sigma_x \bar{\xi} \\ \gamma_\tau \epsilon_k \sigma_x \xi \end{pmatrix} \begin{pmatrix} \gamma_\tau \epsilon_k \bar{\xi}^\intercal \sigma_x & -\gamma_\tau \epsilon_k \xi^\intercal \sigma_x \end{pmatrix}, \quad (4.3.12a)$$

$$= \frac{\gamma_\tau^2 \epsilon_k^2}{P_k + V_k''(\phi)} \begin{pmatrix} -\sigma_x \bar{\xi} \bar{\xi}^\intercal \sigma_x & \sigma_x \bar{\xi} \xi^\intercal \sigma_x \\ \sigma_x \xi \bar{\xi}^\intercal \sigma_x & -\sigma_x \xi \xi^\intercal \sigma_x \end{pmatrix}. \quad (4.3.12b)$$

Terms coming from $\xi \xi^\intercal$ and $\bar{\xi} \bar{\xi}^\intercal$ cannot affect our truncation, and hence we may ignore them. We then write

$$CA^{-1}B = \frac{\gamma_\tau^2 \epsilon_k^2}{P_k + V_k''(\phi)} \begin{pmatrix} 0 & \sigma_x \bar{\xi} \xi^\intercal \sigma_x \\ \sigma_x \xi \bar{\xi}^\intercal \sigma_x & 0 \end{pmatrix} + \cdots, \quad (4.3.13)$$

where the dots stand for terms outside of our truncation. We can then compute $D^{-1}CA^{-1}BD^{-1}$ by employing Eqs. (4.3.10) and (4.3.13). We shall find that

$$D^{-1}CA^{-1}BD^{-1} = \alpha \begin{pmatrix} 0 & f \\ h & 0 \end{pmatrix} + \cdots, \quad (4.3.14)$$

where the coefficient is

$$\alpha = \frac{\epsilon_k^2}{\left[ P_k + V_k''(\phi) \right] \left[ (u^a \partial_a)^2 - \left( R_k^d + \frac{\Omega_k}{2} \right)^2 - \epsilon_k^2 \phi^2 \right]^2} \quad (4.3.15)$$

and the entries are

$$\begin{aligned}
f = {} & \sigma_x \xi \bar{\xi}^\intercal \sigma_x (u^a \partial_a)^2 - i \left( R_k + \frac{\Omega_k}{2} \right) \sigma_x \xi \bar{\xi}^\intercal \sigma_y u^a \partial_a + \epsilon_k \phi \sigma_x \xi \bar{\xi}^\intercal u^a \partial_a \\
& + i \left( R_k + \frac{\Omega_k}{2} \right) \sigma_y \xi \bar{\xi}^\intercal \sigma_x u^a \partial_a + \left( R_k + \frac{\Omega_k}{2} \right)^2 \sigma_y \xi \bar{\xi}^\intercal \sigma_y + i \epsilon_k \phi \left( R_k + \frac{\Omega_k}{2} \right) \sigma_y \xi \bar{\xi}^\intercal \\
& + \epsilon_k \phi \xi \bar{\xi}^\intercal \sigma_x u^a \partial_a - i \epsilon_k \phi \left( R_k + \frac{\Omega_k}{2} \right) \xi \bar{\xi}^\intercal \sigma_x + \epsilon_k^2 \phi^2 \xi \bar{\xi}^\intercal \quad (4.3.16)
\end{aligned}$$

and

$$h = \sigma_x \bar{\xi}\xi^{\mathsf{T}}\sigma_x(u^a\partial_a)^2 + i\left(R_k + \frac{\Omega_k}{2}\right)\sigma_x\bar{\xi}\xi^{\mathsf{T}}\sigma_y u^a\partial_a - \epsilon_k\phi\sigma_x\bar{\xi}\xi^{\mathsf{T}}u^a\partial_a$$

$$- i\left(R_k + \frac{\Omega_k}{2}\right)\sigma_y\bar{\xi}\xi^{\mathsf{T}}\sigma_x u^a\partial_a + \left(R_k + \frac{\Omega_k}{2}\right)^2\sigma_y\bar{\xi}\xi^{\mathsf{T}}\sigma_y + i\epsilon_k\phi\left(R_k + \frac{\Omega_k}{2}\right)\sigma_y\bar{\xi}\xi^{\mathsf{T}}$$

$$- \epsilon_k\phi\bar{\xi}\xi^{\mathsf{T}}\sigma_x u^a\partial_a - i\epsilon_k\phi\left(R_k + \frac{\Omega_k}{2}\right)\bar{\xi}\xi^{\mathsf{T}}\sigma_x + \epsilon_k^2\phi^2\bar{\xi}\xi^{\mathsf{T}}. \quad (4.3.17)$$

Notice that $f$ and $h$ are terms on the matrices F and H shown on Eq. (4.3.1) on page 104. As we see on Eq. (4.3.2) on page 104, in the Wetterich equation they shall occur only through terms with the form

$$\mathrm{Tr}\left[\alpha f\gamma_\tau\partial_t R_k^d\sigma_z\right] \quad \text{and} \quad \mathrm{Tr}\left[\alpha h\gamma_\tau\partial_t R_k^d\sigma_z\right]. \quad (4.3.18)$$

Hence, we only need to keep the terms of $f$ and $h$ that are part of our truncation after being multiplied by $\sigma_z$ and traced over. This means we can write

$$f = \sigma_x\xi\bar{\xi}^{\mathsf{T}}\sigma_x(u^a\partial_a)^2 + \left(R_k + \frac{\Omega_k}{2}\right)^2\sigma_y\xi\bar{\xi}^{\mathsf{T}}\sigma_y + i\epsilon_k\phi\left(R_k + \frac{\Omega_k}{2}\right)\sigma_y\xi\bar{\xi}^{\mathsf{T}} + \cdots \quad (4.3.19)$$

and

$$h = \sigma_x\bar{\xi}\xi^{\mathsf{T}}\sigma_x(u^a\partial_a)^2 + \left(R_k + \frac{\Omega_k}{2}\right)^2\sigma_y\bar{\xi}\xi^{\mathsf{T}}\sigma_y + i\epsilon_k\phi\left(R_k + \frac{\Omega_k}{2}\right)\sigma_y\bar{\xi}\xi^{\mathsf{T}} + \cdots. \quad (4.3.20)$$

We are now in position to give the relevant blocks shown in Eq. (4.3.1) on page 104. From Eqs. (4.3.9) and (4.3.11) on page 105 and on the facing page, we find that

$$\mathbf{A} = \frac{1}{P_k + V_k''(\phi)} + \frac{2\gamma_\tau\epsilon_k^2\left[\epsilon_k\phi\bar{\xi}^{\mathsf{T}}\sigma_x\xi - \left(R_k^d + \frac{\Omega_k}{2}\right)\bar{\xi}^{\mathsf{T}}\sigma_z\xi\right]}{\left[P_k + V_k''(\phi)\right]^2\left[(u^a\partial_a)^2 - \left(R_k^d + \frac{\Omega_k}{2}\right)^2 - \epsilon_k^2\phi^2\right]}. \quad (4.3.21)$$

Eqs. (4.3.10), (4.3.14), (4.3.15) and (4.3.19) on pages 106–107 yield

$$\mathbf{F} = \frac{u^a\partial_a + \left(R_k^d + \frac{\Omega_k}{2}\right)\sigma_z + \epsilon_k\phi\sigma_x}{\gamma_\tau\left[(u^a\partial_a)^2 - \left(R_k^d + \frac{\Omega_k}{2}\right)^2 - \epsilon_k^2\phi^2\right]}$$

$$+ \frac{\epsilon_k^2\left[\sigma_x\xi\bar{\xi}^{\mathsf{T}}\sigma_x(u^a\partial_a)^2 + \left(R_k + \frac{\Omega_k}{2}\right)^2\sigma_y\xi\bar{\xi}^{\mathsf{T}}\sigma_y + i\epsilon_k\phi\left(R_k + \frac{\Omega_k}{2}\right)\sigma_y\xi\bar{\xi}^{\mathsf{T}}\right]}{\left[P_k + V_k''(\phi)\right]\left[(u^a\partial_a)^2 - \left(R_k^d + \frac{\Omega_k}{2}\right)^2 - \epsilon_k^2\phi^2\right]^2} + \cdots, \quad (4.3.22)$$

and these same equations in addition to Eq. (4.3.20) give

$$\mathbf{H} = \frac{u^a\partial_a - \left(R_k^d + \frac{\Omega_k}{2}\right)\sigma_z - \epsilon_k\phi\sigma_x}{\gamma_\tau\left[(u^a\partial_a)^2 - \left(R_k^d + \frac{\Omega_k}{2}\right)^2 - \epsilon_k^2\phi^2\right]}$$

$$+ \frac{\epsilon_k^2\left[\sigma_x\bar{\xi}\xi^{\mathsf{T}}\sigma_x(u^a\partial_a)^2 + \left(R_k + \frac{\Omega_k}{2}\right)^2\sigma_y\bar{\xi}\xi^{\mathsf{T}}\sigma_y + i\epsilon_k\phi\left(R_k + \frac{\Omega_k}{2}\right)\sigma_y\bar{\xi}\xi^{\mathsf{T}}\right]}{\left[P_k + V_k''(\phi)\right]\left[(u^a\partial_a)^2 - \left(R_k^d + \frac{\Omega_k}{2}\right)^2 - \epsilon_k^2\phi^2\right]^2} + \cdots. \quad (4.3.23)$$

### Algebraic Traces

We can now use the expressions we just obtained in Eq. (4.3.2) on page 104 to simplify our expression for $\partial_t \Gamma_k$. For now, let us compute the necessary algebraic traces and reduce the problem of finding $\partial_t \Gamma_k$ to that of computing some functional traces.

Eq. (4.3.21) on the previous page tells us that

$$\text{Tr}\left[ A \partial_t R_k^\phi \right] = \text{Tr}\left[ \frac{\partial_t R_k^\phi}{P_k + V_k''(\phi)} \right] + \text{Tr}\left[ \frac{2\gamma_\tau \epsilon_k^2 \left[ \epsilon_k \phi \bar{\xi}^\dagger \sigma_x \xi - \left( R_k^d + \frac{\Omega_k}{2} \right) \bar{\xi}^\dagger \sigma_z \xi \right] \partial_t R_k^\phi}{\left[ P_k + V_k''(\phi) \right]^2 \left[ (u^a \partial_a)^2 - \left( R_k^d + \frac{\Omega_k}{2} \right)^2 - \epsilon_k^2 \phi^2 \right]} \right]. \tag{4.3.24}$$

Notice, however, that the $\epsilon_k^2 \phi^2$ term on the denominator of the second term can only generate terms that lie outside of our truncation. Indeed, it will only contribute to terms of the form $\phi^n \bar{\xi}^\dagger \sigma_i \xi$ with $n \geq 2$. Hence, we can drop it. We simply get

$$\text{Tr}\left[ A \partial_t R_k^\phi \right] = \text{Tr}\left[ \frac{\partial_t R_k^\phi}{P_k + V_k''(\phi)} \right] + \text{Tr}\left[ \frac{2\gamma_\tau \epsilon_k^2 \left[ \epsilon_k \phi \bar{\xi}^\dagger \sigma_x \xi - \left( R_k^d + \frac{\Omega_k}{2} \right) \bar{\xi}^\dagger \sigma_z \xi \right] \partial_t R_k^\phi}{\left[ P_k + V_k''(\phi) \right]^2 \left[ (u^a \partial_a)^2 - \left( R_k^d + \frac{\Omega_k}{2} \right)^2 \right]} \right]. \tag{4.3.25}$$

We cannot use the same argument to get rid of $V_k''(\phi)$, because it may include terms that do either not depend on the field or depend on it linearly. These would arise from a mass term or from a cubic interaction.

Since we are not considering the running of any derivative couplings, we can treat the fields as constants and write

$$\text{Tr}\left[ A \partial_t R_k^\phi \right] = \text{Tr}\left[ \frac{\partial_t R_k^\phi}{P_k + V_k''(\phi)} \right] - 2\,\text{Tr}\left[ \frac{\gamma_\tau \epsilon_k^2 \left( R_k^d + \frac{\Omega_k}{2} \right) \partial_t R_k^\phi}{\left[ P_k + V_k''(\phi) \right]^2 \left[ (u^a \partial_a)^2 - \left( R_k^d + \frac{\Omega_k}{2} \right)^2 \right]} \right] \bar{\xi}^\dagger \sigma_z \xi$$

$$+ 2\,\text{Tr}\left[ \frac{\gamma_\tau \epsilon_k^3 \partial_t R_k^\phi}{\left[ P_k + V_k''(\phi) \right]^2 \left[ (u^a \partial_a)^2 - \left( R_k^d + \frac{\Omega_k}{2} \right)^2 \right]} \right] \phi \bar{\xi}^\dagger \sigma_x \xi. \tag{4.3.26}$$

Using Eq. (4.3.22) on the preceding page and the fact that Pauli matrices are traceless, we can find

$$\text{Tr}\left[ \mathbb{F} \gamma_\tau \partial_t R_k^d \sigma_z \right] = \text{Tr}\left[ \frac{\left( R_k^d + \frac{\Omega_k}{2} \right) \partial_t R_k^d}{(u^a \partial_a)^2 - \left( R_k^d + \frac{\Omega_k}{2} \right)^2 - \epsilon_k^2 \phi^2} \right]$$

$$+ \text{Tr}\left[ \frac{\gamma_\tau \epsilon_k^2 \left[ \sigma_x \xi \bar{\xi}^\dagger \sigma_x (u^a \partial_a)^2 + \left( R_k + \frac{\Omega_k}{2} \right)^2 \sigma_y \xi \bar{\xi}^\dagger \sigma_y + i \epsilon_k \phi \left( R_k + \frac{\Omega_k}{2} \right) \sigma_y \xi \bar{\xi}^\dagger \right] \partial_t R_k^d \sigma_z}{\left[ P_k + V_k''(\phi) \right] \left[ (u^a \partial_a)^2 - \left( R_k^d + \frac{\Omega_k}{2} \right)^2 - \epsilon_k^2 \phi^2 \right]^2} \right] + \cdots. \tag{4.3.27}$$

Using the cyclic property of the trace and the properties of Pauli matrices, we can simplify this expression to

$$\text{Tr}\left[ \mathbb{F} \gamma_\tau \partial_t R_k^d \sigma_z \right] = \text{Tr}\left[ \frac{\left( R_k^d + \frac{\Omega_k}{2} \right) \partial_t R_k^d}{(u^a \partial_a)^2 - \left( R_k^d + \frac{\Omega_k}{2} \right)^2 - \epsilon_k^2 \phi^2} \right]$$

$$+ \operatorname{Tr}\left[\frac{\gamma_\tau \epsilon_k^2\left[-\xi\tilde{\xi}^\intercal\sigma_z\left(u^a\partial_a\right)^2 - \left(R_k + \frac{\Omega_k}{2}\right)^2\xi\tilde{\xi}^\intercal\sigma_z + \epsilon_k\phi\left(R_k + \frac{\Omega_k}{2}\right)\xi\tilde{\xi}^\intercal\sigma_x\right]\partial_t R_k^d}{\left[P_k + V_k''(\phi)\right]\left[\left(u^a\partial_a\right)^2 - \left(R_k^d + \frac{\Omega_k}{2}\right)^2 - \epsilon_k^2\phi^2\right]^2}\right] + \cdots. \quad (4.3.28)$$

Using that the fields are constant, we can get to

$$\operatorname{Tr}\left[\mathbb{F}\gamma_\tau\partial_t R_k^d\sigma_z\right] = \operatorname{Tr}\left[\frac{\left(R_k^d + \frac{\Omega_k}{2}\right)\partial_t R_k^d}{\left(u^a\partial_a\right)^2 - \left(R_k^d + \frac{\Omega_k}{2}\right)^2 - \epsilon_k^2\phi^2}\right]$$

$$- \operatorname{Tr}\left[\frac{\gamma_\tau\epsilon_k^2\left[\left(u^a\partial_a\right)^2 + \left(R_k + \frac{\Omega_k}{2}\right)^2\right]\partial_t R_k^d}{\left[P_k + V_k''(\phi)\right]\left[\left(u^a\partial_a\right)^2 - \left(R_k^d + \frac{\Omega_k}{2}\right)^2 - \epsilon_k^2\phi^2\right]^2}\right]\operatorname{tr}\left[\xi\tilde{\xi}^\intercal\sigma_z\right]$$

$$+ \operatorname{Tr}\left[\frac{\gamma_\tau\epsilon_k^3\left(R_k + \frac{\Omega_k}{2}\right)\partial_t R_k^d}{\left[P_k + V_k''(\phi)\right]\left[\left(u^a\partial_a\right)^2 - \left(R_k^d + \frac{\Omega_k}{2}\right)^2 - \epsilon_k^2\phi^2\right]^2}\right]\phi\operatorname{tr}\left[\xi\tilde{\xi}^\intercal\sigma_x\right] + \cdots. \quad (4.3.29)$$

If we notice that $\operatorname{tr}\left[\xi\tilde{\xi}^\intercal\sigma_z\right] = -\tilde{\xi}^\intercal\sigma_z\xi$ and $\operatorname{tr}\left[\xi\tilde{\xi}^\intercal\sigma_x\right] = -\tilde{\xi}^\intercal\sigma_x\xi$, then we get to

$$\operatorname{Tr}\left[\mathbb{F}\gamma_\tau\partial_t R_k^d\sigma_z\right] = \operatorname{Tr}\left[\frac{\left(R_k^d + \frac{\Omega_k}{2}\right)\partial_t R_k^d}{\left(u^a\partial_a\right)^2 - \left(R_k^d + \frac{\Omega_k}{2}\right)^2 - \epsilon_k^2\phi^2}\right]$$

$$+ \operatorname{Tr}\left[\frac{\gamma_\tau\epsilon_k^2\left[\left(u^a\partial_a\right)^2 + \left(R_k + \frac{\Omega_k}{2}\right)^2\right]\partial_t R_k^d}{\left[P_k + V_k''(\phi)\right]\left[\left(u^a\partial_a\right)^2 - \left(R_k^d + \frac{\Omega_k}{2}\right)^2 - \epsilon_k^2\phi^2\right]^2}\right]\tilde{\xi}^\intercal\sigma_z\xi$$

$$- \operatorname{Tr}\left[\frac{\gamma_\tau\epsilon_k^3\left(R_k + \frac{\Omega_k}{2}\right)\partial_t R_k^d}{\left[P_k + V_k''(\phi)\right]\left[\left(u^a\partial_a\right)^2 - \left(R_k^d + \frac{\Omega_k}{2}\right)^2 - \epsilon_k^2\phi^2\right]^2}\right]\phi\tilde{\xi}^\intercal\sigma_x\xi + \cdots. \quad (4.3.30)$$

We can ignore the $\epsilon_k^2\phi^2$ contribution in the two latter terms. Hence, our final expression is

$$\operatorname{Tr}\left[\mathbb{F}\gamma_\tau\partial_t R_k^d\sigma_z\right] = \operatorname{Tr}\left[\frac{\left(R_k^d + \frac{\Omega_k}{2}\right)\partial_t R_k^d}{\left(u^a\partial_a\right)^2 - \left(R_k^d + \frac{\Omega_k}{2}\right)^2 - \epsilon_k^2\phi^2}\right]$$

$$+ \operatorname{Tr}\left[\frac{\gamma_\tau\epsilon_k^2\left[\left(u^a\partial_a\right)^2 + \left(R_k + \frac{\Omega_k}{2}\right)^2\right]\partial_t R_k^d}{\left[P_k + V_k''(\phi)\right]\left[\left(u^a\partial_a\right)^2 - \left(R_k^d + \frac{\Omega_k}{2}\right)^2\right]^2}\right]\tilde{\xi}^\intercal\sigma_z\xi$$

$$- \operatorname{Tr}\left[\frac{\gamma_\tau\epsilon_k^3\left(R_k + \frac{\Omega_k}{2}\right)\partial_t R_k^d}{\left[P_k + V_k''(\phi)\right]\left[\left(u^a\partial_a\right)^2 - \left(R_k^d + \frac{\Omega_k}{2}\right)^2\right]^2}\right]\phi\tilde{\xi}^\intercal\sigma_x\xi + \cdots. \quad (4.3.31)$$

Finally, the same procedure using Eq. (4.3.23) on page 107 gives

$$
\begin{aligned}
\mathrm{Tr}\left[\mathsf{H}\gamma_\tau\partial_t R_k^d\sigma_z\right] = &-\mathrm{Tr}\left[\frac{\left(R_k^d+\frac{\Omega_k}{2}\right)\partial_t R_k^d}{(u^a\partial_a)^2-\left(R_k^d+\frac{\Omega_k}{2}\right)^2-\epsilon_k^2\phi^2}\right] \\
&-\mathrm{Tr}\left[\frac{\gamma_\tau\epsilon_k^2\left[(u^a\partial_a)^2+\left(R_k+\frac{\Omega_k}{2}\right)^2\right]\partial_t R_k^d}{\left[P_k+V_k''(\phi)\right]\left[(u^a\partial_a)^2-\left(R_k^d+\frac{\Omega_k}{2}\right)^2\right]^2}\right]\bar{\xi}^\dagger\sigma_z\xi \\
&+\mathrm{Tr}\left[\frac{\gamma_\tau\epsilon_k^3\left(R_k+\frac{\Omega_k}{2}\right)\partial_t R_k^d}{\left[P_k+V_k''(\phi)\right]\left[(u^a\partial_a)^2-\left(R_k^d+\frac{\Omega_k}{2}\right)^2\right]^2}\right]\phi\bar{\xi}^\dagger\sigma_x\xi+\cdots. \quad (4.3.32)
\end{aligned}
$$

If we bring all of these expressions back to Eq. (4.3.2) on page 104, we find that

$$
\begin{aligned}
\partial_t\Gamma_k = &\frac{1}{2}\mathrm{Tr}\left[\frac{\partial_t R_k^\phi}{P_k+V_k''(\phi)}\right]-\mathrm{Tr}\left[\frac{\left(R_k^d+\frac{\Omega_k}{2}\right)\partial_t R_k^d}{(u^a\partial_a)^2-\left(R_k^d+\frac{\Omega_k}{2}\right)^2-\epsilon_k^2\phi^2}\right] \\
&-\mathrm{Tr}\left[\frac{\gamma_\tau\epsilon_k^2\left(R_k^d+\frac{\Omega_k}{2}\right)\partial_t R_k^\phi}{\left[P_k+V_k''(\phi)\right]^2\left[(u^a\partial_a)^2-\left(R_k^d+\frac{\Omega_k}{2}\right)^2\right]}\right]\bar{\xi}^\dagger\sigma_z\xi \\
&-\mathrm{Tr}\left[\frac{\gamma_\tau\epsilon_k^2\left[(u^a\partial_a)^2+\left(R_k+\frac{\Omega_k}{2}\right)^2\right]\partial_t R_k^d}{\left[P_k+V_k''(\phi)\right]\left[(u^a\partial_a)^2-\left(R_k^d+\frac{\Omega_k}{2}\right)^2\right]^2}\right]\bar{\xi}^\dagger\sigma_z\xi \\
&+\mathrm{Tr}\left[\frac{\gamma_\tau\epsilon_k^3\partial_t R_k^\phi}{\left[P_k+V_k''(\phi)\right]^2\left[(u^a\partial_a)^2-\left(R_k^d+\frac{\Omega_k}{2}\right)^2\right]}\right]\phi\bar{\xi}^\dagger\sigma_x\xi \\
&+\mathrm{Tr}\left[\frac{\gamma_\tau\epsilon_k^3\left(R_k+\frac{\Omega_k}{2}\right)\partial_t R_k^d}{\left[P_k+V_k''(\phi)\right]\left[(u^a\partial_a)^2-\left(R_k^d+\frac{\Omega_k}{2}\right)^2\right]^2}\right]\phi\bar{\xi}^\dagger\sigma_x\xi+\cdots. \quad (4.3.33)
\end{aligned}
$$

Notice that the $\epsilon_k^2\phi^2$ term on the second trace will generate all of the even self-interactions of the quantum field. It is interesting to notice that even something as simple as a particle detector is sufficient to initiate the flow of infinitely many coupling constants.

### The Taylor Trick

Our next goal is to actually compute the functional traces occurring on Eq. (4.3.33). One may notice some of them involve derivatives in the denominator. This means we will need to be clever when dealing with them.

While we expect the following techniques to work on more general cases, we shall make a few simplifying assumptions to keep the calculations more treatable. Namely, we will assume:

i. that the coupling between quantum field and detector is time-independent, *i.e.*, we assume $\epsilon_k(\tau) = \epsilon_k$;

ii. that the detector is moving inertially so that its proper time corresponds to an inertial global time function.

So far, these assumptions were not necessary, and they likely can be dropped in the following discussions. Nevertheless, since this is (to my knowledge) the first ever treatment of the FRG flow for a two-level system, it is interesting to first make the computations for the simplest possible case.

The difficult traces on Eq. (4.3.33) on the facing page have the general form

$$\text{Tr}\left[\frac{W(\Delta)f(x)(u^a\partial_a)^{2n}}{\left[(u^a\partial_a)^2 - A(\Delta)\right]^m}\right], \tag{4.3.34}$$

where $n$ and $m$ are non-negative integers, $W(\Delta)$ and $A(\Delta)$ are some functions of the covariant Laplacian $\Delta = -\partial^2$, and $f(x)$ is some spacetime function (such as $\chi_\tau$). Neither $W$ nor $A$ depend on spacetime events. If Eq. (4.3.34) only had derivatives in the numerator, we would be able to compute this functional trace by simply employing off-diagonal heat kernel techniques (see Section 3.B). However, this is not the case. How can we proceed?

We shall approach this calculation by performing a "trick" with a Taylor series. Hence, we will refer to this technique as the "Taylor trick" from here onward. The main idea is to perform a formal Taylor series expansion in $u^a\partial_a$, so that all derivatives end up in the numerator. Once this happens we can use off-diagonal heat kernel techniques and then resum the series.

More explicitly, we write

$$\text{Tr}\left[\frac{W(\Delta)f(x)(u^a\partial_a)^{2n}}{\left[(u^a\partial_a)^2 - A(\Delta)\right]^m}\right] = (-1)^m \sum_{l=0}^{+\infty} \frac{(m)_l}{l!} \text{Tr}\left[\frac{W(\Delta)f(x)(u^a\partial_a)^{2n+2l}}{A(\Delta)^{m+l}}\right], \tag{4.3.35}$$

where we used the series expansion

$$\frac{1}{(x^2 - a)^m} = \frac{(-1)^m}{a^m} \sum_{l=0}^{+\infty} \frac{(m)_l x^{2l}}{l! a^l}, \tag{4.3.36}$$

where $(m)_l$ is the (rising) Pochhammer symbol (see Arfken, Weber, and Harris 2013, Eq. (1.72)). This series expansion converges for $|x|^2 < |a|$, and hence the expression on Eq. (4.3.35) should have a non-vanishing radius of convergence as long as $A(\Delta)$ does not vanish identically. Comparing Eqs. (4.3.33) and (4.3.35) on the preceding page and on this page, we see that typically $A(\Delta) = \left(R_k^d + \frac{\Omega_k}{2}\right)^2$, and hence we aim to use the regulator $R_k^d$ to keep $A(\Delta)$ finite even in the gapless limit. Notice, however, that this will fail in the ultraviolet (UV) region, since the regulator is small for large values of its argument. We shall look at this more closely in Section 4.4.

The traces inside the sum of Eq. (4.3.35) have the form given on Eq. (3.B.27) on page 90, and hence we can now proceed with off-diagonal heat-kernel techniques. Namely, let us consider a trace with the form

$$\text{Tr}\left[W(\Delta)f(x)(u^a\partial_a)^{2n}\right]. \tag{4.3.37}$$

We can deal with it by using the expressions in Section 3.B. In flat spacetime we have

$$\text{Tr}\big[W(\Delta)f(x)(u^a\partial_a)^{2n}\big] = \frac{(-1)^n(2n)!}{(4\pi)^{\frac{d}{2}}4^n n!}\int_0^{+\infty}\widetilde{W}(s)s^{-\frac{d}{2}-n}\int f(x)u^{a_1}\cdots u^{a_{2n}}\partial_{(a_1 a_2}\cdots\partial_{a_{2n-1}a_{2n})}\,\mathrm{d}^d x\,\mathrm{d}s\,.$$
$$(4.3.38)$$

The formulae in Section 3.B also tell us the trace would be zero if there was an odd number of insertions of $u^a\partial_a$. On Eq. (4.3.38) we used the fact that the detector is inertial to be able to write

$$(u^a\partial_a)^{2n} = u^{a_1}\cdots u^{a_{2n}}\partial_{a_1}\cdots\partial_{a_{2n}}\,. \qquad (4.3.39)$$

Since $\eta_{ab}u^a u^b = -1$ in Lorentzian signature, we will have $\delta_{ab}u^a u^b = +1$ in Euclidean signature. Hence,

$$\text{Tr}\big[W(\Delta)f(x)(u^a\partial_a)^{2n}\big] = \frac{(-1)^n(2n)!}{(4\pi)^{\frac{d}{2}}4^n n!}\int_0^{+\infty}\widetilde{W}(s)s^{-\frac{d}{2}-n}\int f(x)\,\mathrm{d}^d x\,\mathrm{d}s\,. \qquad (4.3.40)$$

If we recall the definition of the $Q$-functionals given on Eq. (3.B.21) on page 89, we see that

$$\text{Tr}\big[W(\Delta)f(x)(u^a\partial_a)^{2n}\big] = \frac{(-1)^n(2n)!}{(4\pi)^{\frac{d}{2}}4^n n!}\int f(x)\,\mathrm{d}^d x\,Q_{n+\frac{d}{2}}[W]\,. \qquad (4.3.41)$$

Notice the integral over $f$ can be used to consider, for example, the effects of the detector being turned on for only a finite time.

Using Eqs. (4.3.35) and (4.3.41) on the previous page and on this page, we find that

$$\text{Tr}\left[\frac{W(\Delta)f(x)(u^a\partial_a)^{2n}}{\big[(u^a\partial_a)^2 - A(\Delta)\big]^m}\right] = (-1)^m\sum_{l=0}^{+\infty}\frac{(-1)^{n+l}(m)_l(2(n+l))!}{(4\pi)^{\frac{d}{2}}4^{n+l}(n+l)!l!}\int f(x)\,\mathrm{d}^d x\,Q_{n+l+\frac{d}{2}}\left[\frac{W}{A^{m+l}}\right]. \qquad (4.3.42)$$

We are only interested in the cases with $d > 0$ and $n, l \geq 0$. Hence, we can use Eq. (3.B.22) on page 90 when computing the $Q$-functionals and write

$$\text{Tr}\left[\frac{W(\Delta)f(x)(u^a\partial_a)^{2n}}{\big[(u^a\partial_a)^2 - A(\Delta)\big]^m}\right] = (-1)^m\sum_{l=0}^{+\infty}\frac{(-1)^{n+l}(m)_l(2(n+l))!\int f(x)\,\mathrm{d}^d x}{(4\pi)^{\frac{d}{2}}4^{n+l}(n+l)!l!\Gamma\!\left(n+l+\frac{d}{2}\right)}\int_0^{+\infty}\frac{W(z)z^{n+l+\frac{d}{2}-1}}{A(z)^{m+l}}\,\mathrm{d}z\,. \qquad (4.3.43)$$

Assuming we can exchange the limits between the series and the $z$ integral, and using the identities

$$\Gamma\!\left(n+l+\frac{d}{2}\right) = \Gamma\!\left(n+\frac{d}{2}\right)\!\left(n+\frac{d}{2}\right)_l \quad\text{and}\quad \frac{(2(n+l))!}{(n+l)!} = \frac{(2n)!}{n!}\!\left(n+\frac{1}{2}\right)_l 4^l\,, \qquad (4.3.44)$$

we can write

$$\text{Tr}\left[\frac{W(\Delta)f(x)(u^a\partial_a)^{2n}}{\big[(u^a\partial_a)^2 - A(\Delta)\big]^m}\right] = \frac{(-1)^{m+n}(2n)!\int f(x)\,\mathrm{d}^d x}{(4\pi)^{\frac{d}{2}}4^n n!\Gamma\!\left(n+\frac{d}{2}\right)}\int_0^{+\infty}\frac{W(z)z^{n+\frac{d}{2}-1}}{A(z)^m}\sum_{l=0}^{+\infty}\frac{(m)_l\!\left(n+\frac{1}{2}\right)_l}{\left(n+\frac{d}{2}\right)_l l!}\!\left(-\frac{z}{A(z)}\right)^l\,\mathrm{d}z\,. \qquad (4.3.45)$$

We can now see the series evaluates to a hypergeometric function (see Arfken, Weber, and Harris 2013, Sec. 18.5). Therefore, we may write

$$\text{Tr}\left[\frac{W(\Delta)f(x)(u^a\partial_a)^{2n}}{\big[(u^a\partial_a)^2 - A(\Delta)\big]^m}\right] = \frac{(-1)^{m+n}(2n)!\int f(x)\,\mathrm{d}^d x}{(4\pi)^{\frac{d}{2}}4^n n!\Gamma\!\left(n+\frac{d}{2}\right)}\int_0^{+\infty}\frac{W(z)z^{n+\frac{d}{2}-1}F\!\left(m, n+\frac{1}{2}; n+\frac{d}{2}; -\frac{z}{A(z)}\right)}{A(z)^m}\,\mathrm{d}z\,. \qquad (4.3.46)$$

Using this expression, we can write the beta functional on Eq. (4.3.33) on page 110 in terms of integrals of hypergeometric functions. While these integrals may be complicated, in the worst case scenario we can still evaluate them numerically. This allows us to at least obtain a portrait of how the RG flow behaves in parameter space.

### Functional Traces

We can now use the Taylor trick to deal with each of the functional traces. To further simplify the integrals that Eq. (4.3.46) on the preceding page yields us, it is convenient to introduce a few new definitions. Firstly, we will write the cutoffs $R_k^\phi$ and $R_k^d$ in terms of cutoff profiles as shown on Eq. (4.2.16) on page 103. To avoid writing $\frac{z}{k^2}$ all the time, we will also introduce a new variable $u = \frac{z}{k^2}$. Notice that

$$\partial_t R_k^\phi(z) = 2k^2\big[r_\phi(u) - ur_\phi'(u)\big] \quad \text{and} \quad \partial_t R_k^d(z) = k\big[r_d(u) - 2ur_d'(u)\big]. \tag{4.3.47}$$

We will also introduce dimensionless couplings. $\Omega_k$ always has dimension of energy regardless of the spacetime dimension, so we shall define its dimensionless version through

$$\tilde{\Omega}_k = k^{-1}\Omega_k. \tag{4.3.48}$$

The dimensions of $\epsilon_k$ do depend on spacetime dimension. From Eq. (4.1.26) on page 100, we can tell that the dimensionless version of $\epsilon_k$ should be defined as

$$\tilde{\epsilon}_k = k^{\frac{d}{2}-2}\epsilon_k. \tag{4.3.49}$$

The potential's second derivative, $V_k''$, is also dimensionful. We will define

$$U_k(\phi) = k^{-2}V_k(\phi) \tag{4.3.50}$$

so that $U_k''$ is dimensionless.

With these new definitions in mind, let us move on to the traces.

### Field Self-Interactions

There are two traces contributing to the field's self-interactions. The first of them is

$$\mathrm{Tr}\left[\frac{\partial_t R_k^\phi}{P_k + V_k''(\phi)}\right] = \frac{V_d}{(4\pi)^{\frac{d}{2}}}Q_{\frac{d}{2}}\left[\frac{\partial_t R_k^\phi}{P_k + V_k''(\phi)}\right] \tag{4.3.51}$$

which we had already computed on Eq. (3.3.14) on page 76. Hence, we will not further develop this expression. We recall that $V_d$ denotes the (infinite) volume of $d$-dimensional spacetime.

The new trace is generated by the detector. To read it as a series of terms being generated for the field, it is useful to make a series expansion in $\phi$ before computing it. We have

$$\mathrm{Tr}\left[\frac{\left(R_k^d + \frac{\Omega_k}{2}\right)\partial_t R_k^d}{\left(u^a\partial_a\right)^2 - \left(R_k^d + \frac{\Omega_k}{2}\right)^2 - \epsilon_k^2\phi^2}\right] = \sum_{l=0}^{+\infty}\epsilon_k^{2l}\mathrm{Tr}\left[\frac{\left(R_k^d + \frac{\Omega_k}{2}\right)\partial_t R_k^d}{\left[\left(u^a\partial_a\right)^2 - \left(R_k^d + \frac{\Omega_k}{2}\right)^2\right]^{l+1}}\right]\phi^{2l}. \tag{4.3.52}$$

We can compute the traces in the series by using Eq. (4.3.46) on page 112 with $m = l + 1$, $n = 0$, $f(x) = 1$,

$$W(\Delta) = \left(R_k^d(\Delta) + \frac{\Omega_k}{2}\right)\partial_t R_k^d(\Delta), \quad \text{and} \quad A(\Delta) = \left(R_k^d(\Delta) + \frac{\Omega_k}{2}\right)^2. \tag{4.3.53}$$

We find

$$\mathrm{Tr}\left[\frac{\left(R_k^d + \frac{\Omega_k}{2}\right)\partial_t R_k^d}{\left[(u^a\partial_a)^2 - \left(R_k^d + \frac{\Omega_k}{2}\right)^2\right]^{l+1}}\right]$$
$$= \frac{(-1)^{l+1}V_d}{(4\pi)^{\frac{d}{2}}\Gamma\left(\frac{d}{2}\right)}\int_0^{+\infty}\frac{\left(R_k^d(z) + \frac{\Omega_k}{2}\right)\partial_t R_k^d(z)z^{\frac{d}{2}-1}}{\left(R_k^d(z) + \frac{\Omega_k}{2}\right)^{2l+2}}F\left(l+1, \frac{1}{2}; \frac{d}{2}; -\frac{z}{\left(R_k^d(z) + \frac{\Omega_k}{2}\right)^2}\right)\mathrm{d}z. \tag{4.3.54}$$

Using Eqs. (4.2.16) and (4.3.47) to (4.3.49) on page 103 and on the previous page, we see that

$$\mathrm{Tr}\left[\frac{\left(R_k^d + \frac{\Omega_k}{2}\right)\partial_t R_k^d}{\left[(u^a\partial_a)^2 - \left(R_k^d + \frac{\Omega_k}{2}\right)^2\right]^{l+1}}\right]$$
$$= \frac{(-1)^{l+1}V_d k^{d-2l}}{(4\pi)^{\frac{d}{2}}\Gamma\left(\frac{d}{2}\right)}\int_0^{+\infty}\frac{\left[r_d(u) - 2ur_d'(u)\right]u^{\frac{d}{2}-1}}{\left(r_d(u) + \frac{\tilde{\Omega}_k}{2}\right)^{2l+1}}F\left(l+1, \frac{1}{2}; \frac{d}{2}; -\frac{u}{\left(r_d(u) + \frac{\tilde{\Omega}_k}{2}\right)^2}\right)\mathrm{d}u. \tag{4.3.55}$$

When we compute the RG flow, we will need to fix

i. $d$, to select the spacetime dimension we are interested in;

ii. $r_d$, to select the cutoff we will be working with;

iii. $l$, to select the term whose beta function we are computing.

Hence, the only free parameter in the integral shown on Eq. (4.3.55) is the dimensionless gap $\tilde{\Omega}_k$. We can then solve the integral numerically, for example, for some range of $\tilde{\Omega}_k$ and obtain the RG flow numerically. In some cases, it may be possible to solve the integral exactly.

Similar statements hold for all of the following integrals. They can be put in a form in which they depend solely on the dimensionless couplings, at which stage we can solve the integrals numerically for some different values of couplings and understand the flow.

### Gap Term

Our next step is to consider the traces for the gap term. The first of them is

$$\mathrm{Tr}\left[\frac{\gamma_\tau\left(R_k^d + \frac{\Omega_k}{2}\right)\partial_t R_k^\phi}{\left[P_k + V_k''(\phi)\right]^2\left[(u^a\partial_a)^2 - \left(R_k^d + \frac{\Omega_k}{2}\right)^2\right]}\right]. \tag{4.3.56}$$

We can compute it with Eq. (4.3.46) on page 112 by setting $m = 1$, $n = 0$, $f(x) = \gamma_\tau$,

$$W(\Delta) = \frac{\left(R_k^d(\Delta) + \frac{\Omega_k}{2}\right)\partial_t R_k^\phi(\Delta)}{\left[P_k(\Delta) + V_k''(\phi)\right]^2}, \quad \text{and} \quad A(\Delta) = \left(R_k^d(\Delta) + \frac{\Omega_k}{2}\right)^2. \tag{4.3.57}$$

Therefore,

$$\text{Tr}\left[\frac{\gamma_\tau\left(R_k^d + \frac{\Omega_k}{2}\right)\partial_t R_k^\phi}{\left[P_k + V_k''(\phi)\right]^2\left[(u^a\partial_a)^2 - \left(R_k^d + \frac{\Omega_k}{2}\right)^2\right]}\right]$$

$$= \frac{-\int \gamma_\tau \, \mathrm{d}^d x}{(4\pi)^{\frac{d}{2}}\Gamma\left(\frac{d}{2}\right)}\int_0^{+\infty}\frac{\partial_t R_k^\phi(z) z^{\frac{d}{2}-1}}{\left[P_k(z) + V_k''(\phi)\right]^2\left(R_k^d(z) + \frac{\Omega_k}{2}\right)}F\left(1, \frac{1}{2}; \frac{d}{2}; -\frac{z}{\left(R_k^d(z) + \frac{\Omega_k}{2}\right)^2}\right)\mathrm{d}z. \tag{4.3.58}$$

Notice that the definition of $\gamma_\tau$, Eq. (4.2.9) on page 102, implies

$$\int \gamma_\tau \, \mathrm{d}^d x = \int \tilde{\partial}^{(d-1)}(\vec{x} - \vec{\tilde{z}}(t))\frac{\mathrm{d}\tau}{\mathrm{d}t} \, \mathrm{d}^d x, \tag{4.3.59a}$$

$$= \int \frac{\mathrm{d}\tau}{\mathrm{d}t} \, \mathrm{d}t, \tag{4.3.59b}$$

$$= \int \mathrm{d}\tau, \tag{4.3.59c}$$

$$= V_1, \tag{4.3.59d}$$

where $V_1$ is the (infinite) length of the detector's worldline.

Using this fact and the dimensionless parameters we introduced on Eqs. (4.2.16) and (4.3.47) to (4.3.50) on page 103 and on page 113, we can write

$$\text{Tr}\left[\frac{\gamma_\tau\left(R_k^d + \frac{\Omega_k}{2}\right)\partial_t R_k^\phi}{\left[P_k + V_k''(\phi)\right]^2\left[(u^a\partial_a)^2 - \left(R_k^d + \frac{\Omega_k}{2}\right)^2\right]}\right]$$

$$= \frac{-2V_1 k^{d-3}}{(4\pi)^{\frac{d}{2}}\Gamma\left(\frac{d}{2}\right)}\int_0^{+\infty}\frac{\left[r_\phi(u) - ur_\phi'(u)\right]u^{\frac{d}{2}-1}}{\left[u + r_\phi(u) + U_k''(\phi)\right]^2\left(r_d(u) + \frac{\tilde{\Omega}_k}{2}\right)}F\left(1, \frac{1}{2}; \frac{d}{2}; -\frac{u}{\left(r_d(u) + \frac{\tilde{\Omega}_k}{2}\right)^2}\right)\mathrm{d}u. \tag{4.3.60}$$

The second relevant trace is

$$\text{Tr}\left[\frac{\gamma_\tau\left[(u^a\partial_a)^2 + \left(R_k + \frac{\Omega_k}{2}\right)^2\right]\partial_t R_k^d}{\left[P_k + V_k''(\phi)\right]\left[(u^a\partial_a)^2 - \left(R_k + \frac{\Omega_k}{2}\right)^2\right]^2}\right]. \tag{4.3.61}$$

We will split it into two separate traces in order to be able to apply Eq. (4.3.46) on page 112. Namely,

$$
\text{Tr}\left[\frac{\gamma_\tau\left[(u^a\partial_a)^2 + \left(R_k + \frac{\Omega_k}{2}\right)^2\right]\partial_t R_k^d}{\left[P_k + V_k'''(\phi)\right]\left[(u^a\partial_a)^2 - \left(R_k^d + \frac{\Omega_k}{2}\right)^2\right]^2}\right] = \text{Tr}\left[\frac{\gamma_\tau\partial_t R_k^d(u^a\partial_a)^2}{\left[P_k + V_k'''(\phi)\right]\left[(u^a\partial_a)^2 - \left(R_k^d + \frac{\Omega_k}{2}\right)^2\right]^2}\right]
$$
$$
+ \text{Tr}\left[\frac{\gamma_\tau\left(R_k + \frac{\Omega_k}{2}\right)^2\partial_t R_k^d}{\left[P_k + V_k'''(\phi)\right]\left[(u^a\partial_a)^2 - \left(R_k^d + \frac{\Omega_k}{2}\right)^2\right]^2}\right]. \quad (4.3.62)
$$

We then find

$$
\text{Tr}\left[\frac{\gamma_\tau\partial_t R_k^d(u^a\partial_a)^2}{\left[P_k + V_k'''(\phi)\right]\left[(u^a\partial_a)^2 - \left(R_k^d + \frac{\Omega_k}{2}\right)^2\right]^2}\right]
$$
$$
= \frac{-V_1}{2(4\pi)^{\frac{d}{2}}\Gamma\left(1 + \frac{d}{2}\right)}\int_0^{+\infty}\frac{\partial_t R_k^d(z)z^{\frac{d}{2}}}{\left[P_k + V_k'''(\phi)\right]\left(R_k^d + \frac{\Omega_k}{2}\right)^4}F\left(2, \frac{3}{2}; 1 + \frac{d}{2}; -\frac{z}{\left(R_k^d + \frac{\Omega_k}{2}\right)^2}\right)\mathrm{d}z \quad (4.3.63)
$$

and

$$
\text{Tr}\left[\frac{\gamma_\tau\left(R_k + \frac{\Omega_k}{2}\right)^2\partial_t R_k^d}{\left[P_k + V_k'''(\phi)\right]\left[(u^a\partial_a)^2 - \left(R_k^d + \frac{\Omega_k}{2}\right)^2\right]^2}\right]
$$
$$
= \frac{V_1}{(4\pi)^{\frac{d}{2}}\Gamma\left(\frac{d}{2}\right)}\int_0^{+\infty}\frac{\partial_t R_k^d(z)z^{\frac{d}{2}-1}}{\left[P_k + V_k'''(\phi)\right]\left(R_k^d + \frac{\Omega_k}{2}\right)^2}F\left(2, \frac{1}{2}; \frac{d}{2}; -\frac{z}{\left(R_k^d + \frac{\Omega_k}{2}\right)^2}\right)\mathrm{d}z. \quad (4.3.64)
$$

If we add both terms and change to dimensionless quantities, we find

$$
\text{Tr}\left[\frac{\gamma_\tau\left[(u^a\partial_a)^2 + \left(R_k + \frac{\Omega_k}{2}\right)^2\right]\partial_t R_k^d}{\left[P_k + V_k'''(\phi)\right]\left[(u^a\partial_a)^2 - \left(R_k^d + \frac{\Omega_k}{2}\right)^2\right]^2}\right]
$$
$$
= \frac{-V_1 k^{d-3}}{2(4\pi)^{\frac{d}{2}}\Gamma\left(1 + \frac{d}{2}\right)}\int_0^{+\infty}\frac{\left[r_d(u) - 2ur_d'(u)\right]u^{\frac{d}{2}}}{\left[u + r_\phi(u) + U_k'''(\phi)\right]\left(r_d(u) + \frac{\tilde{\Omega}_k}{2}\right)^4}F\left(2, \frac{3}{2}; 1 + \frac{d}{2}; -\frac{u}{\left(r_d(u) + \frac{\tilde{\Omega}_k}{2}\right)^2}\right)\mathrm{d}u
$$
$$
+ \frac{V_1 k^{d-3}}{(4\pi)^{\frac{d}{2}}\Gamma\left(\frac{d}{2}\right)}\int_0^{+\infty}\frac{\left[r_d(u) - 2ur_d'(u)\right]u^{\frac{d}{2}-1}}{\left[u + r_\phi(u) + U_k'''(\phi)\right]\left(r_d(u) + \frac{\tilde{\Omega}_k}{2}\right)^2}F\left(2, \frac{1}{2}; \frac{d}{2}; -\frac{u}{\left(r_d(u) + \frac{\tilde{\Omega}_k}{2}\right)^2}\right)\mathrm{d}u. \quad (4.3.65)
$$

### Interaction Term

Lastly we need to compute the traces associated with the interaction term. The first of them is

$$\text{Tr}\left[\frac{\gamma_\tau \partial_t R_k^\phi}{\left[P_k + V_k''(\phi)\right]^2 \left[(u^a \partial_a)^2 - \left(R_k^d + \frac{\Omega_k}{2}\right)^2\right]}\right]$$

$$= \frac{-V_1}{(4\pi)^{\frac{d}{2}} \Gamma\left(\frac{d}{2}\right)} \int_0^{+\infty} \frac{\partial_t R_k^\phi(z) z^{\frac{d}{2}-1}}{\left[P_k + V_k''(\phi)\right]^2 \left(R_k^d + \frac{\Omega_k}{2}\right)^2} F\left(1, \frac{1}{2}; \frac{d}{2}; -\frac{z}{\left(R_k^d + \frac{\Omega_k}{2}\right)^2}\right) \mathrm{d}z. \quad (4.3.66)$$

In terms of the dimensionless quantities

$$\text{Tr}\left[\frac{\gamma_\tau \partial_t R_k^\phi}{\left[P_k + V_k''(\phi)\right]^2 \left[(u^a \partial_a)^2 - \left(R_k^d + \frac{\Omega_k}{2}\right)^2\right]}\right]$$

$$= \frac{-2V_1 k^{d-4}}{(4\pi)^{\frac{d}{2}} \Gamma\left(\frac{d}{2}\right)} \int_0^{+\infty} \frac{\left[r_\phi(u) - u r_\phi'(u)\right] u^{\frac{d}{2}-1}}{\left[u + r_\phi(u) + U_k''(\phi)\right]^2 \left(r_d(u) + \frac{\tilde{\Omega}_k}{2}\right)^2} F\left(1, \frac{1}{2}; \frac{d}{2}; -\frac{u}{\left(r_d(u) + \frac{\tilde{\Omega}_k}{2}\right)^2}\right) \mathrm{d}u. \quad (4.3.67)$$

The second and last trace is

$$\text{Tr}\left[\frac{\gamma_\tau \left(R_k + \frac{\Omega_k}{2}\right) \partial_t R_k^d}{\left[P_k + V_k''(\phi)\right] \left[(u^a \partial_a)^2 - \left(R_k^d + \frac{\Omega_k}{2}\right)^2\right]^2}\right]$$

$$= \frac{V_1}{(4\pi)^{\frac{d}{2}} \Gamma\left(\frac{d}{2}\right)} \int_0^{+\infty} \frac{\partial_t R_k^d(z) z^{\frac{d}{2}-1}}{\left[P_k + V_k''(\phi)\right] \left(R_k^d + \frac{\Omega_k}{2}\right)^3} F\left(2, \frac{1}{2}; \frac{d}{2}; -\frac{z}{\left(R_k^d + \frac{\Omega_k}{2}\right)^2}\right) \mathrm{d}z. \quad (4.3.68)$$

Or, in terms of dimensionless quantities,

$$\text{Tr}\left[\frac{\gamma_\tau \left(R_k + \frac{\Omega_k}{2}\right) \partial_t R_k^d}{\left[P_k + V_k''(\phi)\right] \left[(u^a \partial_a)^2 - \left(R_k^d + \frac{\Omega_k}{2}\right)^2\right]^2}\right]$$

$$= \frac{V_1 k^{d-4}}{(4\pi)^{\frac{d}{2}} \Gamma\left(\frac{d}{2}\right)} \int_0^{+\infty} \frac{\left[r_d(u) - 2u r_d'(u)\right] u^{\frac{d}{2}-1}}{\left[u + r_\phi(u) + U_k''(\phi)\right] \left(r_d(u) + \frac{\tilde{\Omega}_k}{2}\right)^3} F\left(2, \frac{1}{2}; \frac{d}{2}; -\frac{u}{\left(r_d(u) + \frac{\tilde{\Omega}_k}{2}\right)^2}\right) \mathrm{d}u. \quad (4.3.69)$$

### Beta Functions

Let us bring Eqs. (4.3.51), (4.3.52), (4.3.55), (4.3.60), (4.3.65), (4.3.67) and (4.3.69) on pages 113–117 together, to reduce Eq. (4.3.33) on page 110 to integrals. Recalling Eq. (4.3.49) on page 113, we see

that

$$\partial_t \Gamma_k = \frac{V_d}{2(4\pi)^{\frac{d}{2}}} Q_{\frac{d}{2}} \left[ \frac{\partial_t R_k^\phi}{P_k + V_k''(\phi)} \right]$$

$$+ \sum_{l=0}^{+\infty} \frac{(-1)^l \epsilon_k^{2l} V_d k^{d+(d-2)l} \phi^{2l}}{(4\pi)^{\frac{d}{2}} \Gamma\left(\frac{d}{2}\right)} \int_0^{+\infty} \frac{\left[ r_d(u) - 2u r_d'(u) \right] u^{\frac{d}{2}-1}}{\left( r_d(u) + \frac{\tilde{\Omega}_k}{2} \right)^{2l+1}} F\left( l+1, \frac{1}{2}; \frac{d}{2}; -\frac{u}{\left( r_d(u) + \frac{\tilde{\Omega}_k}{2} \right)^2} \right) \mathrm{d}u$$

$$+ \frac{2\epsilon_k^2 V_1 k \tilde{\xi}^\dagger \sigma_z \xi}{(4\pi)^{\frac{d}{2}} \Gamma\left(\frac{d}{2}\right)} \int_0^{+\infty} \frac{\left[ r_\phi(u) - u r_\phi'(u) \right] u^{\frac{d}{2}-1}}{\left[ u + r_\phi(u) + U_k''(\phi) \right]^2 \left( r_d(u) + \frac{\tilde{\Omega}_k}{2} \right)} F\left( 1, \frac{1}{2}; \frac{d}{2}; -\frac{u}{\left( r_d(u) + \frac{\tilde{\Omega}_k}{2} \right)^2} \right) \mathrm{d}u$$

$$+ \frac{\epsilon_k^2 V_1 k \tilde{\xi}^\dagger \sigma_z \xi}{2(4\pi)^{\frac{d}{2}} \Gamma\left(1 + \frac{d}{2}\right)} \int_0^{+\infty} \frac{\left[ r_d(u) - 2u r_d'(u) \right] u^{\frac{d}{2}}}{\left[ u + r_\phi(u) + U_k''(\phi) \right] \left( r_d(u) + \frac{\tilde{\Omega}_k}{2} \right)^4} F\left( 2, \frac{3}{2}; 1 + \frac{d}{2}; -\frac{u}{\left( r_d(u) + \frac{\tilde{\Omega}_k}{2} \right)^2} \right) \mathrm{d}u$$

$$- \frac{\epsilon_k^2 V_1 k \tilde{\xi}^\dagger \sigma_z \xi}{(4\pi)^{\frac{d}{2}} \Gamma\left(\frac{d}{2}\right)} \int_0^{+\infty} \frac{\left[ r_d(u) - 2u r_d'(u) \right] u^{\frac{d}{2}-1}}{\left[ u + r_\phi(u) + U_k''(\phi) \right] \left( r_d(u) + \frac{\tilde{\Omega}_k}{2} \right)^2} F\left( 2, \frac{1}{2}; \frac{d}{2}; -\frac{u}{\left( r_d(u) + \frac{\tilde{\Omega}_k}{2} \right)^2} \right) \mathrm{d}u$$

$$- \frac{2\epsilon_k^3 V_1 k^{2-\frac{d}{2}} \phi \tilde{\xi}^\dagger \sigma_z \xi}{(4\pi)^{\frac{d}{2}} \Gamma\left(\frac{d}{2}\right)} \int_0^{+\infty} \frac{\left[ r_\phi(u) - u r_\phi'(u) \right] u^{\frac{d}{2}-1}}{\left[ u + r_\phi(u) + U_k''(\phi) \right]^2 \left( r_d(u) + \frac{\tilde{\Omega}_k}{2} \right)^2} F\left( 1, \frac{1}{2}; \frac{d}{2}; -\frac{u}{\left( r_d(u) + \frac{\tilde{\Omega}_k}{2} \right)^2} \right) \mathrm{d}u$$

$$+ \frac{\epsilon_k^3 V_1 k^{2-\frac{d}{2}} \phi \tilde{\xi}^\dagger \sigma_z \xi}{(4\pi)^{\frac{d}{2}} \Gamma\left(\frac{d}{2}\right)} \int_0^{+\infty} \frac{\left[ r_d(u) - 2u r_d'(u) \right] u^{\frac{d}{2}-1}}{\left[ u + r_\phi(u) + U_k''(\phi) \right] \left( r_d(u) + \frac{\tilde{\Omega}_k}{2} \right)^3} F\left( 2, \frac{1}{2}; \frac{d}{2}; -\frac{u}{\left( r_d(u) + \frac{\tilde{\Omega}_k}{2} \right)^2} \right) \mathrm{d}u + \cdots .$$

$$(4.3.70)$$

Eq. (4.3.70) completely describes the RG flow for the field-detector system. Our current goal is to extract the beta functions from it, since they are more convenient to work with.

Since the argument of the hypergeometric functions always has the same form, let us define

$$\rho(u, \tilde{\Omega}_k) \equiv \frac{u}{\left( r_d(u) + \frac{\tilde{\Omega}_k}{2} \right)^2}. \qquad (4.3.71)$$

From Eq. (4.2.10) on page 102, and from the assumption that the fields are constant throughout spacetime, we can notice that

$$\partial_t \Gamma_k = V_d \partial_t V_k(\phi) + \frac{V_1}{2} \beta_\Omega \tilde{\xi}^\dagger \sigma_z \xi + V_1 \beta_c \phi \tilde{\xi}^\dagger \sigma_x \xi. \qquad (4.3.72)$$

Using Eqs. (3.3.4) and (4.3.70) to (4.3.72) on page 72 and on the current page, we find for the potential that

$$\partial_t V_k(\phi) = \frac{1}{2(4\pi)^{\frac{d}{2}}} Q_{\frac{d}{2}} \left[ \frac{\partial_t R_k^\phi}{P_k + V_k''(\phi)} \right]$$

$$+ \sum_{l=0}^{+\infty} \frac{(-1)^l \tilde{\xi}_k^{2l} k^{d+(d-2)l} \phi^{2l}}{(4\pi)^{\frac{d}{2}} \Gamma\left(\frac{d}{2}\right)} \int_0^{+\infty} \frac{\left[r_d(u) - 2u r_d'(u)\right] u^{\frac{d}{2}-1} F\left(l+1, \frac{1}{2}; \frac{d}{2}; -\rho(u, \tilde{\Omega}_k)\right)}{\left(r_d(u) + \frac{\tilde{\Omega}_k}{2}\right)^{2l+1}} \, \mathrm{d}u \, . \quad (4.3.73)$$

For the gap term, we get the dimensionless beta function

$$\tilde{\beta}_\Omega = -\tilde{\Omega}_k + \frac{\tilde{\xi}_k^2}{(4\pi)^{\frac{d}{2}} \Gamma\left(\frac{d}{2}\right)} \int_0^{+\infty} \frac{\left[r_\varphi(u) - u r_\varphi'(u)\right] u^{\frac{d}{2}-1} F\left(1, \frac{1}{2}; \frac{d}{2}; -\rho(u, \tilde{\Omega}_k)\right)}{\left[u + r_\varphi(u) + U_k'''(\phi)\right]^2 \left(r_d(u) + \frac{\tilde{\Omega}_k}{2}\right)} \, \mathrm{d}u$$

$$+ \frac{\tilde{\xi}_k^2}{4(4\pi)^{\frac{d}{2}} \Gamma\left(1 + \frac{d}{2}\right)} \int_0^{+\infty} \frac{\left[r_d(u) - 2u r_d'(u)\right] u^{\frac{d}{2}} F\left(2, \frac{3}{2}; 1 + \frac{d}{2}; -\rho(u, \tilde{\Omega}_k)\right)}{\left[u + r_\varphi(u) + U_k'''(\phi)\right]^2 \left(r_d(u) + \frac{\tilde{\Omega}_k}{2}\right)^4} \, \mathrm{d}u$$

$$- \frac{\tilde{\xi}_k^2}{2(4\pi)^{\frac{d}{2}} \Gamma\left(\frac{d}{2}\right)} \int_0^{+\infty} \frac{\left[r_d(u) - 2u r_d'(u)\right] u^{\frac{d}{2}-1} F\left(2, \frac{1}{2}; \frac{d}{2}; -\rho(u, \tilde{\Omega}_k)\right)}{\left[u + r_\varphi(u) + U_k'''(\phi)\right] \left(r_d(u) + \frac{\tilde{\Omega}_k}{2}\right)^2} \, \mathrm{d}u \, . \quad (4.3.74)$$

At last, for the interaction term, we get

$$\tilde{\beta}_\ell = \left(\frac{d}{2} - 2\right) \tilde{\xi}_k - \frac{2\tilde{\xi}_k^3}{(4\pi)^{\frac{d}{2}} \Gamma\left(\frac{d}{2}\right)} \int_0^{+\infty} \frac{\left[r_\varphi(u) - u r_\varphi'(u)\right] u^{\frac{d}{2}-1} F\left(1, \frac{1}{2}; \frac{d}{2}; -\rho(u, \tilde{\Omega}_k)\right)}{\left[u + r_\varphi(u) + U_k'''(\phi)\right]^2 \left(r_d(u) + \frac{\tilde{\Omega}_k}{2}\right)^2} \, \mathrm{d}u$$

$$+ \frac{\tilde{\xi}_k^3}{(4\pi)^{\frac{d}{2}} \Gamma\left(\frac{d}{2}\right)} \int_0^{+\infty} \frac{\left[r_d(u) - 2u r_d'(u)\right] u^{\frac{d}{2}-1} F\left(2, \frac{1}{2}; \frac{d}{2}; -\rho(u, \tilde{\Omega}_k)\right)}{\left[u + r_\varphi(u) + U_k'''(\phi)\right] \left(r_d(u) + \frac{\tilde{\Omega}_k}{2}\right)^3} \, \mathrm{d}u \, . \quad (4.3.75)$$

Let us quickly recall the simplifications we made to obtain these results. These expressions assume that:

  i. we are working in Minkowski spacetime;

 ii. the detector is inertial and its proper time is an inertial global time function;

iii. the detector's coupling to the quantum field does not change with time.

The assumption of Minkowski spacetime simplifies the heat kernel expressions. The assumption of inertial motion is a hypothesis of our Taylor trick formulae. The assumption of constant coupling simplifies the functional traces.

## 4.4 Gapless Divergences

We would now like to compute the RG flow for the theory and see how the constants evolve, but there is at least one issue with these expressions: some of them are divergent in the gapless limit $\tilde{\Omega}_k \to 0$. Consider, for example, the integral

$$I(\tilde{\Omega}_k) \equiv \int_0^{+\infty} \frac{\left[r_d(u) - 2u r_d'(u)\right] u^{\frac{d}{2}-1} F\left(2, \frac{1}{2}; \frac{d}{2}; -\rho(u, \tilde{\Omega}_k)\right)}{\left[u + r_\varphi(u)\right] \left(r_d(u) + \frac{\tilde{\Omega}_k}{2}\right)^3} \, \mathrm{d}u \, , \quad (4.4.1)$$

which occurs on Eq. (4.3.75) on the preceding page in the particular case of a free massless field. For both $d = 4$ and $d = 2$ this integral will diverge at $\tilde{\Omega}_k$ for the usual cutoffs we discussed in Section 3.2. For example, Figs. 4.1 and 4.2 on the next page illustrate the values of the integral for different values of $\tilde{\Omega}_k$ for each of the cutoffs illustrated in Fig. 3.3 on page 66 (the same profile was chosen for the field and for the detector).

Since the divergence occurs in the gapless limit, it appears to be an infrared (IR) divergence. Nevertheless, it is interesting to plot the integrand of Eq. (4.4.1) on the previous page for fixed values of $\tilde{\Omega}_k$. This is shown in Fig. 4.3 on page 122 for the exponential cutoff $r(x) = x(e^x - 1)^{-1}$ and $d = 4$. Notice that as $\tilde{\Omega}_k$ decreases, the integrand grows for large values of $u$. This corresponds to the UV region, not to the IR region. Hence, it seems the divergence might somehow be related to UV effects.

To understand this behavior, it is useful to look at the denominator of the integrand on Eq. (4.4.1) on the previous page. In particular, notice the term

$$\left( r_d(u) + \frac{\tilde{\Omega}_k}{2} \right)^3. \tag{4.4.2}$$

In the UV, $r_d(u)$ should be small enough so that it does not suppress the high energy modes in the path integral. However, if $\tilde{\Omega}_k$ is also small, then the denominator is small in the UV and we end up with a divergent expression.

This does not happen for the scalar field because the kinetic term contributes with a term $u$ that grows in the UV, and hence keeps the denominator large enough to prevent divergences from happening.

The absence of such a "control term" in the detector comes from the fact that we were not able to solve the functional traces that occurred on Eq. (4.3.33) on page 110 in a straightforward manner, but rather had to resort to the Taylor trick. This removes the kinetic term contribution from the denominator and leads to this issue. Therefore, we should question the validity of Eq. (4.3.35) on page 111 in the UV region, since Eq. (4.3.36) on page 111 only converges for $|x|^2 < |a|$, and in the gapless UV limit $A(\Delta)$ will be really small. At the same time, the UV region is energetic, and hence $u^a \partial_a$ is expected to be large.

This poses a difficulty to the problem: without relying on the Taylor trick, it is not clear how to compute the traces of the form given on Eq. (4.3.34) on page 111, which are essential for our purposes.

Given this conundrum, let us try to investigate possible origins of the issues we are facing. For example, two things we may try are[*]

  i. to exploit the fact we are working in Minkowski spacetime and attempt to compute the traces using Fourier analysis;

 ii. to try to compute the one-loop effective action using standard quantum field theory (QFT) techniques and see whether we run into similar problems with the gapless limit even when we are using other regularization schemes.

Each of these routes brings different advantages. Fourier analysis allows us to bypass the need for heat kernel techniques and we can see whether avoiding the Taylor trick still leads to the same issues. On the other hand, a one-loop computation allows us to notice whether the difficulties we have been facing are somehow related to our choice of cutoff $\mathcal{R}_k$.

---

[*] I thank Prof. Antônio Duarte Pereira Junior for these suggestions.

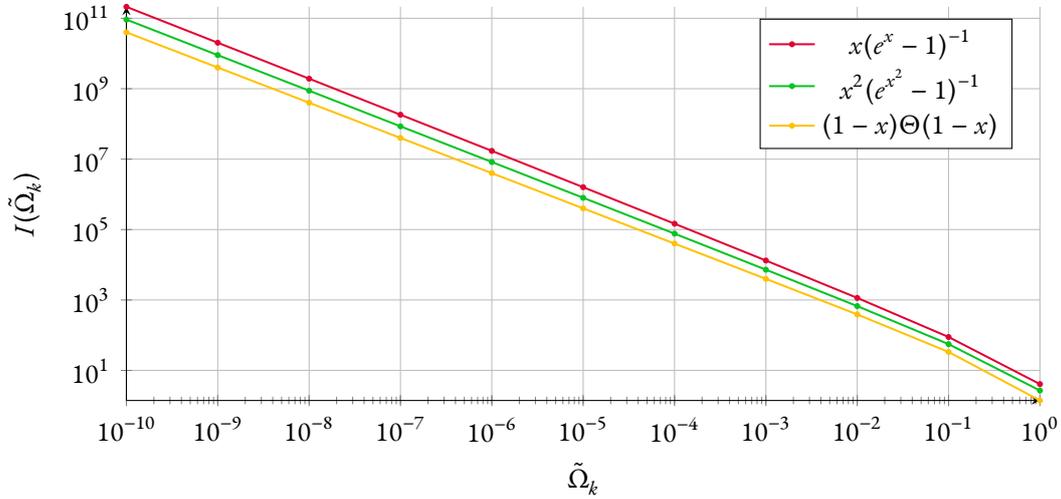

Figure 4.1: Numerical solutions of the integral on Eq. (4.4.1) on page 119 for the cutoffs shown on Fig. 3.3 on page 66 and $d = 4$. Notice how the integral grows as $\tilde{\Omega}_k$ vanishes. The values plotted were obtained using Mathematica's (Wolfram Research 2021) NIntegrate function with $\tilde{\Omega}_k \geq 0.01$.

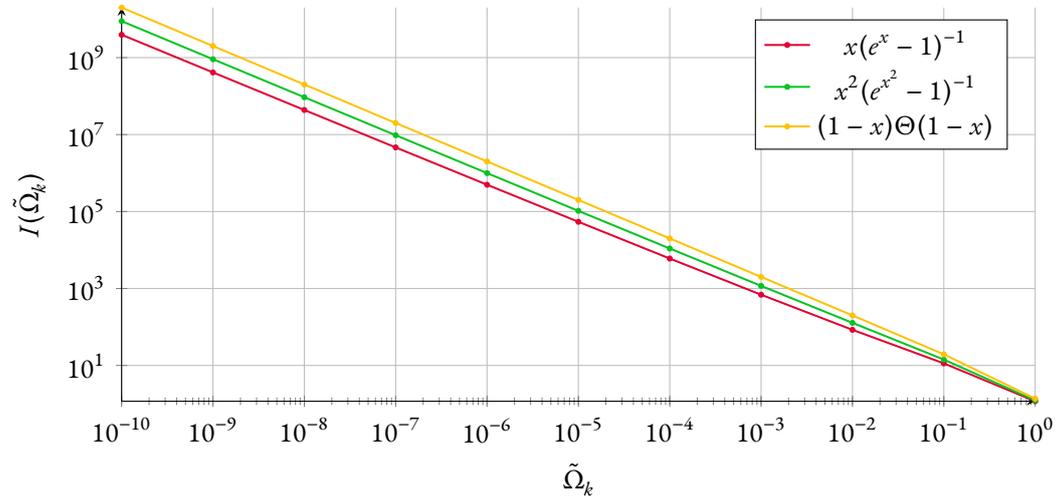

Figure 4.2: Numerical solutions of the integral on Eq. (4.4.1) on page 119 for the cutoffs shown on Fig. 3.3 on page 66 and $d = 2$. Notice how the integral grows as $\tilde{\Omega}_k$ vanishes. The values plotted were obtained using Mathematica's (Wolfram Research 2021) NIntegrate function with $\tilde{\Omega}_k \geq 0.01$.

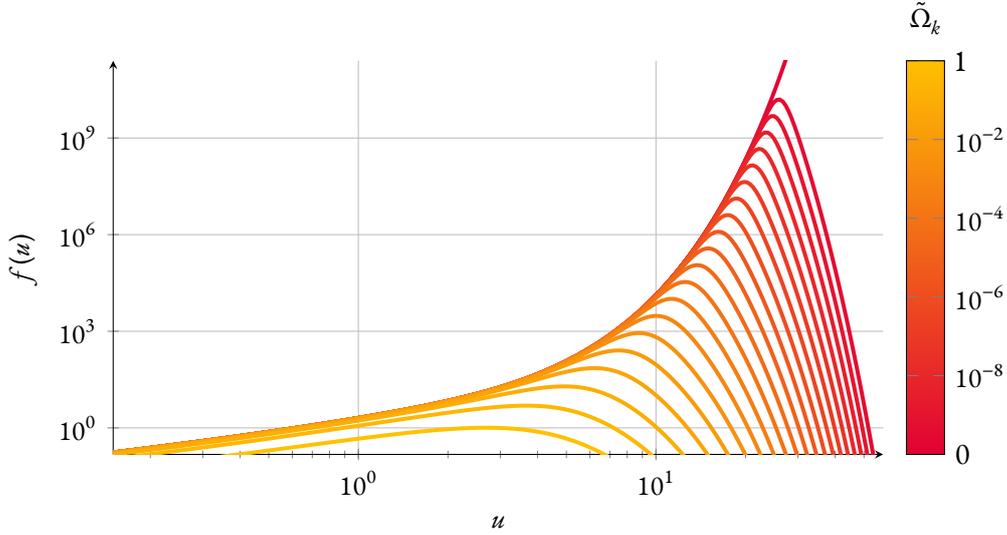

Figure 4.3: Plots of the integrand $f(u)$ of Eq. (4.4.1) on page 119 for the cutoff profile $r(u) = u(e^u - 1)^{-1}$ and $d = 4$. While the color scale is mostly logarithmic, the smallest value $10^{-10}$ was manually replaced by 0. Notice how the integrand grows as $\tilde{\Omega}_k$ vanishes. It is interesting to notice that this growth happens for large values of $u$, which are associated with UV modes. The values plotted were obtained using Mathematica (Wolfram Research 2021).

## Fourier Analysis

Let us begin by computing some of the divergent traces using Fourier techniques. This might not be generalized in a straightforward manner for other spacetimes, and might even be too difficult to be used in the case of an uniformly accelerated detector, for example. Hence, our goal with this calculation is mostly to try to understand if the Taylor trick is problematic. Therefore, we will perform this calculation with a single functional trace with the sole goal of learning whether its divergence depends or not on the Taylor trick.

We shall compute a functional trace associated with the field self-interactions. The results for the traces of the self-interactions obtained by means of the Taylor trick were given on Eqs. (4.3.51) and (4.3.52) on page 113. Some of the field self-interactions involve divergent traces. Indeed, the integral

$$I(\tilde{\Omega}_k) = \int_0^{+\infty} \frac{\left[r_d'(u) - 2u r_d'(u)\right] u}{\left(r_d(u) + \frac{\tilde{\Omega}_k}{2}\right)^3} F\left(2, \frac{1}{2}; 2; -\frac{u}{\left(r_d(u) + \frac{\tilde{\Omega}_k}{2}\right)^2}\right) \mathrm{d}u, \qquad (4.4.3)$$

which occurs on Eq. (4.3.51) on page 113 for $d = 4$ and $l = 1$, diverges in the gapless limit for the three cutoffs we have been considering, as illustrated in Fig. 4.4 on the facing page.

Let us then compute the same functional trace by using Fourier analysis. We will choose a Litim cutoff to keep the calculations feasible. Computing a functional trace by working in momentum space is a trick often used in standard QFT textbooks, such as those by Peskin and Schroeder (1995,

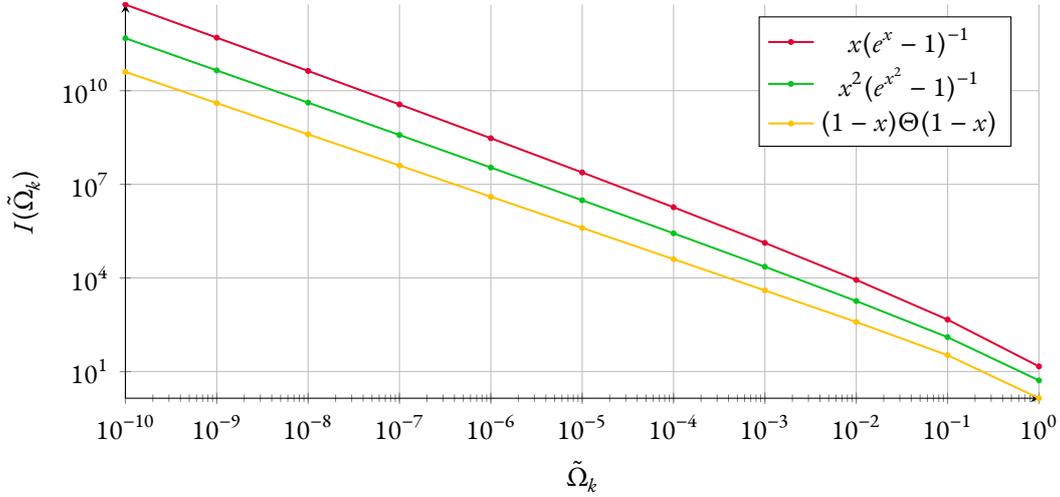

Figure 4.4: Numerical solutions of the integral on Eq. (4.4.3) on the preceding page for the cutoffs shown on Fig. 3.3 on page 66. Notice how the integral grows as $\tilde{\Omega}_k$ vanishes. The values plotted were obtained using Mathematica's (Wolfram Research 2021) NIntegrate function with $\tilde{\Omega}_k \geq 0.01$.

p. 374) and Weinberg (1996, p. 70), and hence we will not go into much detail. We have

$$\mathrm{Tr}\left[\frac{\left(R_k^d + \frac{\Omega_k}{2}\right)\partial_t R_k^d}{\left[\left(\iota^a \partial_a\right)^2 - \left(R_k^d + \frac{\Omega_k}{2}\right)^2\right]^2}\right] = \frac{V_4}{(2\pi)^4}\int \frac{\left(R_k^d(p^2) + \frac{\Omega_k}{2}\right)\partial_t R_k^d(p^2)}{\left[\left(\iota^a p_a\right)^2 + \left(R_k^d(p^2) + \frac{\Omega_k}{2}\right)^2\right]^2}\,\mathrm{d}^4 p\,, \tag{4.4.4a}$$

$$= \frac{V_4}{(2\pi)^4}\int_{\|p\| \leq k} \frac{\left(k - \frac{p^2}{k} + \frac{\Omega_k}{2}\right)\left(k + \frac{p^2}{k}\right)}{\left[\left(p_0\right)^2 + \left(k - \frac{p^2}{k} + \frac{\Omega_k}{2}\right)^2\right]^2}\,\mathrm{d}^4 p\,. \tag{4.4.4b}$$

Let us define spherical coordinates on the 3-sphere through

$$\begin{cases} p_0 = p\cos\psi, \\ p_1 = p\sin\psi\cos\theta, \\ p_2 = p\sin\psi\sin\theta\cos\phi, \\ p_3 = p\sin\psi\sin\theta\sin\phi, \end{cases} \tag{4.4.5}$$

where $p \geq 0$, $\psi, \theta \in [0, \pi]$, and $\phi \in [0, 2\pi)$. Furthermore, let us define dimensionless variables through

$$p = kq \quad \text{and} \quad \Omega_k = k\tilde{\Omega}_k. \tag{4.4.6}$$

We can then write

$$\mathrm{Tr}\left[\frac{\left(R_k^d + \frac{\Omega_k}{2}\right)\partial_t R_k^d}{\left[\left(\iota^a \partial_a\right)^2 - \left(R_k^d + \frac{\Omega_k}{2}\right)^2\right]^2}\right] = \frac{V_4 k^2}{(2\pi)^4}\int_{q \leq 1} \frac{\left(1 - q^2 + \frac{\tilde{\Omega}_k}{2}\right)\left(1 + q^2\right)q^3\,\mathrm{d}q\,\mathrm{d}^3\Omega}{\left[q^2\cos^2\psi + \left(1 - q^2 + \frac{\tilde{\Omega}_k}{2}\right)^2\right]^2}\,, \tag{4.4.7a}$$

$$= \frac{V_4 k^2}{4\pi^3} \int_0^\pi \int_0^1 \frac{\left(1 - q^2 + \frac{\tilde{\Omega}_k}{2}\right)\left(1 + q^2\right)q^3 \sin^2 \psi \, \mathrm{d}q \, \mathrm{d}\psi}{\left[q^2 \cos^2 \psi + \left(1 - q^2 + \frac{\tilde{\Omega}_k}{2}\right)^2\right]^2}. \tag{4.4.7b}$$

While solving this integral is not trivial, we can once again resort to numerical methods. For the purpose of comparison, notice that Eq. (4.3.51) on page 113 would lead to

$$\mathrm{Tr}\left[\frac{\left(R_k^d + \frac{\Omega_k}{2}\right)\partial_t R_k^d}{\left[\left(u^a \partial_a\right)^2 - \left(R_k^d + \frac{\Omega_k}{2}\right)^2\right]^2}\right] = \frac{V_4 k^2}{16\pi^2} \int_0^1 \frac{(1+u)u}{\left(1 - u + \frac{\tilde{\Omega}_k}{2}\right)^3} F\left(2, \frac{1}{2}; 2; -\frac{u}{\left(1 - u + \frac{\tilde{\Omega}_k}{2}\right)^2}\right) \mathrm{d}u. \tag{4.4.8}$$

Eqs. (4.4.7) and (4.4.8) on the preceding page and on the current page invite us to consider the integrals

$$F(\tilde{\Omega}_k) = \frac{1}{4\pi^3} \int_0^\pi \int_0^1 \frac{\left(1 - q^2 + \frac{\tilde{\Omega}_k}{2}\right)\left(1 + q^2\right)q^3 \sin^2 \psi \, \mathrm{d}q \, \mathrm{d}\psi}{\left[q^2 \cos^2 \psi + \left(1 - q^2 + \frac{\tilde{\Omega}_k}{2}\right)^2\right]^2} \tag{4.4.9}$$

and

$$T(\tilde{\Omega}_k) = \frac{1}{16\pi^2} \int_0^1 \frac{(1+u)u}{\left(1 - u + \frac{\tilde{\Omega}_k}{2}\right)^3} F\left(2, \frac{1}{2}; 2; -\frac{u}{\left(1 - u + \frac{\tilde{\Omega}_k}{2}\right)^2}\right) \mathrm{d}u, \tag{4.4.10}$$

where $F$ stands for "Fourier" and $T$ for "Taylor".

These integrals are actually equal. Indeed, if one defines $u = q^2$ and $\xi = \cos \psi$, it can be shown that

$$F(\tilde{\Omega}_k) = \frac{1}{4\pi^3} \int_0^1 \int_0^1 \frac{(1+u)u\sqrt{1 - \xi^2} \, \mathrm{d}u \, \mathrm{d}\xi}{\left(1 - u + \frac{\tilde{\Omega}_k}{2}\right)^3 \left[\rho(u, \tilde{\Omega}_k)\xi^2 + 1\right]^2}, \tag{4.4.11}$$

with

$$\rho(u, \tilde{\Omega}_k) = \frac{u}{\left(1 - u + \frac{\tilde{\Omega}_k}{2}\right)^2}. \tag{4.4.12}$$

Notice that

$$\int_0^1 \frac{\sqrt{1 - \xi^2}}{\left[\rho\xi^2 + 1\right]^2} \, \mathrm{d}\xi = \sum_{n=0}^{+\infty} \frac{(2)_n(-\rho)^n}{n!} \int_0^1 \sqrt{1 - \xi^2}\,\xi^{2n} \, \mathrm{d}\xi, \tag{4.4.13a}$$

$$= \frac{\pi}{4} \sum_{n=0}^{+\infty} \frac{(2)_n\left(\frac{1}{2}\right)_n}{(2)_n n!}(-\rho)^n, \tag{4.4.13b}$$

$$= \frac{\pi}{4} F\left(2, \frac{1}{2}; 2; -\rho\right). \tag{4.4.13c}$$

From Eqs. (4.4.10) to (4.4.13), we can conclude that the Taylor trick and Fourier analysis lead to the same result.

### One-Loop Quantum Action and Beta Functions

A second analysis we may perform in an attempt to understand what is going on is to consider the one-loop effective action for this detector. This will let us check whether the issue lies on our choice of cutoffs. We shall then compute the functional trace given on Eq. (3.A.33) on page 86. Therefore, our goal is to compute

$$\Gamma^{\text{one-loop}} = S + \frac{1}{2} \operatorname{STr}\big[\log S^{(2)}\big] \tag{4.4.14}$$

for the Unruh–DeWitt detector.

Notice that our goal is not to obtain a complete and general result, but rather just to check what happens in the gapless limit. Since the gapless divergences were occurring even for a detector coupled to a free field, we shall focus on the structure due to the detector and ignore the one-loop contributions to the field. We will also overlook the contributions to terms that lie outside the truncation we are interested in.

Our previous ansatz for the EAA was based on the classical action for the detector, and hence we have actually already computed $S^{(2)}$. From Eq. (4.2.14) on page 103 we know that

$$S^{(2)}(x, x') = \begin{pmatrix} -\partial^2 & \gamma_\tau \epsilon \tilde{\xi}^\dagger \sigma_x & -\gamma_\tau \epsilon \xi^\dagger \sigma_x \\ -\gamma_\tau \epsilon \sigma_x \tilde{\xi} & 0 & \gamma_\tau \big[u^a \partial_a - \frac{\Omega}{2} \sigma_z - \epsilon \phi \sigma_x\big] \\ \gamma_\tau \epsilon \sigma_x \xi & \gamma_\tau \big[u^a \partial_a + \frac{\Omega}{2} \sigma_z + \epsilon \phi \sigma_x\big] & 0 \end{pmatrix} \delta^{(d)}(x - x'), \tag{4.4.15}$$

where $S$ is given by (cf. Eq. (4.2.10) on page 102)

$$S = \int \frac{1}{2} \partial_a \phi \partial^a \phi + \gamma_\tau \Big[\tilde{\xi}^\dagger u^a \partial_a \xi + \frac{\Omega}{2} \tilde{\xi}^\dagger \sigma_z \xi + \epsilon \phi \tilde{\xi}^\dagger \sigma_x \xi\Big] \, \mathrm{d}^d x. \tag{4.4.16}$$

We already dropped the time dependence of $\epsilon$ and the quantum field potential, for simplicity.

We begin by splitting $S^{(2)}$ in blocks according to (cf. Eq. (4.3.4) on page 105)

$$\begin{pmatrix} A & B \\ C & D \end{pmatrix} \equiv \left( \begin{array}{c|cc} -\partial^2 + V''(\phi) & \gamma_\tau \epsilon \tilde{\xi}^\dagger \sigma_x & -\gamma_\tau \epsilon \xi^\dagger \sigma_x \\ \hline -\gamma_\tau \epsilon \sigma_x \tilde{\xi} & 0 & \gamma_\tau \big[u^a \partial_a - \frac{\Omega}{2} \sigma_z - \epsilon \phi \sigma_x\big] \\ \gamma_\tau \epsilon \sigma_x \xi & \gamma_\tau \big[u^a \partial_a + \frac{\Omega}{2} \sigma_z + \epsilon \phi \sigma_x\big] & 0 \end{array} \right). \tag{4.4.17}$$

Using this decomposition and Eq. (3.A.32) on page 86, we can notice that the one-loop correction to the quantum action is given by

$$\Gamma_1 = \frac{1}{2} \operatorname{STr}\big[\log S^{(2)}\big], \tag{4.4.18a}$$

$$= \frac{1}{2} \log \big(\operatorname{SDet} S^{(2)}\big), \tag{4.4.18b}$$

$$= \frac{1}{2} \log \big[\operatorname{Det}\big(A - BD^{-1}C\big)\big] - \frac{1}{2} \log \big[\operatorname{Det} D\big], \tag{4.4.18c}$$

$$= \frac{1}{2} \operatorname{Tr}\big[\log\big(A - BD^{-1}C\big)\big] - \frac{1}{2} \operatorname{Tr}\big[\log D\big], \tag{4.4.18d}$$

where we used known properties of the superdeterminant to decompose it in terms of standard functional determinants (see DeWitt 2003, Eq. (A.81)).

As we can see from Eq. (4.4.17), $D$ does not contain any instances of the detector variables, and hence will not contribute to the pieces of the action we are interested in. Hence, we shall ignore the term $\mathrm{Tr}\,[\log D]$, although a more complete analysis of the theory's one-loop structure would need to take it into account. We are left with

$$\Gamma_1 = \frac{1}{2}\,\mathrm{Tr}\,\big[\log\big(A - BD^{-1}C\big)\big] + \cdots, \tag{4.4.19}$$

where the dots denote terms we are not interested in. To compute the logarithm we shall expand it in a series and notice that

$$\Gamma_1 = \frac{1}{2}\,\mathrm{Tr}\,[\log A] - \frac{1}{2}\sum_{n=1}^{+\infty}\frac{1}{n}\,\mathrm{Tr}\,\big[\big(A^{-1}BD^{-1}C\big)^n\big] + \cdots. \tag{4.4.20}$$

As we can see from Eq. (4.4.17) on the previous page, $A$ does not involve the detector variables, and the terms $(A^{-1}BD^{-1}C)^n$ with $n \geq 2$ will either vanish due to involving squares of Grassmann variables or will lie outside the truncation we are interested in. Therefore, we write

$$\Gamma_1 = -\frac{1}{2}\,\mathrm{Tr}\,\big[A^{-1}BD^{-1}C\big] + \cdots. \tag{4.4.21}$$

With some algebraic work, we can get to (*cf.* Eq. (4.3.11) on page 106)

$$\Gamma_1 = -\frac{1}{2}\,\mathrm{Tr}\left[\frac{2\gamma_\tau \epsilon^2\big[\epsilon\phi\bar\xi^\dagger\sigma_x\xi - \frac{\Omega}{2}\bar\xi^\dagger\sigma_z\xi\big]}{[-\partial^2]\big[(u^a\partial_a)^2 - \frac{\Omega^2}{4} - \epsilon^2\phi^2\big]}\right] + \cdots. \tag{4.4.22}$$

Since we are not interested in the $\phi$ contributions that might come from the denominator (they do not contribute to our original truncation), we will ignore them and write simply

$$\Gamma_1 = -\,\mathrm{Tr}\left[\frac{\gamma_\tau \epsilon^2\big[\epsilon\phi\bar\xi^\dagger\sigma_x\xi - \frac{\Omega}{2}\bar\xi^\dagger\sigma_z\xi\big]}{[-\partial^2]\big[(u^a\partial_a)^2 - \frac{\Omega^2}{4}\big]}\right] + \cdots. \tag{4.4.23}$$

By treating the fields as constants, we find

$$\Gamma_1 = -\epsilon^2\,\mathrm{Tr}\left[\frac{\gamma_\tau}{[-\partial^2]\big[(u^a\partial_a)^2 - \frac{\Omega^2}{4}\big]}\right]\left(\epsilon\phi\bar\xi^\dagger\sigma_x\xi - \frac{\Omega}{2}\bar\xi^\dagger\sigma_z\xi\right) + \cdots, \tag{4.4.24}$$

Eq. (4.3.46) on page 112 now implies that

$$\mathrm{Tr}\left[\frac{\gamma_\tau}{[-\partial^2]\big[(u^a\partial_a)^2 - \frac{\Omega^2}{4}\big]}\right] = \frac{-4V_1}{(4\pi)^{\frac{d}{2}}\Gamma\big(\frac{d}{2}\big)\Omega^2}\int_0^{+\infty}\frac{z^{\frac{d}{2}-1}F\big(1, \frac{1}{2}; \frac{d}{2}; -\frac{4z}{\Omega^2}\big)}{z}\,\mathrm{d}z. \tag{4.4.25}$$

We may now use the fact that (*cf.* Bateman 1954, p. 336, Eq. (6.9.3))

$$\int_0^{+\infty} z^{\frac{d}{2}-2}F\big(1, \frac{1}{2}; \frac{d}{2}; -\frac{4z}{\Omega^2}\big)\,\mathrm{d}z = \Omega^{d-2}\Gamma(3-d)\Gamma\big(\frac{d}{2}-1\big)\Gamma\big(\frac{d}{2}\big), \tag{4.4.26}$$

which holds for $2 < d < 3$. We are, of course, aware of the fact that the expression diverges when we take $d \to 4$, and hence we need to regularize it. We then have

$$\mathrm{Tr}\left[\frac{\gamma_\tau}{[-\partial^2]\big[(u^a\partial_a)^2 - \frac{\Omega^2}{4}\big]}\right] = \frac{-4V_1\Omega^{d-4}}{(4\pi)^{\frac{d}{2}}}\Gamma(3-d)\Gamma\big(\frac{d}{2}-1\big). \tag{4.4.27}$$

If we define $\varepsilon = 4 - d$ and perform a series expansion in $\varepsilon$, we get to

$$\text{Tr}\left[\frac{\gamma_\tau}{[-\partial^2]\left[(u^a\partial_a)^2 - \frac{\Omega^2}{4}\right]}\right] = \frac{V_1}{8\pi^2}\left[\frac{2}{\varepsilon} - \gamma_E + \log 4\pi + 2 - \log \Omega^2\right] + \mathcal{O}(\varepsilon), \tag{4.4.28}$$

where $\gamma_E$ is the Euler–Mascheroni constant. Notice the difference between the coupling $\varepsilon$ and the parameter $\varepsilon$.

Notice that Eq. (4.4.28) still has a divergence in the gapless limit, in spite of the fact that we regularized the expression by means of dimensional regularization this time. Nevertheless, it is worth pointing out that similar divergences occur in the massless limit for a scalar theory (see, *e.g.*, Peskin and Schroeder 1995, Eq. (11.77)).

In the modified minimal subtraction scheme (Peskin and Schroeder 1995, p. 377; Weinberg 1996, Sec. 18.6), we can write Eq. (4.4.28) as

$$\text{Tr}\left[\frac{\gamma_\tau}{[-\partial^2]\left[(u^a\partial_a)^2 - \frac{\Omega^2}{4}\right]}\right] \to \frac{V_1}{8\pi^2}\left[2 - \log\left(\frac{\Omega^2}{k^2}\right)\right], \tag{4.4.29}$$

where $k$ is a mass scale introduced to keep dimensional consistency. It can be interpreted as the RG sliding scale.

From Eqs. (4.4.16), (4.4.24) and (4.4.29) on pages 125–127, we find the effective potential

$$V_{\text{eff}}(\phi, \bar{\xi}, \xi) = \frac{\gamma_\tau}{2}\left(\Omega + \frac{\left[2 - \log\left(\frac{\Omega^2}{k^2}\right)\right]\epsilon^2\Omega}{8\pi^2}\right)\bar{\xi}^i\sigma_x\xi + \gamma_\tau\left(\epsilon - \frac{\left[2 - \log\left(\frac{\Omega^2}{k^2}\right)\right]\epsilon^3}{8\pi^2}\right)\phi\bar{\xi}^i\sigma_x\xi + \cdots, \tag{4.4.30}$$

where the dots indicate the terms we chose to ignore for simplicity during our calculations.

Eq. (4.4.30) invites us to define the running couplings

$$\Omega_k = \Omega + \frac{\left[2 - \log\left(\frac{\Omega^2}{k^2}\right)\right]\epsilon^2\Omega}{8\pi^2} \quad \text{and} \quad \epsilon_k = \epsilon - \frac{\left[2 - \log\left(\frac{\Omega^2}{k^2}\right)\right]\epsilon^3}{8\pi^2}, \tag{4.4.31}$$

and differentiation leads us to

$$\beta_\Omega = k\partial_k\Omega_k = \frac{\Omega\epsilon^2}{4\pi^2} \quad \text{and} \quad \beta_\epsilon = k\partial_k\epsilon_k = -\frac{\epsilon^3}{4\pi^2}. \tag{4.4.32}$$

To one-loop, we can invert Eq. (4.4.31) and insert it on Eq. (4.4.32) to obtain

$$\beta_\Omega = \frac{\Omega_k\epsilon_k^2}{4\pi^2} \quad \text{and} \quad \beta_\epsilon = -\frac{\epsilon_k^3}{4\pi^2}. \tag{4.4.33}$$

Therefore, to one-loop we get beta functions that do not diverge in the gapless limit, although the intermediate calculations did involve gapless divergences. However, these divergences seemed to have the same nature of the infrared divergences we would run into in the intermediate calculations for a scalar field. This suggests our difficulties might indeed be arising from our choice of cutoff $\mathcal{R}_k$.

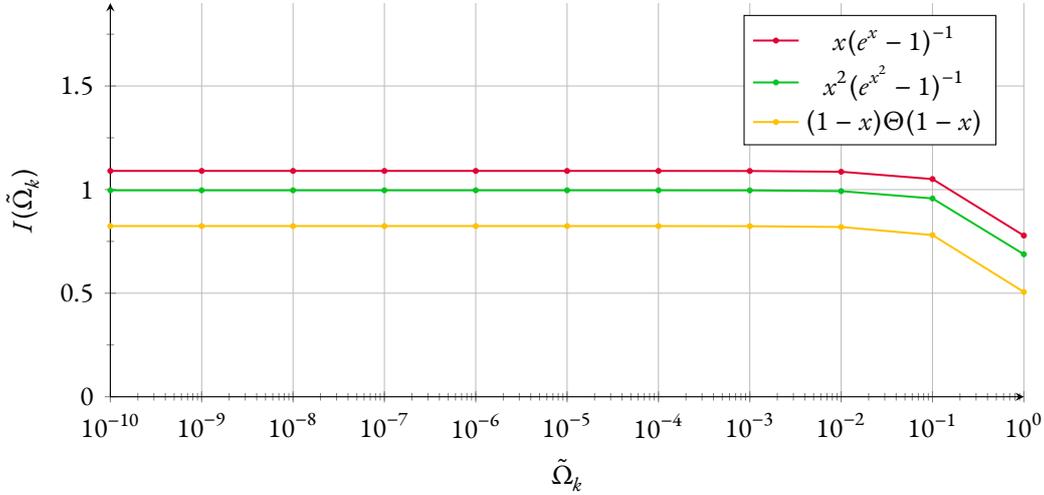

Figure 4.5: Numerical solutions of the integral on Eq. (4.4.34) for the cutoffs shown on Fig. 3.3 on page 66.
Notice how the integral seems to remain finite as $\tilde{\Omega}_k$ vanishes. The values plotted were obtained
using Mathematica's (Wolfram Research 2021) NIntegrate function with $\tilde{\Omega}_k \geq 0.01$.

## Conclusions and Other Possibilities

Given the previous discussions, let us recall the information we have and other possible approaches
we could try to understand what is going wrong with this calculation.

Firstly, we must discuss the Taylor trick. Since we see at least one of the divergences it predicts is
also obtained by means of Fourier analysis, it seems abandoning it would not cure the issues we are
having. Furthermore, it seems that the Taylor trick still works for some other traces. Consider, for
example, the integral

$$I(\tilde{\Omega}_k) = \int_0^{+\infty} \frac{\left[r_\varphi(u) - u r_\varphi'(u)\right] u}{\left(u + r_\varphi(u)\right)^2 \left(r_d(u) + \frac{\tilde{\Omega}_k}{2}\right)} F\left(1, \frac{1}{2}; 2; -\frac{u}{\left(r_d(u) + \frac{\tilde{\Omega}_k}{2}\right)^2}\right) \mathrm{d}u, \qquad (4.4.34)$$

which occurs on Eq. (4.3.60) on page 115. Plotting the numerical values for this integral, as done in
Fig. 4.5, we see that the integral remains finite even as the gap vanishes. Hence, if the Taylor trick is
an issue, it does not cause trouble on every single trace.

Secondly we raise the question of whether our cutoffs have been chosen adequately. The one-loop
beta functions we computed are finite in the gapless limit, which suggests it might be a good idea to
choose different cutoffs and try to regulate the expressions in different manners. A possibility is trying
to mimic the kinetic term instead of the gap term on Eq. (4.2.15) on page 103, or to try a combination
of both approaches.

A third possible issue is our formal manipulation of functional matrices. Rigorously, we should
see the multiplication of functional matrices as given on Eq. (4.3.3) on page 105, but we chose to just
manipulate them formally. This it typically avoided by diagonalizing the operators one will trace over,
but the complexity of Eq. (4.3.4) on page 105 suggests this might not be feasible on this calculation.
Each entry of the matrix is in principle diagonalized on a different functional basis, and hence we
would need to find a manner of diagonalizing both the algebraic and functional matrices at the same time. Since the algebraic matrix already includes Grassmann variables, this is not an easy task, if possible.

In any case, do notice the same formal manipulations entered our one-loop calculations when going from Eq. (4.4.21) on page 126 to Eq. (4.4.22) on page 126, since we employed Eq. (4.3.11) on page 106, which already involved formal manipulation of functional matrices. Hence, if there is something wrong with these procedures, they do not seem to be responsible for our difficulties, or at least they are not the sole culprits.

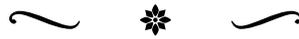

# Five

# Conclusions

*Our previous discussions are recapitulated and we discuss what possibilities lie ahead of us.*

Hic sunt dracones.

Latin for "here be dragons". Generally, but falsely, believed to be an old cartographer inscription to denote uncharted and possibly dangerous regions in maps.

In this thesis we studied one of the oldest physics questions known to humankind: why do things fall? Simple questions often have difficult answers. Hence, our approach was specified to a very particular attack route. Let us recall it.

At first, we reviewed the main ideas and concepts lying behind quantum field theory in curved spacetime (QFTCS). This was done through multiple different approaches, so we could always focus on the aspects that interested us the most at each stage of the thesis. We formulated QFTCS within the algebraic approach, which is particularly clear from a conceptual point of view, and later employed Euclidean techniques to be able to perform renormalization group (RG) computations. We also noticed how fragile the current knowledge about the relationship between these two approaches seems to be. Through these discussions we filled a pedagogical gap in the physics literature, since most (if not all) texts concerning the algebraic approach are written exclusively with mathematics-oriented readers in mind.

A particularly interesting prediction of QFTCS is the Unruh effect, which states that different observers on the same spacetime might interpret the vacuum state of a quantum field theory (QFT) in very different ways. Namely, while an inertial observer might be freezing in the absence of particles, an accelerated observer might be burning in a thermal bath. We derived this conclusion through four different approaches and, in particular, noticed how particle detectors can be used to probe these effects.

Using Euclidean techniques we were able to study the RG in a nonperturbative formulation, when it is often referred to as the functional renormalization group (FRG). Our discussion was mainly based on the notion of effective average action (EAA), which is a "scale-dependent action", and the functional differential equation it satisfies—the Wetterich equation. Our derivation of the Wetterich equation considered both bosonic and fermionic fields from the start, which is uncommon in pedagogical expositions.

Employing the techniques and ideas developed in the previous chapter, we then tackled for the first time ever the problem of describing the RG flow of a particle detector. This required us to reformulate the Unruh–DeWitt particle detector in terms of an action so that we could use the Wetterich equation machinery on it. This led us to difficult functional traces that do not occur on most problems in QFT. We dealt with them by means of the Taylor trick, but got to beta functions that diverge in the gapless limit. Nevertheless, even without the beta functions, we were able to notice, for example, that even the interaction with a particle detector is sufficient to generate infinitely many interactions in the EAA. We also analyzed some alternative calculations and noticed the gapless divergences still occurred on some of the traces even without the Taylor trick, and found that the one-loop calculation has well-behaved beta functions.

At this stage it is natural to ask ourselves which other possibilities lie ahead for applications of the FRG to QFTCS, both within the problem of understanding Unruh–DeWitt detectors in a nonperturbative fashion and for other eventual applications.

## 5.1   On Particle Detectors

Our explorations with particle detectors led to difficulties in evaluating the beta functions for small gaps, which we noticed both with the Taylor trick and with Fourier analysis techniques. By studying the one-loop approximation, we found beta functions that were well-behaved, but in the process we found expressions that were infrared (IR) divergent. Nevertheless, these IR divergences were similar to what one would find when performing the same computations for a scalar field.

The Taylor trick seems to be reliable, since it predicts the same divergences we obtained with Fourier analysis. Furthermore, it does not lead to divergences on all traces. Hence, while it might be interesting to search for a more straightforward manner of computing the traces we obtained, it seems we should first attempt at other solutions.

Since the one-loop solution gave finite beta functions in the gapless limit, it seems possible that our regularization scheme is not working. In other words, our choice of cutoff made on Eq. (4.2.15) on page 103 might not be adequate. In this case, it is interesting to make other choices and see whether the results improve. An immediate option is to mimic a kinetic term rather than a gap term on Eq. (4.2.15) on page 103, or to attempt at a combination of both ideas that allows for a regulation of both kinetic and gap terms simultaneously.

Another point of stress in the calculations are the formal manipulations we carried out with functional matrices. Abandoning this simplification seems considerably difficult at first, and hence it seems more productive to first attempt at using different cutoffs and see whether this is enough to obtain sensible results. If not, a manner of avoiding these manipulations entirely could be to refrain from promoting the detector variables to spacetime fields. In such a manner, it might be that the functional matrices will all be trivially diagonalized in a momentum-like basis, and hence the manipulation will be much more clean. Nevertheless, computing the functional traces by means of heat kernel techniques might no longer be possible, in which case it is not immediately clear whether we can actually complete the calculations at all.

## 5.2   On Compact Interactions

One of the difficulties of computing the renormalization flow of a particle detector is the fact that this interaction happens only on a specific region of spacetime. This is unusual, since most QFTs are typically considered with interaction terms that hold for the whole spacetime. Since this new feature lies at the heart of the particle detector problem, it might be productive to further study these sorts of uncommon interactions. While they might in principle occur on infinite regions of spacetime (as is the case of a particle detector that is always turned on), I will refer to these interactions as "compact interactions" for the remainder of this section.

Since the Unruh–DeWitt detector has a complicated field structure due to its dependence on four Grassmann fields, a compactly interacting scalar field might provide a useful toy model to investigate similar issues. One might study how the renormalization flow of a scalar field is affected by allowing the interaction to happen only on a worldline, or only on some subset of spacetime. This would provide a simpler field model that can still lead, for example, to non-trivial functional matrices similar to the ones we had to deal with. At the end of the calculation we should somehow be able to recover the typical results for a scalar theory by taking the interaction to happen on the whole spacetime.

This might be an interesting model to also investigate whether the flow for a detector that is turned on and off depends on how often this detector is turned on and off. During our previous calculations, we obtained the beta functions for a single detector that is always turned on, and hence we interpreted Eq. (4.3.72) on page 118 as meaning the beta functions referred to the exact terms we would like them to refer to. Nevertheless, it could be that the renormalization flow would happen only on a certain region of spacetime. Our present formalism does not allow to discuss this and considers the flow as something completely global, in spite of the interactions being allowed to be compact.

A possible application of these ideas is the following. Consider a detector that, classically, was prescribed to be turned on and off between two given times. If the renormalization flow does not depend on spacetime, then is it possible that quantum effects will turn on the particle detector later? Is it possible they would turn it on earlier?

Due to our analytic continuations, the function describing how the detector turns on and off is not allowed to vanish on an open set, since analyticity would then force it to vanish on the whole complex plane. Nevertheless, even in this situation, we can ask whether the RG flow can force a detector to be strongly coupled, for example.

While interesting, it is important to notice these remarks are not a challenge on causality of spacetime. Even if the RG flow turns on a detector earlier than planned, this can hardly be seem as a violation of causality, since even tiny interactions seem to be enough to turn on an RG flow. Hence, it would be impossible to tell whether the detector has been turned on because it would be turned on in the future or just because it was not possible to isolate it from external noise in the first place.

Similarly, one might wonder whether the RG flow of field self-interactions due to the detector could be seem as a violation of causality. After all, our calculations suggest that turning on a detector at late times could be sufficient to render the scalar field self-interacting at early times. While beta functions are not observables, this would have observable consequences, since it would alter the cross sections of scattering experiments, for example. Indeed, the cross section for a $\phi\phi \rightarrow \phi\phi$ process with a self-interacting field and with a free field are expected to behave very differently. Nevertheless, do notice that measuring such a process would require to probe the quantum field somehow, which means the field would need to be coupled to some sort of detector. While this detector might not be a qubit, we would then also expect it to be sufficient to cause a RG flow on the field, which would be rendered interacting simply by being probed. This prevents anyone from claiming causality violations based on RG flows, since the only way a field would manage to remain free would be if it has no interactions with anything else, in which case it is completely unobservable.

Another interesting application of compact interactions to particle detectors is to consider whether two different detectors can affect each other's RG flow. Consider the situation in which two detectors follow trajectories on which they never cross. In this case, we can promote the degrees of freedom of both detectors to the same set of four spacetime Grassmann variables, rather than promoting each detector to a different set of variables. This is possible because the theory's action will only consider the values of the variables on the detectors' worldlines, and hence it is just a notational trick. This choice is not ambiguous, because the only degrees of freedom that enter the theory are the ones written in the action, and hence if we used each spacetime field to model a single detector we would be throwing away degrees of freedom that we might also use to model other detectors. Since the FRG calculation involve manipulations of matrices in field space, this can greatly simplify the operations, for it allows us to deal with only five spacetime fields instead of nine.

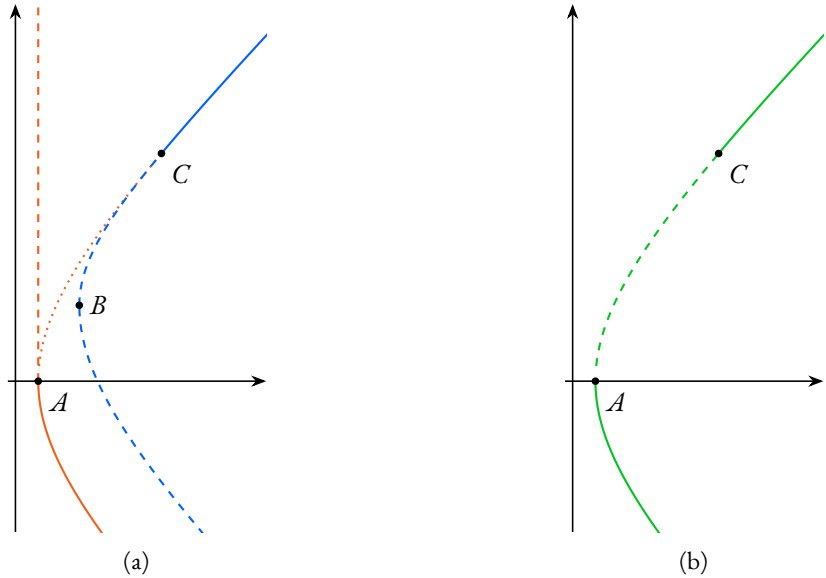

Figure 5.1: **(a)**: Two observers, Alice and Bob, are moving with the same acceleration at first. At the event $A$, Alice stops accelerating and continues in inertial motion. A little after, at event $B$, Bob increases his acceleration to reach event $C$. Bob's acceleration between $B$ and $C$ was carefully chosen so that, at event $C$, he can now return to his original acceleration and continue his motion on the exact trajectory Alice would perform had she never turned off her acceleration. The dashed lines represent when the observers keep their detectors turned off, and the solid lines represent when the observers keep their detectors turned on. The dotted line depicts the worldline Alice would have followed if she had not changed her acceleration. Hence, Alice has her detector turned on only before reaching event $A$ and Bob has his detector turned on only after reaching event $C$. **(b)**: A single observer with constant acceleration turns their detector on and off in such a way that the detector is turned on at the very same events Alice and Bob would have their detectors turned on.

Let us then prescribe worldlines to the detectors, such as the ones shown in Fig. 5.1. In that figure, we consider two observers, Alice and Bob. Alice and Bob coordinated their motions and the functions describing how their detectors are turned on and off in such a manner that at every Cauchy surface for the spacetime there is a single detector turned on. Furthermore, the detector that is turned on is always following a trajectory of constant acceleration. Hence, from a purely mathematical perspective, there is no way to distinguish these two detectors from a single detector undergoing constant acceleration and being turned on and off.

Under these conditions it is interesting to wonder whether Alice's detector can influence the RG flow of Bob's detector. If so, then applications of nonperturbative flows to quantum communication might require a more careful analysis that considers the impacts of one detector onto different ones.

Once again, it is relevant to recall that the detectors cannot be turned off in an absolute sense, since the Euclidean techniques require us to analytically continue their trajectories to imaginary time. Therefore, there are still traces of Alice and Bob in the action, even though they might be faint. Even so, it is interesting to wonder how these faint tails can affect the calculations, since this restricts the simplications one can make when modelling the behavior of a detector. If the results are sensible to these faint effects, then models for a physical situation might need to be excessively complicated to match experimental results.

In order to distinguish between the flows of both detectors when describing them with the same set of spacetime fields, one might use, for example, the spacetime cutoffs used on perturbative algebraic quantum field theory (see Rejzner 2016, Sec. 4.1). The Wetterich equation would then not reduce to some expression times the spacetime volume, but rather to some integral involving the spacetime cutoff. By changing cutoffs one might be able to probe different regions of spacetime and figure out how different detectors respond to the flow.

## 5.3   On Higher Spins

Our calculations were performed with a scalar field, which is customary within QFTCS. Scalar fields are typically simpler to deal with and allow us to obtain a general view of how more complicated phenomena will happen.

While our calculations were not fully completed, they already suggest a few results. For example, Eq. (4.3.73) on page 119 suggests the interaction with a detector will cause the generation of a mass term on a scalar field, even though the field could have initially been massless. This appearance of a mass term would surely be unexpected on gauge theories and it might be interesting to wonder what happens to it on quantum electrodynamics, for example. Since we have detected the photon and know it is massless to a good precision, verifying the photon mass beta function vanishes could provide a consistency check for these computations.

Do notice that higher spins require more complex models for a particle detector.

## 5.4   Final Remarks

While this thesis has focused on the study of an Unruh–DeWitt detector, it is also interesting to wonder in which other ways the functional renormalization group might allow us to slide from quantum field theory in curved spacetime toward quantum gravity.

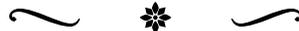

# A

# Notation and Conventions

## A.1  Units

We assume units with $k_B = \hbar = c = 1$ throughout the whole text, although some expressions might restore these quantities to increase clarity. In Chapters 1 and 2, we also take $G = 1$. Nevertheless, since when working with quantum field theory (QFT) it is often convenient to discuss mass dimensions, we understand $G$ as being dimensional in Chapters 3 and 4.

Assume a physical quantity has dimensions of $[\text{mass}]^D$ in units with $k_B = \hbar = c = 1$. The value $D$ is often referred to as the "mass dimension" of the physical quantity.

## A.2  Conventions

We work in the $+ + +$ relativity convention as defined by Misner, Thorne, and Wheeler (1973). This also coincides with the conventions used by Hawking and Ellis (1973), Percacci (2017), Reuter and Saueressig (2018), and Wald (1984, 1994) up to a few different choices on the positions of indices. Since these choices mostly affect only how one writes the Christoffel symbols of the second kind—*viz.*, $\Gamma^a{}_{bc}$, $\Gamma_{bc}{}^a$, and so on—these differences do not affect this thesis. These conventions imply that we use the "mostly pluses" metric convention—*i.e.*, the line element in Minkowski spacetime is given in Cartesian coordinates as

$$\mathrm{d}s^2 = -\mathrm{d}t^2 + \mathrm{d}x^2 + \mathrm{d}y^2 + \mathrm{d}z^2 \,. \tag{A.2.1}$$

When working in Euclidean signature, the metric is taken to be positive-definite.

Many QFT textbooks often use the "mostly minuses" sign convention. Nevertheless, Năstase (2020) and Weinberg (1995, 1996), for example, work in the "mostly pluses" convention.

We take the physicist's convention on Hermitian inner products, meaning that if $\psi, \phi \in \mathscr{H}$ and $\mu, \lambda \in \mathbb{C}$, then

$$\langle \mu\psi | \lambda\phi \rangle = \bar{\mu}\lambda \langle \psi | \phi \rangle \,, \tag{A.2.2}$$

where $\bar{\mu}$ denotes the complex conjugate of $\mu$.

No conventions are chosen with respect to the Fourier transform. While it occurs a few times throughout this thesis, all expressions are written in a manner that holds for any choice of sign and normalization.

The Einstein summation convention is in effect at all times, unless stated otherwise or clear from context. Sometimes the summation is written explicitly to make its range clear or avoid ambiguity.

## A.3   Notation

### Differential Geometry

Spacetime is denoted by $\mathcal{M}$, Cauchy surfaces are denoted by $\Sigma$. The space of compactly supported smooth functions on a manifold $X$ is denoted by $\mathscr{C}_0^\infty(X)$. supp denotes the support and $J^\pm(X)$ denote the causal past and future of some set $X \subseteq \mathcal{M}$. For more information on the causal structure of spacetime, including the definitions of Cauchy surface and causal past and future, see, *e.g.*, the references due to Geroch (1970), Hawking and Ellis (1973), Pfäffle (2009), and Wald (1984).

Tensorial expressions are written in abstract index notation (Geroch 2013, Chap. 4; Wald 1984, Sec. 2.4). Hence, tensors written with Latin indices such as $R_{abc}{}^d$ are meant as abstract tensors, while expressions with Greek indices such as $R_{\mu\nu\rho}{}^\sigma$ denote tensor components on some coordinate system. Sometimes we also write $x^\mu x_\mu$ to denote the invariant interval in Minkowski spacetime.

A few times we work in a $3 + 1$ decomposition of the spacetime manifold. In these cases, we might write $\vec{x}$ (with an arrow) to denote points on the spacelike manifold. Nevertheless, it is important to point out that such manifolds might not always have a vector space structure. Since I find it unlikely that this will lead to confusion, I consider it convenient to use the notation common in Minkowski spacetime. Within this thesis, calculations in which $\vec{x}$ could be mistaken for a vector occur only in Minkowski spacetime, and hence there is not really much loss.

The volume element on a manifold is written as $\sqrt{|g|}\, \mathrm{d}^d x$, where $g_{ab}$ is the metric and $g$ its determinant in the coordinate system of consideration. Typically we will write $\sqrt{-g}$ in Lorentzian signature and $\sqrt{g}$ in Euclidean. Technically this notation makes reference to a coordinate system, but we use it even when a coordinate system is not strictly necessary. One could alternatively write $\mathrm{d}\mu_g(x)$ or write the volume form explicitly, for example.

We might write symmetrizations as

$$T_{(a_1 \cdots a_n)} = \frac{1}{n!} \sum_\pi T_{\pi(a_1)\cdots\pi(a_1)}, \tag{A.3.1}$$

where the sum runs over permutations of $n$ elements.

### Distributions

When working with distributions, we sometimes write w-lim to denote the "weak limit". This means the limit should be taken only after integrating the distribution against some test function. For example, the distributions $\varphi_\pm$ defined on Eq. (2.B.12) on page 51 act on functions according to

$$\varphi_\pm(f) = \lim_{\varepsilon \to 0^+} \int \frac{f(x)}{x \pm i\varepsilon}\, \mathrm{d}x. \tag{A.3.2}$$

The term "weak limit" refers to the fact this corresponds to a limit in the weak topology generated by the space of test functions (*cf.* Simon 2015, Sec. 5.7).

The Dirac delta in a $d$-dimensional flat manifold is given in Cartesian coordinates as $\delta^{(d)}(x, y)$. Hence, one has, for any function $f$,

$$f(x) = \int \delta^{(d)}(x, y)\, \mathrm{d}^d y = \int \frac{\delta^{(d)}(x, y)}{\sqrt{-g(y)}} \sqrt{-g(y)}\, \mathrm{d}^d y. \tag{A.3.3}$$

### Field Theory

When writing path integrals, we put a hat on fields being integrated over, while their classical counterparts do not have a hat. For example, $\Phi = \langle \hat{\Phi} \rangle$. This is overlooked in Chapter 2, in which fields being integrated over can be clearly understood from context.

In Chapter 3, we occasionally employ DeWitt's condensed notation (DeWitt 1964), where a field's internal indices and spacetime dependence are written in terms of a single index. For example, $\alpha$ in $\Phi_\alpha$ labels the field and stands for both the internal indices of the field in, *e.g.*, gauge space and for the spacetime dependency. Einstein's summation convention then also implies integrals over the spacetime variables. When writing integrals explicitly for these indices (as we sometimes do in Section 3.A) a summation over the discrete variables is implied. The correct integration measure is understood.

Chapter 3 also employs an abstract notation for functional matrices. For example, we often manipulate $\Phi$ by understanding it as a vector with components $\Phi_\alpha$. We also introduce the inner product $\langle \cdot, \cdot \rangle$ to denote

$$\langle J, \Phi \rangle = J_\alpha \Phi_\alpha. \tag{A.3.4}$$

This notation is adapted from the one used by Kopietz, Bartosch, and Schütz (2010, Chaps. 6 and 7).

### Linear Algebra and Functional Analysis

$\mathcal{H}$ denotes a Hilbert space. $\mathcal{L}(\mathcal{H})$ denotes the space of linear operators from $\mathcal{H}$ to itself. Similarly, $\mathcal{B}(\mathcal{H})$ denotes the space of bounded linear operators from $\mathcal{H}$ to itself, *i.e.*, the space of linear operators with eigenvalues of a bounded set. More details can be found on the book by Reed and Simon (1980).

We often need to consider many different sorts of traces. Traces over finite-dimensional spaces are denoted as tr. Functional traces are denoted by Tr and functional determinants by Det. The supertrace (see DeWitt 2003, p. 989) is denoted by STr—while in principle it can be used in finite-dimensional spaces, all of the occurrences of the supertrace in this thesis involve functional traces. Similarly, the superdeterminant (DeWitt 2003, p. 989), also known as Berezinian, is denoted by SDet.

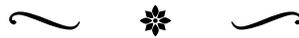